\documentclass[reprint,amsmath,amssymb,nofootinbib,aps]{revtex4-1}
\usepackage{subfig}
\usepackage{graphicx}
\usepackage{dcolumn}
\usepackage{bm}
\usepackage{amsfonts}
\usepackage{slashed}
\usepackage{mathtools}
\usepackage{amssymb}
\usepackage{amsmath}
\usepackage{xcolor}
\usepackage{soul}
\usepackage[bottom]{footmisc}

\def\be{\begin{equation}}
\def\ee{\end{equation}}

\def\bwt{\begin{widetext}}
	\def\ewt{\end{widetext}}

\def\Xint#1{\mathchoice
	{\XXint\displaystyle\textstyle{#1}}%
	{\XXint\textstyle\scriptstyle{#1}}%
	{\XXint\scriptstyle\scriptscriptstyle{#1}}%
	{\XXint\scriptscriptstyle\scriptscriptstyle{#1}}%
	\!\int}
\def\XXint#1#2#3{{\setbox0=\hbox{$#1{#2#3}{\int}$}
		\vcenter{\hbox{$#2#3$}}\kern-.5\wd0}}

\def\dashint{\Xint-}

\begin{document}
	\title{Interpolating 't Hooft model between instant and front forms}

 \author{Bailing Ma and Chueng-Ryong Ji}
\affiliation{Department of Physics, North Carolina State University,
Raleigh, North Carolina  27695-8202, USA} 

\begin{abstract}

The 't Hooft model, i.e. the two-dimensional quantum chromodynamics in the limit of infinite number of colors, is interpolated
by an angle parameter $\delta$ between $\delta=0$ for the instant form dynamics (IFD) and $\delta=\pi/4$ for the light-front dynamics (LFD). 
With this parameter $\delta$, we formulate the interpolating mass gap equation which takes into account the non-trivial vacuum effect on the bare fermion mass
to find the dressed fermion mass.
Our interpolating mass gap solutions not only reproduce the previous IFD result at $\delta=0$ as well as the previous LFD result at $\delta=\pi/4$ but also link them together between the IFD and LFD results with the
$\delta$ parameter. 
We find the interpolation angle independent characteristic 
energy function which satisfies the energy-momentum dispersion relation of
the dressed fermion, identifying the renormalized fermion mass function and the wave function renormalization factor. The renormalized fermion condensate is also found independent of $\delta$, indicating the persistence of
the non-trivial vacuum structure even in the LFD.
Using the dressed fermion propagator interpolating
between IFD and LFD, we derive the corresponding quark-antiquark bound-state
equation in the interpolating formulation verifying its agreement with the previous bound-state equations in the IFD and LFD at $\delta=0$ and $\delta=\pi/4$, respectively.
The mass spectra of mesons bearing the feature of the Regge trajectories are found independent of the $\delta$-parameter reproducing the previous results in LFD and IFD for the equal mass quark and antiquark bound-states. The Gell-Mann - Oakes - Renner relation for
the pionic ground-state in the zero fermion mass limit is confirmed indicating
that the spontaneous breaking of the chiral symmetry occurs in the 't Hooft model
regardless of the quantization for $0 \le \delta \le \pi/4$.
We obtain the corresponding bound-state wave functions and discuss 
their reference frame dependence with respect to the frame independent 
LFD result.  Applying them 
for the computation of the so-called quasi parton distribution functions now in the interpolating formulation between IFD and LFD, 
we note a possibility of utilizing not only the reference frame dependence
but also the interpolation angle dependence to get an alternative effective approach to the LFD-like results.  

\end{abstract}
\maketitle

\section{\label{sec:intro}Introduction}
The two-dimensional quantum chromodynamics ($\text{QCD}_{\text{2}} $) with the number of colors $N_c\to\infty$ has served as a theoretical laboratory for the study of strong interactions. In 't Hooft's seminal paper in 1974~\cite{tHooft}, the power of $ 1/N_c $ expansion~\cite{largeN} was demonstrated in solving $ \text{QCD}_{\text{2}}$ in the limit of $N_c \to \infty$, which was then widely studied also in relation
to the string model and dual theories with the idea of $1/N_c$ expansion as a topological expansion in 
the motion of physical strings (e.g. by Witten~\cite{largeNWitten}). Under the large $N_c$ approximation, non-planar diagrams are negligible and 
thus, for example, only the rainbow diagrams need to be summed over for the computation of the quark's 
self-mass. The two other parameters in $\text{QCD}_{\text{2}}$ besides $N_c$, are the dimensionful coupling constant $g$ and the quark mass $m$. 
Sharing the same mass dimension, $g$ and $m$ play an important role in determining the phase of
$\text{QCD}_{\text{2}}$~\cite{Zhicon}. Depending on the value of the dimensionless coupling $g^2 N_c /m^2$,
it is known that there are at least two phases in $\text{QCD}_{\text{2}}$~\cite{boseform}. 
While the regime of the strong coupling phase which doesn't require the finiteness condition on the dimensionless coupling $g^2 N_c /m^2$~\cite{equiv} can be studied by the bosonization method~\cite{bosonization},
the regime of the weak coupling phase which keeps the so-called ``'t Hooft coupling" $\lambda \sim g^2 N_c$ finite in the limit of not only $N_c \to \infty$ but also $g \to 0$ is investigated typically in $\text{QCD}_{\text{2}}$. 
Although the strong coupling regime of $\text{QCD}_{\text{2}}$ is interesting and deserves further study,  
the scope of the present work is limited to the weak coupling regime of $\text{QCD}_{\text{2}}$. 
Yet, we notice that solving $\text{QCD}_{\text{2}}$ in the weak coupling 
regime, i.e. 't Hooft model, is still highly nontrivial
as the theory captures the property of quark confinement and involves the infrared-cutoff procedures 
discussed in two-dimensional gauge field theories~\cite{consistency,inconsistency}.

In particular, we notice that the 't Hooft model was originally formulated and solved in the Light Front Dynamics (LFD)~\cite{tHooft} well before it was re-derived and discussed in the Instant Form Dynamcis (IFD)~\cite{BG}. The numerical solution of the 't Hooft model in the IFD was presented in the rest frame of the meson~\cite{Li} and more recently also in the moving frames~\cite{mov}.
 While a particular family of the axial gauges interpolated between the IFD and the LFD was explored, the principal-value prescription for regulating the infrared divergences 
was shown to be inconsistent with the interpolated general axial gauges~\cite{inconsistency}. 
Since then, however, this issue involving the interpolation between the IFD and the LFD has not yet been examined any further although the nontrivial vacua in two-dimensional models including $\text{QED}_{\text{2}}$ have been extensively discussed~\cite{Inter}.
It thus motivates us to explore the interpolation of 
the 't Hooft model between the IFD and the LFD fully beyond the gauge sector and discuss the outcome of the full interpolation which naturally remedies the previous issue on novel inconsistency with the interpolating gauge~\cite{inconsistency}.

Before we get into the details and specific discussions on the 1+1 dimensional nature of the 't Hooft model, we first briefly summarize the general remarks on 
distinguished features of the IFD and the LFD proposed originally by Dirac in 1949~\cite{Dirac} 
and the efforts of interpolating them together~\cite{Inter,Op,Sca,Ele,Spi,InterQED,Hiller,Ilderton}.
The LFD has the advantage of having 
the maximum number (seven) of the kinematic operators among the ten Poincar\`{e} operators. 
More kinematic operators provide more symmetries that effectively save the efforts of solving dynamic equations. The conversion of the dynamic operator in one form of the dynamics into the kinematic operator in another form of the dynamics can be traced by introducing an interpolation angle parameter spanning between the two different forms of the dynamics. This in fact motivates the study of the interpolation between the IFD and the LFD.

In our previous works, we have applied the interpolation method to the scattering amplitude of two scalar particles~\cite{Sca}, the electromagnetic gauge fields~\cite{Ele}, as well as the helicity spinors~\cite{Spi} and established the interpolating QED theory between the IFD and LFD~\cite{InterQED}. In particular, we presented~\cite{InterQED} the formal derivation of the interpolating QED in the canonical field theory approach and discussed the constraint fermion degrees of freedom, which appear uniquely in the LFD. The constraint component of the fermion degrees of freedom in LFD results in the instantaneous contribution to the fermion propagator, which is genuinely distinguished from the ordinary equal-time forward and backward propagation of the relativistic fermion degrees of freedom. The helicity of the on-mass-shell fermion spinors in LFD is also distinguished from the ordinary Jacob-Wick helicity in the IFD with respect to whether the helicity depends on the reference frame or not. Our analyses clarified any conceivable confusion in the prevailing notion of the equivalence between the infinite momentum frame (IMF) approach and the LFD.

To link the $1+1$ dimensional IFD space-time coordinates with the LFD ones, we introduce the ``hat notation'' for the interpolating variables, as we have done in Refs.~\cite{Sca,Ele,Spi,InterQED}:
\begin{align}\label{eqn:interpolangle}
\left( \begin{array}{c}
x^{\widehat{+}}  \\
x^{\widehat{-}}
\end{array}
\right) = \left( \begin{array}{cc}\cos\delta & \sin\delta \\
\sin\delta & -\cos\delta \end{array} \right)
\left(\begin{array}{c} x^0 \\ x^1  \end{array}\right),
\end{align} 
where the interpolation angle $\delta$ is allowed to be in the region of $0\le \delta \le \frac{\pi}{4}$. When $ \delta=0 $, we recover the IFD coordinates $ (x^0,-x^1) $, and when $ \delta=\pi/4 $, we arrive at the LFD coordinates denoted typically by $x^{\pm}= (x^0\pm x^1)/\sqrt{2}$ without the ``hat''.

In this coordinate system, the metric becomes
\begin{equation}\label{eqn:metric}
g^{\hat{\mu}\hat{\nu}}=	g_{\hat{\mu}\hat{\nu}}=\left( \begin{array}{cc}
\mathbb{C} & \mathbb{S} \\
\mathbb{S} & -\mathbb{C} \end{array}\right) ,
\end{equation}
where we use the short-hand notation $ \mathbb{C}=\cos2\delta $ and $ \mathbb{S}=\sin2\delta $. 
Apparently, the interpolating $ g^{\hat{\mu}\hat{\nu}} $ goes to the IFD metric
$ \left( \begin{array}{cc}
1 & 0 \\
0 & -1 \end{array}\right)  $ when $ \delta=0 $, and the LFD metric $ \left( \begin{array}{cc}
0 & 1 \\
1 & 0 \end{array}\right)  $ when $ \delta=\pi/4 $. 

The components of covariant and contravariant two-vector $a$ are then related with each other by
\begin{align}
a_{\hat{+}}&=\mathbb{C}a^{\hat{+}}+\mathbb{S}a^{\hat{-}};\ \ \ \ a^{\hat{+}}=\mathbb{C}a_{\hat{+}}+\mathbb{S}a_{\hat{-}};\notag\\
a_{\hat{-}}&=\mathbb{S}a^{\hat{+}}-\mathbb{C}a^{\hat{-}};\ \ \ \
a^{\hat{-}}=\mathbb{S}a_{\hat{+}}-\mathbb{C}a_{\hat{-}},\label{eqn:upper_lower}
\end{align}
and the inner product of two vectors $a$ and $b$ can be written as
\begin{equation}\label{inner_product}
a^{\hat{\mu}}b_{\hat{\mu}}=\mathbb{C}\left(a^{\hat{+}}b^{\hat{+}}-a^{\hat{-}}b^{\hat{-}} \right)+\mathbb{S}\left( a^{\hat{+}}b^{\hat{-}}+a^{\hat{-}}b^{\hat{+}}\right)  .
\end{equation}
The same transformation as shown in Eq.(\ref{eqn:interpolangle}) applies to momentum variables as well, i.e.
\begin{align}
p^{\hat{+}}=p^0\cos\delta+p^1\sin\delta;\notag\\
p^{\hat{-}}=p^0\sin\delta-p^1\cos\delta.
\end{align}
According to Eq.(\ref{eqn:upper_lower}), we also have
\begin{align}
p_{\hat{+}}=p^0\cos\delta-p^1\sin\delta;\notag\\
p_{\hat{-}}=p^0\sin\delta+p^1\cos\delta.
\end{align}
A useful relationship for the energy-momentum of the on-mass-shell particle with mass $m$ and two-momentum vector $p^{\hat{\mu}}$ can be found as follows
\begin{equation}\label{eqn:energy_momentum}
(p^{\hat{+}})^2=(p_{\hat{-}})^2+\mathbb{C}m^2.
\end{equation}

With the ``hat notation", the theory of $\text{QCD}_{1+1}$ in the interpolating quantization is then given by the Lagrangian density~\footnote{It is worth noting that our definition of $g$ is the same with that of Ref.~\cite{BG}, but differs with that of Ref.~\cite{tHooft} by a factor of $\frac{1}{\sqrt{2}}$, i.e.,  when $\delta\to \frac{\pi}{4}$, (quantities with a superscript ``t'' denote the notation used in Ref.~\cite{tHooft}, and the ones without are ours)
	\begin{equation}\label{eqn:not_diff}
	A_{\mu}^{\mathrm{t}}=-i\sqrt{2}A_{\mu}^at_a,\ \ 
	g^{\mathrm{t}}=\frac{1}{\sqrt{2}}g,\ \ {\rm and}\ \ 
	\gamma_{\mu}^{\mathrm{t}}=-i\gamma^{\mu}.
	\end{equation}
}
\begin{equation}\label{eqn:Lag}
\mathcal{L}=-\frac{1}{4}F_{\hat{\mu}\hat{\nu}}^aF^{\hat{\mu}\hat{\nu} a}+\bar{\psi}(i\gamma^{\hat{\mu}}D_{\hat{\mu}}-m)\psi,
\end{equation}
where
\begin{equation}\label{eqn:Dmu}
D_{\hat{\mu}}=\partial_{\hat{\mu}}-i g A_{\hat{\mu}}^a t_a
\end{equation}
and
\begin{equation}\label{eqn:Fmunu}
F_{\hat{\mu}\hat{\nu}}^a=\partial_{\hat{\mu}}A_{\hat{\nu}}^a-\partial_{\hat{\nu}}A_{\hat{\mu}}^a+gf^{abc}A_{\hat{\mu}}^bA_{\hat{\nu}}^c.
\end{equation}
\noindent
In this work, we start from the interpolating Lagrangian density, Eq.(\ref{eqn:Lag}), 
derive the corresponding Hamiltonian and solve the mass gap equation which interpolates
between the IFD and the LFD.  
We then apply the solutions of mass gap equation to the calculations of chiral condensates and 
the quark-antiquark bound-states to find the meson mass spectra and the corresponding wavefunctions. As expected for any physical observables, the meson mass spectra are found to be independent of the interpolation angle parameter. Since we obtain the meson wavefunctions in terms of the interpolation angle parameter 
$\delta$, we use these $\delta$-dependent wavefunctions to compute the corresponding parton distribution functions (PDFs), comparing them with the PDFs in the LFD
and the so-called quasi-PDFs based on the IMF approach in IFD~\cite{pdf}.  

The paper is organized as follows. In Section~\ref{sec:self}, we derive
the fermion mass gap equation in $ {\rm QCD}_{\rm 1+1}(N_c\to\infty) $ in the quantization interpolating between the IFD and the LFD, using a couple of different methods, namely, the Hamiltonian method and the Feynman-diagram method. In Section~\ref{sec:sol}, we present the solutions of the mass gap equation numerically spanning the interpolation angle between 
$\delta=0$ (IFD) and $\delta=\pi/4$ (LFD). In Section~\ref{sec:implications}, we apply the mass gap solutions to the calculations of the chiral condensates and the constituent quark mass defined in the full fermion propagator. In Section~\ref{sec:bound}, we derive the quark-antiquark bound-state equations in the interpolating dynamics and present their solutions in Section~\ref{sec:solbound}, including the meson mass spectra, wavefunctions, and (quasi-)PDFs in subsections \ref{sub:spec}, \ref{sub:wavefunc} and \ref{sub:quasipdf}, respectively.
The summary and conclusions follow in Section~\ref{sec:conclusion}. In Appendix~\ref{app:Bogo}, we describe in detail the derivation of the interacting
quark/anti-quark spinor representation using the Bogoliubov transformation. In Appendix~\ref{app:Var}, we show the method of minimizing the vacuum energy with respect to the Bogoliubov angle in getting the mass gap equation. In Appendix~\ref{app:FreeComparison}, we discuss 
the interpolating mass gap equation and solution in terms of the rescaled variables with respect to
the mass dimension $\sqrt{2\lambda}$
and its treatment associated with the $\lambda = 0$ (Free) case vs.  
the $\lambda \neq 0$ (Interacting) case. In 
Appendix~\ref{app:figures},
we present additional numerical solutions of the mesonic wavefunction for a few different quark masses beyond the ones presented in subsection~\ref{sub:wavefunc}.
The corresponding quasi-PDFs are discussed in Appendix \ref{quasi-PDF-other-masses}.
In Appendix \ref{app:rest}, we present the quark-antiquark bound-state equations and solutions in the rest frame of the meson.

\section{\label{sec:self}The mass gap equation}
In this section, we will derive the quark self-energy equation in $ {\rm QCD}_{\rm 1+1}(N_c\to\infty) $ in the interpolating dynamics between the IFD and the LFD. 
While we use two different methods, i.e. the Hamiltonian method in Sec.~\ref{sub:Ham} and the Feynman-diagram method in Sec.~\ref{sub:dia}, we show that  both methods provide exactly the same set of equations. 
When $ \delta\to0 $, $ \mathbb{C}\to1 $ and $ p_{\hat{-}}\to p^1 $, these equations become the 
IFD mass gap equations presented in Ref.~\cite{BG} (i.e. Eqs.(3.18) and (3.19) of Ref~\cite{BG}).
The agreement to Ref.~\cite{tHooft} of the $  \delta\to\frac{\pi}{4} $ limit is discussed in Sec.~\ref{sub:behaviorLF}.

\subsection{\label{sub:Ham}The Hamiltonian method}


Before we start, we need to choose a gauge as in the case of any gauge field theory.
We adopt here the interpolating axial gauge, i.e. $ A_{\hat{-}}^a=0 $, 
as explored previously in Ref.~\cite{inconsistency}.
 In this gauge, the gluon self-couplings are absent. With the gauge condition, Eq.~(\ref{eqn:Lag}) reduces to
\begin{equation}\label{eqn:Lagred}
\mathcal{L}=\frac{1}{2}\left( \partial_{\hat{-}}A_{\hat{+}}^a\right)^2+\bar{\psi}(i\gamma^{\hat{+}}D_{\hat{+}}+i\gamma^{\hat{-}}\partial_{\hat{-}}-m)\psi.
\end{equation}
As no interpolation-time derivative of $ A_{\hat{+}}^a $, i.e. $\partial_{\hat{+}}A_{\hat{+}}^a$, appears
in Eq.~(\ref{eqn:Lagred}), $A_{\hat{+}}^a$ is not a dynamical variable but a constrained degree of freedom. We substitute this constrained degree of freedom  using the equation of motion for the gluon field $A_{\hat{+}}^a$ given by
\begin{equation}\label{eqn:gluon_field_eom}
\partial_{\hat{-}}^2A_{\hat{+}}^a =\psi^{\dagger}\gamma^0\gamma^{\hat{+}}gt^a\psi\equiv \rho^a=J^{\hat{+}a}.
\end{equation}
The general solution of Eq.(\ref{eqn:gluon_field_eom}) is given by
\begin{eqnarray}\label{eqn:sol_A_eom}
&&A_{\hat{+}}^a(x^{\hat{+}},x^{\hat{-}})   \\
&=&\frac{1}{2}\int dy^{\hat{-}}|x^{\hat{-}}-y^{\hat{-}}|\rho^a(x^{\hat{+}},y^{\hat{-}})-x^{\hat{-}}F^a(x^{\hat{+}})+B^a(x^{\hat{+}}), \nonumber
\end{eqnarray}
where $ F^a $ and $ B^a $ are constants. While $B^a $ is irrelevant as it can always be eliminated by a gauge transformation, $F^a$ is a background electric field which can provide some interesting physical effect such as the axial anomaly in Abelian gauge field theory~\cite{AxialAnomaly}. 
For the color-singlet sector in the non-Abelian gauge field theory, however,  
the background field $F^a$ has no effect, e.g. on the spectrum of hadrons in the $ q\bar{q} $ channel. For this reason, we drop the background field $F^a$ and take the first term of Eq.(\ref{eqn:sol_A_eom}) as the solution of 
$A_{\hat{+}}^a(x^{\hat{+}},x^{\hat{-}})$ in this work.
More details of the discussion on the effect from dropping the background field in the 't Hooft model can 
be found in Ref.~\cite{BG}. 

The energy-momentum tensor in the interpolation form is
\begin{align}\label{eqn:Tmunu}
{T^{\hat{\mu}}}_{\hat{\nu}}
&=-F^{\hat\mu\hat\lambda a}{F_{\hat\nu\hat\lambda}}^a+i\bar{\psi}\gamma^{\hat\mu} D_{\hat\nu}\psi-{g^{\hat\mu}}_{\hat\nu}\mathcal{L} .
\end{align}
Thus, the interpolating Hamiltonian is 
\begin{align}\label{eqn:hamilt}
H\equiv P_{\hat{+}}&=\int dx^{\hat{-}} {T^{\hat{+}}}_{\hat{+}}=\int dx^{\hat{-}}\left(  \frac{1}{2}(\partial_{\hat{-}}A_{\hat+}^a)^2\right.\notag\\
&+\left.\psi^{\dagger}(x^{\hat{-}})\left( -i\gamma^0\gamma^{\hat{-}}\partial_{\hat{-}}+\gamma^0m\right) \psi(x^{\hat{-}})\right) .
\end{align}
As we have shown in~\cite{InterQED}, all components of the $ \psi $ field are dynamical degrees of freedom for $ 0\leq \delta<\pi/4 $, while half of the components become constrained for $ \delta=\pi/4 $.
The field operator conjugate to $ \psi(x) $ is
\begin{equation}\label{eqn:conjugatefield}
\Pi(x)=\frac{\partial\mathcal{L}}{\partial\left( \partial_{\hat{+}}\psi(x)\right) }=i\gamma^0\gamma^{\hat{+}}\psi^{\dagger}(x).
\end{equation}
The anti-commutation relation at $x^{\hat{+}}=x'^{\hat{+}}$ is
\begin{align}\label{eqn:anticommutation}
\left\lbrace \Pi(x),\psi(x')\right\rbrace_{x^{\hat{+}}=x'^{\hat{+}}} &=i\gamma^0\gamma^{\hat{+}}\left\lbrace \psi^{\dagger}(x^{\hat{-}}),\psi(x^{'\hat{-}})\right\rbrace\notag\\ &=i\delta(x^{\hat{-}}-x^{'\hat{-}}).
\end{align}
Consequently,
\begin{equation}\label{eqn:anti-commu}\left\lbrace \psi^{\dagger}(x^{\hat{-}}),\psi(x^{'\hat{-}})\right\rbrace =\left( \gamma^0\gamma^{\hat{+}}\right) ^{-1}\delta(x^{\hat{-}}-x^{'\hat{-}}).
\end{equation}
The Dirac field $ \psi $ can be expanded in terms of the quark creation and annihilation operators
\begin{align}\label{eqn:psi_field}
&\psi(x^{\hat{-}})\\
=&\int \frac{dp_{\hat{-}}}{2\pi\sqrt{2p^{\hat{+}}}} [b(p_{\hat{-}})u(p_{\hat{-}})+d^{\dagger}(-p_{\hat{-}})v(-p_{\hat{-}})] \ {\rm e}^{-ip_{\hat{-}}x^{\hat{-}}}.\notag
\end{align}
When the form goes to the limit of IFD, i.e., $\delta\to 0$, $p_{\hat{-}}\to -p_1=p^1 $, and $x^{\hat{-}}\to -x^1$, and Eq.(\ref{eqn:psi_field}) becomes the ordinary field operator expansion 
in IFD~\footnote{It differs from the expression in Ref.~\cite{BG} by a normalization factor $ \frac{1}{\sqrt{2p^{\hat{+}}}} $, which was inserted in order to be consistent with standard textbook~\cite{text}.}

The non-trivial vacuum, $|\Omega>$, which is defined by
\begin{equation}
\label{non-trivial-vacuum}
b^i|\Omega>=0 , \ d^i|\Omega>=0,
\end{equation} 
is different from the trivial vacuum $|0>$ defined by
\begin{equation}
b^{i(0)}|0>=0,\  d^{i(0)}|0>=0.
\end{equation}
The trivial and non-trivial sets of creation and annihilation operators are related by a Bogoliubov transformation
\begin{align}\label{transf}
\left( \begin{array}{c}
b^i(p_{\hat{-}})\\
d^{i\dagger}(-p_{\hat{-}})
\end{array}\right) =& \left( \begin{array}{cc}
\cos\zeta(p_{\hat{-}}) & -\sin\zeta(p_{\hat{-}}) \\
\sin\zeta(p_{\hat{-}}) & \cos\zeta(p_{\hat{-}})
\end{array}\right) \notag \\ 
&\cdot \left( \begin{array}{c}
b^{i(0)}(p_{\hat{-}})\\
d^{i(0)\dagger}(-p_{\hat{-}})
\end{array}\right) .
\end{align}

The non-trivial set of operators, just like the trivial ones, satisfy the canonical anti-commutation relations 
at $x^{\hat{+}}=x'^{\hat{+}}$
\begin{equation}
\left\lbrace b^i(p_{\hat{-}},x^{\hat{+}}),b^{\dagger j}(p'_{\hat{-}},x'^{\hat{+}})\right\rbrace_{x^{\hat{+}}=x'^{\hat{+}}} =2\pi\delta(p_{\hat{-}}-p'_{\hat{-}})\delta^{ij},
\end{equation}
\begin{equation}
\left\lbrace d^i(-p_{\hat{-}},x^{\hat{+}}),d^{\dagger j}(-p'_{\hat{-}},x'^{\hat{+}})\right\rbrace_{x^{\hat{+}}=x'^{\hat{+}}} =2\pi\delta(p_{\hat{-}}-p'_{\hat{-}})\delta^{ij},
\end{equation}
and all others are zero.

The spinors can be defined through a combination of boost and Bogoliubov transformation, which can be represented by
\begin{equation}
\theta(p_{\hat{-}})=\theta_f(p_{\hat{-}})+2\zeta(p_{\hat{-}}) ,
\end{equation} 
where $\theta_f(p_{\hat{-}})$ is the boost part given by 
\begin{equation}\label{eqn:thetaf}
\theta_f(p_{\hat{-}})=\arctan\frac{p_{\hat{-}}}{\sqrt{\mathbb{C}}m} ,
\end{equation}
and $\zeta(p_{\hat{-}}) $ is the Bogoliubov angle defined in Eq.(\ref{transf}). 

While the details of the derivation are given in Appendix~\ref{app:Bogo},
the results of the spinors are given by
\begin{equation}\label{eqn:uspinor}
u(p_{\hat{-}})=\sqrt{2p^{\hat{+}}}\left( \begin{array}{c}
\sqrt{\frac{1-\sin\theta(p_{\hat{-}})}{2(\cos\delta-\sin\delta)}}\\
\sqrt{\frac{1+\sin\theta(p_{\hat{-}})}{2(\cos\delta+\sin\delta)}}
\end{array}\right)
\end{equation}
and
\begin{equation}\label{eqn:vspinor}
v(-p_{\hat{-}})=\sqrt{2p^{\hat{+}}}\left( \begin{array}{c}
\sqrt{\frac{1+\sin\theta(p_{\hat{-}})}{2(\cos\delta-\sin\delta)}}\\
-\sqrt{\frac{1-\sin\theta(p_{\hat{-}})}{2(\cos\delta+\sin\delta)}}
\end{array}\right).
\end{equation}
In the case of free particles, Eqs.~(\ref{eqn:uspinor}) and (\ref{eqn:vspinor}) 
simplify with $\theta(p_{\hat{-}})=\theta_f(p_{\hat{-}})$ given by Eq.~(\ref{eqn:thetaf}) as
\begin{equation}
u^{(0)}(p_{\hat{-}})=\left( \begin{array}{c}
\sqrt{\frac{p^{\hat{+}}-p_{\hat{-}}}{\cos\delta-\sin\delta}}\\
\sqrt{\frac{p^{\hat{+}}+p_{\hat{-}}}{\cos\delta+\sin\delta}}
\end{array}\right)\label{u_inter_free_}
\end{equation}
and
\begin{equation}
v^{(0)}(-p_{\hat{-}})=\left( \begin{array}{c}
\sqrt{\frac{p^{\hat{+}}+p_{\hat{-}}}{\cos\delta-\sin\delta}}\\
-\sqrt{\frac{p^{\hat{+}}-p_{\hat{-}}}{\cos\delta+\sin\delta}}
\end{array}\right)\label{v_inter_free_}.
\end{equation}

Now, plugging the gluon field solution Eq.~(\ref{eqn:sol_A_eom}) without the 
background field into Eq.~(\ref{eqn:hamilt}), we obtain the interpolating Hamiltonian as
\begin{align}\label{eqn:InterpolHamil}
H&=\int dx^{\hat{-}}\psi^{\dagger}(x^{\hat{-}})\left(-i\gamma^0\gamma^{\hat{-}}\partial_{\hat{-}}+\gamma^0m \right)\psi(x^{\hat{-}})\notag\\ &-\frac{1}{4}\int dx^{\hat{-}}\int dy^{\hat{-}}\rho^a(x^{\hat{-}})|x^{\hat{-}}-y^{\hat{-}}|\rho^a(y^{\hat{-}})
\notag \\
&=T+V,
\end{align}
where the kinetic energy
\begin{equation}
T=\int dx^{\hat{-}}\psi^{\dagger}(x^{\hat{-}})\left( -i\gamma^0\gamma^{\hat{-}}\partial_{\hat{-}}+\gamma^0m\right) \psi(x^{\hat{-}}),
\end{equation}
and the potential energy
\begin{align}
V &=-\frac{1}{4}\int dx^{\hat{-}}\int dy^{\hat{-}}\rho^a(x^{\hat{-}})|x^{\hat{-}}-y^{\hat{-}}|\rho^a(y^{\hat{-}})\notag\\
&=-\frac{g^2}{4}\int dx^{\hat{-}}\int dy^{\hat{-}}|x^{\hat{-}}-y^{\hat{-}}|\psi^{\dagger}(x^{\hat{-}})\gamma^0\gamma^{\hat{+}}t^a\psi(x^{\hat{-}})\notag\\
&\times\psi^{\dagger}(y^{\hat{-}})\gamma^0\gamma^{\hat{+}}t^a\psi(y^{\hat{-}}).
\end{align}

Now, if we define the 't Hooft coupling as
\begin{equation}
\label{lambda-def}
\lambda=\frac{g^2\left(N_c- 1/N_c\right) }{4\pi},
\end{equation}
then $\lambda$ has the dimension of mass squared in 1+1 dimension.
By normal-ordering the Hamiltonian, we can write it in three pieces
\begin{equation}
\label{Hamiltonian}
H=L N_c \mathcal{E}_v +:H_2:+:H_4:\ .
\end{equation}
Here, the vacuum energy density $\mathcal{E}_v$ for the one-dimensional volume $L$ is given by

\begin{widetext}
\begin{align}
\mathcal{E}_v&=\int\frac{dp_{\hat{-}}}{(2\pi)(2p^{\hat{+}})}\text{Tr}\left[ \left(-\gamma^0\gamma^{\hat{-}}p_{\hat{-}}+m\gamma^0 \right) v(-p_{\hat{-}})v^{\dagger}(-p_{\hat{-}})\right] \notag\\
&+\frac{\lambda}{4\pi}\int\frac{dp_{\hat{-}}}{2p^{\hat{+}}}\int\frac{dk_{\hat{-}}}{2k^{\hat{+}}}\frac{1}{(p_{\hat{-}}-k_{\hat{-}})^2} \text{Tr}\left[\gamma^0\gamma^{\hat{+}}u(k_{\hat{-}})u^{\dagger}(k_{\hat{-}})\gamma^0\gamma^{\hat{+}} v(-p_{\hat{-}})v^{\dagger}(-p_{\hat{-}})\right],\label{eqn:vacuum_energy}
	\end{align}
the two-body interaction term including the kinetic energy is given by
	\begin{equation}
	:H_2:=:T:+:V_2:\ ,\label{eqn:H2}
	\end{equation}
with
	\begin{align}
	:T:=\int\frac{dp_{\hat{-}}}{(2\pi)(2p^{\hat{+}})^2}&\left\lbrace \text{Tr}\left[ \left(-\gamma^0\gamma^{\hat{-}}p_{\hat{-}}+m\gamma^0 \right)u(p_{\hat{-}})u^{\dagger}(p_{\hat{-}})\right]  b^{\dagger}(p_{\hat{-}})b(p_{\hat{-}})\right. \notag\\
	&+\text{Tr}\left[ \left(-\gamma^0\gamma^{\hat{-}}p_{\hat{-}}+m\gamma^0 \right)v(-p_{\hat{-}})u^{\dagger}(p_{\hat{-}})\right]  b^{\dagger}(p_{\hat{-}})d^{\dagger}(-p_{\hat{-}})\notag\\
	&+\text{Tr}\left[ \left(-\gamma^0\gamma^{\hat{-}}p_{\hat{-}}+m\gamma^0 \right)u(p_{\hat{-}})v^{\dagger}(-p_{\hat{-}})\right]  d(-p_{\hat{-}})b(p_{\hat{-}})\notag\\
	&\left.-\text{Tr}\left[ \left(-\gamma^0\gamma^{\hat{-}}p_{\hat{-}}+m\gamma^0 \right)v(-p_{\hat{-}})v^{\dagger}(-p_{\hat{-}})\right]  d^{\dagger}(-p_{\hat{-}})d(-p_{\hat{-}})
	\right\rbrace ,\label{equationT}
	\end{align}
	and
	\begin{align}
	:V_2:=\frac{\lambda}{2}\int\frac{dp_{\hat{-}}}{(2\pi)(2p^{\hat{+}})^2}&\int\frac{dk_{\hat{-}}}{(2k^{\hat{+}})\left( p_{\hat{-}}-k_{\hat{-}}\right)^2 }\notag\\
	&\times\left\lbrace \text{Tr}\left[\gamma^0\gamma^{\hat{+}}\left( u(k_{\hat{-}})u^{\dagger}(k_{\hat{-}})-v(-k_{\hat{-}})v^{\dagger}(-k_{\hat{-}})\right)\gamma^0\gamma^{\hat{+}}u(p_{\hat{-}})u^{\dagger}(p_{\hat{-}}) \right]  b^{\dagger}(p_{\hat{-}})b(p_{\hat{-}})\right. \notag\\
	&+\text{Tr}\left[\gamma^0\gamma^{\hat{+}}\left( u(k_{\hat{-}})u^{\dagger}(k_{\hat{-}})-v(-k_{\hat{-}})v^{\dagger}(-k_{\hat{-}})\right)\gamma^0\gamma^{\hat{+}}v(-p_{\hat{-}})u^{\dagger}(p_{\hat{-}}) \right] b^{\dagger}(p_{\hat{-}})d^{\dagger}(-p_{\hat{-}})\notag\\
	&+\text{Tr}\left[\gamma^0\gamma^{\hat{+}}\left( u(k_{\hat{-}})u^{\dagger}(k_{\hat{-}})-v(-k_{\hat{-}})v^{\dagger}(-k_{\hat{-}})\right)\gamma^0\gamma^{\hat{+}}u(p_{\hat{-}})v^{\dagger}(-p_{\hat{-}}) \right] d(-p_{\hat{-}})b(p_{\hat{-}})\notag\\
	&\left.-\text{Tr}\left[\gamma^0\gamma^{\hat{+}}\left( u(k_{\hat{-}})u^{\dagger}(k_{\hat{-}})-v(-k_{\hat{-}})v^{\dagger}(-k_{\hat{-}})\right)\gamma^0\gamma^{\hat{+}}v(-p_{\hat{-}})v^{\dagger}(-p_{\hat{-}}) \right] d^{\dagger}(-p_{\hat{-}})d(-p_{\hat{-}})
	\right\rbrace ,\label{equationV2}
	\end{align}
and the four-body interaction term is given by
	\begin{align}
	:H_4:=-\frac{g^2}{4}\int dx^{\hat{-}}\int dy^{\hat{-}}|x^{\hat{-}}-y^{\hat{-}}|:\psi^{\dagger}(x^{\hat{-}})\gamma^0\gamma^{\hat{+}}t^a\psi(x^{\hat{-}})\psi^{\dagger}(y^{\hat{-}})\gamma^0\gamma^{\hat{+}}t^a\psi(y^{\hat{-}}):.
	\end{align}
Although the mass gap equation can be obtained either by minimizing $\mathcal{E}_v$ with respect to the Bogoliubov angle or by requiring $:H_2:$ to be diagonal in the quark anti-quark creation and annihilation operator basis, both methods provide the same resulting equations. While we present the derivation of  
minimizing $\mathcal{E}_v$ in Appendix~\ref{app:Var}, we derive here the mass gap equations by requiring
$:H_2:$ to be diagonal. 
The requirement of $:H_2:$ to be diagonal means that it must take the form
\begin{align}
\int\frac{dp_{\hat{-}}}{(2\pi)(2p^{\hat{+}})}[E_u(p_{\hat{-}})b^{\dagger}(p_{\hat{-}})b(p_{\hat{-}})
-E_v(p_{\hat{-}})d^{\dagger}(-p_{\hat{-}})d(-p_{\hat{-}})].
\end{align}
The divergent piece that comes out during the normal-ordering process is regulated removing the infinite energy~\cite{BG}, and using the principal value prescription as was done in Ref.~\cite{Li}, 
	\begin{align}\label{eqn:PV}
	\int \frac{dy}{(x-y)^2}f(y) &\to \int \frac{dy}{(x-y)^2}\left[f(y)-f(x)-(y-x)\frac{df(x)}{dx} \right]\equiv \dashint \frac{dy}{(x-y)^2}f(y).
	\end{align}
Thus, the eigenvalue conditions on the spinors are given by
\begin{subequations}
\begin{align}
E_u(p_{\hat{-}})&=\text{Tr}\left[ \left(-\gamma^0\gamma^{\hat{-}}p_{\hat{-}}+m\gamma^0 \right)\Gamma^+(p_{\hat{-}})+\frac{\lambda}{2}\dashint\frac{dk_{\hat{-}}}{(p_{\hat{-}}-k_{\hat{-}})^2}\gamma^0\gamma^{\hat{+}}\left( \Gamma^+(k_{\hat{-}})-\Gamma^-(k_{\hat{-}})\right)\gamma^0\gamma^{\hat{+}}\Gamma^+(p_{\hat{-}}) \right] \label{Eu},\\
E_v(p_{\hat{-}})&=\text{Tr}\left[ \left(-\gamma^0\gamma^{\hat{-}}p_{\hat{-}}+m\gamma^0 \right)\Gamma^-(p_{\hat{-}})+\frac{\lambda}{2}\dashint\frac{dk_{\hat{-}}}{(p_{\hat{-}}-k_{\hat{-}})^2}\gamma^0\gamma^{\hat{+}}\left( \Gamma^+(k_{\hat{-}})-\Gamma^-(k_{\hat{-}})\right)\gamma^0\gamma^{\hat{+}}\Gamma^-(p_{\hat{-}}) \right] \label{Ev},
\end{align}
\end{subequations}
where $\Gamma^{\pm}$ is defined by
\begin{align}
\Gamma^+(p_{\hat{-}})\equiv\frac{u(p_{\hat{-}})u^{\dagger}(p_{\hat{-}})}{2p^{\hat{+}}}
=\left( \begin{array}{cc}
\frac{1-\sin\theta(p_{\hat{-}})}{2(\cos\delta-\sin\delta)} & \frac{\cos\theta(p_{\hat{-}})}{2\sqrt{\mathbb{C}}}\\
\frac{\cos\theta(p_{\hat{-}})}{2\sqrt{\mathbb{C}}} & \frac{1+\sin\theta(p_{\hat{-}})}{2(\cos\delta+\sin\delta)}
\end{array}\right) ,\label{Gamma^+`}
\end{align}
\begin{align}
\Gamma^-(p_{\hat{-}})\equiv\frac{v(-p_{\hat{-}})v^{\dagger}(-p_{\hat{-}})}{2p^{\hat{+}}}
=\left( \begin{array}{cc}
\frac{1+\sin\theta(p_{\hat{-}})}{2(\cos\delta-\sin\delta)} & -\frac{\cos\theta(p_{\hat{-}})}{2\sqrt{\mathbb{C}}}\\
-\frac{\cos\theta(p_{\hat{-}})}{2\sqrt{\mathbb{C}}} & \frac{1-\sin\theta(p_{\hat{-}})}{2(\cos\delta+\sin\delta)}
\end{array}\right) .\label{Gamma^-`}
\end{align}

By using Eqs.(\ref{eqn:uspinor}) and (\ref{eqn:vspinor}), one may see that the matrices on the right hand side of Eqs.(\ref{Gamma^+`}) and (\ref{Gamma^-`}) can be obtained by direct computation. 
Now, let us define
\begin{equation}\label{eqn:defineE}
E(p_{\hat{-}})\equiv\frac{\mathbb{C}}{2}\left[ E_u(p_{\hat{-}})-E_v(p_{\hat{-}})\right]. 
\end{equation}
Then, by subtracting Eqs.(\ref{Eu}) and (\ref{Ev}) as well as plugging in Eqs.(\ref{Gamma^+`}) and (\ref{Gamma^-`}), we arrive at	

\begin{equation}\label{gap_eq_inter_E}
E(p_{\hat{-}})=p_{\hat{-}}\sin\theta(p_{\hat{-}})+\sqrt{\mathbb{C}}m\cos\theta(p_{\hat{-}})
+\frac{\mathbb{C}\lambda}{2}\dashint\frac{dk_{\hat{-}}}{(p_{\hat{-}}-k_{\hat{-}})^2}\cos\left( \theta(p_{\hat{-}})-\theta(k_{\hat{-}})\right).
\end{equation}
On the other hand, by adding them, we get
\begin{equation}\label{eqn:defineEother} E_u(p_{\hat{-}})+E_v(p_{\hat{-}})=-\frac{2\mathbb{S}}{\mathbb{C}}p_{\hat{-}}. \end{equation}
Also, we know that the off-diagonal elements of $ :H_2: $ have to vanish
\begin{subequations} \label{off-diag}
	\begin{align} 
	0&=\text{Tr}\left[ \left(-\gamma^0\gamma^{\hat{-}}p_{\hat{-}}+m\gamma^0 \right)\frac{v(-p_{\hat{-}})u^{\dagger}(p_{\hat{-}})}{2p^{\hat{+}}}+\frac{\lambda}{2}\dashint\frac{dk_{\hat{-}}}{(p_{\hat{-}}-k_{\hat{-}})^2}\gamma^0\gamma^{\hat{+}}\left( \Gamma^+(k_{\hat{-}})-\Gamma^-(k_{\hat{-}})\right)\gamma^0\gamma^{\hat{+}} \frac{v(-p_{\hat{-}})u^{\dagger}(p_{\hat{-}})}{2p^{\hat{+}}}\right]  \label{off-diaga}, \\
	0&=\text{Tr}\left[ \left(-\gamma^0\gamma^{\hat{-}}p_{\hat{-}}+m\gamma^0 \right)\frac{u(p_{\hat{-}})v^{\dagger}(-p_{\hat{-}})}{2p^{\hat{+}}}+\frac{\lambda}{2}\dashint\frac{dk_{\hat{-}}}{(p_{\hat{-}}-k_{\hat{-}})^2}\gamma^0\gamma^{\hat{+}}\left( \Gamma^+(k_{\hat{-}})-\Gamma^-(k_{\hat{-}})\right)\gamma^0\gamma^{\hat{+}}\frac{u(p_{\hat{-}})v^{\dagger}(-p_{\hat{-}})}{2p^{\hat{+}}} \right] \label{off-diagb} .
	\end{align}
\end{subequations}
From either Eq.(\ref{off-diaga}) or Eq.(\ref{off-diagb}), we get
\begin{equation}\label{gap_eq_inter_theta}
\frac{p_{\hat{-}}}{\mathbb{C}}\cos\theta(p_{\hat{-}})-\frac{m}{\sqrt{\mathbb{C}}}
\sin\theta(p_{\hat{-}})=\frac{\lambda}{2}\dashint\frac{dk_{\hat{-}}}{(p_{\hat{-}}-k_{\hat{-}})^2}
\sin\left( \theta(p_{\hat{-}})-\theta(k_{\hat{-}})\right) .
\end{equation}
\end{widetext}
%
%
%
Eqs.(\ref{gap_eq_inter_E}) and (\ref{gap_eq_inter_theta}) are the mass gap equations in the interpolating dynamics. The same set of mass gap equations can be derived using the Feynman diagram method 
as we present in the following subsection~\ref{sub:dia}.

\subsection{\label{sub:dia}The Feynman diagram method}

\begin{figure}
	\centering
	\includegraphics[width=1.0\linewidth]{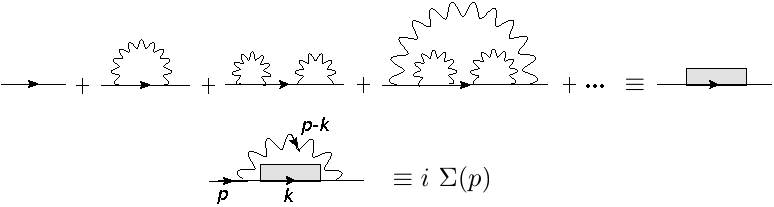}
	\caption{Self-energy equation}
	\label{fig:ds}
\end{figure}

\begin{figure}
	\centering
	\includegraphics[width=0.8\linewidth]{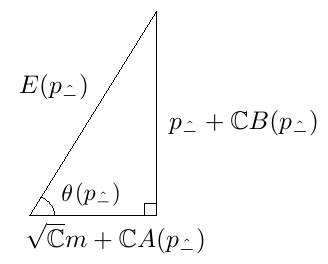}
	\caption{Geometrical representation of mass gap equations representing Eqs.(\ref{energy-square})-(\ref{gapeq_dia2}).}
	\label{fig:righttriangle}
\end{figure}

The self-energy equation in the large $N_c$ approximation is drawn pictorially 
in Fig.~\ref{fig:ds}. Following the Feynman rules for the gluon propagator, the free quark propagator and 
the vertex as $\frac{1}{k_{\hat{-}}^2} $, $ \frac{1}{\slashed{k}-m+i\epsilon} $ and $g\gamma^{\hat{+}}t^a$,
respectively, with the momentum assignment shown in Fig.~\ref{fig:ds}, 
we have
\begin{equation}\label{Sigma}
\Sigma(p_{\hat{-}})=i\frac{\lambda}{2\pi}\dashint\frac{dk_{\hat{-}}dk_{\hat{+}}}{\left( p_{\hat{-}}-k_{\hat{-}}\right)^2}\gamma^{\hat{+}} \frac{1}{\slashed{k}-m-\Sigma(k_{\hat{-}})+i\epsilon}\gamma^{\hat{+}}.
\end{equation}
Writing the self-energy as
\begin{eqnarray}\label{SigmaAB}
\Sigma(p_{\hat{-}})&=&\sqrt{\mathbb{C}}A(p_{\hat{-}})+\gamma_{\hat{-}}B(p_{\hat{-}}) \nonumber \\
&=&\sqrt{\mathbb{C}}A(p_{\hat{-}})+(\mathbb{S}\gamma^{\hat{+}}-\mathbb{C}\gamma^{\hat{-}})B(p_{\hat{-}}) ,
\end{eqnarray}
we express the dressed quark propagator as
\begin{widetext}
\begin{align}\label{dressed_prop}
S(k)=&\left[\slashed{k}-m-\Sigma(k_{\hat{-}})+i\epsilon\right]^{-1}\notag\\
=&\left[\gamma^{\hat{+}}\left( k_{\hat{+}}-\mathbb{S}B(k_{\hat{-}})\right)+\gamma^{\hat{-}}\left( k_{\hat{-}}+\mathbb{C}B(k_{\hat{-}})\right)-\left( m+\sqrt{\mathbb{C}}A(k_{\hat{-}})\right)+i\epsilon\right]^{-1} .
\end{align}
This dressed quark propagator can be obtained from the bare quark propagator 
with the replacement given by
\begin{eqnarray}\label{eqn:replace}
\left\{
\begin{array}{l}
k_{\hat{+}}\to k_{\hat{+}}-\mathbb{S}B(k_{\hat{-}})\\
k_{\hat{-}}\to k_{\hat{-}}+\mathbb{C}B(k_{\hat{-}})\\
m\to m+\sqrt{\mathbb{C}}A(k_{\hat{-}})
\end{array} .
\right.
\end{eqnarray}
Then, Eq.~(\ref{Sigma}) becomes
\begin{align}\label{Sigma2}
\Sigma(p_{\hat{-}})
=&i\frac{\lambda}{2\pi}\dashint\frac{dk_{\hat{-}}dk_{\hat{+}}}{\left( p_{\hat{-}}-k_{\hat{-}}\right)^2}\gamma^{\hat{+}}\frac{1}{\gamma^{\hat{+}}\left( k_{\hat{+}}-\mathbb{S}B(k_{\hat{-}})\right)+\gamma^{\hat{-}}\left( k_{\hat{-}}+\mathbb{C}B(k_{\hat{-}})\right)-\left( m+\sqrt{\mathbb{C}}A(k_{\hat{-}})\right)+i\epsilon}\gamma^{\hat{+}}
\notag\\
=&i\frac{\lambda}{2\pi}\dashint\frac{dk_{\hat{-}}dk_{\hat{+}}}{\left( p_{\hat{-}}-k_{\hat{-}}\right)^2}\notag\\
&\times\frac{\mathbb{C}\gamma^{\hat{+}}\left(k_{\hat{+}}-\mathbb{S}B(k_{\hat{-}}) \right) +\left( 2\mathbb{S}\gamma^{\hat{+}}-\mathbb{C}\gamma^{\hat{-}}\right)\left( k_{\hat{-}}+\mathbb{C}B(k_{\hat{-}})\right) +\mathbb{C}\left(m+\sqrt{\mathbb{C}}A(k_{\hat{-}}) \right)  }{\mathbb{C}\left( k_{\hat{+}}-\mathbb{S}B(k_{\hat{-}})\right)^2+2\mathbb{S}\left(k_{\hat{+}}-\mathbb{S}B(k_{\hat{-}}) \right)\left(k_{\hat{-}}+\mathbb{C}B(k_{\hat{-}}) \right) -\mathbb{C}\left(k_{\hat{-}}+\mathbb{C}B(k_{\hat{-}}) \right)^2-\left( m+\sqrt{\mathbb{C}}A(k_{\hat{-}})\right)^2+i\epsilon},
\end{align}
\end{widetext}
where we have used the algebra for the interpolating $\gamma$ matrices $ \left( \gamma^{\hat{+}}\right)^2=\mathbb{C}\cdot\mathbf{I}_{2\times 2}  $, $ \left( \gamma^{\hat{-}}\right)^2=-\mathbb{C} \cdot\mathbf{I}_{2\times 2} $, and $ \left\lbrace\gamma^{\hat{+}},\gamma^{\hat{-}} \right\rbrace=2\mathbb{S}\cdot\mathbf{I}_{2\times 2} $.
Now, the two poles of $ k_{\hat{+}} $ in the denominator of Eq.~(\ref{Sigma2}) are given by
\begin{align}\label{k_+poles}
-\frac{\mathbb{S}}{\mathbb{C}}k_{\hat{-}}\pm\sqrt{\left(\frac{k_{\hat{-}}}{\mathbb{C}} +B(k_{\hat{-}})\right)^2+\left(\frac{m}{\sqrt{\mathbb{C}}}+A(k_{\hat{-}}) \right)^2  }\mp i\epsilon'.
\end{align} 
After doing the $k_{\hat{+}}$ pole integration using Cauchy's theorem, we get
\begin{widetext}
\begin{align}\label{Sigma_result}
\Sigma(p_{\hat{-}})
=&\frac{\lambda}{2}\dashint\frac{dk_{\hat{-}}}{\left( p_{\hat{-}}-k_{\hat{-}}\right)^2 } \frac{\sqrt{\mathbb{C}}\left(\sqrt{\mathbb{C}} m+\mathbb{C}A(k_{\hat{-}})\right) +\gamma_{\hat{-}}\left(k_{\hat{-}} +\mathbb{C}B(k_{\hat{-}})\right) }{\sqrt{\left(k_{\hat{-}}+\mathbb{C}B(k_{\hat{-}}) \right)^2+\left( \sqrt{\mathbb{C}}m+\mathbb{C}A(k_{\hat{-}})\right)^2}}\notag\\
=&\frac{\lambda}{2}\dashint\frac{dk_{\hat{-}}}{\left( p_{\hat{-}}-k_{\hat{-}}\right)^2 }\left( \sqrt{\mathbb{C}}\cos\theta(k_{\hat{-}})+\gamma_{\hat{-}}\sin\theta(k_{\hat{-}})\right) ,
\end{align}
\end{widetext}
where $ \theta(k_{\hat{-}}) $ is defined by 
\begin{equation}
\theta(k_{\hat{-}})=\tan^{-1}\left[ \frac{k_{\hat{-}}+\mathbb{C}B(k_{\hat{-}})}{\sqrt{\mathbb{C}}m+\mathbb{C}A(k_{\hat{-}})}\right].\label{Esin}
\end{equation}
By comparing Eq.(\ref{Sigma_result}) with Eq.(\ref{SigmaAB}), we can identify
\begin{align}\label{Ap_}
A(p_{\hat{-}})&=\frac{\lambda}{2}\dashint\frac{dk_{\hat{-}}}{\left( p_{\hat{-}}-k_{\hat{-}}\right)^2 }\cos\theta(k_{\hat{-}})
\end{align}
and
\begin{align}
\label{Bp_}
B(p_{\hat{-}})&=\frac{\lambda}{2}\dashint\frac{dk_{\hat{-}}}{\left( p_{\hat{-}}-k_{\hat{-}}\right)^2 }\sin\theta(k_{\hat{-}}).
\end{align}
From Eq.(\ref{Esin}), we may geometrically represent the effective mass and longitudinal momentum of the particle 
moving in non-trivial vacuum by drawing the triangle picture shown in Fig.~\ref{fig:righttriangle}.  
From Fig.~\ref{fig:righttriangle}, we identify the energy $E(p_{\hat{-}})$ as
\begin{equation}
\label{energy-square}
E(p_{\hat{-}})^2 = \left(p_{\hat{-}}+\mathbb{C}B(p_{\hat{-}}) \right)^2+\left( \sqrt{\mathbb{C}}m+\mathbb{C}A(p_{\hat{-}})\right)^2, 
\end{equation}
and find the mass gap equations
\begin{align}\label{gapeq_dia1}
&E(p_{\hat{-}})\cos\theta(p_{\hat{-}})=\sqrt{\mathbb{C}}m+\mathbb{C}\cdot\frac{\lambda}{2}\dashint\frac{dk_{\hat{-}}}{\left( p_{\hat{-}}-k_{\hat{-}}\right)^2 }\cos\theta(k_{\hat{-}}),\\
\label{gapeq_dia2}
&E(p_{\hat{-}})\sin\theta(p_{\hat{-}})=p_{\hat{-}}+\mathbb{C}\cdot\frac{\lambda}{2}\dashint\frac{dk_{\hat{-}}}{\left( p_{\hat{-}}-k_{\hat{-}}\right)^2 }\sin\theta(k_{\hat{-}}),
\end{align}
where we used Eqs.~(\ref{Ap_}) and (\ref{Bp_}) for $A(p_{\hat{-}})$ and $B(p_{\hat{-}})$.
Now, multiplying Eq.~(\ref{gapeq_dia1}) by $\sin\theta(p_{\hat{-}})$ and Eq.~(\ref{gapeq_dia2}) by $\cos\theta(p_{\hat{-}})$ and subtracting them, we get the exact same equation in subsection~\ref{sub:Ham} as given by Eq.~(\ref{gap_eq_inter_theta}).
Also, by multiplying Eq.~(\ref{gapeq_dia1}) by $\cos\theta(p_{\hat{-}})$ and Eq.~(\ref{gapeq_dia2}) by $\sin\theta(p_{\hat{-}})$ and adding them, we get the same equation as Eq.~(\ref{gap_eq_inter_E}) in subsection~\ref{sub:Ham}.

In Sec.~\ref{sec:sol}, we numerically solve Eq.~(\ref{gap_eq_inter_theta}) to obtain $\theta(p_{\hat{-}})$ and 
plug it into Eq.~(\ref{gap_eq_inter_E}) to find $E(p_{\hat{-}})$. We note that the solution of $E(p_{\hat{-}})$ is 
not always positive and thus the geometric interpretation of Fig.~\ref{fig:righttriangle} should be regarded as an pictorial device to represent Eqs.~(\ref{energy-square})-(\ref{gapeq_dia2}) without assuming that the lengths of 
the triangle sides are positive. 
We also recognize Eq.~(\ref{k_+poles}) as $E_u $ and $E_v$ mentioned in the previous subsection,
Sec.~\ref{sub:Ham}, i.e.
\begin{equation}\label{EuEv}
E_u(p_{\hat{-}})=-\frac{\mathbb{S}}{\mathbb{C}}p_{\hat{-}}+\frac{E(p_{\hat{-}})}{\mathbb{C}},\ E_v(p_{\hat{-}})=-\frac{\mathbb{S}}{\mathbb{C}}p_{\hat{-}}-\frac{E(p_{\hat{-}})}{\mathbb{C}},
\end{equation}
where we note $E_u(-p_{\hat{-}}) = -E_v(p_{\hat{-}})$ and $E_v(-p_{\hat{-}}) = -E_u(p_{\hat{-}})$
due to 
the evenness of $E(p_{\hat{-}})$ under $p_{\hat{-}} \leftrightarrow -p_{\hat{-}}$,
i.e. $E(p_{\hat{-}})=E(-p_{\hat{-}})$.
When $E(p_{\hat{-}})$ is positive, $E_u $ is the first energy pole corresponding to the plus sign in Eq.~(\ref{k_+poles}), and $E_v$ to the minus sign
as written in Eq.(\ref{EuEv}). When $E(p_{\hat{-}})$ is negative, however, $ E_u $ is the pole with minus sign, and $ E_v $ the plus sign.
Moreover, one can naturally obtain Eqs.~(\ref{eqn:defineE}) and (\ref{eqn:defineEother}) by adding and subtracting $E_u $ and $E_v$ in Eq.~(\ref{EuEv}).

\subsection{Behavior of the gap equation when approaching the light front\label{sub:behaviorLF}}
We note that the interpolating mass gap equations are greatly simplified in the limit to the LFD,
i.e. $ \mathbb{C}\to0$. In this limit, Eq.~(\ref{gap_eq_inter_theta}) becomes
\begin{equation}
\label{LFmassgap}
p^+ \cos\theta(p^+)=0,
\end{equation}
where one should note $p_{\hat{-}}=p_-=p^+$ as $ \mathbb{C}\to0$.
The solution of Eq.~(\ref{LFmassgap}) is analytically given by 
\begin{equation}\label{lfthetasol}
\theta(p^+)=\frac{\pi}{2}\mathrm{sgn} (p^+).
\end{equation}
Likewise, Eq.~(\ref{gap_eq_inter_E}) is simplified as
\begin{equation}\label{lftenergysol}
E(p^+)=p^+\sin\theta(p^+).
\end{equation}
Moreover, Eqs.~(\ref{Ap_}) and (\ref{Bp_}) are now given by 
\begin{align}
A(p^+)&=0\ ({\rm except}\ p^+=0),\label{eqn:Alf}\\
B(p^+)&=\frac{\lambda}{2}\dashint\frac{dk^+}{(p^+-k^+)^2}{\rm sgn}(k^+)\ ,\label{LF_Bpplus}
\end{align}
where the light-front zero-mode $p^+=0$ contribution 
should be taken into account separately
with the form of $A(p^+) = A(0) \delta (p^+)$ solution in mind. Besides the $p^+=0$ contribution,
it is interesting to note a remarkable simplification of the self-energy in the LFD given by
$\Sigma(p^+)=\gamma^+B(p^+)$ due to the absence of the scalar part, i.e. $A(p^+)=0$. 
For the computation of $B(p^+)$ in Eq.~(\ref{LF_Bpplus}), 't Hooft~\cite{tHooft} didn't use the principal value
but discussed how to make the infrared region finite by introducing the infrared cutoff parameter $\varepsilon$ 
as summaried below for $p^+>0$ and $p^+<0$, respectively, i.e. in the limit $\varepsilon \to 0$,
for $p^+>0$,
\begin{align}
B(p^+)&=\frac{\lambda}{2}\left(\int_{-\infty}^{0}-\frac{dk^+}{(p^+-k^+)^2}+ \int_{0}^{p^+-\varepsilon}\frac{dk^+}{(p^+-k^+)^2}\right.\notag\\
&+\left.\int_{p^++\varepsilon}^{+\infty}\frac{dk^+}{(p^+-k^+)^2}\right) \notag\\
&=-\lambda\left( \frac{1}{p^+}-\frac{1}{\varepsilon}\right)\ ,
\end{align}
while for $p^+<0$,
\begin{align}
B(p^+)&=\frac{\lambda}{2}\left(\int_{-\infty}^{p^+-\varepsilon}-\frac{dk^+}{(p^+-k^+)^2}+ \int_{p^++\varepsilon}^{0}-\frac{dk^+}{(p^+-k^+)^2}\right.\notag\\
&+\left.\int_{0}^{+\infty}\frac{dk^+}{(p^+-k^+)^2}\right) \notag\\
&=-\lambda\left( \frac{1}{p^+}+\frac{1}{\varepsilon}\right)\ .
\end{align}
Thus, the 't Hooft's solution for $B(p^+)$ is given by
\begin{equation}
\label{B-sol}
B(p^+)=\lambda\left( \frac{{\rm sgn}(p^+)}{\varepsilon}-\frac{1}{p^+}\right) \ .
\end{equation}
With these solutions, the replacement given by Eq.~(\ref{eqn:replace}) in the LFD limit ($\delta\to\frac{\pi}{4}$) becomes 
\begin{equation}
\left\{
\begin{array}{r}
p_{\hat{+}}\to p_{\hat{+}}-\mathbb{S}B\\
p_{\hat{-}}\to p_{\hat{-}}+\mathbb{C}B\\
m\to m+\sqrt{\mathbb{C}}A
\end{array}
\right. \, \, \, \overset{\mathbb C \to 0} {\longrightarrow} 
\left\{
\begin{array}{l}
p^-\to p^- +\frac{\lambda}{p^+}-\lambda\frac{{\rm sgn}(p^+)}{\varepsilon}\\
p^+\to p^+\\
m\to m \, ,
\end{array}
\right.
\end{equation}
and it leads to the dressed fermion propagator given by
\begin{align}
&S(p)=\frac{\gamma^+\left( p^-+\frac{\lambda}{p^+}-\lambda\frac{{\rm sgn}(p^+)}{\varepsilon}\right) 
+\gamma^-p^++m}{2 p^+p^- - m^2+2\lambda -2\lambda\frac{|p^+|}{\varepsilon} +i\epsilon }\notag\\
&=
\frac{\gamma^+\left( p^--p^-_{\rm on}+\frac{\lambda}{p^+}-\lambda\frac{{\rm sgn}(p^+)}{\varepsilon}\right)+\gamma^+p^-_{\rm on} +\gamma^-p^++m}{2p^+\left( p^--p^-_{\rm on}+\frac{\lambda}{p^+}-\lambda\frac{{\rm sgn}(p^+)}{\varepsilon}  \right) +i\epsilon}\notag\\
&=\frac{\gamma^+}{2p^+}+\frac{\slashed{p}_{\rm on}+m}{p^2-m^2+2\lambda-2\lambda\frac{|p^+|}{\varepsilon} +i\epsilon},
\end{align}
where $p_{\rm on}^- = \frac{m^2}{2 p^+}$ and $p_{\rm on}^+ = p^+$.
Here, we note the splitting of the instantaneous contribution ($\sim \gamma^+$) and 
the so-called on-mass-shell part $\sim (\slashed{p}_{\rm on}+m$) of the fermion propagator in LFD. 
While the interpolating fermion propagator can split into the forward moving part with the energy denominator $1/(p_{\hat{+}}-E_u(p_{\hat{-}}))$ and the backward moving part with the energy denominator $1/(p_{\hat{+}}-E_v(p_{\hat{-}}))$ as we discuss the details of the fermion propagator in Sec.~\ref{FPandCM}, one can notice the behaviors of $E_u(p_{\hat{-}})$ and $E_v(p_{\hat{-}})$ in the limit $\mathbb{C} \to 0$ as $E_u(p_{\hat{-}}) \to B(p^+)+\frac{m^2}{2p^+}$ and $E_v(p_{\hat{-}}) \to -(\frac{2p^+}{\mathbb{C}}+B(p^+)+\frac{m^2}{2p^+})$ for $p^+>0$ while 
$E_u(-p_{\hat{-}}) \to +(\frac{2p^+}{\mathbb{C}}+B(p^+)+\frac{m^2}{2p^+})$
and $E_v(-p_{\hat{-}}) \to -(B(p^+)+\frac{m^2}{2p^+})$  for $p^+<0$
using Eqs.~(\ref{energy-square}) and (\ref{EuEv}). 
Taking $p^+>0$ for the sake of discussion, 
one can rather easily identify that the on-mass-shell part $\sim (\slashed{p}_{\rm on}+m$) of the dressed fermion propagator corresponds to the forward moving part
from the 't Hooft solution for $B(p^+)$ given by Eq.(\ref{B-sol}) as well as the correspondence $\slashed{p}_{\rm on}+m$ with the spinor biproduct $u(p^+) {\bar u}(p^+)$ of the dressed fermion. Likewise, one can identify the instantaneous contribution ($\sim \gamma^+$) to the backward moving part by noting the cancellation of $1/{\mathbb{C}}$ factor in $E_v(p_{\hat{-}}) \sim -\frac{2p^+}{\mathbb{C}}$ with 
the $1/{\mathbb{C}}$ factor in the backward moving spinor biproduct $v(-p^+) {\bar v}(-p^+) \sim \frac{2p^+ \gamma^+}{\mathbb{C}}$. 
However, the instantaneous contribution effectively vanishes in the actual calculation of
the rainbow and ladder diagrams as every quark line is multiplied by the vertex factor 
$ g\gamma^+ $ from both sides and $ \left( \gamma^+\right) ^2=0 $. Moreover, as 
$ \left\lbrace \gamma^+,\gamma^-\right\rbrace=2  $, the dressed quark propagator can be effectively given by 
\begin{equation}
S(p)=\frac{p^+}{2 p^+p^- - m^2+2\lambda -2\lambda\frac{|p^+|}{\varepsilon} +i\epsilon}\label{eqn:tHooftprop_noPV},
\end{equation}  
with the effective vertex factor $2g$.
We can now see that, due to the $1/\varepsilon$ infrared divergence, the on-mass-shell pole of this dressed quark propagator moves towards the infinity from the on-mass-shell pole $p^-_{\rm on}=\frac{m^2}{2p^+}$. This disappearance of the on-mass-shell pole due to the infrared cutoff term was interpreted as the confinement of the fermions in the 't Hooft model~\cite{tHooft}.

As shown in Ref.~\cite{tHooft}, however, the infrared cutoff terms cancel themselves in the bound-state spectroscopy calculation. Thus, one can use the principal value prescription as defined in Eq.~(\ref{eqn:PV}) to regulate the infinite piece. With the same principal value prescription,
$B(p^+)$ in the limit of $\varepsilon\to 0$ is given by  
\begin{align}\label{eqn:Bp-}
B(p^+)=&\frac{\lambda}{2}\int\frac{dk^+}{(p^+-k^+)^2}\left( {\rm sgn}(k^+)-{\rm sgn}(p^+)\right) \notag\\
=&\lambda\left( \frac{{\rm sgn}(p^+)}{\varepsilon}-\frac{1}{p^+}\right)-\frac{\lambda}{2}{\rm sgn}(p^+)\notag\\
&\times\left( \int_{-\infty}^{p^+-\varepsilon}\frac{dk^+}{(p^+-k^+)^2}+\int_{p^++\varepsilon}^{+\infty}\frac{dk^+}{(p^+-k^+)^2}\right) \notag\\
=&-\frac{\lambda}{p^+}\ .
\end{align}
%
The reduced quark propagator without the $1/\varepsilon$ infrared divergence factor in the denominator is then given by  
\begin{equation}
S(p)=\frac{p^+}{2 p^+p^- -m^2+2\lambda +i\epsilon }\label{eqn:tHooftprop},
\end{equation}  
which leads to the same effective planar Feynman rule of the fermion propagator in the light-front gauge  presented in Ref.~\cite{tHooft} besides the notation difference explained in the footnote in Sec.~\ref{sec:intro}.

\section{\label{sec:sol} The mass gap solution}

\begin{figure}
	\centering
	\subfloat[]{
		\includegraphics[width=1.0\columnwidth]{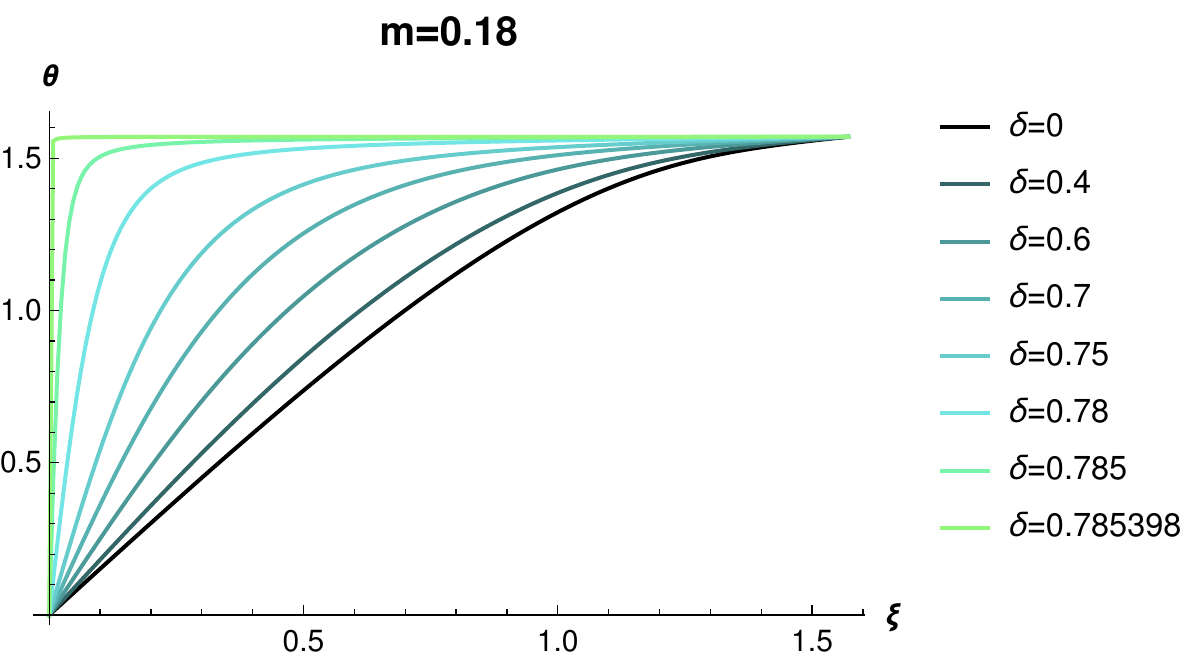}
		\label{fig:m018_basic_result_theta}}
	\\
	\centering
	\subfloat[]{
		\includegraphics[width=1.0\columnwidth]{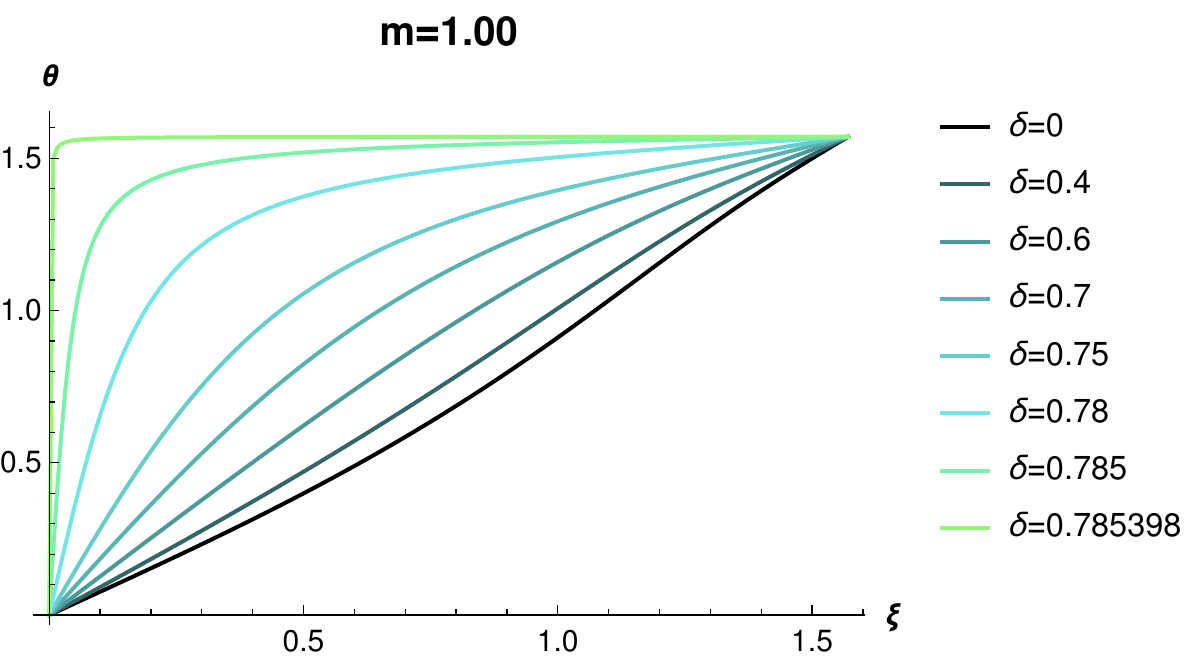}
		\label{fig:m100_basic_result_theta}}
	\\
	\centering
	\subfloat[]{
		\includegraphics[width=1.0\columnwidth]{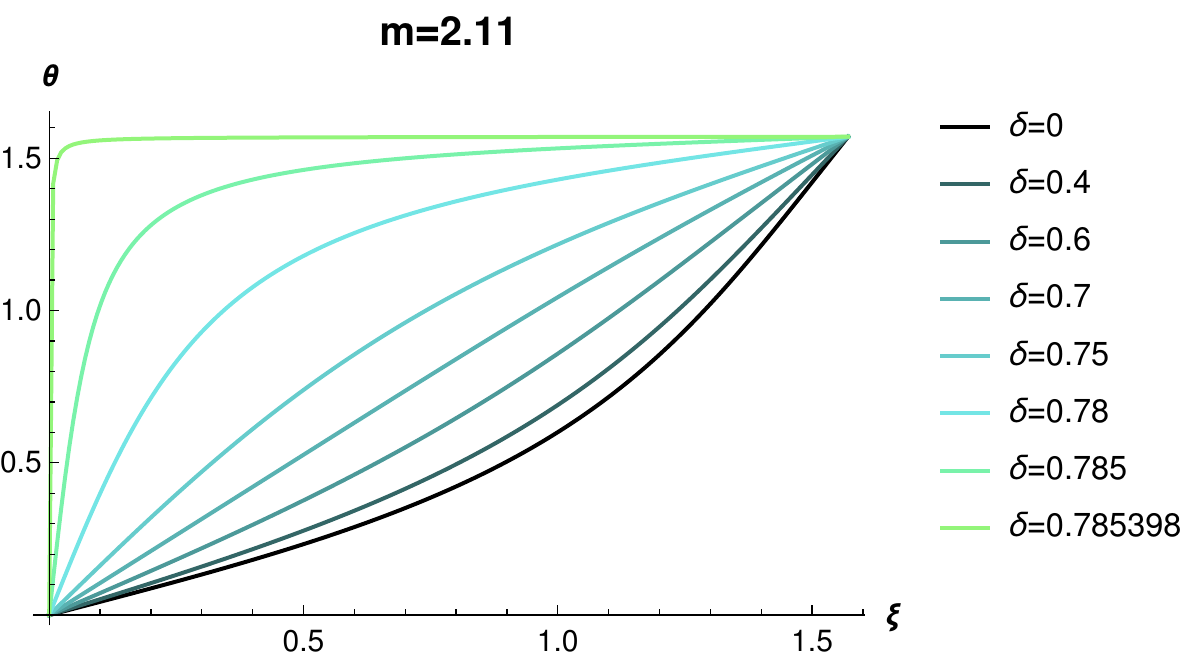}
		\label{fig:m211_basic_result_theta}}
	\caption{\label{fig:solutiontheta}The numerical solutions of $ \theta(p_{\hat{-}}) $ for several interpolation angles, corresponding to the quark mass values in Figure 4 of Ref.~\cite{Li}. All the mass values are in the unit of 
		$ \sqrt{2\lambda}$. Note that $ \theta(p_{\hat{-}}) $ is an odd function of $ p_{\hat{-}} $ and only the positive 
		$ p_{\hat{-}} $ range is plotted with the variable $ \xi = \tan^{-1} p_{\hat{-}}$.}
\end{figure}
\begin{figure}
	\centering
	\subfloat[]{
		\includegraphics[width=1.0\columnwidth]{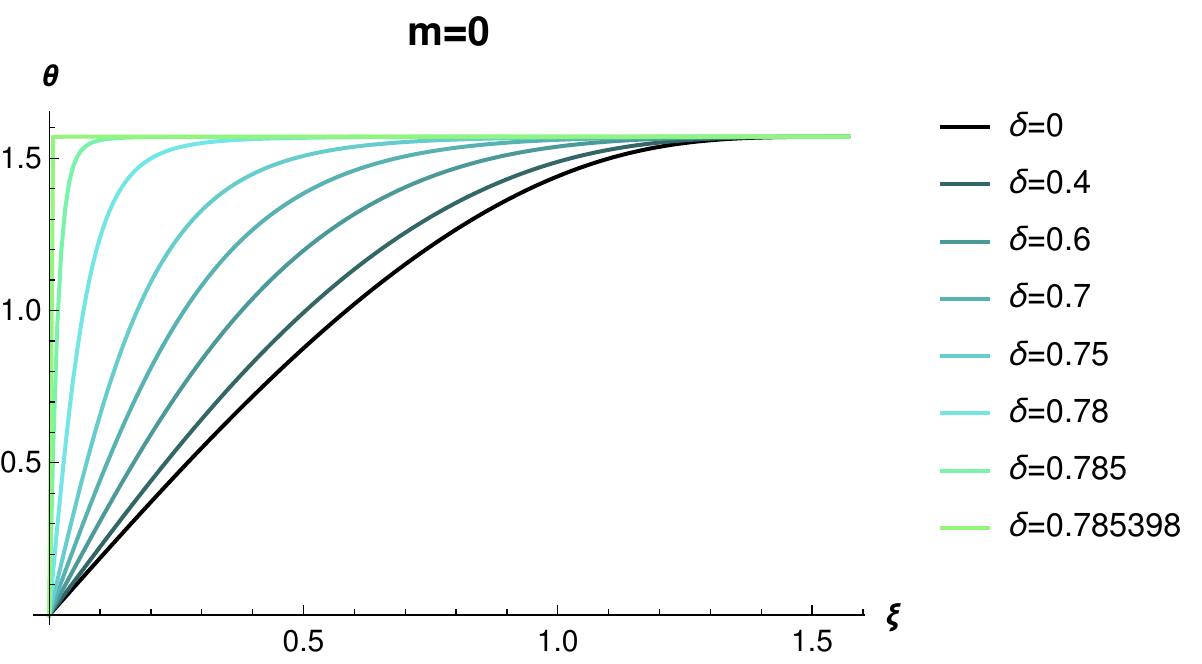}	 \label{fig:m0_basic_result_theta}}\\
	\centering
	\subfloat[]{
		\includegraphics[width=1.0\columnwidth]{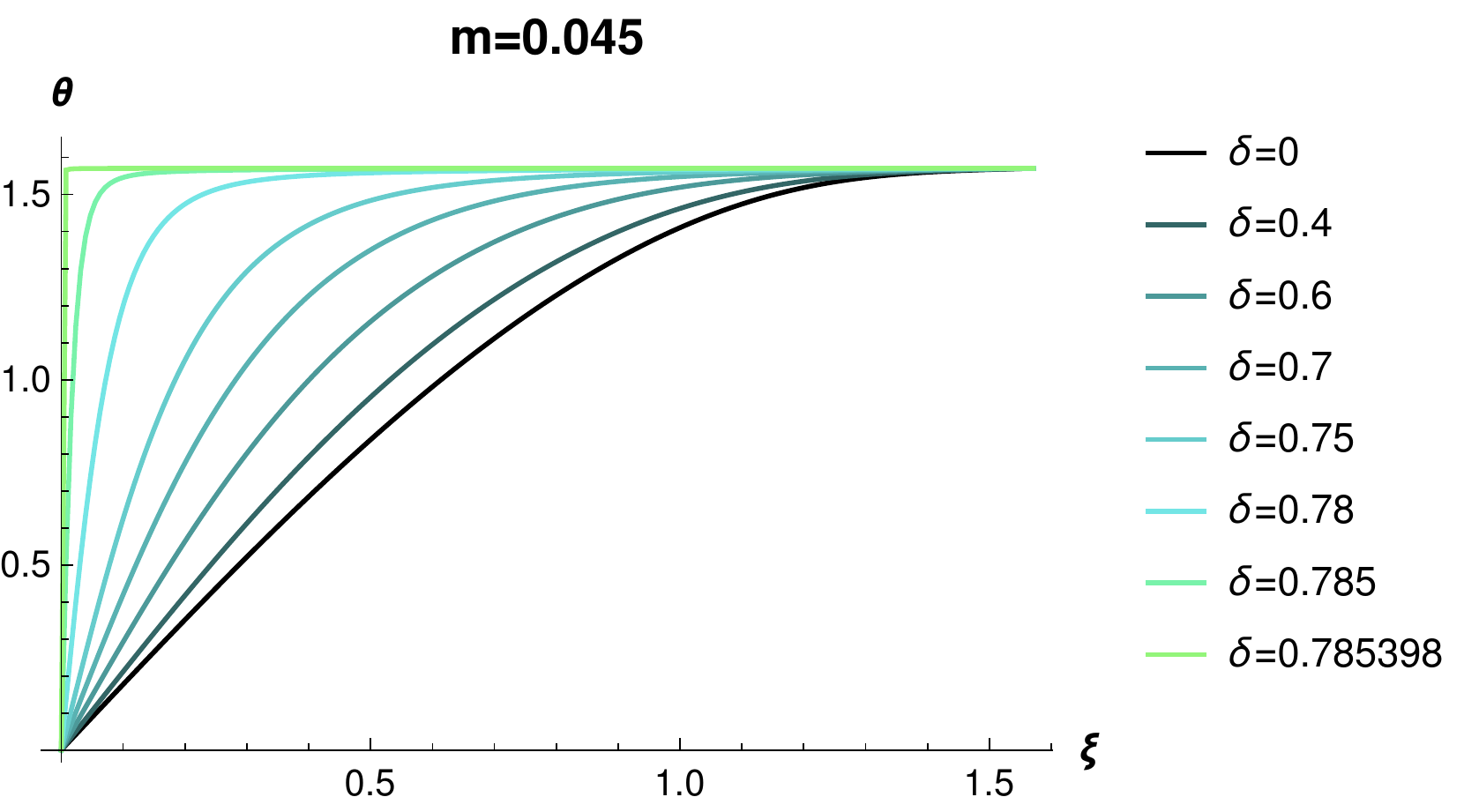}
		\label{fig:m0045_basic_result_theta}}\\
	\centering
	\subfloat[]{
		\includegraphics[width=1.0\columnwidth]{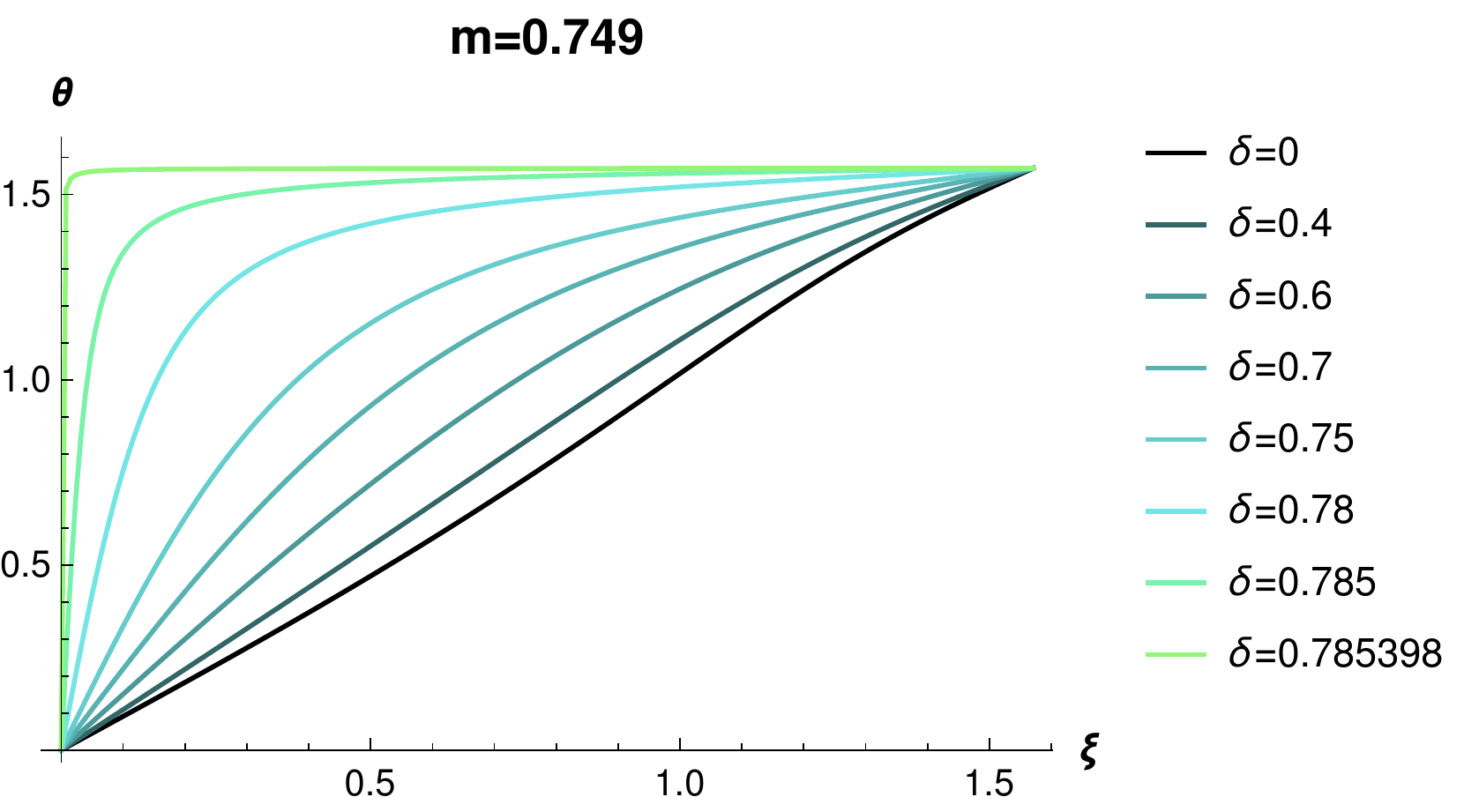}
		\label{fig:m0749_basic_result_theta}}\\
	\centering
	\subfloat[]{
		\includegraphics[width=1.0\columnwidth]{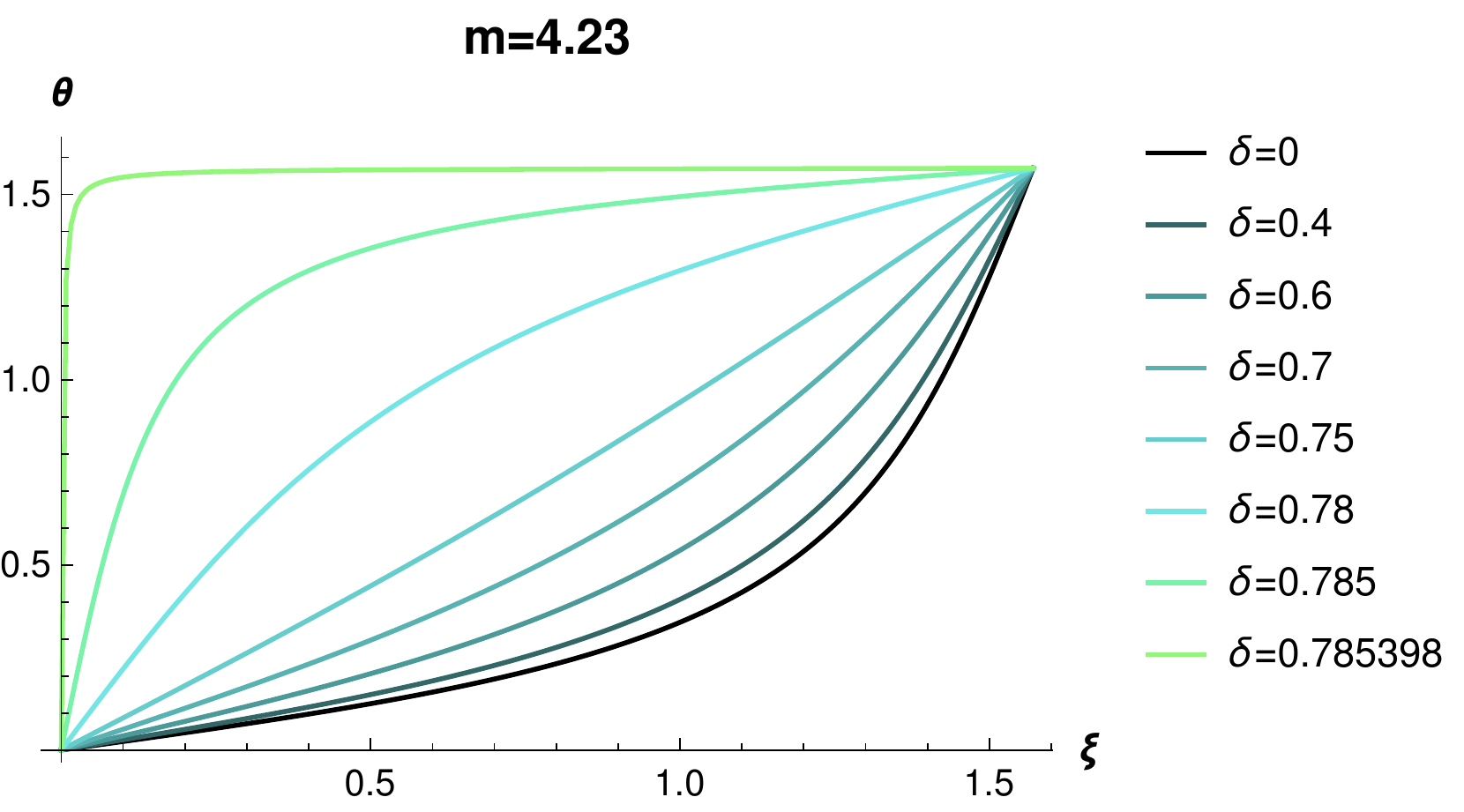}
		\label{fig:m423_basic_result_theta}}\\
	\caption{\label{fig:solutiontheta2}The numerical solutions of $ \theta(p_{\hat{-}}) $ for several interpolation angles, corresponding to the quark mass values in Figure 2 of Ref.~\cite{mov}. All the mass values are in the unit of 
		$ \sqrt{2\lambda}$. Note that $ \theta(p_{\hat{-}}) $ is an odd function of $ p_{\hat{-}} $ and only the positive 
		$ p_{\hat{-}} $ range is plotted with the variable $ \xi = \tan^{-1} p_{\hat{-}}$.}
\end{figure}

\begin{figure}
	\centering
	\subfloat[]{
		\includegraphics[width=1.0\columnwidth]{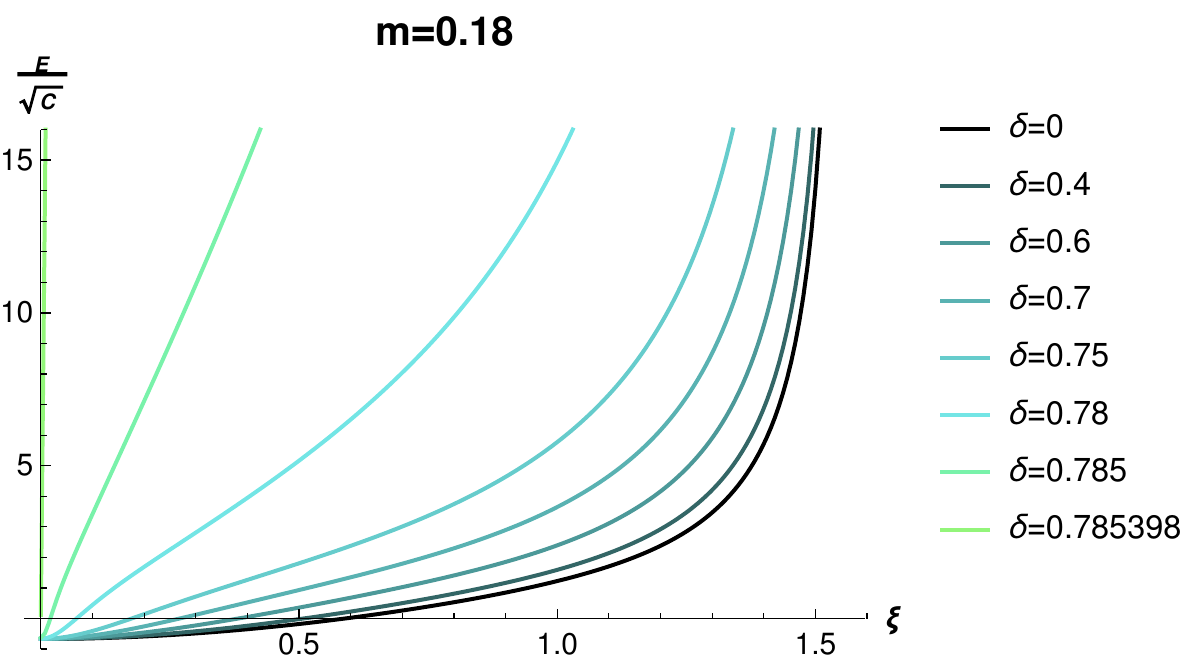}
		\label{fig:m018_basic_result_EC}}
	\\
	\centering
	\subfloat[]{
		\includegraphics[width=1.0\columnwidth]{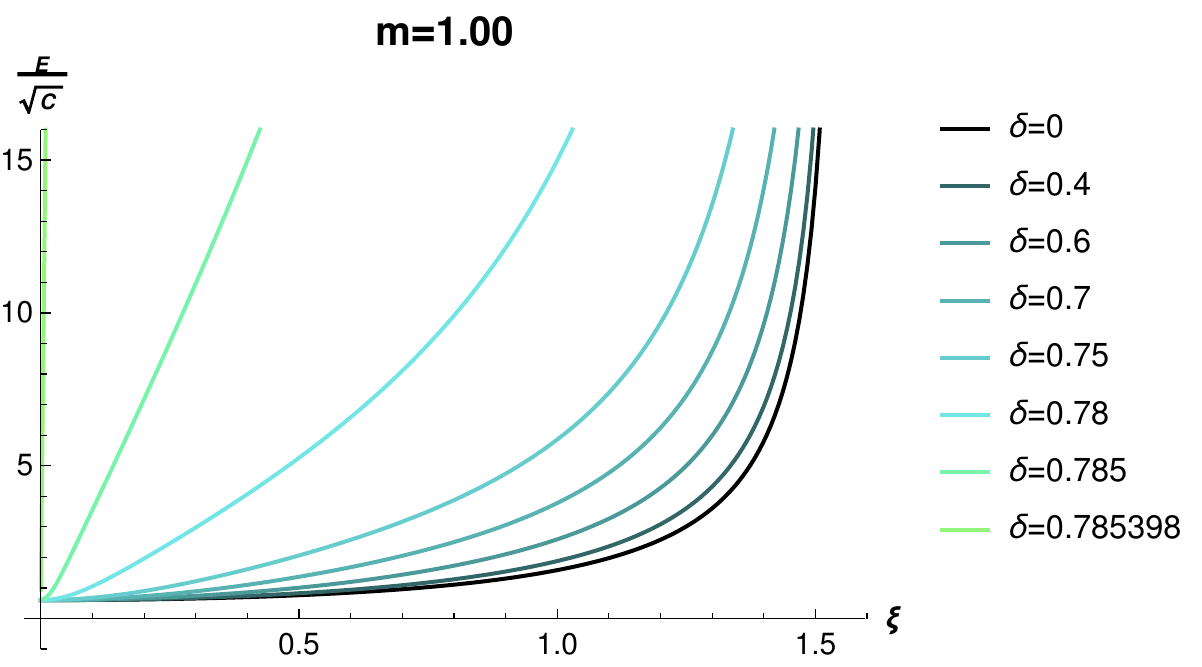}
		\label{fig:m100_basic_result_EC}}
	\\
	\centering
	\subfloat[]{
		\includegraphics[width=1.0\columnwidth]{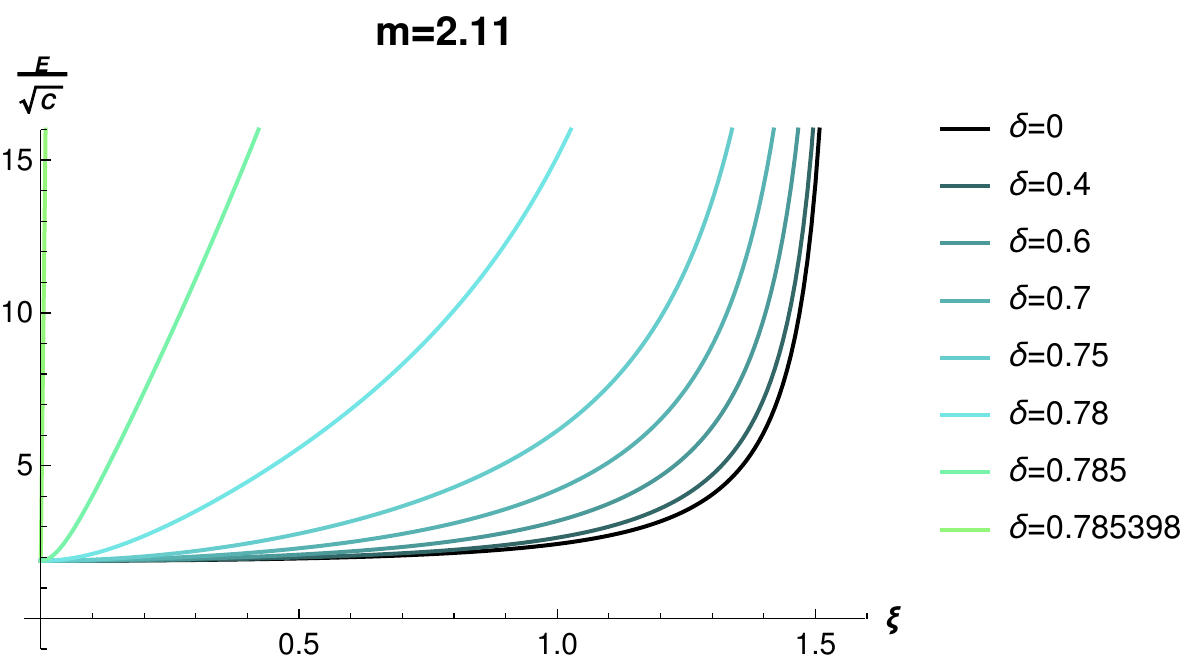}
		\label{fig:m211_basic_result_EC}}
	\caption{\label{fig:solutionEC}The solutions of $ E(p_{\hat{-}})/\sqrt{\mathbb{C}} $ for several interpolation angles for different choices of quark mass corresponding to the quark mass values in Figure 5 of Ref.~\cite{Li}. All quantities are in proper units of $ \sqrt{2\lambda} $. $ E(p_{\hat{-}}) $ is an even function of $ p_{\hat{-}} $. We plot only for positive $ p_{\hat{-}} $ with the variable $ \xi = \tan^{-1} p_{\hat{-}}$.}
\end{figure}
\begin{figure}
	\centering
	\subfloat[]{
		\includegraphics[width=1.0\columnwidth]{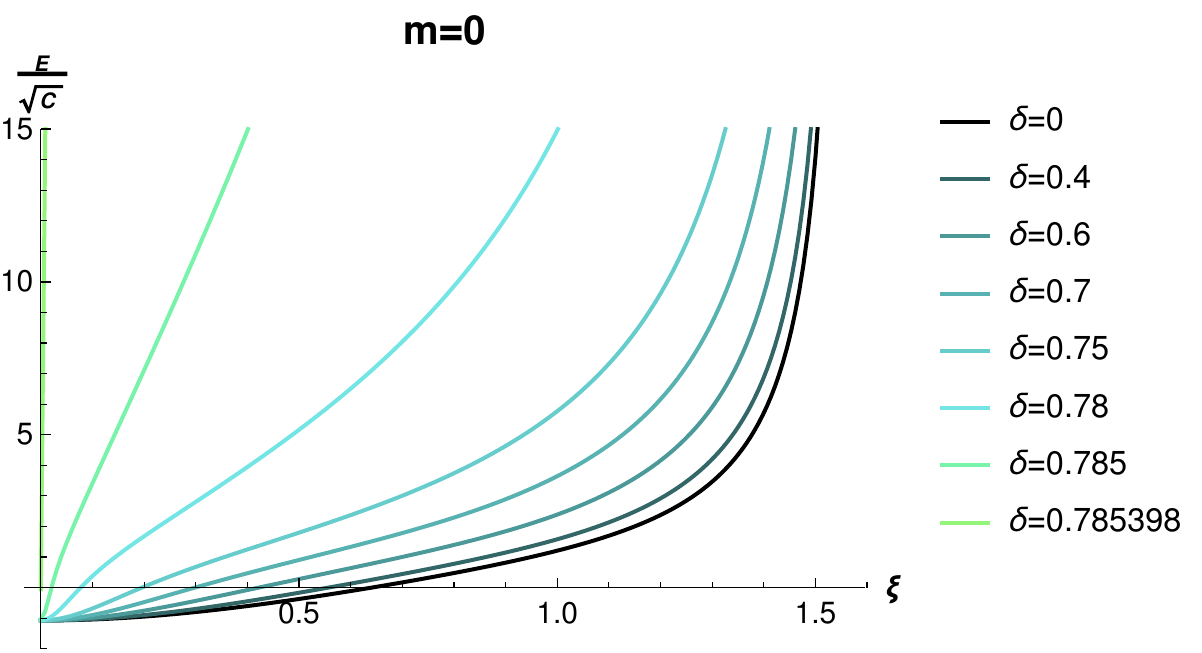}
		\label{fig:m0_basic_result_EC}
	}\\
	\centering
	\subfloat[]{
		\includegraphics[width=1.0\columnwidth]{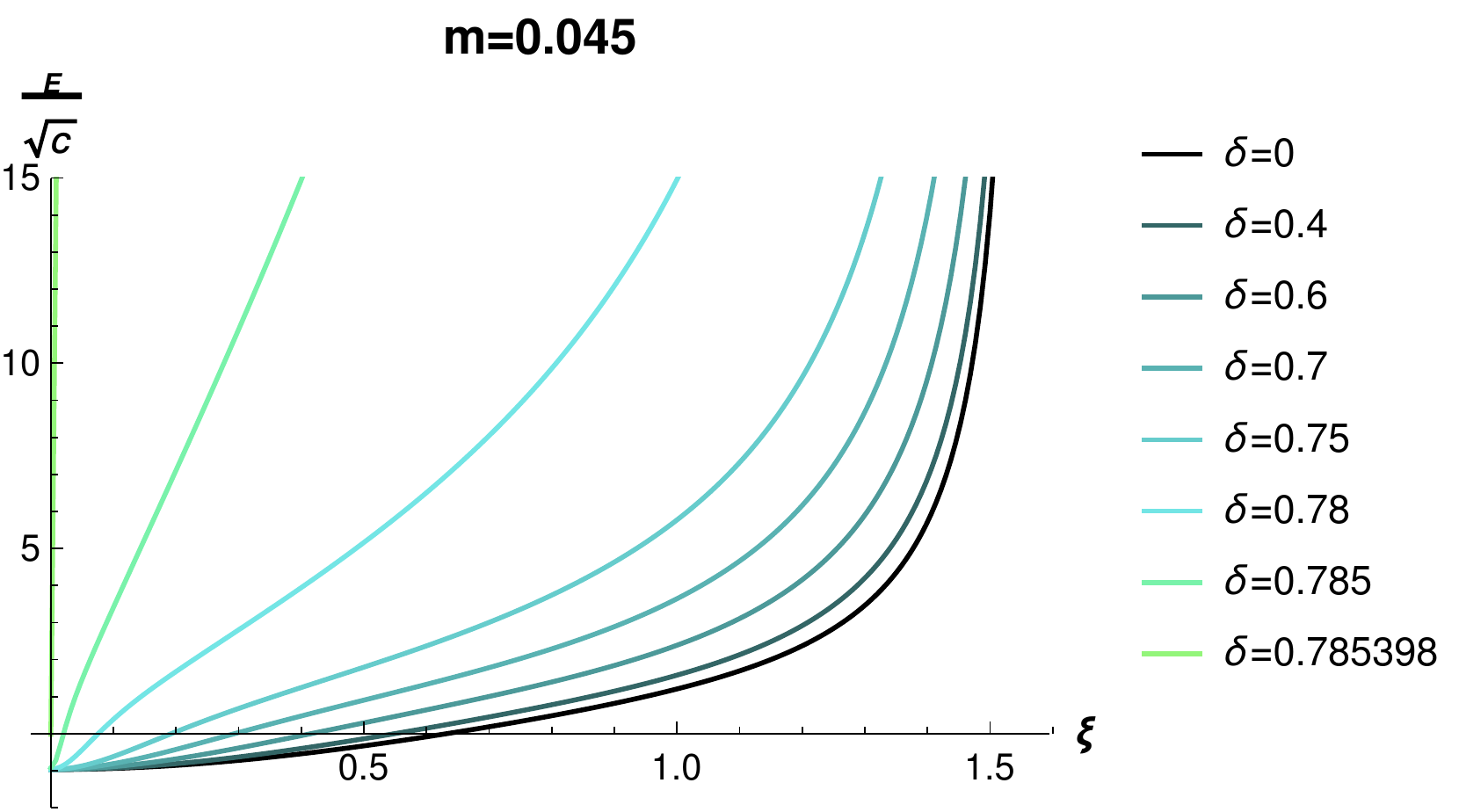}
		\label{fig:m0045_basic_result_EC}
	}\\
	\centering
	\subfloat[]{
		\includegraphics[width=1.0\columnwidth]{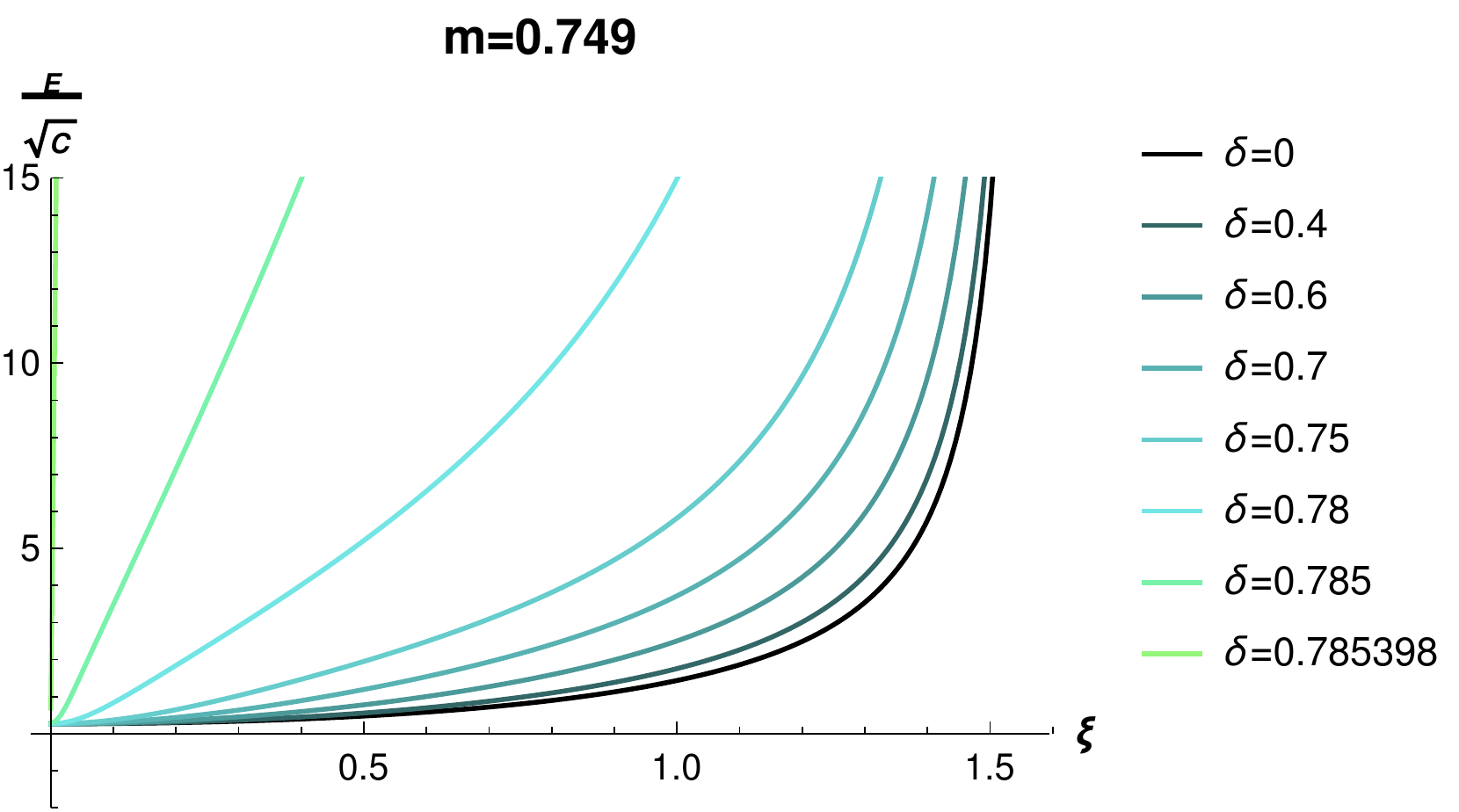}
		\label{fig:m0749_basic_result_EC}
	}\\
	\centering
	\subfloat[]{
		\includegraphics[width=1.0\columnwidth]{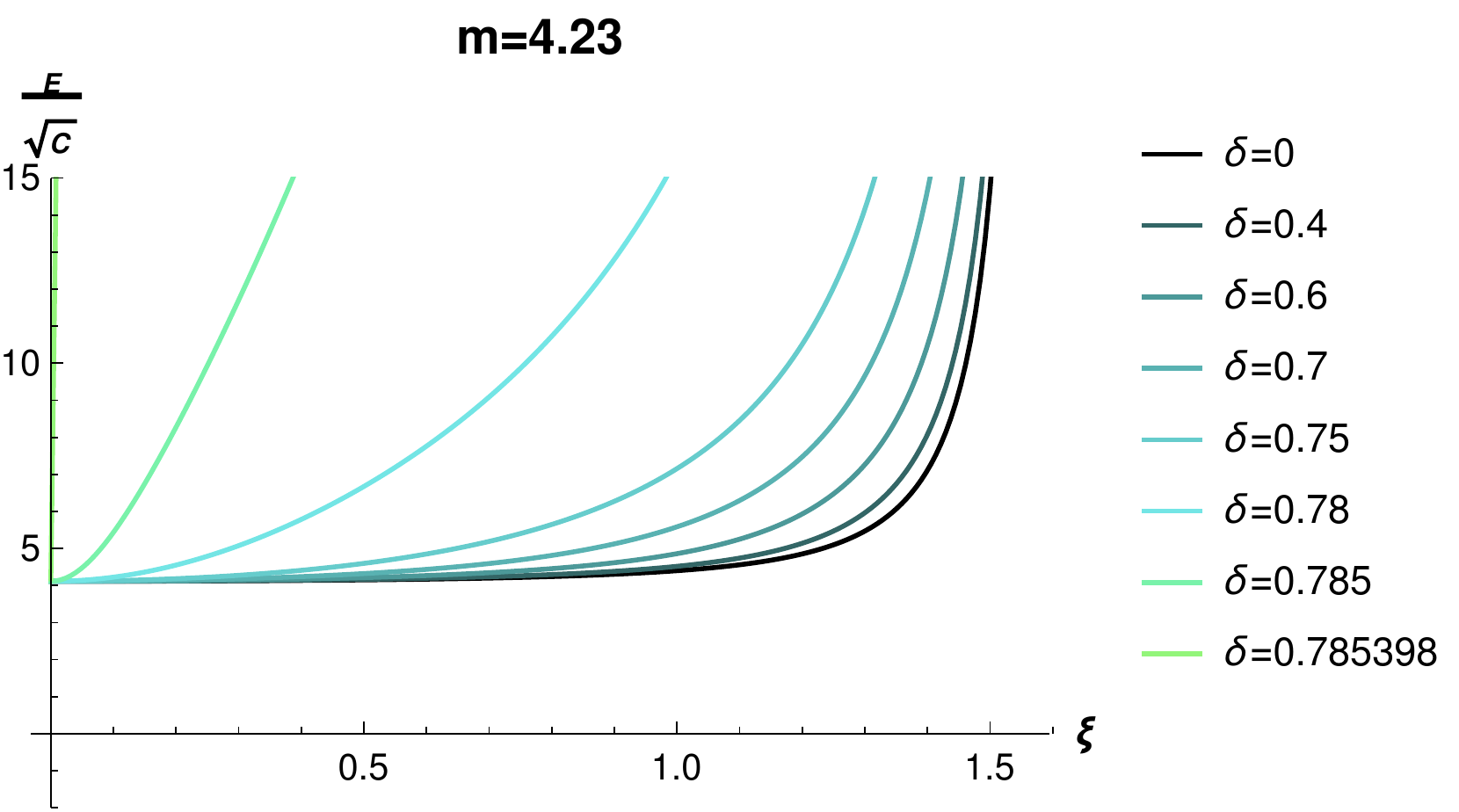}
		\label{fig:m423_basic_result_EC}
	}\\
	\caption{\label{fig:solutionEC2}The solutions of $ E(p_{\hat{-}})/\sqrt{\mathbb{C}} $ for several interpolation angles for different choices of quark mass, corresponding to the quark mass values in Figure 2 of Ref.~\cite{mov}. All quantities are in proper units of $ \sqrt{2\lambda} $. $ E(p_{\hat{-}}) $ is an even function of $ p_{\hat{-}} $. We plot only for positive $ p_{\hat{-}} $ with the variable $ \xi = \tan^{-1} p_{\hat{-}}$.}
\end{figure}

The mass gap equation given by Eq.~(\ref{gap_eq_inter_theta}) is solved numerically using a generalized Newton method 
mentioned in Ref.~\cite{Li} as well as in Ref.~\cite{mov}, and we use the same numerical method 
elaborated in Refs.~\cite{Li} and \cite{mov}. Using the same $ 200 $ grid points and choosing the same quark mass 
values (in unit of $ \sqrt{2\lambda} $) as in Ref.~\cite{Li} and in Ref.~\cite{mov}, we obtain the numerical solutions of Eq.~(\ref{gap_eq_inter_theta}). To cover the entire interpolating longitudinal momentum range, $-\infty <p_{\hat{-}} <+\infty$, we use  the variable $ \xi = \tan^{-1} p_{\hat{-}} $, where $ \xi\in (-\frac{\pi}{2},\frac{\pi}{2}) $. As the solutions of $ \theta(p_{\hat{-}}) $ are antisymmetric under the transformation of $p_{\hat{-}} \to - p_{\hat{-}}$, i.e. $ \theta(-p_{\hat{-}}) = -\theta(p_{\hat{-}})$,
we present the results for the region $0< p_{\hat{-}}<+\infty$ only, i.e. $0<\xi<\frac{\pi}{2}$,
for several quark mass values in Figs.~\ref{fig:solutiontheta} and \ref{fig:solutiontheta2}.
In Fig.~\ref{fig:solutiontheta}, the numerical solutions of $ \theta(p_{\hat{-}}) $ for several interpolation angles are plotted for the quark mass values presented in Ref.~\cite{Li}. The profiles of $\theta(\xi)$ for $\delta=0$ coincide with the IFD results provided in Ref.~\cite{Li}'s Figure 4. Also, the profiles of $\theta(\xi)$ for $\delta \rightarrow \pi/4$ approach to
the analytic solution given by Eq.~(\ref{lfthetasol}) in LFD.
It is interesting to note that the IFD results exhibit a convex profile for the lighter
quark mass ($m = 0.18$), and as the quark mass increases ($m =$ 1.00 and 2.11) the profile gets
more and more concave. 
For the quark mass values presented in Ref.~\cite{mov}, 
the numerical solutions of $ \theta(p_{\hat{-}}) $ for several interpolation angles are plotted in Fig.~\ref{fig:solutiontheta2}.
The authors of Ref.~\cite{mov} chose the 't Hooft coupling $\lambda$ as $\pi\lambda = 0.18$ GeV$^2$ in conformity to the value of the string tension in the realistic $\text{QCD}_{\text{4}}$, and then 
determined the light quark mass $m_{u/d} = 0.045$ to obtain
the physical pion mass $M_\pi = 0.41$ in the unit of $\sqrt{2\lambda}$ by solving the bound-state equation, which we present in Sec.~\ref{sec:bound}. Likewise, 
the heavy quark mass $m_c = 4.23$ was determined to get the physical $J/\psi$ mass 
$M_{J/\psi} = 9.03 $ in the same unit. The strange quark mass $m_s = 0.749$ was taken to provide
a threshold, below (above) which is called light (heavy) flavor, by minimizing the relative distance between
the $\theta(\xi)$ solution and the straight line $\theta(\xi) = \xi$ in IFD as the profile of $\theta(\xi)$ in IFD (i.e. $\delta = 0$) passes from the convex to concave with the increasing quark mass as discussed
in Fig.~\ref{fig:solutiontheta}. The chiral limit with $m_u = 0$ was also considered and the Gell-Mann - Oakes - Renner relation (GOR) was discussed in Ref.~\cite{mov}. For those quark masses, $m= 0, 0.045, 0.749$
and 4.23, we obtain the interpolating mass gap solutions for several $\delta$ values between 
$\delta=0$ (IFD) and $\delta=\pi/4$ (LFD) as plotted in Fig.~\ref{fig:solutiontheta2}.
Our numerical solutions for $\delta=0$ coincide with the IFD results provided in Ref.~\cite{mov}'s Figure 2 (left panel). Again, the profiles of $\theta(\xi)$ for $\delta \rightarrow \pi/4$ approach to
the analytic solution given by Eq.(\ref{lfthetasol}) in LFD.

We then obtain the solutions for $ E(p_{\hat{-}}) $ by plugging the mass gap solutions
presented in Figs.~\ref{fig:solutiontheta} and ~\ref{fig:solutiontheta2} into Eq.~(\ref{gap_eq_inter_E}).
As the plots of $E(p_{\hat{-}})$ themselves for different interpolation angles are too close to each other to observe clearly, we present the solutions of $ E(p_{\hat{-}})/\sqrt{\mathbb{C}} $ in Figs.~\ref{fig:solutionEC} and \ref{fig:solutionEC2}. The quark mass values in Figs.~\ref{fig:solutionEC} and \ref{fig:solutionEC2} correspond to those in Figs.~\ref{fig:solutiontheta} and~\ref{fig:solutiontheta2}, respectively.
At $\delta=0$, the profiles of $E(p_{\hat{-}})/\sqrt{\mathbb{C}} = E(p^1)$ in Figs.~\ref{fig:solutionEC} and~\ref{fig:solutionEC2} coincide with the IFD results provided in Ref.~\cite{Li}'s Figure 5 and
Ref.~\cite{mov}'s Figure 2 (right panel), respectively. 
Examining the results of $E(p_{\hat{-}})$ themselves in the limit $\delta \to \pi/4$, 
we also find that all of them approach to the analytic result of $E(p^+)$ given by Eq.~(\ref{lftenergysol})
independent of quark mass $m$.  Moreover, the LFD result of $E(p^+)$ given by Eq.~(\ref{lftenergysol}) is always positive regardless of $p^+$ while the IFD ($\delta=0$) results of $E(p^1)$ for small quark mass values, e.g. 
$m=0.18$ in Fig.~\ref{fig:m018_basic_result_EC} and $m=0$ and 0.045 in Figs.~\ref{fig:m0_basic_result_EC} and~\ref{fig:m0045_basic_result_EC}, respectively,
are negative for small momentum regions. Indeed, we notice that the region of small momentum
for the negative value of $E(p^1)$ gets shrunken as $m$ gets larger but persists up to $m \approx 0.56$. It was argued in Ref.~\cite{BG} that the existence of negative quark self-energy does not cause any concern though, due to 
the lack of observability for the energy of a confined single quark. While similar
aspect of the negative quark energy exists for other interpolation angle values unless $\delta = \pi/4$, the corresponding
range of the small momentum gets reduced as $\delta$ approaches to $\pi/4$ as 
depicted in Figs.~\ref{fig:m018_basic_result_EC},~\ref{fig:m0_basic_result_EC} and 
~\ref{fig:m0045_basic_result_EC}. It is interesting to note that our observation of
diminishing the negative quark self-energy region in the limit $\delta \to \pi/4$ (LFD) 
appears consistent with the result
that the single particle energies do not change sign due to 
the additional constant term appearing in converting the light-front longitudinal momentum sum 
to a principal value integral discussed in Ref.~\cite{LenzThies} with  
the formulation of a finite light-front $x^-$ coordinate interval.

\section{Chiral Condensate and Constituent Quark Mass} 
\label{sec:implications}

As we have obtained the mass gap solutions, we now utilize them to discuss the chiral condensate and 
the constituent quark mass in this section. 

\subsection{The chiral condensate}\label{sub:condensate}

While the Coleman's theorem~\cite{Col} prohibits the SBCS in two-dimensional theories with a finite number of degrees of freedom, the large $N_c$ limit of the 't Hooft model does not contradict with the Coleman's theorem. 
The exact result for the chiral condensate 
was found in the chiral limit ($m\to 0$) for the weak coupling regime of QCD$_2$ ($m >> g \sim 1/\sqrt{N_c}$)~\cite{Zhicon}
as 
\begin{equation}
\label{analytic-condensation}
<\bar{\psi}\psi>=-N_c/\sqrt{12}
\end{equation}
in the unit of the mass dimension $\sqrt{2\lambda}$ with the definition of $\lambda$ given by 
Eq.(\ref{lambda-def}).
This indicates that the spontaneous breaking of the chiral symmetry (SBCS) occurs in the 't Hooft model, in contrast to the strong coupling regime of QCD$_2$ in which the SBCS does not occur according to the Coleman's theorem~\cite{Col}.
Ramifications of the nontrivial chiral condensates with respect to the vacuum in the LFD as well as its non-analytic behavior were discussed for nonzero quark masses and its chiral limit~\cite{Burkardtcon}. 
For nonzero quark masses, the renormalized quark condensation was defined by subtracting
the free field expectation value to render the quark condensation finite~\cite{Burkardtcon}, 
\begin{align}\label{con_mno0}
<\bar{\psi}\psi>|_{\rm ren}&\equiv<\bar{\psi}\psi>-<\bar{\psi}\psi>|_{g=0}.
\end{align}
In the chiral limit, $m \to 0$, $<\bar{\psi}\psi>|_{g=0} = 0$ and thus $<\bar{\psi}\psi>|_{\rm ren} = <\bar{\psi}\psi>$. 
The numerical computation verifying Eq.(\ref{analytic-condensation}) was presented 
in IFD~\cite{Licon} substituting the mass gap solution for $\theta(p^1)$ in the chiral limit $m \to 0$ 
as 
\begin{equation}
<\bar{\psi}\psi>|_{\rm ren}=-N_c\int_{-\infty}^{+\infty}\frac{dp^1}{2\pi}\cos\theta(p^1) \approx -0.29 N_c .
\end{equation}

With the interpolating quark field between IFD and LFD given by Eq.~(\ref{eqn:psi_field}), we find  
\begin{align}\label{con_ren}
&<\bar{\psi}\psi>|_{\rm ren}=N_c\int_{-\infty}^{+\infty}\frac{dp_{\hat{-}}}{(2\pi)(2p^{\hat{+}})}\notag\\
&\times\bigg[{\rm Tr}\left\lbrace v(-p_{\hat{-}})\bar{v}(-p_{\hat{-}})\right\rbrace
-{\rm Tr}\lbrace v^{(0)}(-p_{\hat{-}})\bar{v}^{(0)}(-p_{\hat{-}})\rbrace\biggr]\notag\\
&=-\frac{N_c}{2\pi\sqrt{\mathbb{C}}}\int_{-\infty}^{+\infty}dp_{\hat{-}}
\left[ \cos \theta(p_{\hat{-}})-\cos \theta_f(p_{\hat{-}})\right] ,
\end{align}
where we have used the interpolating spinors given by Eqs.~(\ref{eqn:vspinor}) 
and (\ref{v_inter_free_}).
We numerically compute Eq.~(\ref{con_ren}) in the chiral limit, $m\to 0$, for different interpolation angles, and the results are shown in Table~\ref{tab:conden}. We observe that the closer one gets to the LFD ($\delta=\frac{\pi}{4}$), the higher accuracy one needs for the numerical computation and thus we list the results obtained by increasing the number of grid points used. 
One can see that when $\delta$ is away from $\pi/4$, the coarser grid is already good enough to obtain an accurate result. For $\delta$ closer to $\pi/4$, the number of grid points 
should be increased to improve the numerical accuary. 
Table~\ref{tab:conden} indicates an eventual agreement to the analytical value $-1/\sqrt{12}\approx -0.29$ regardless of the interpolation angle $\delta$ with the enhancement of the numerical accuracy.

\begin{table}
	\caption{\label{tab:conden}The numerically calculated condensation values in the chiral limit with different interpolation angles and computational accuracy. All quantities are in proper units of $ \sqrt{2\lambda} $.}
	\begin{ruledtabular}
		\begin{tabular}{c|c|c}
			\centering
			$\delta$ &  number of grid points & $ <\bar{\psi}\psi>|_{m\to 0}/N_c $\\ \hline
			$	0$&$ 200$&$ -0.285209$\\
			& $600$ &$ -0.287508$\\
			\hline
			$0.4$&$ 200$& $-0.285164$ \\
			& $600$& $-0.287496$\\
			\hline
			$0.6$& $200$& $-0.284792$\\
			&$ 600 $& $ -0.287375 $\\
			\hline
			$0.7$& $200$& $-0.283836$\\
			&$ 600 $ & $ -0.287059 $\\
			\hline
			$0.75$& $200$& $-0.281837$\\
			&$ 600 $ & $ -0.286396 $\\
			\hline
			$0.78$& $200$& $-0.296575$ \\
			& $600$& $-0.291334$\\
			\hline
			$0.785$& $600$& $-0.298104$\\
			& $ 1000 $ & $ -0.294377 $\\
			& $ 3000 $ & $  -0.290590 $\\
			\hline
			$0.78535$
			& $ 1000 $ & $  -0.304659 $\\
			& $ 3000 $ & $   -0.294134$\\
			& $ 5000 $ & $ 	-0.291964$\\
		\end{tabular}
	\end{ruledtabular}
\end{table}

\begin{figure}
	\centering
	\includegraphics[width=1.0\linewidth]{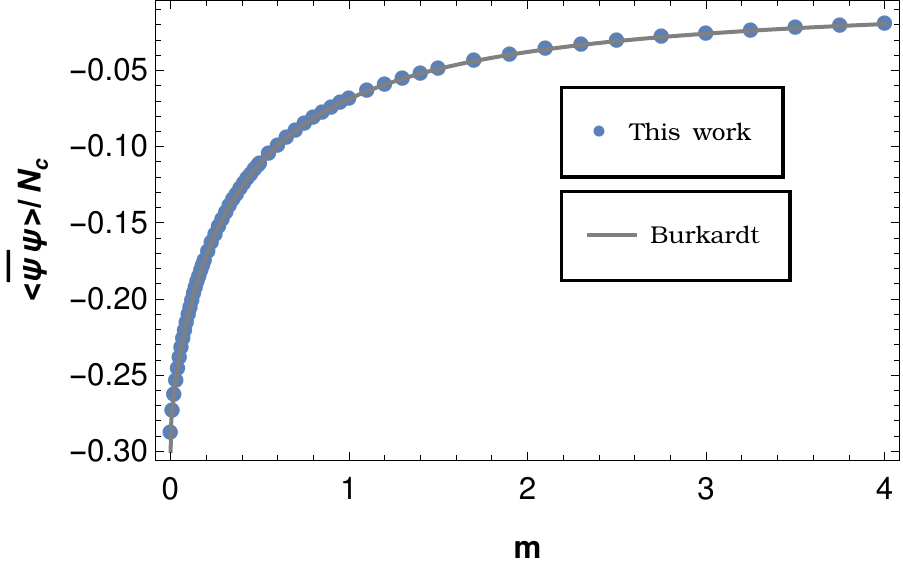}
	\caption{Numerical results of the condensation $ <\bar{\psi}\psi>|_{\rm ren} $ as a function of $m$
		in comparison with the analytic result in Ref.~\cite{Burkardtcon}. All quantities are in proper units of $ \sqrt{2\lambda} $.}
	\label{fig:Con_IFD}
\end{figure}

Even if the chiral limit is not taken, we notice that the renormalized chiral condensate must be independent of the interpolating angle $\delta$ as it must be the characteristic quantity 
determining the vacuum property for a given phase of the theory. 
In fact, the interpolating longitudinal momentum variable
$p_{\hat{-}}$ can be scaled out by the interpolating parameter $\sqrt{\mathbb{C}}$ and Eq.~(\ref{con_ren}) can be given by the rescaled variable 
$p'_{\hat{-}}=p_{\hat{-}}/\sqrt{\mathbb{C}}$, i.e. 
\begin{align}\label{con_rescale}
<\bar{\psi}\psi>|_{\rm ren}
&=-\frac{N_c}{2\pi} \int_{-\infty}^{+\infty}dp'_{\hat{-}}\left[ \cos \theta(p'_{\hat{-}})-\cos\theta_f(p'_{\hat{-}})\right] .
\end{align}
The same variable change can be applied to the interpolating mass gap equation given by
Eq.(\ref{gap_eq_inter_theta}) as we illustrate in Appendix~\ref{app:FreeComparison} to obtain 
the rescaled mass gap equation without any apparent interpolation
angle dependence as given by Eq.~(\ref{gap_eq_inter_theta_rescaled}).
With $\theta(p'_{\hat{-}})$ being the solution of Eq.~(\ref{gap_eq_inter_theta_rescaled})  
as well as $\theta_f(p'_{\hat{-}}) = \tan^{-1}(p'_{\hat{-}}/m)$ from Eq.(\ref{eqn:thetaf}), 
we can confirm that the chiral condensate is indeed independent of the interpolation angle $\delta$.
Thus, the result of the chiral condensate must be identical whichever dynamics is 
chosen for the computation between $\delta=0$ (IFD) and $\delta=\pi/4$ (LFD).
Computing $<\bar{\psi}\psi>|_{\rm ren}$ in Eq.(\ref{con_ren}) for any given interpolation
angle between $\delta=0$ and $\delta=\pi/4$, one can numerically verify the uniqueness of the result for each and 
every quark mass values that one takes. 
Varying the quark masses between $m = 0$ and $m = 4$ in the unit of $\sqrt{2\lambda}$, we 
computed $<\bar{\psi}\psi>|_{\rm ren}$ from Eq.(\ref{con_rescale}) and 
obtained the numerical result shown in Fig.~\ref{fig:Con_IFD}.
Our numerical result for $m=0$ in Fig.~\ref{fig:Con_IFD} confirms the results obtained in 
Table~\ref{tab:conden} reproducing the analytic value $-1/\sqrt{12} \approx -0.29$. 
Likewise, our numerical results for $0 < m \leq 4$ coincide with the analytic result
given by Eq.(2.19) of Ref.~\cite{Burkardtcon} which was also numerically confirmed in 
Ref.~\cite{mov}.

In the formulation of a finite light-front $x^-$ coordinate interval~\cite{LenzThies},
a phase transition was reported in the weak coupling limit as a 
dramatic change of the quark condensate value to zero was observed. 
In the continuum limit, however, there is
no such phase transition and the non-zero condensate value is intact regardless of
the form of dynamics between $\delta =0$ (IFD) and $\delta= \pi/4$ (LFD)
as discussed in this subsection. In this respect, it is important to realize that
the phase characterized by the SBCS is uniquely viable in the continuum 't Hooft model. While there is a simple analytic step function solution of Eq.(\ref{gap_eq_inter_theta}) in IFD ($\delta=0$) given by $\theta(p^1)=\frac{\pi}{2}\mathrm{sgn} (p^1)$ in the chiral limit~\cite{BG}, it should be clearly distinguished from the step function solution  in LFD given by Eq.(\ref{lfthetasol}).
The step function solution in IFD leads to the zero chiral condensate $<\bar{\psi}\psi>=0$ in contrast to the non-trivial chiral condensate given by Eq.(\ref{analytic-condensation}). It was indeed demonstrated in Ref.~\cite{Bicudo} that this chirally symmetric step function solution in IFD reveals the possession of infinite vacuum energy compared to the vacuum energy of the SBCS solution. 
Thus, the step function solution in IFD should be clearly distinguished from the physically viable solution characterized by the SBCS~\cite{review} discussed in this subsection.

\subsection{The Fermion Propagator and Constituent Mass}
\label{FPandCM}

The interpolating free fermion propagator has been discussed at length in our previous work~\cite{InterQED} and 
can be readily applied for the interpolating bare quark propagator in 1+1 dimension with the notation of
the on-mass-shell two-momenta $ P_{a\hat{\mu}}=(P_{a\hat{+}},p_{\hat{-}}) $, $ P_{b\hat{\mu}}=(-P_{b\hat{+}},-p_{\hat{-}}) $ taking $ P_{a\hat{+}} $ and $ -P_{b\hat{+}} $ as the positive and negative on-shell 
interpolating energies of the bare quark, i.e. 
\begin{equation}\label{Pa+}
P_{a\hat{+}}=-\frac{\mathbb{S}}{\mathbb{C}}p_{\hat{-}}+\frac{\sqrt{p_{\hat{-}}^2+\mathbb{C}m^2}}{\mathbb{C}}=\frac{-\mathbb{S}p_{\hat{-}}+p^{\hat{+}}}{\mathbb{C}},
\end{equation}
and
\begin{equation}\label{Pb+}
-P_{b\hat{+}}=-\frac{\mathbb{S}}{\mathbb{C}}p_{\hat{-}}-\frac{\sqrt{p_{\hat{-}}^2+\mathbb{C}m^2}}{\mathbb{C}}=\frac{-\mathbb{S}p_{\hat{-}}-p^{\hat{+}}}{\mathbb{C}}.
\end{equation}
The interpolating bare quark propagator is then given by 
\begin{equation}\label{Sfree}
S(p)_{\rm f}=\frac{1}{2p^{\hat{+}}}\left( \frac{\slashed{P}_a+m}{p_{\hat{+}}-P_{a\hat{+}}+i\epsilon}+\frac{\slashed{P}_b-m}{p_{\hat{+}}+P_{b\hat{+}}-i\epsilon}\right) ,
\end{equation}
where $\left(\slashed{P}_a+m \right) = u^{(0)}(p_{\hat{-}})\bar{u}^{(0)}(p_{\hat{-}})$
and $\left(\slashed{P}_b-m \right) = v^{(0)}(-p_{\hat{-}})\bar{v}^{(0)}(-p_{\hat{-}})$
can be easily verified using the interpolating free spinors given by Eqs.(\ref{u_inter_free_}) and 
(\ref{v_inter_free_}). In Eq.(\ref{Sfree}), we call $m$ as the bare quark mass.
As discussed in Sec.~\ref{sub:dia}, the interpolating dressed quark propagator can be obtained from 
the interpolating bare quark propagator with the replacement given by Eq.(\ref{eqn:replace}). 
Effectively, $P_{a\hat{+}}$ and $P_{b\hat{+}}$ are replaced by $E_u$ and $E_v$ in Eq.(\ref{EuEv}) 
as well as the free spinors, $u^{(0)}(p_{\hat{-}})$ and $v^{(0)}(p_{\hat{-}})$, are replaced by the spinors given by Eqs.(\ref{eqn:uspinor}) and (\ref{eqn:vspinor}), $u(p_{\hat{-}})$ and $v(p_{\hat{-}})$, respectively. 
The dressed quark propagator is then obtained as
\begin{equation}\label{Sdressed}
S(p)=\frac{1}{2p^{\hat{+}}}\left[ \frac{u(p_{\hat{-}})\bar{u}(p_{\hat{-}})}{p_{\hat{+}}-E_u(p_{\hat{-}})+i\epsilon}+\frac{v(-p_{\hat{-}})\bar{v}(-p_{\hat{-}})}{p_{\hat{+}}-E_v(p_{\hat{-}})-i\epsilon}\right],
\end{equation}
where the first and second terms correspond to the forward and backward moving propagators,
respectively. 
The equivalence of Eq.(\ref{Sdressed}) to Eq.(\ref{dressed_prop}) can be verified using
the relations 
\begin{equation}\label{uubar}
\frac{u(p_{\hat{-}})\bar{u}(p_{\hat{-}})}{2p^{\hat{+}}}=\Gamma^+\gamma^0=\left( \begin{array}{cc}
\frac{\cos\theta(p_{\hat{-}})}{2\sqrt{\mathbb{C}}} & \frac{1-\sin\theta(p_{\hat{-}})}{2(\cos\delta-\sin\delta)} \\
\frac{1+\sin\theta(p_{\hat{-}})}{2(\cos\delta+\sin\delta)} & \frac{\cos\theta(p_{\hat{-}})}{2\sqrt{\mathbb{C}}}
\end{array}\right),\nonumber
\end{equation}
\begin{equation}\label{vvbar}
\frac{v(-p_{\hat{-}})\bar{v}(-p_{\hat{-}})}{2p^{\hat{+}}}=\Gamma^-\gamma^0=\left( \begin{array}{cc}
-\frac{\cos\theta(p_{\hat{-}})}{2\sqrt{\mathbb{C}}} & \frac{1+\sin\theta(p_{\hat{-}})}{2(\cos\delta-\sin\delta)} \\
\frac{1-\sin\theta(p_{\hat{-}})}{2(\cos\delta+\sin\delta)} & -\frac{\cos\theta(p_{\hat{-}})}{2\sqrt{\mathbb{C}}} 
\end{array}\right),\nonumber
\end{equation}
where $\Gamma^+$ and $\Gamma^-$ are given by Eqs.(\ref{Gamma^+`}) and (\ref{Gamma^-`}), respectively.  
As discussed in Sec.~\ref{sub:behaviorLF}, one can verify that the forward and backward moving parts in the limit $\mathbb{C} \to 0$ correspond to the on-mass-shell part $\sim( \slashed{p}_{\rm on}+m)$ and the instantaneous contribution ($\sim \gamma^+$) in LFD, respectively.

\begin{figure}
	\centering
	\subfloat[]{
		\includegraphics[width=1.0\linewidth]{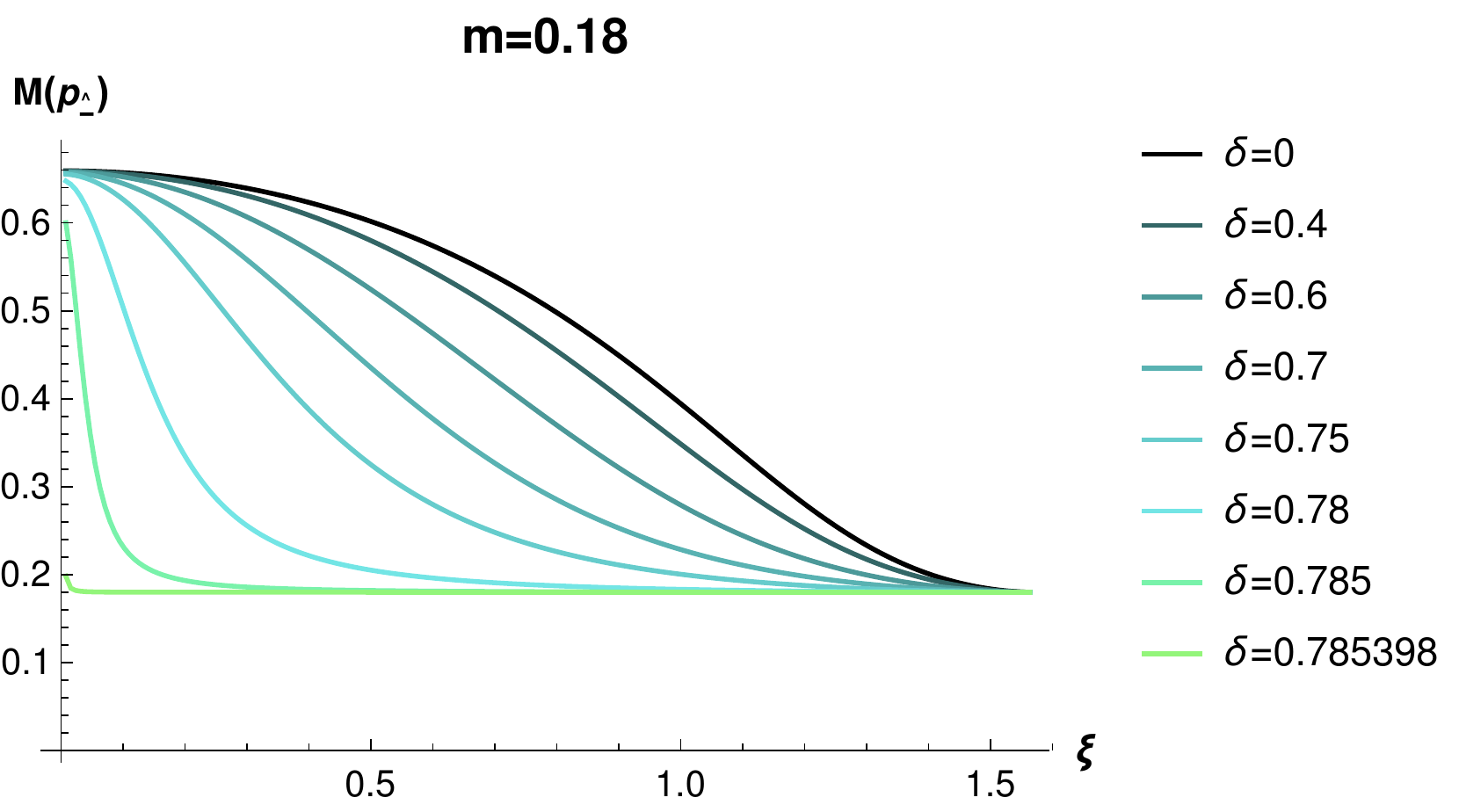}
		\label{fig:m018_constituentmass}
	}\\
	\subfloat[]{
		\includegraphics[width=1.0\linewidth]{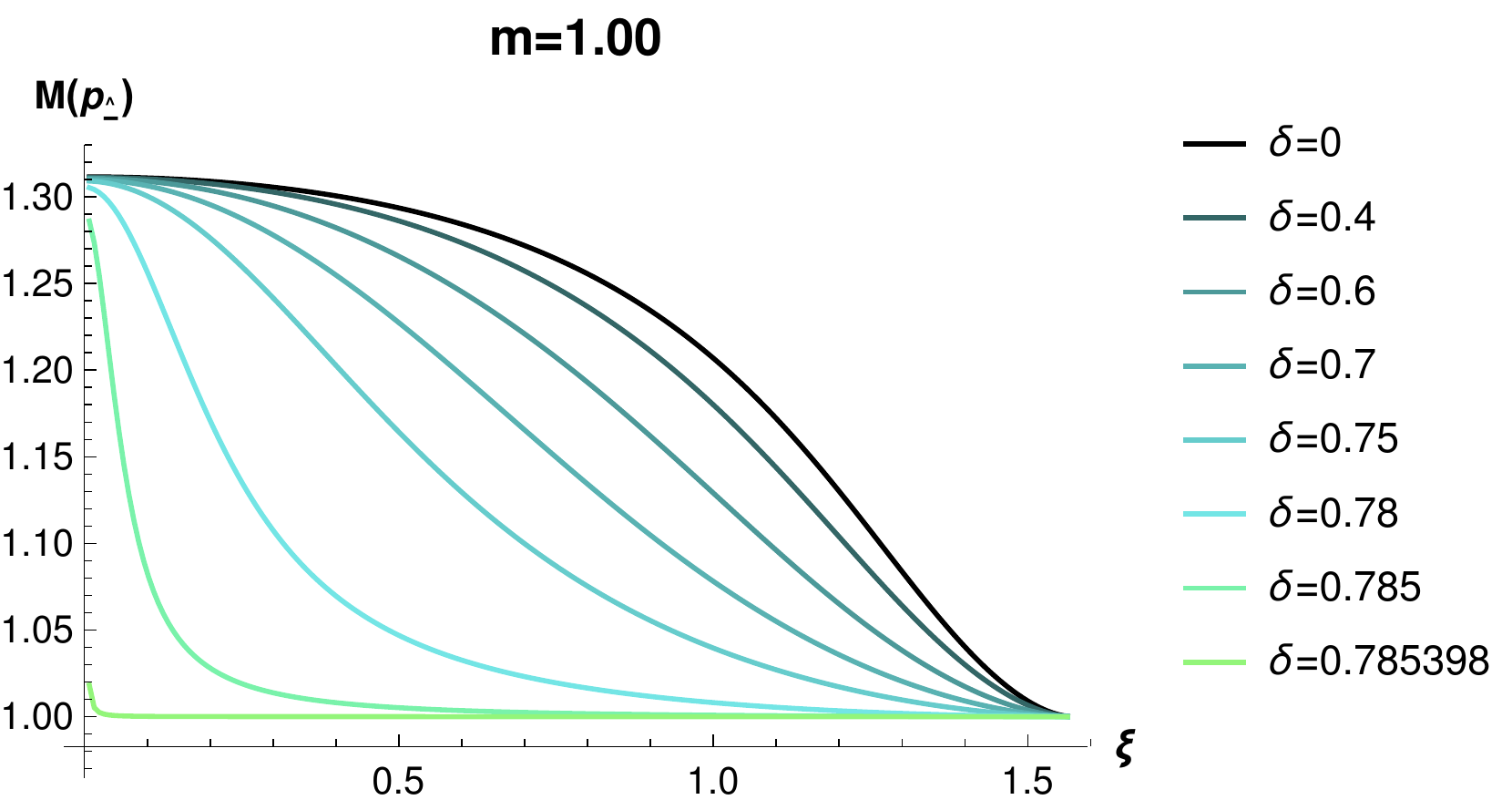}
		\label{fig:m100_constituentmass}
	}\\
	\subfloat[]{
		\includegraphics[width=1.0\linewidth]{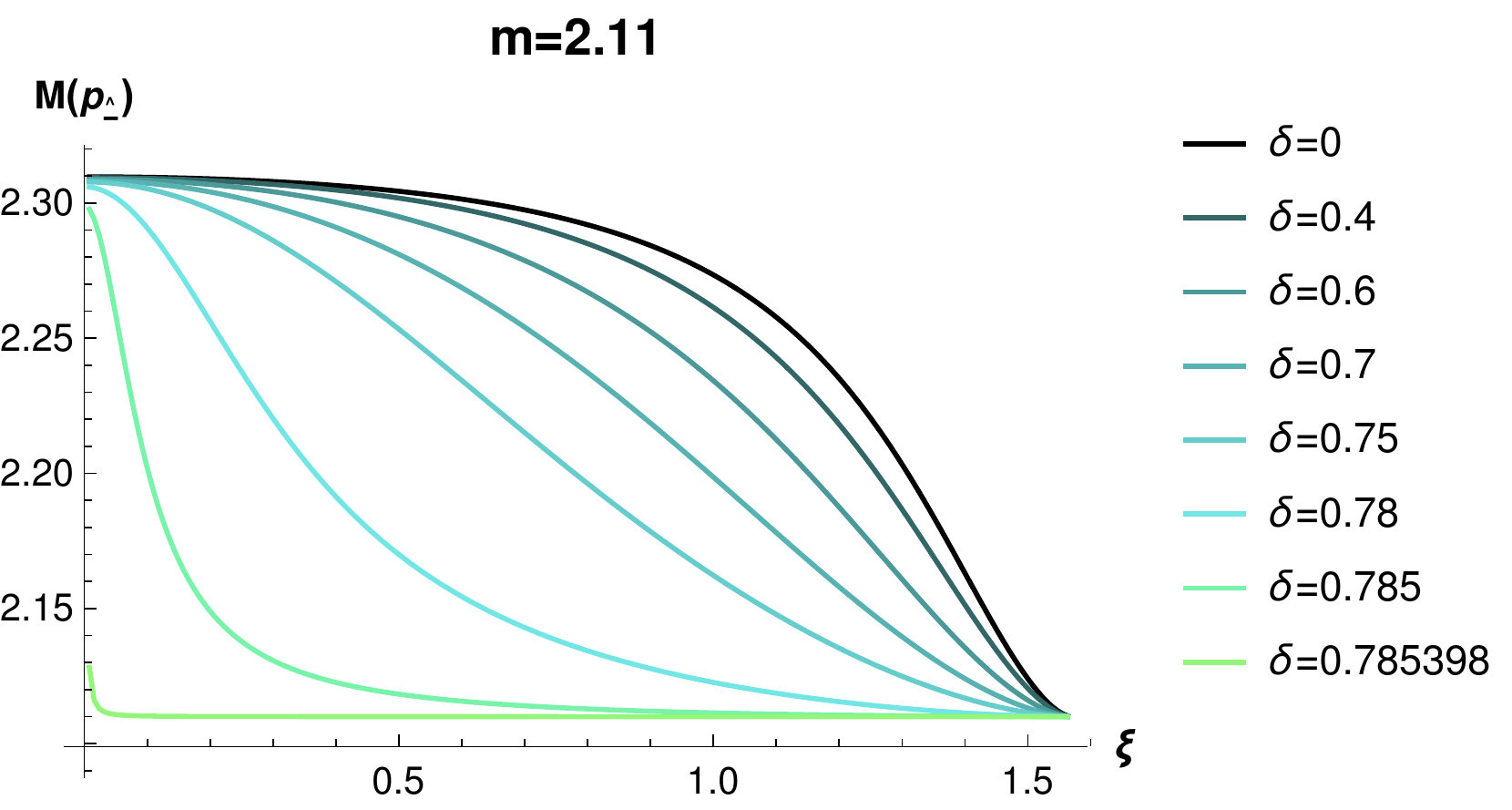}
		\label{fig:m211_constituentmass}
	}
	\caption{Constituent mass as a function of 
	$\xi=\tan^{-1} p_{\hat{-}}$
	for (a) $m=0.18$, (b) $m=1.00$ and (c) $m=2.11$. All quantities are in proper units of $ \sqrt{2\lambda} $.\label{fig:ms_constituentmass}}
\end{figure}
\begin{figure}
	\centering
	\subfloat[]{
		\includegraphics[width=1.0\linewidth]{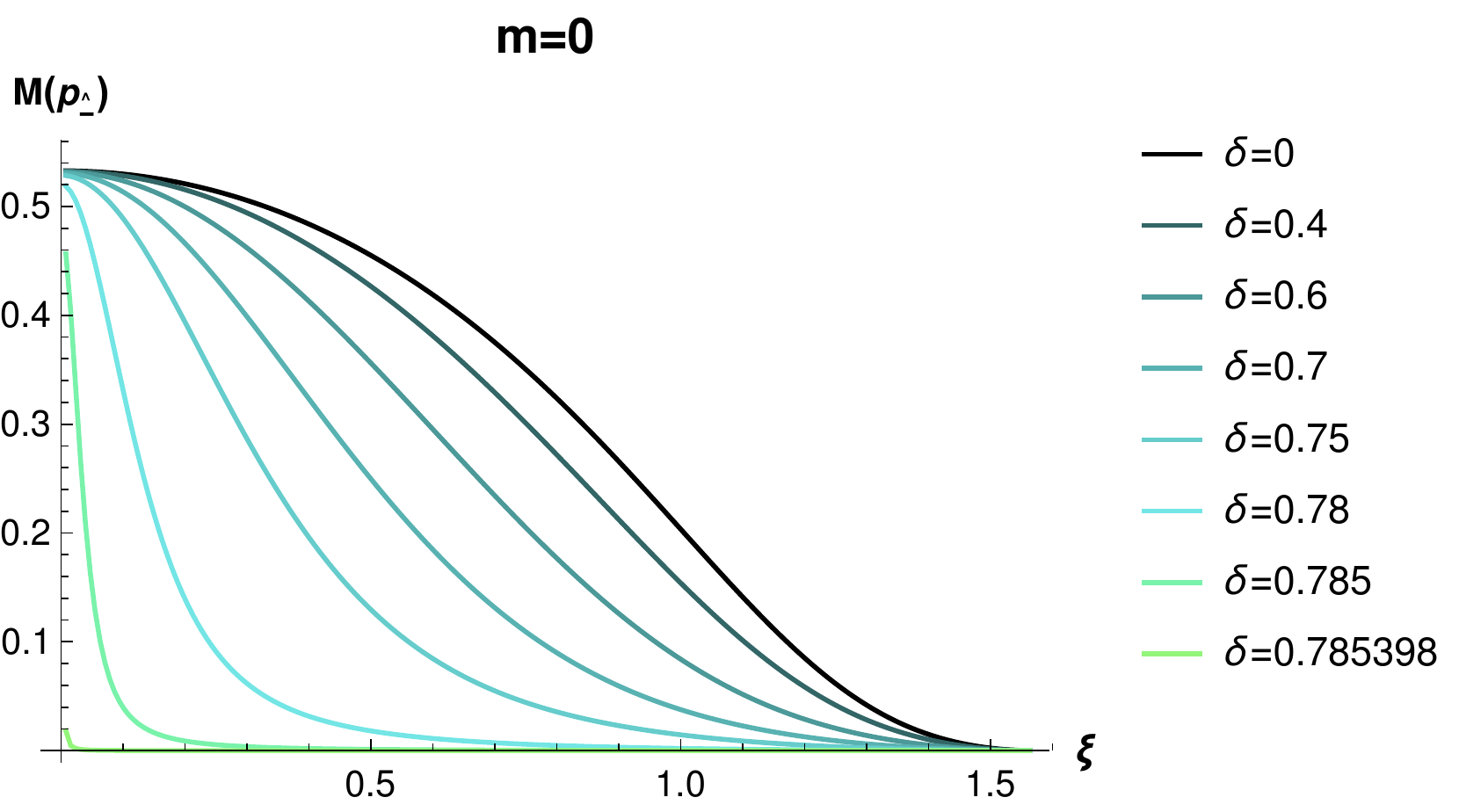}
		\label{fig:m0_constituentmass}
	}\\
	\centering
	\subfloat[]{
		\includegraphics[width=1.0\linewidth]{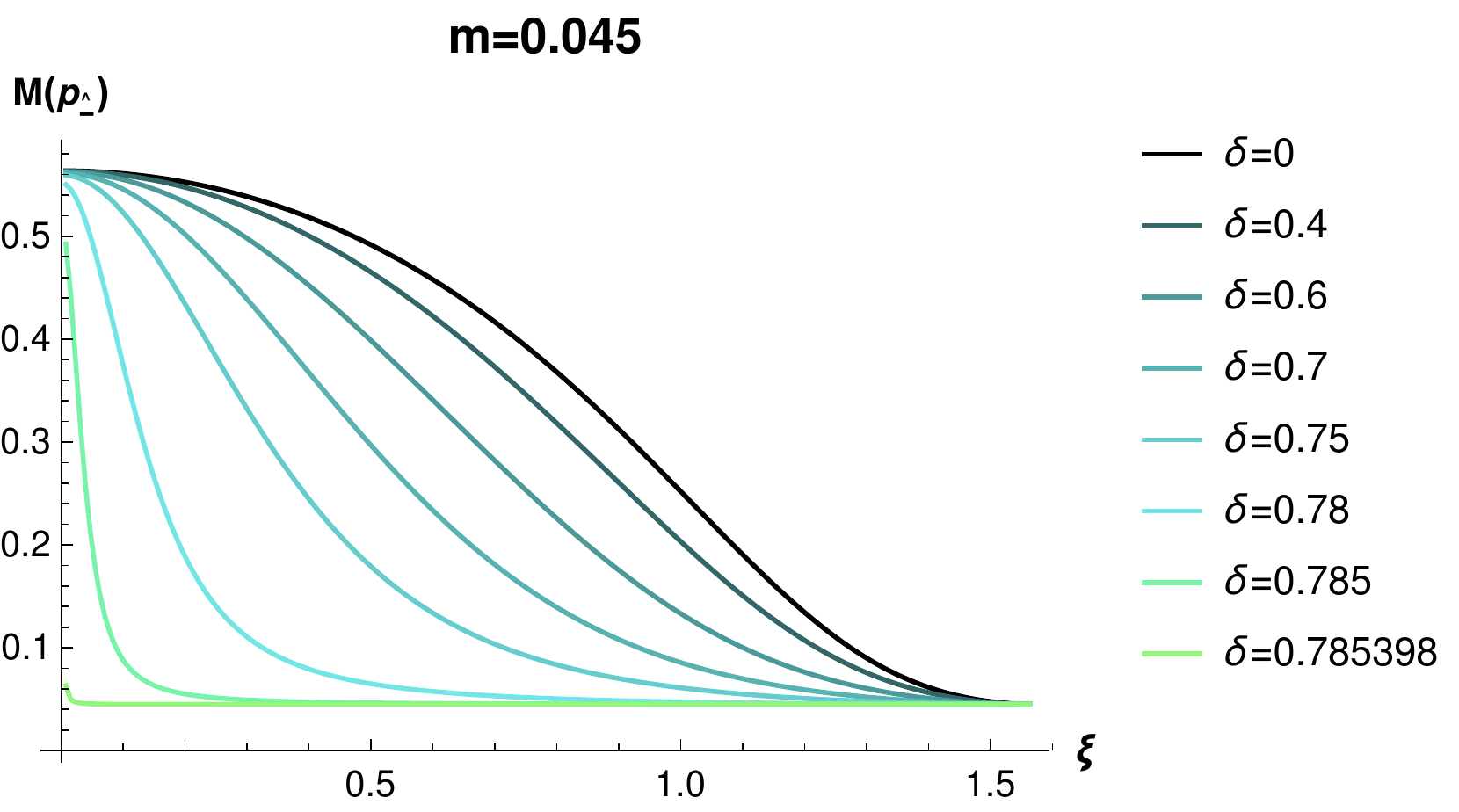}
		\label{fig:m0045_constituentmass}
	}\\
	\centering
	\subfloat[]{
		\includegraphics[width=1.0\linewidth]{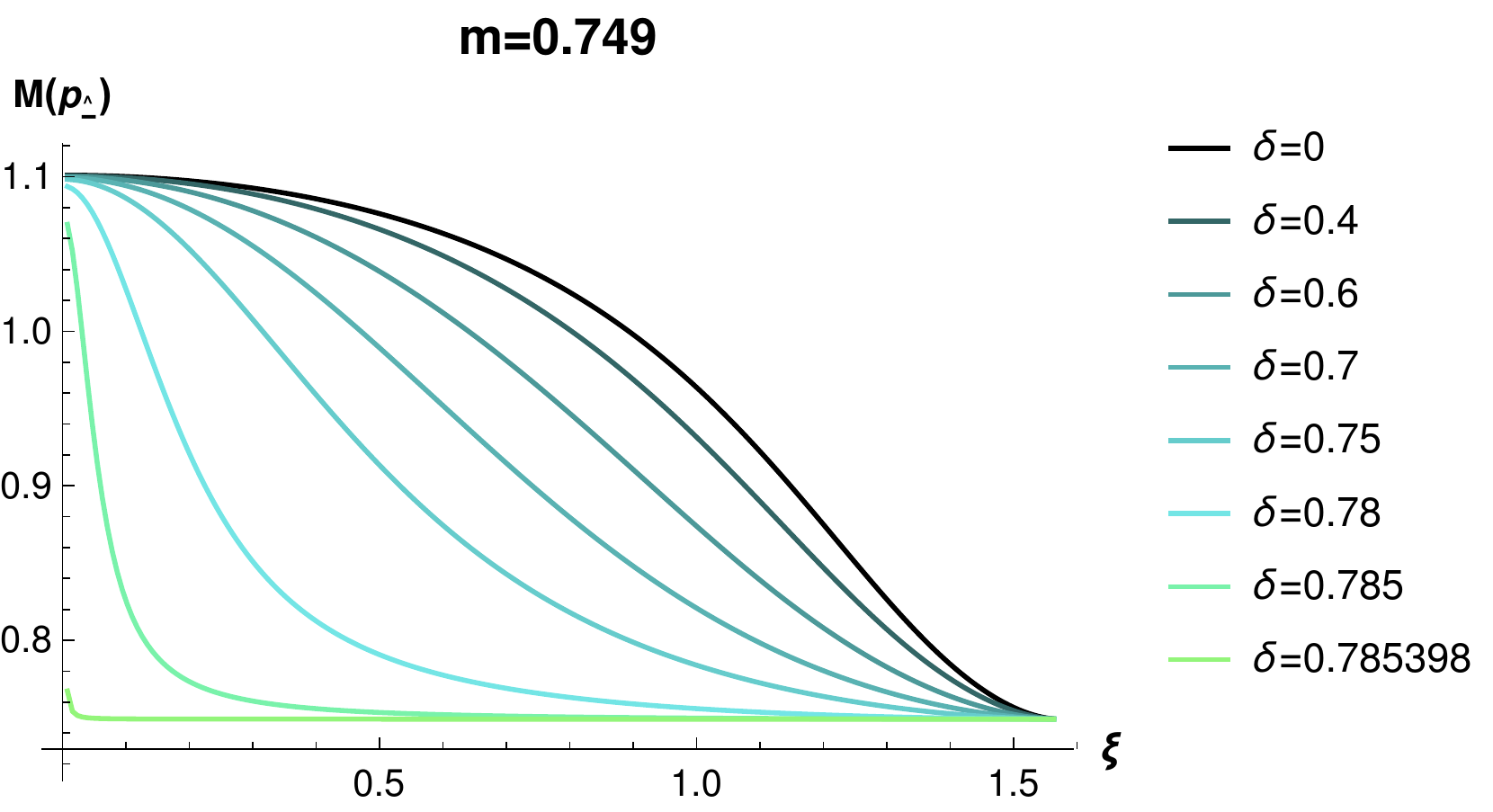}
		\label{fig:m0749_constituentmass}
	}\\
	\subfloat[]{
		\includegraphics[width=1.0\linewidth]{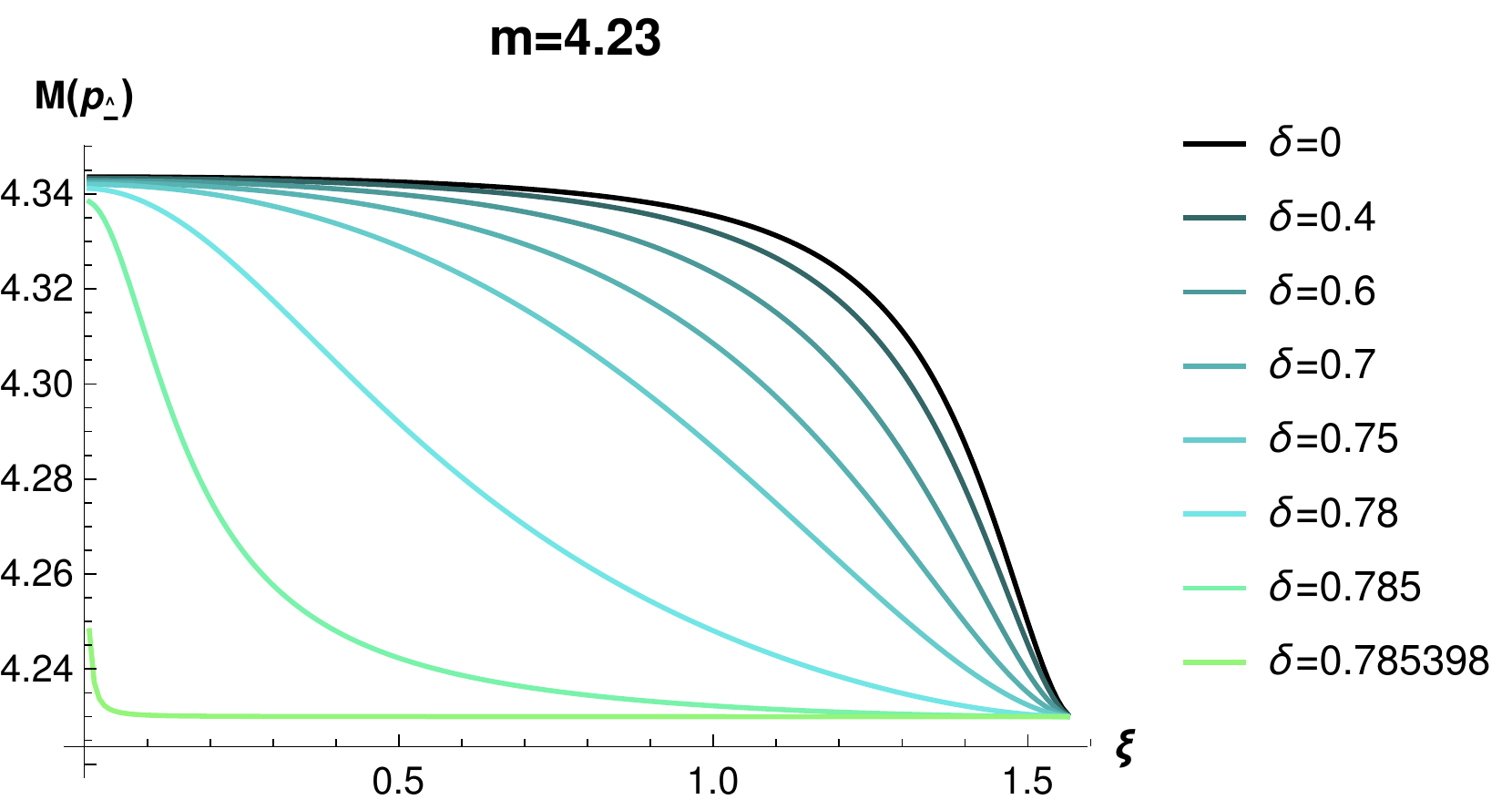}
		\label{fig:m423_constituentmass}
	}\\
	\caption{Constituent mass as a function of 
	$\xi=\tan^{-1} p_{\hat{-}}$ for (a) $m=0$, (b) $m=0.045$, (c) $m=0.749$ and (d) $m=4.23$. All quantities are in proper units of $ \sqrt{2\lambda} $.\label{fig:ms_constituentmass2}}
\end{figure}

To discuss the dressed fermion propagator in more physical terms~\cite{Mph}, one can express 
the dressed quark propagator given by Eq.(\ref{dressed_prop})
in terms of the mass function $M(p_{\hat{-}})$ and 
the wave function renormalization factor $F(p_{\hat{-}})$, i.e. 
\begin{equation}\label{eqn:Sponeway}
S(p)=\frac{F(p_{\hat{-}})}{\slashed{p}-M(p_{\hat{-}})},
\end{equation}
and identify $M(p_{\hat{-}})$ and $F(p_{\hat{-}})$ respectively as 
\begin{equation}\label{eqn:Mph}
M(p_{\hat{-}})=p_{\hat{-}}\frac{m+\sqrt{\mathbb{C}}A(p_{\hat{-}})}{p_{\hat{-}}+\mathbb{C}B(p_{\hat{-}})}=\frac{p_{\hat{-}}}{\sqrt{\mathbb{C}}}\cot\theta(p_{\hat{-}})
\end{equation}
and
\begin{equation}\label{eqn:Fofp}
F(p_{\hat{-}})=\left( 1+\frac{\mathbb{C}B(p_{\hat{-}})}{p_{\hat{-}}}\right)^{-1} =\frac{p_{\hat{-}}}{E(p_{\hat{-}})\sin\theta(p_{\hat{-}})}.
\end{equation}
We then numerically compute $M(p_{\hat{-}})$ and $F(p_{\hat{-}})$
using the mass gap solutions $\theta(p_{\hat{-}})$ and $E(p_{\hat{-}})$ obtained in Sec.~\ref{sec:sol}.

In Figs.~\ref{fig:ms_constituentmass} and \ref{fig:ms_constituentmass2}, the results of the mass function 
$M(p_{\hat{-}})$ are shown as a function of the variable $\xi = \tan^{-1}p_{\hat{-}}$ for the bare quark mass values in Ref.~\cite{Li} ($m=0.18, 1.00, 2.11$) and Ref.~\cite{mov} ($m=0, 0.045, 0.749, 4.23$), respectively.
As we can see in Figs.~\ref{fig:ms_constituentmass} and \ref{fig:ms_constituentmass2}, 
the mass function $M(p_{\hat{-}})$ approaches the respective bare quark mass value $m$
as $\xi \to \pi/2$ or $p_{\hat{-}}\to \infty$ while it gets to the respective characteristic mass value $M(0)$
at $\xi=0$ or $p_{\hat{-}}=0$ regardless of the interpolation angle $\delta$, although the profile of $M(p_{\hat{-}})$ in $0<\xi<\pi/2$ does depend on the value of $\delta$. The characteristic 
mass value $M(0)$ may be regarded as the dressed quark mass acquired from the dynamical 
self-energy interaction depicted in Fig.~\ref{fig:ds}. We note that the profile of $M(p_{\hat{-}})$
as $\delta \to \pi/4$ approaches to the shape of the step-function which drops from $M(0)$ 
to $m$ away from $p^+=0$. 

\begin{figure}
	\centering
	\subfloat[]{
		\includegraphics[width=1.0\linewidth]{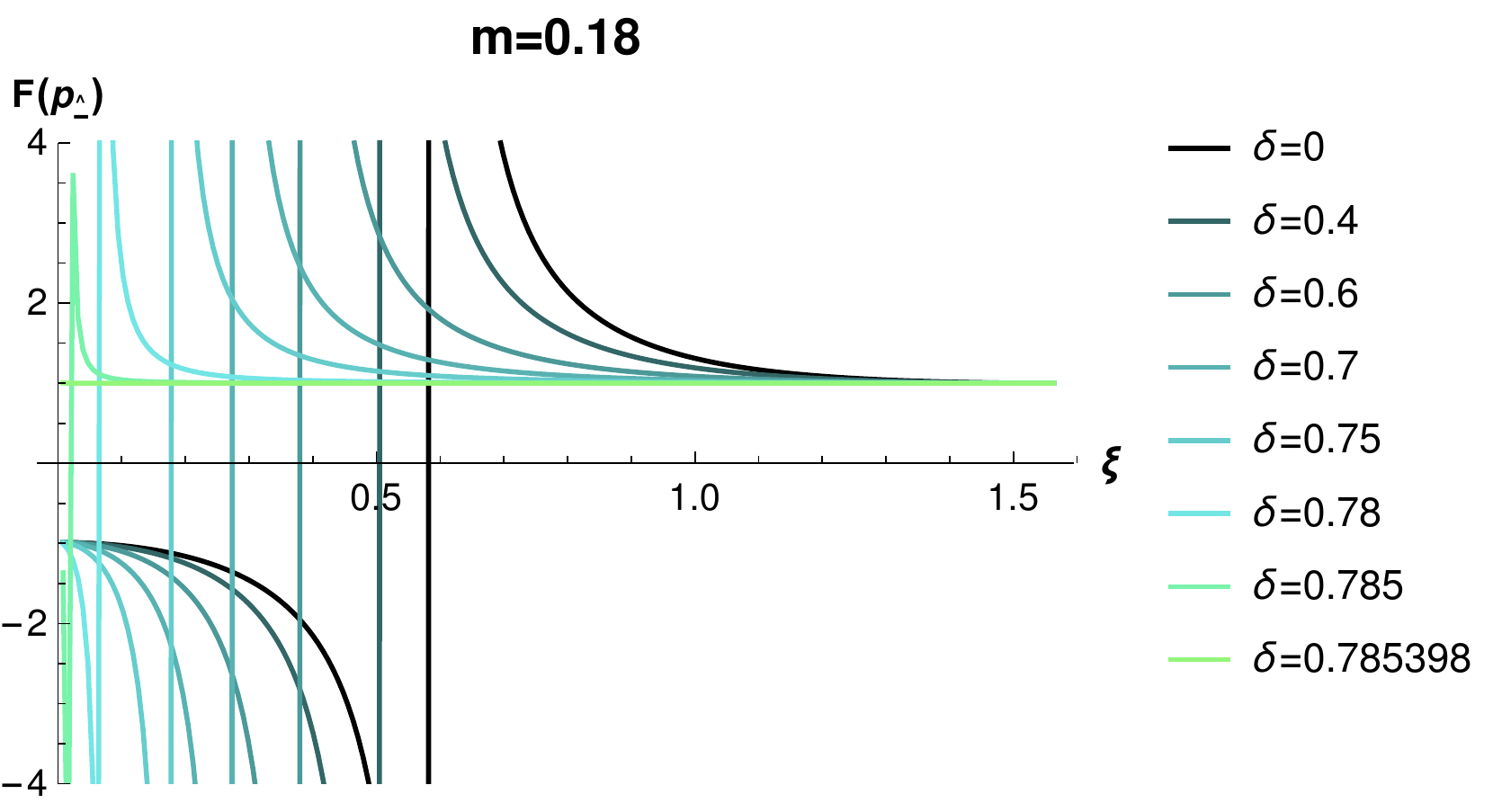}
		\label{fig:m018_Fofp}
	}\\
	\subfloat[]{
		\includegraphics[width=1.0\linewidth]{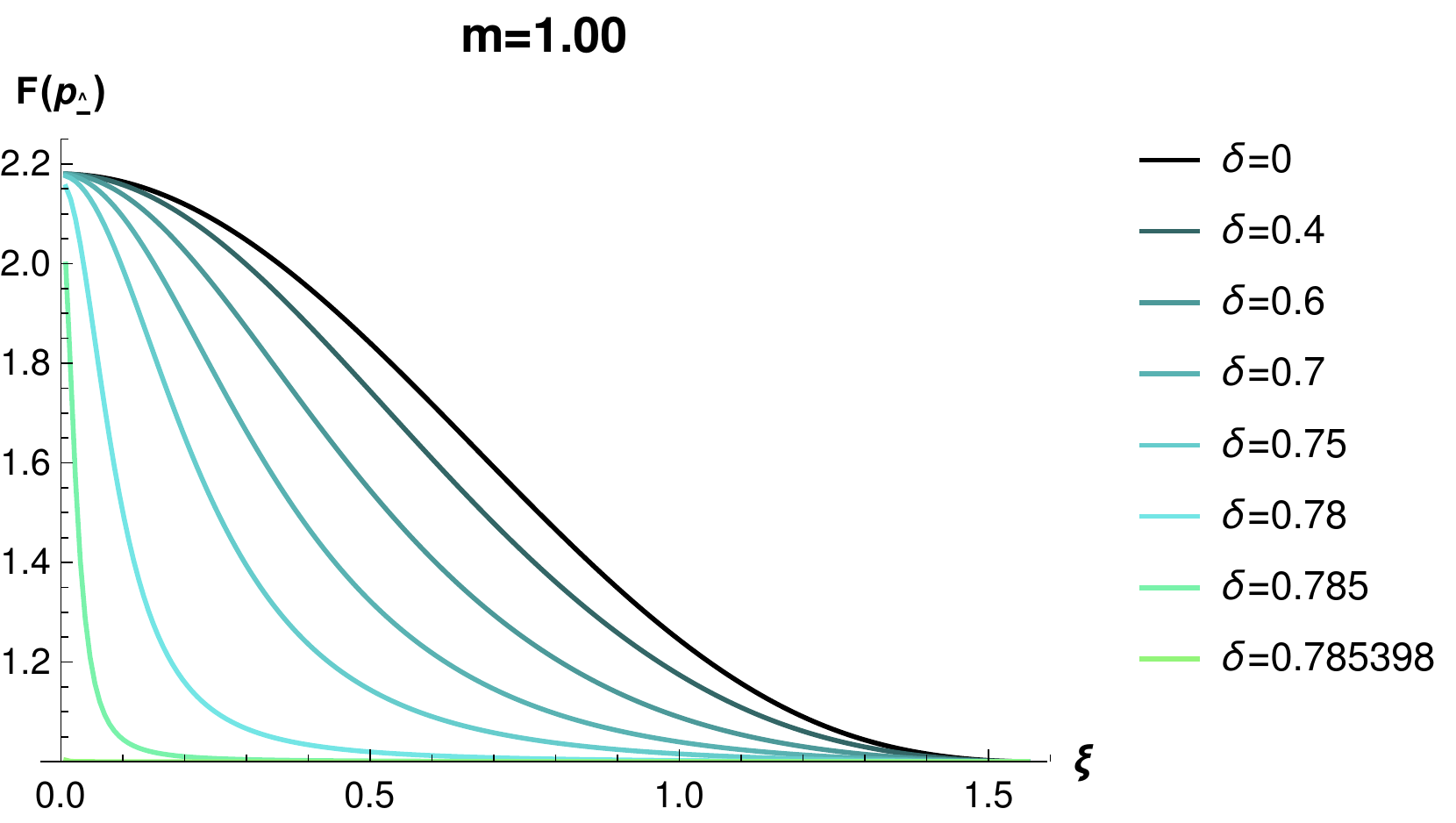}
		\label{fig:m100_Fofp}
	}\\
	\subfloat[]{
		\includegraphics[width=1.0\linewidth]{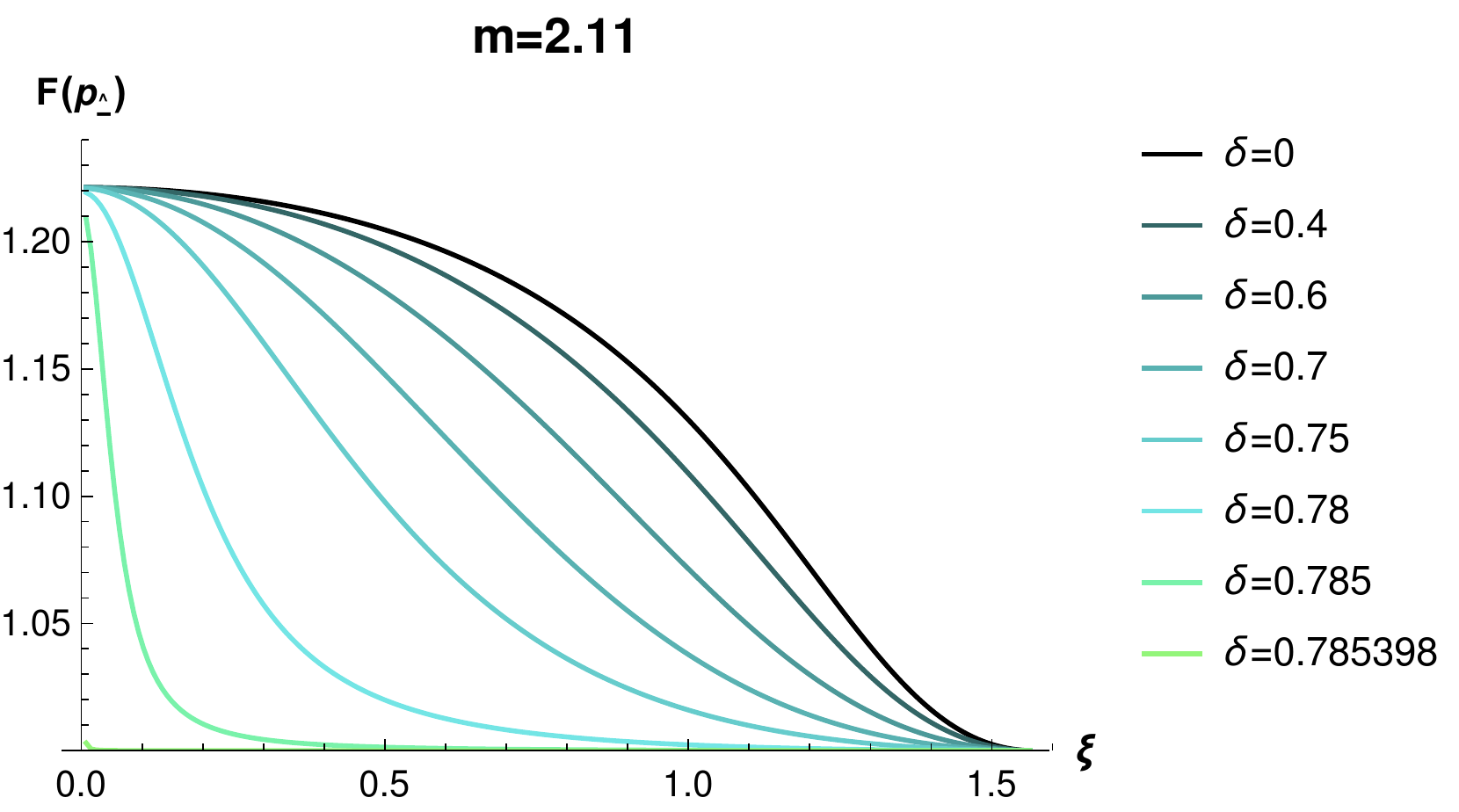}
		\label{fig:m211_Fofp}
	}
	\caption{Wavefunction renormalization as a function of 
	$\xi=\tan^{-1} p_{\hat{-}}$ for (a) $m=0.18$, (b) $m=1.0$ and (c)
		$m=2.11$. All quantities are in proper units of $ \sqrt{2\lambda} $.\label{fig:ms_Fofp}}
\end{figure}
\begin{figure}
	\centering
	\subfloat[]{
		\includegraphics[width=1.0\linewidth]{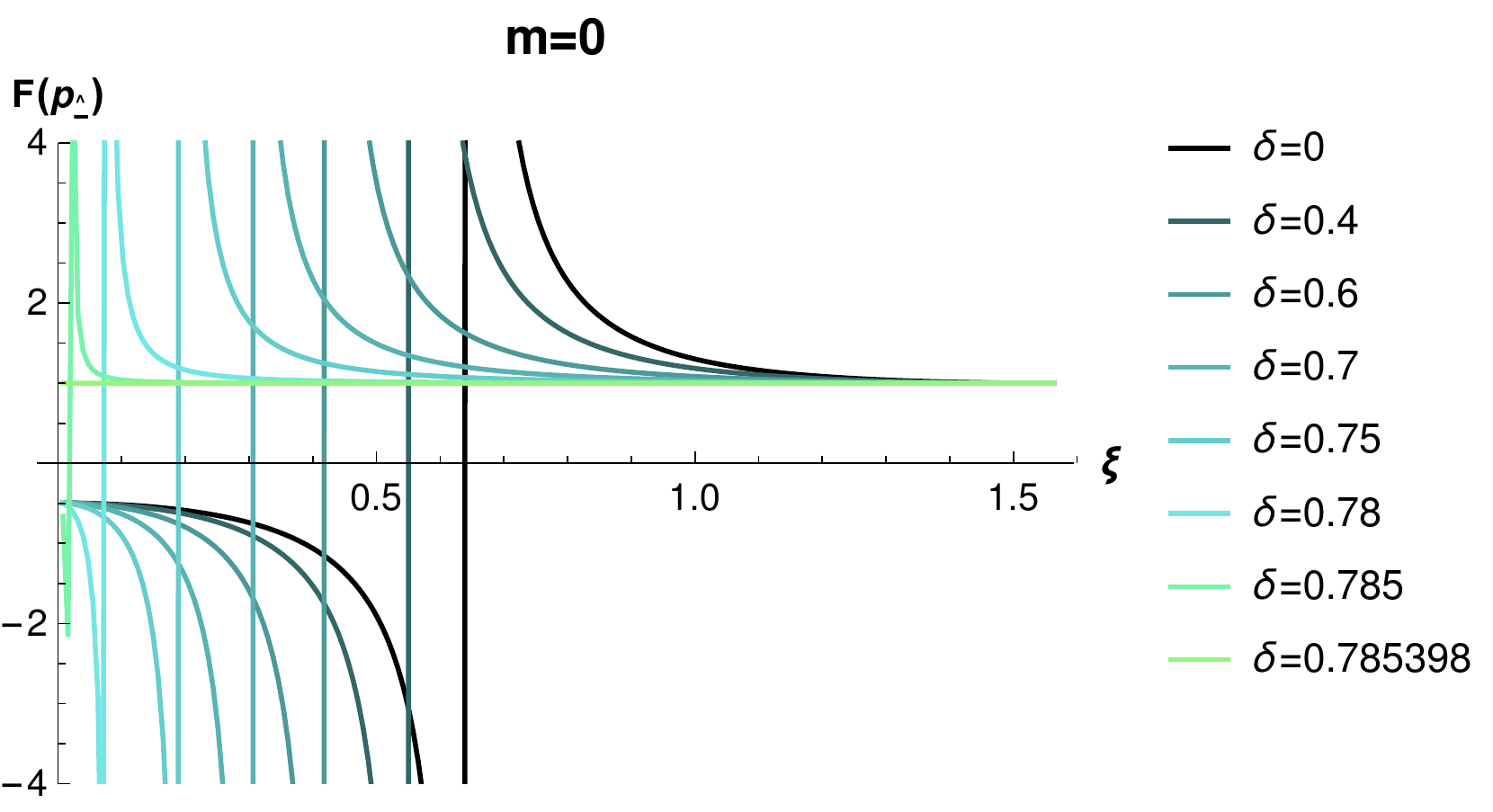}
		\label{fig:m0_Fofp}
	}\\
	\centering
	\subfloat[]{
		\includegraphics[width=1.0\linewidth]{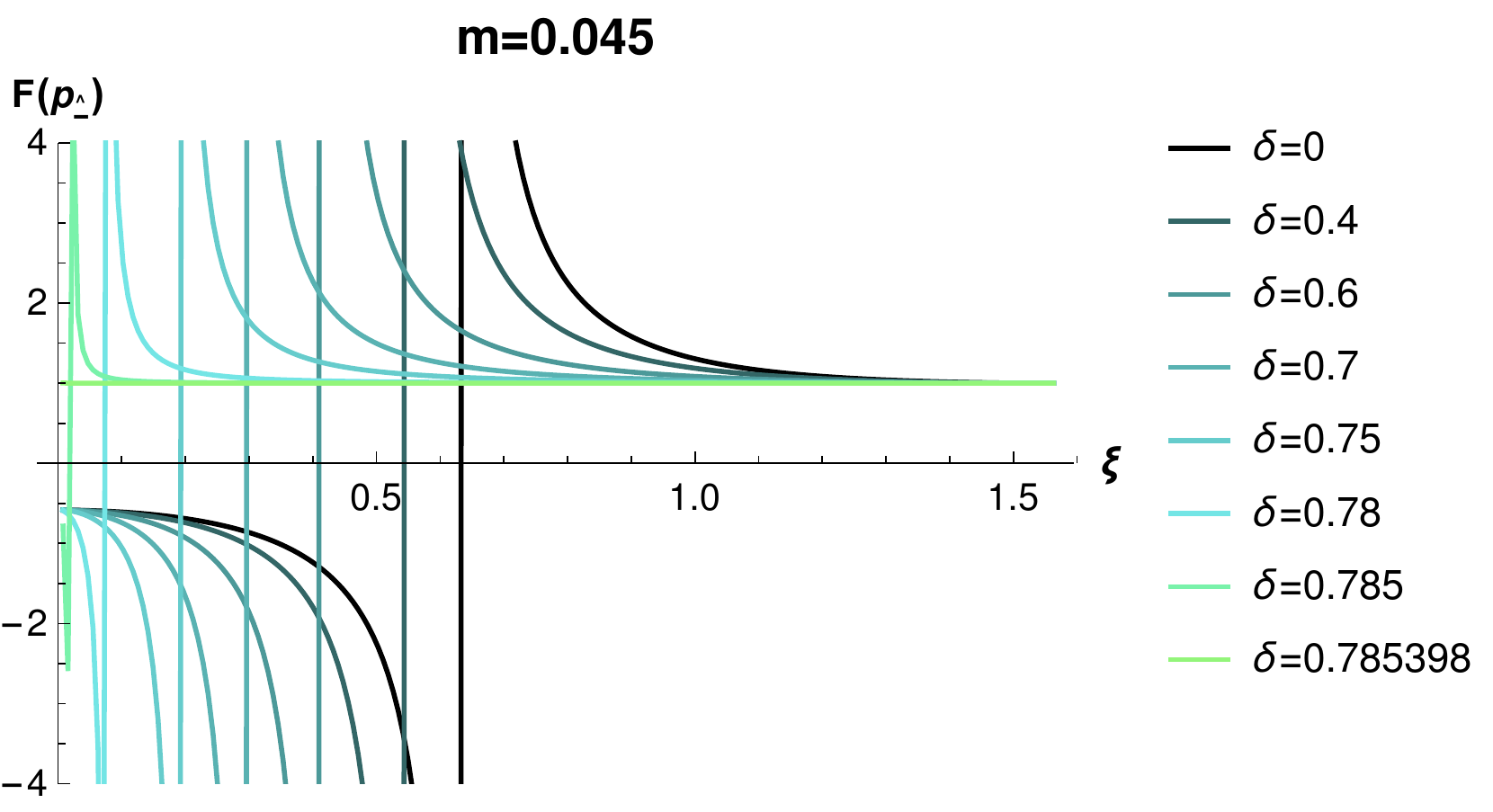}
		\label{fig:m0045_Fofp}
	}\\
	\subfloat[]{
		\includegraphics[width=1.0\linewidth]{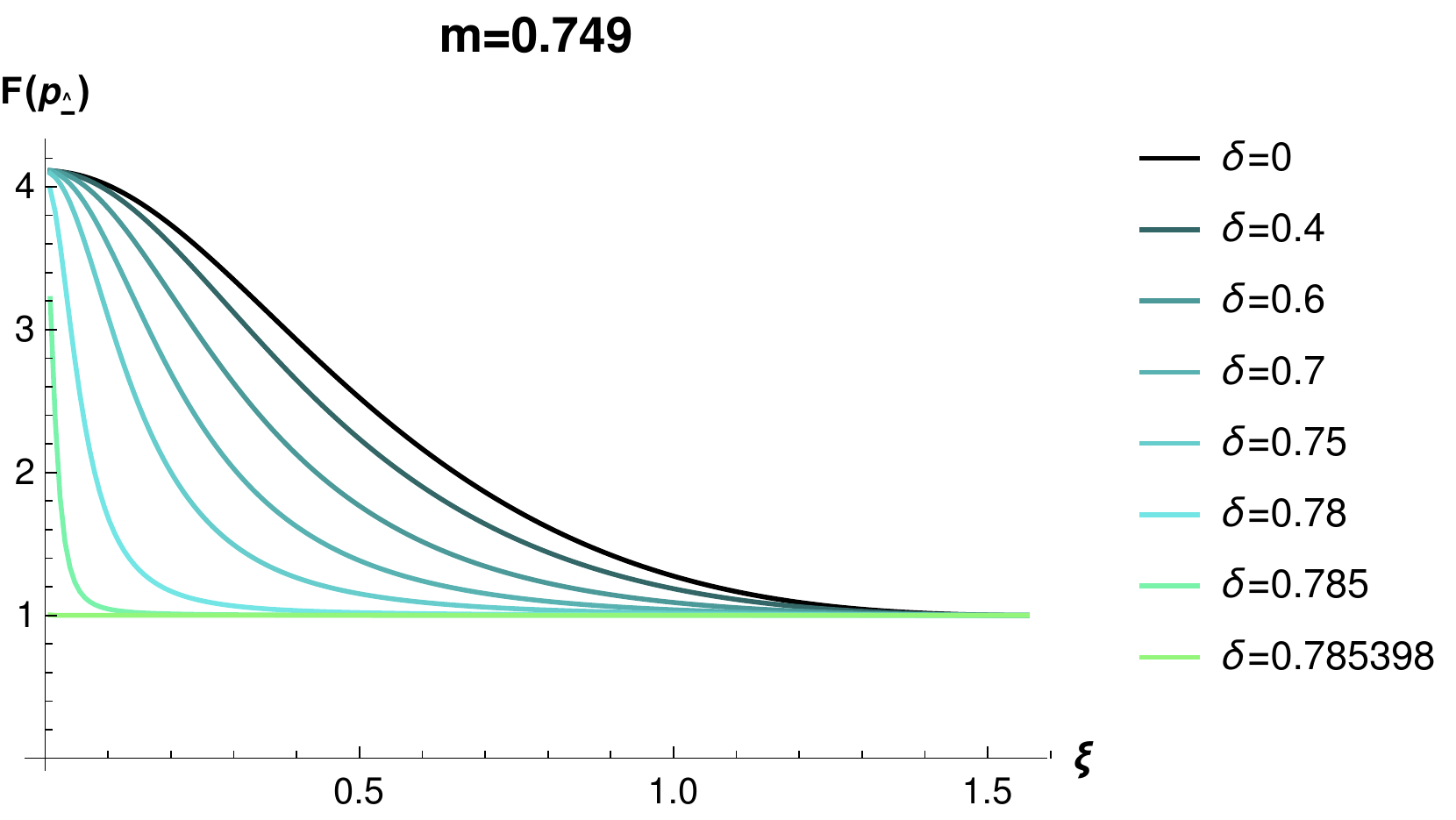}
		\label{fig:m0749_Fofp}
	}\\
	\subfloat[]{
		\includegraphics[width=1.0\linewidth]{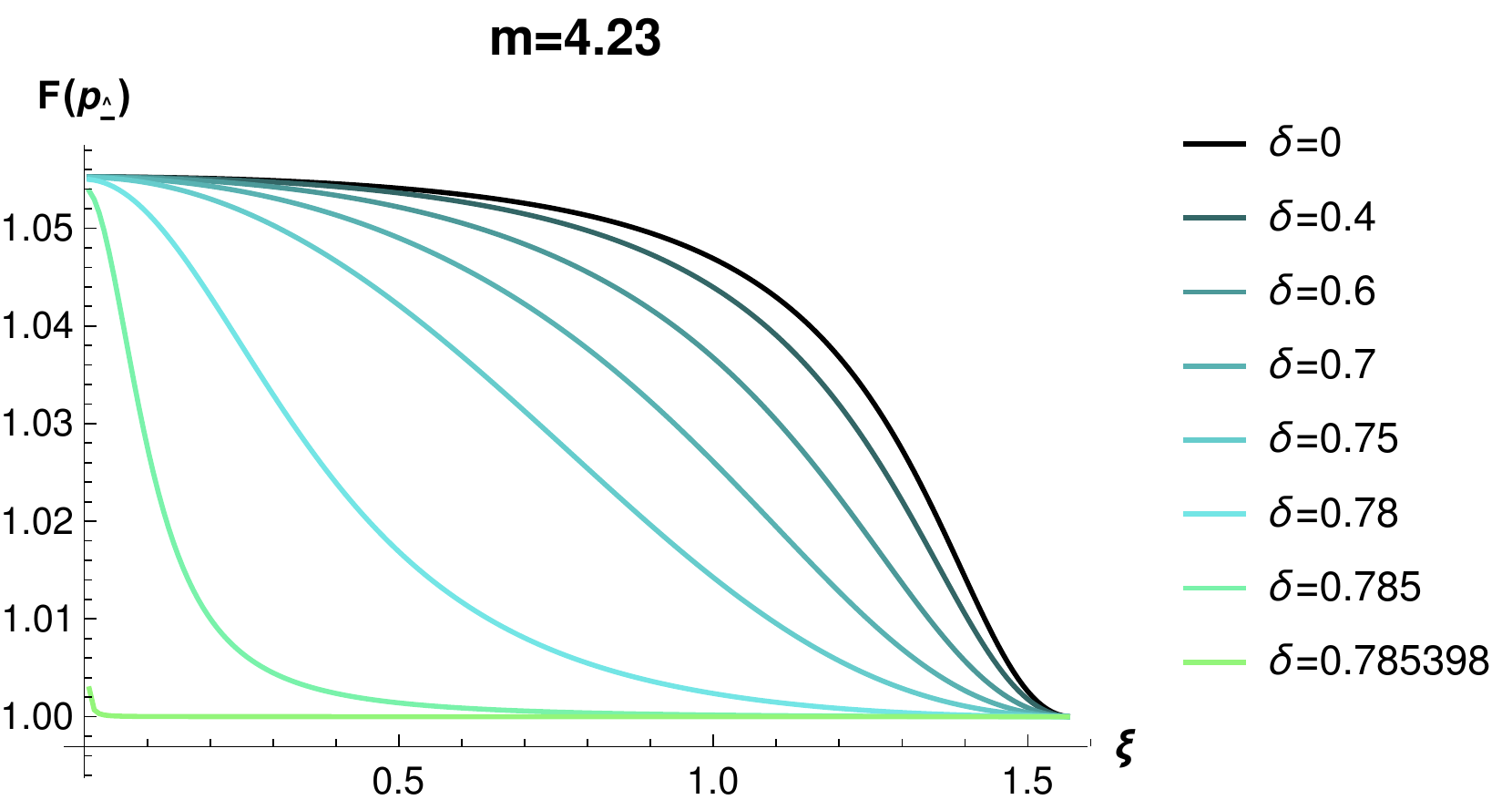}
		\label{fig:m423_Fofp}
	}
	\caption{Wavefunction renormalization as a function of 
	$\xi=\tan^{-1} p_{\hat{-}}$
	for (a) $m=0$, (b) $m=0.045$, (c) $m=0.749$ and (d)
		$m=4.23$. All quantities are in proper units of $ \sqrt{2\lambda} $.\label{fig:ms_Fofp2}}
\end{figure}

In Figs.~\ref{fig:ms_Fofp} and \ref{fig:ms_Fofp2}, the results of the wavefunction renormalization 
factor $F(p_{\hat{-}})$ are plotted with the same variable $\xi$ for the bare quark mass values in Ref.~\cite{Li} ($m=0.18, 1.00, 2.11$) and Ref.~\cite{mov} ($m=0, 0.045, 0.749, 4.23$), respectively. Rather immediately, we notice a dramatic difference in the results of $F(p_{\hat{-}})$ for the lower values of bare quark mass $m=0.18$ in Fig.~\ref{fig:ms_Fofp} as well as $m=0$ and 0.045 in Fig.~\ref{fig:ms_Fofp2} due to the negative values for small 
$ \xi = \tan^{-1} p_{\hat{-}}$ region. Interestingly, the appearance of the negative values 
in $F(p_{\hat{-}})$ for those bare quark mass values is correlated with the negative values
of $ E(p_{\hat{-}})/\sqrt{\mathbb{C}}$ discussed previously for  Figs.~\ref{fig:m018_basic_result_EC} ($m=0.18$), ~\ref{fig:m0_basic_result_EC} ($m=0$)
and ~\ref{fig:m0045_basic_result_EC} ($m=0.045$). To comprehend the sign correlation
between $F(p_{\hat{-}})$ and $E(p_{\hat{-}})$, we write $\cos\theta(p_{\hat{-}})$
and $\sin\theta(p_{\hat{-}})$ in terms of $M(p_{\hat{-}})$ and $F(p_{\hat{-}})$ 
from Eqs.(\ref{eqn:Mph}) and (\ref{eqn:Fofp}), as
\begin{equation}
\cos\theta(p_{\hat{-}})=\frac{\sqrt{\mathbb{C}} M(p_{\hat{-}})}{F(p_{\hat{-}})E(p_{\hat{-}})},
\end{equation}
and 
\begin{equation}
\label{sinp'}
\sin\theta(p_{\hat{-}})=\frac{p_{\hat{-}}}{F(p_{\hat{-}})E(p_{\hat{-}})},
\end{equation}
so that we may rewrite Eq.(\ref{energy-square}) associated with the triangle diagram 
shown in Fig.~\ref{fig:righttriangle} as 
\begin{equation}
\label{eqn:Fofp2}
F(p_{\hat{-}})E(p_{\hat{-}})= \sqrt{\mathbb{C} M(p_{\hat{-}})^2 + p_{\hat{-}}^2},
\end{equation}
or 
\begin{equation}
E(p_{\hat{-}})= \frac{\sqrt{\mathbb{C} M(p_{\hat{-}})^2 + p_{\hat{-}}^2}}{F(p_{\hat{-}})}.
\end{equation}
In contrast to Eq.(\ref{energy-square}), we can now express $E(p_{\hat{-}})$ itself without squaring it as $E(p_{\hat{-}})^2$ with the support from the wavefunction renormalization factor $F(p_{\hat{-}})$ as well as the mass function $M(p_{\hat{-}})$. This is rather remarkable because the issue of $E(p_{\hat{-}})$ not being always positive for $m \lesssim 0.56$, which was discussed 
in Sec.~\ref{sec:sol}, is now resolved by expressing the dressed quark propagator $S(p)$
in terms of $F(p_{\hat{-}})$ and $M(p_{\hat{-}})$ as given by Eq.(\ref{eqn:Sponeway}). 
While $E(p_{\hat{-}})$ can be negative, $F(p_{\hat{-}})E(p_{\hat{-}})$ is always positive
due to the sign correlation between $E(p_{\hat{-}})$ and $F(p_{\hat{-}})$ as one can 
see from Eq.(\ref{eqn:Fofp}) or equivalently from Eq.(\ref{sinp'}) due to the 
sign correlation between $\theta(p_{\hat{-}})$ and $p_{\hat{-}}$.
This allows us more physically transparent interpretation of the energy-momentum
dispersion relation for the interpolating dressed quark with its self-energy.
Moreover, using the rescaled variable $p'_{\hat{-}}=p_{\hat{-}}/\sqrt{\mathbb{C}}$ introduced in Eq.(\ref{con_rescale}) for the renormalized chiral condensate, we can 
assert the interpolation angle independence of the rescaled energy-momentum dispersion given by 
\begin{equation}
\label{reduced-energy-momentum-DR}
\frac{F(p'_{\hat{-}})E(p'_{\hat{-}})}{\sqrt{\mathbb{C}}}=
\sqrt{M(p'_{\hat{-}})^2 + {p'_{\hat{-}}}^2} \equiv {\tilde E}(p'_{\hat{-}}),
\end{equation}
where we define the interpolation angle independent energy function ${\tilde E}(p'_{\hat{-}})$
which extends the interpolating energy-momentum dispersion relation
of the on-mass-shell particle given by Eq.(\ref{eqn:energy_momentum}) 
to the case of the dressed quark with the rescaled variable, i.e. 
\begin{equation}
\label{eqn:energy_momentum_dressed}
{{\tilde E}(p'_{\hat{-}})}^2 ={p'_{\hat{-}}}^2 + M(p'_{\hat{-}})^2.
\end{equation}
As the solution of ${\tilde E}(p'_{\hat{-}})$ is always positive in contrast to 
\begin{figure}
	\centering
	\includegraphics[width=0.6\linewidth]{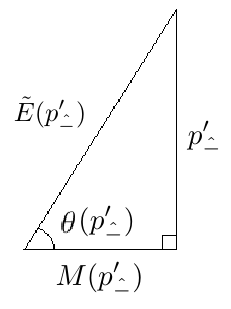}
	\caption{Geometrical representation of mass gap equations with the interpolation angle independent energy function ${\tilde E}(p'_{\hat{-}})$ defined in Eq.(\ref{reduced-energy-momentum-DR}).}
	\label{fig:righttriangle2}
\end{figure}
$E(p_{\hat{-}})$, we can now promote the mere pictorial device of     
geometric interpretation depicted in Fig.~\ref{fig:righttriangle} 
to the more physically meaningful geometric interpretation 
with ${\tilde E}(p'_{\hat{-}}), M(p'_{\hat{-}})$ and $p'_{\hat{-}}$
as shown in Fig.~\ref{fig:righttriangle2} with all the positive lengths of
the triangle sides.  
From Eqs.(\ref{eqn:energy_momentum}) and (\ref{eqn:energy_momentum_dressed}),
we also note the correspondence $m \leftrightarrow M(p'_{\hat{-}})$ and 
$p^{\hat{+}}/\sqrt{\mathbb{C}} \leftrightarrow {\tilde E}(p'_{\hat{-}})$ 
between the bare quark and the dressed quark. As an illustration of this correspondence,
we plot the profiles of ${\tilde E}$ as a function of $p'_{\hat{-}}$ for the two cases of $m=0$ and $m=0.18$ in Fig.~\ref{fig:m0018fepppplot}. It is evident that 
${\tilde E}(p'_{\hat{-}}) \to p^{\hat{+}}/\sqrt{\mathbb{C}}$ as $p'_{\hat{-}} \to \infty$,
which is consistent with the result that the mass function $M(p_{\hat{-}})$ approaches the bare quark mass value $m$ as $\xi \to \pi/2$ or $p_{\hat{-}}\to \infty$ 
(See Figs.~\ref{fig:ms_constituentmass} and \ref{fig:ms_constituentmass2}). 
As $p'_{\hat{-}} \to 0$, however, ${\tilde E}(p'_{\hat{-}})$ approaches the characteristic 
mass value $M(0)$ as shown in Fig.~\ref{fig:m0018fepppplot}. 
Indeed we note ${\tilde E}(0)=\frac{F(0)E(0)}{\sqrt{\mathbb{C}}}=M(0)$, confirming
the sign correlation between $E(p_{\hat{-}})$ and $F(p_{\hat{-}})$ mentioned earlier, 
i.e. the negativity of $E(p_{\hat{-}})$ for the small $p_{\hat{-}}$ region is compensated
by the corresponding negativity of $F(p_{\hat{-}})$  to yield the mass function $M(p_{\hat{-}})$
positive always for any kinematic region of $p_{\hat{-}}$. In Table~\ref{tab:M0F0}, 
\begin{table*}
	\caption{\label{tab:M0F0}The numerical values of $M(0)$ and $F(0)$ for several different quark mass values. All quantities are in proper units of $ \sqrt{2\lambda} $.}
	\begin{ruledtabular}
		\begin{tabular}{c|c|c|c|c|c|c|c}
			\centering
			$m$ &  $0$ & $ 0.045 $ & $0.18$ & $0.749$ & $1.00$ & $2.11$ & $4.23$\\ \hline
			$	M(0)$ & $0.532778$ & $0.563644 $ & $0.659112$ & $1.10105$ & $1.31167$ & $2.30969$ & $4.34358$\\ \hline
			$F(0)$ & $ -0.495173$ & $-0.584175$ & $-0.987673$ & $4.11079$ & $2.17976$ &$1.22134$ & $1.05526$ \\ 
		\end{tabular}
	\end{ruledtabular}
\end{table*}
the numerical values of $M(0)$ and $F(0)$ are tabulated for the quark mass values shown in 
Figs.~\ref{fig:ms_constituentmass} and \ref{fig:ms_constituentmass2} as well as 
in Figs.~\ref{fig:ms_Fofp} and \ref{fig:ms_Fofp2}.
As expected, $F(0)$ values are negative for the small bare quark mass values ($m\lesssim 0.56$) to compensate the corresponding negative values of $E(0)$ while $M(0)$ values are all positive.   
\begin{figure}
		\centering
	\subfloat[]{
		\includegraphics[width=1.0\linewidth]{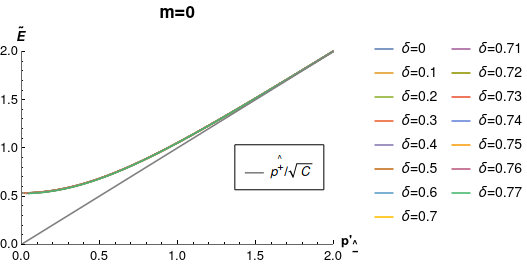}
		\label{fig:m0fepppplot}
	}\\
		\centering
\subfloat[]{
	\includegraphics[width=1.0\linewidth]{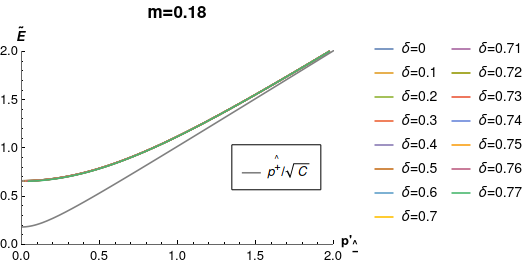}
	\label{fig:m018fepppplot}
}\\
	\caption{	\label{fig:m0018fepppplot}The profiles of $ {\tilde E}(p'_{\hat{-}})$ for several interpolation angles for two example small quark masses (a) $m=0$ and (b) $m=0.18$, in comparison with $p^{\hat{+}}/\sqrt{\mathbb{C}}$ for a free particle of the corresponding same masses, as a function of $p'_{\hat{-}}$. All quantities are in proper units of $ \sqrt{2\lambda} $.}
\end{figure}

Now, between the two limits (${p'}_{\hat -} \to 0$ and ${p'}_{\hat -} \to \infty$),
both ${\tilde E}(p'_{\hat{-}})$ and $M(p'_{\hat{-}})$ are running with the variable $p'_{\hat{-}}$
and their profiles of the $p'_{\hat{-}}$-dependence are completely independent of 
the interpolation angle $\delta$. The invariance of ${\tilde E}(p'_{\hat{-}})$ and $M(p'_{\hat{-}})$
under the interpolation between IFD and LFD indicates their universal nature as physically meaningful quantities. In this respect, one may call ${\tilde E}(p'_{\hat{-}})$ and $M(p'_{\hat{-}})$ as the effective energy and the constituent mass of the dressed quark moving with the scaled longitudinal momentum $p'_{\hat{-}}$. In principle, these physical quantities can be also computed in the well-known Euclidean approaches~\cite{TTWu} such as the lattice QCD~\cite{Lattice} and the manifestly covariant Dyson-Schwinger formulation. In the Euclidean formulation, which can in principle be applied to
the interpolation dynamics as far as $0 \le \delta < \pi/4$ except 
$\delta=\pi/4$ due to the light-like nature of LFD, the effective energy and the constituent quark mass would be given in terms of the Lorentz-invariant 
Euclidean variable ${\tilde P}^2 < 0$, i.e. ${\tilde E}({\tilde P}^2)$ and $M({\tilde P}^2)$.
In IFD, one may correspond the Wick rotated energy ${\tilde P}^0$ with the imaginary 
effective energy, i.e. ${\tilde P}^0= i {\tilde E({\tilde P}^2)}$ (purely imaginary value), and the longitudinal momentum ${\tilde P}^1$ with  $p^1$ of the dressed quark, so that ${\tilde P}^2 = ({\tilde P}^0)^2 + ({\tilde P}^1)^2 = - {\tilde E({\tilde P}^2)}^2 + ({p^1})^2 = -(p^1)^2-M({\tilde P}^2)^2+(p^1)^2=
- M({\tilde P}^2)^2 < 0 $.
In the same token, as ${p}^2 = \frac{({p}^{\hat +})^2 - ({p}_{\hat -})^2}{{\mathbb{C}}}$ generally in the interpolating dynamics, the Wick rotated interpolating energy ${\tilde P}^{\hat{+}}$
with the imaginary effective energy, i.e. 
${\tilde P}^{\hat{+}}/\sqrt{\mathbb{C}}= i {\tilde E({p'_{\hat{-}}}^2)}$ (purely imaginary value), and the longitudinal momentum ${\tilde P}_{\hat{-}}/\sqrt{\mathbb{C}}$ with  $p'_{\hat{-}}$ of the dressed quark, so that 
${\tilde P}^2 = \frac{({\tilde P}^{\hat{+}})^2
+ ({\tilde P}_{\hat{-}})^2}{\mathbb{C}}
= - {\tilde E}({p'_{\hat{-}}}^2)^2 + {p'_{\hat{-}}}^2 
= -{p'_{\hat{-}}}^2 - M(p'_{\hat{-}})^2 + {p'_{\hat{-}}}^2 
= - M(p'_{\hat{-}})^2 < 0 $.
Thus, it is natural to correspond the square of the rescaled longitudinal momentum ($p'_{\hat{-}}$) to the Euclidean variable ${\tilde P}^2$
for the space-like region ${\tilde P}^2 <0$ in the interpolating dynamics, i.e.
\begin{equation}\label{eqn:covpsquared}
{p'}_{\hat{-}}^2 \leftrightarrow - {\tilde P}^2 .
\end{equation}
This correspondence is supported not only from the relativistic form invariance of
${\tilde E}(p'_{\hat{-}})$ and $M(p'_{\hat{-}})$ but also from the matching condition 
between the Minkowsky space and the Euclidean space. 
Namely, when the real energy value in the Minkowsky space is converted into the purely imaginary value in the Euclidean space, the matching between the real value and the purely imaginary value occurs precisely 
where the value itself is zero. For instance, in IFD, the Wick rotation of the energy 
$p^0 \to {\tilde P}^0 = i p^0$ has the common energy value $p^0 = {\tilde P}^0 = 0$ so that the Lorentz-invariant momentum-squared value in the Minkowsky space can be matched with the corresponding Euclidean momentum-squared value, ${\tilde P}^2$, by taking simultaneously both the real value $p^0 =0$ in the Minkowsky space and the purely imaginary value ${\tilde P}^0 = i p^0 =0$. 
Likewise, in the interpolation of the relativistic dynamics for $0 \le \delta < \pi/4$, one can 
match the Minkowsky space and the Euclidean space   
by taking  
$p^{\hat{+}}/\sqrt{\mathbb{C}} 
= i p^{\hat{+}}/\sqrt{\mathbb{C}} 
={\tilde P}^{\hat +}/\sqrt{\mathbb{C}} = 0$, 
confirming the correspondence given by
Eq.(\ref{eqn:covpsquared}).

\section{\label{sec:bound}The bound-state equation}
\begin{figure}
	\centering
	\includegraphics[width=1.0\linewidth]{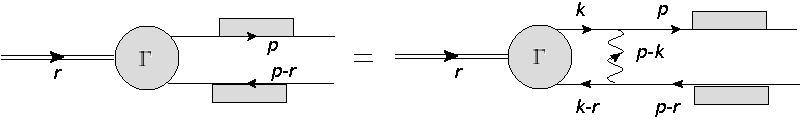}
	\caption{Diagrammatic representation of the Bethe-Salpeter Equation.}
	\label{fig:BSE}
\end{figure}

Having solved the mass gap equation and obtained the dressed quark propagator
interpolating between IFD and LFD, we now derive the quark-antiquark bound-state equation 
in the interpolating form. While we consider the quark-antiquark bound-state with 
the equal bare mass $m$ in this work, one may generalize it rather straightforwardly 
to the unequal quark and antiquark mass cases. Bound-states for the unequal quark and antiquark mass cases were analyzed in Refs.~\cite{tHooft,Brower}, and 
the quark-hadron duality, analytical heavy quark expansion and chiral symmetry breaking effects in the heavy-light mesons have been discussed respectively in Refs.~\cite{GL,BU,GSSW}.
 Denoting the covariant Bethe-Salpeter
amplitude $\Gamma(r,p)$ of the two-body bound-state with the two-momentum 
$r^{\hat\mu}$
of the bound-state and the two-momentum $p^{\hat\mu}$ of one of the two constituents 
depicted in Fig.~\ref{fig:BSE}, we first write down the Bethe-Salpeter equation following 
the Feynman rules of the gluon propagator and the dressed 
quark propagator in the 't Hooft model 
\begin{equation}
\Gamma(r,p)=\frac{i\lambda}{2\pi}\dashint\frac{dk_{\hat{-}}dk_{\hat{+}}}{(p_{\hat{-}}-k_{\hat{-}})^2}S(p)\gamma^{\hat{+}} \Gamma(r,k)\gamma^{\hat{+}} S(p-r) . \label{eqn:BSE}
\end{equation}

To project the covariant Bethe-Salpeter amplitude $\Gamma(r,p)$ into the interpolating equal-time
wavefunction, we first integrate out the interpolating energy $p_{\hat{+}}$ and 
define the wave function $\phi(r_{\hat{-}},p_{\hat{-}})$ as
$\phi(r_{\hat{-}},p_{\hat{-}})=\int dp_{\hat{+}} \Gamma(r,p)$ with the on-mass-shell condition
to fix the external interpolating energy $r_{\hat{+}}$.
Then, Eq.(\ref{eqn:BSE}) is converted to the equation for $\phi(r_{\hat{-}},p_{\hat{-}})$
\begin{widetext} 
\begin{align}
\phi(r_{\hat{-}},p_{\hat{-}})&=\frac{i\lambda}{2\pi}\dashint\frac{dk_{\hat{-}}}{(p_{\hat{-}}-k_{\hat{-}})^2}\int dp_{\hat{+}}S(p)\gamma^{\hat{+}} \phi(r_{\hat{-}},k_{\hat{-}})\gamma^{\hat{+}} S(p-r).\label{inter_BSE}
\end{align}
Now, we also write the fermion propagator $S(p)$ given by Eq.~(\ref{Sdressed}) as 
\begin{equation}\label{Sp_new}
S(p)=\frac{\tilde{T}(p_{\hat{-}})\Lambda^+\tilde{T}^{\dagger}(p_{\hat{-}})\gamma^0}{p_{\hat{+}}-E_u(p_{\hat{-}})+i\epsilon}+\frac{\tilde{T}(p_{\hat{-}})\Lambda^-\tilde{T}^{\dagger}(p_{\hat{-}})\gamma^0}{p_{\hat{+}}-E_v(p_{\hat{-}})-i\epsilon} ,
\end{equation}
using Eqs.(\ref{eqn:Bogo_u}) and (\ref{eqn:TpItp}) to rewrite $u(p_{\hat{-}})\bar{u}(p_{\hat{-}})$
and $v(-p_{\hat{-}})\bar{v}(-p_{\hat{-}})$ as 
\begin{align}
u(p_{\hat{-}})\bar{u}(p_{\hat{-}})=T(p_{\hat{-}})u^{(0)}(0)u^{(0)\dagger}(0)T^{\dagger}(p_{\hat{-}})\gamma^0\label{uubar_inter}
=\tilde{T}(p_{\hat{-}})\Lambda^+\tilde{T}^{\dagger}(p_{\hat{-}})\gamma^0
\end{align}
and
\begin{equation}
v(-p_{\hat{-}})\bar{v}(-p_{\hat{-}})=T(p_{\hat{-}})v^{(0)}(p_{\hat{-}}0)v^{(0)\dagger}(p_{\hat{-}}0)T^{\dagger}(p_{\hat{-}})\gamma^0\label{vvbar_inter}
=\tilde{T}(p_{\hat{-}})\Lambda^-\tilde{T}^{\dagger}(p_{\hat{-}})\gamma^0,
\end{equation}
where we define 
\begin{equation}
\Lambda^+\equiv\frac{u^{(0)}(0)u^{(0)\dagger}(0)}{2\sqrt{\mathbb{C}}m}=\left( \begin{array}{cc}
\frac{1}{2(\cos\delta-\sin\delta)} & \frac{1}{2\sqrt{\mathbb{C}}}\\
\frac{1}{2\sqrt{\mathbb{C}}} &\frac{1}{2(\cos\delta+\sin\delta)}
\end{array}\right)
\end{equation}
\begin{equation}
\Lambda^-\equiv\frac{v^{(0)}(0)v^{(0)\dagger}(0)}{2\sqrt{\mathbb{C}}m}=\left( \begin{array}{cc}
\frac{1}{2(\cos\delta-\sin\delta)} & -\frac{1}{2\sqrt{\mathbb{C}}}\\
-\frac{1}{2\sqrt{\mathbb{C}}} &\frac{1}{2(\cos\delta+\sin\delta)}
\end{array}\right)
\end{equation}
and
\begin{equation}\label{Ttilde}
\tilde{T}(p_{\hat{-}})=\exp[-\frac{1}{2}\theta(p_{\hat{-}})\frac{\gamma_{\hat{-}}}{\sqrt{\mathbb{C}}} ].
\end{equation}
Substituting then Eq.(\ref{Sp_new}) to Eq.(\ref{inter_BSE})
and performing the $ p_{\hat{+}} $ integration, we get
\begin{align}
\phi(r_{\hat{-}},p_{\hat{-}})&=\lambda\dashint\frac{dk_{\hat{-}}}{(p_{\hat{-}}-k_{\hat{-}})^2}\notag\\
&\times\left[ \frac{\tilde{T}(p_{\hat{-}})\Lambda^+\tilde{T}^{\dagger}(p_{\hat{-}})\gamma^0\gamma^{\hat{+}}\phi(r_{\hat{-}},k_{\hat{-}})\gamma^{\hat{+}}\tilde{T}(p_{\hat{-}}-r_{\hat{-}})\Lambda^-\tilde{T}^{\dagger}(p_{\hat{-}}-r_{\hat{-}})\gamma^0}{-r_{\hat{+}}+E_u(p_{\hat{-}})-E_v(p_{\hat{-}}-r_{\hat{-}})}\right.\notag\\
&+\left.\frac{\tilde{T}(p_{\hat{-}})\Lambda^-\tilde{T}^{\dagger}(p_{\hat{-}})\gamma^0\gamma^{\hat{+}}\phi(r_{\hat{-}},k_{\hat{-}})\gamma^{\hat{+}}\tilde{T}(p_{\hat{-}}-r_{\hat{-}})\Lambda^+\tilde{T}^{\dagger}(p_{\hat{-}}-r_{\hat{-}})\gamma^0}{r_{\hat{+}}+E_u(p_{\hat{-}}-r_{\hat{-}})-E_v(p_{\hat{-}})}\right]. \end{align}
Here, we note $\tilde{T}(-p_{\hat{-}}) \tilde{T}(p_{\hat{-}}) = I$ and further define
$\tilde{\phi}(r_{\hat{-}},p_{\hat{-}})=\tilde{T}(-p_{\hat{-}})\phi(r_{\hat{-}},p_{\hat{-}})\tilde{T}(r_{\hat{-}}-p_{\hat{-}})$,
to plug in 
$\phi(r_{\hat{-}},k_{\hat{-}})=\tilde{T}(k_{\hat{-}})\tilde{\phi}(r_{\hat{-}},k_{\hat{-}})\tilde{T}(k_{\hat{-}}-r_{\hat{-}})$
and obtain
\begin{align}
\tilde{\phi}(r_{\hat{-}},p_{\hat{-}})=\lambda\dashint\frac{dk_{\hat{-}}}{(p_{\hat{-}}-k_{\hat{-}})^2}\left[
\frac{\Lambda^+\tilde{T}^{\dagger}(p_{\hat{-}})\gamma^0\gamma^{\hat{+}}\tilde{T}(k_{\hat{-}})\tilde{\phi}(r_{\hat{-}},k_{\hat{-}})\tilde{T}(k_{\hat{-}}-r_{\hat{-}})\gamma^{\hat{+}}\tilde{T}(p_{\hat{-}}-r_{\hat{-}})\Lambda^-\gamma^0}{-r_{\hat{+}}+E_u(p_{\hat{-}})-E_v(p_{\hat{-}}-r_{\hat{-}})}\right.\notag\\
+ \left.\frac{\Lambda^-\tilde{T}^{\dagger}(p_{\hat{-}})\gamma^0\gamma^{\hat{+}}\tilde{T}(k_{\hat{-}})\tilde{\phi}(r_{\hat{-}},k_{\hat{-}})\tilde{T}(k_{\hat{-}}-r_{\hat{-}})\gamma^{\hat{+}}\tilde{T}(p_{\hat{-}}-r_{\hat{-}})\Lambda^+\gamma^0}{r_{\hat{+}}+E_u(p_{\hat{-}}-r_{\hat{-}})-E_v(p_{\hat{-}})}
\right] \label{phi_tilde_inter}.
\end{align}

Examining the general structure of $\tilde{\phi}$ in Eq.(\ref{phi_tilde_inter}),
we realize that it can be split into the forward moving part ${\hat\phi}_{+}$ and the backward 
moving part ${\hat\phi}_{-}$ with the two $2\times2$ matrices ${\hat M}^{+}$ and ${\hat M}^{-}$, i.e.  
$\tilde{\phi}={\hat\phi}_{+} {\hat M}^{+} + {\hat\phi}_{-} {\hat M}^{-}$, 
where
\begin{align}
{\hat M}^{+}&=\frac{\gamma^5 \sqrt{\mathbb{C}} + \gamma_{\hat{-}}}{2}= 
\left( \begin{array}{cc}
-\frac{\sqrt{\mathbb{C}}}{2} & \frac{\cos\delta+\sin\delta}{2}\\
-\frac{\cos\delta-\sin\delta}{2} & \frac{ \sqrt\mathbb{C}}{2}
\end{array}\right) ,\\ {\hat M}^{-}&=\frac{\gamma^5\sqrt{\mathbb{C}} - \gamma_{\hat{-}}}{2} =  \left( \begin{array}{cc}
-\frac{ \sqrt\mathbb{C}}{2} & -\frac{\cos\delta+\sin\delta}{2}\\
\frac{\cos\delta-\sin\delta}{2} & \frac{ \sqrt\mathbb{C}}{2}
\end{array}\right).
\end{align}
We note here that ${\hat M}^{+}$ and ${\hat M}^{-}$ coincide with the two $2\times2$ matrices 
$M^{+}$ and $M^{-}$ in the IFD given by Eq.(4.9) of Ref.~\cite{BG} as 
$\delta\to 0$ or $\mathbb{C}\to 1$, i.e. 
$M^+ =\frac{\gamma^5+\gamma^1}{2} =
\frac{1+\gamma^0}{2}\gamma^5$, 
and 
$M^- =\frac{\gamma^5-\gamma^1}{2} =
\frac{1-\gamma^0}{2}\gamma^5$,
due to $\gamma^0 \gamma^5 =\gamma^1$,
while ${\hat M}^\pm \to M_{\rm LF}^\pm = \pm \frac{\gamma^+}{2}$ in the LFD limit 
$\delta \to \pi/4$ or $\mathbb{C}\to 0$.
For the interpolation angle $0 \le \delta \le \pi/4$ in general, from the direct calculation of the matrix multiplications, we find
\begin{align}\label{M++}
\left[ \Lambda^+\tilde{T}^{\dagger}(p_{\hat{-}})\gamma^0\gamma^{\hat{+}}\tilde{T}(k_{\hat{-}})\right] {\hat M}^+\left[ \tilde{T}(k_{\hat{-}}-r_{\hat{-}})\gamma^{\hat{+}}\tilde{T}(p_{\hat{-}}-r_{\hat{-}})\Lambda^-\gamma^0\right] 
&=C(p_{\hat{-}},k_{\hat{-}},r_{\hat{-}}){\hat M}^+ ,
\end{align}
\begin{align}\label{M-+}
\left[ \Lambda^+\tilde{T}^{\dagger}(p_{\hat{-}})\gamma^0\gamma^{\hat{+}}\tilde{T}(k_{\hat{-}})\right] {\hat M}^-\left[ \tilde{T}(k_{\hat{-}}-r_{\hat{-}})\gamma^{\hat{+}}\tilde{T}(p_{\hat{-}}-r_{\hat{-}})\Lambda^-\gamma^0\right] 
&=-S(p_{\hat{-}},k_{\hat{-}},r_{\hat{-}}){\hat M}^+,
\end{align}
\begin{align}\label{M+-}
\left[ \Lambda^-\tilde{T}^{\dagger}(p_{\hat{-}})\gamma^0\gamma^{\hat{+}}\tilde{T}(k_{\hat{-}})\right] {\hat M}^+\left[ \tilde{T}(k_{\hat{-}}-r_{\hat{-}})\gamma^{\hat{+}}\tilde{T}(p_{\hat{-}}-r_{\hat{-}})\Lambda^+\gamma^0\right] 
&=-S(p_{\hat{-}},k_{\hat{-}},r_{\hat{-}}){\hat M}^-,
\end{align}
\begin{align}\label{M--}
\left[ \Lambda^-\tilde{T}^{\dagger}(p_{\hat{-}})\gamma^0\gamma^{\hat{+}}\tilde{T}(k_{\hat{-}})\right] {\hat M}^-\left[ \tilde{T}(k_{\hat{-}}-r_{\hat{-}})\gamma^{\hat{+}}\tilde{T}(p_{\hat{-}}-r_{\hat{-}})\Lambda^+\gamma^0\right] 
&=C(p_{\hat{-}},k_{\hat{-}},r_{\hat{-}}){\hat M}^-,
\end{align}
where 
\begin{equation}\label{Cpkr}
C(p_{\hat{-}},k_{\hat{-}},r_{\hat{-}})=\cos \left(\frac{\theta(p_{\hat{-}})-\theta(k_{\hat{-}})}{2}\right)\cos \left(\frac{\theta(r_{\hat{-}}-p_{\hat{-}})-\theta(r_{\hat{-}}-k_{\hat{-}})}{2}\right),
\end{equation}
and
\begin{equation}\label{Spkr}
S(p_{\hat{-}},k_{\hat{-}},r_{\hat{-}})=\sin \left(\frac{\theta(p_{\hat{-}})-\theta(k_{\hat{-}})}{2}\right)\sin \left(\frac{\theta(r_{\hat{-}}-p_{\hat{-}})-\theta(r_{\hat{-}}-k_{\hat{-}})}{2}\right) .
\end{equation}
With Eqs.(\ref{M++})-(\ref{M--}), we finally split Eq.(\ref{phi_tilde_inter}) into 
the two coupled bound-state equations of ${\hat\phi}_{+}$ and ${\hat\phi}_{\hat -}$ 
\begin{subequations}\label{boundeq}
	\begin{align}\label{boundeq1}
	\left[ -r_{\hat{+}}+E_u(p_{\hat{-}})-E_v(p_{\hat{-}}-r_{\hat{-}})\right] {\hat\phi}_{+}(r_{\hat{-}},p_{\hat{-}})=\lambda\dashint\frac{dk_{\hat{-}}}{(p_{\hat{-}}-k_{\hat{-}})^2}\left[C(p_{\hat{-}},k_{\hat{-}},r_{\hat{-}}) {\hat\phi}_{+}(r_{\hat{-}},k_{\hat{-}})-S(p_{\hat{-}},k_{\hat{-}},r_{\hat{-}}) {\hat\phi}_{-}(r_{\hat{-}},k_{\hat{-}})\right] ,\\
	\left[ r_{\hat{+}}+E_u(p_{\hat{-}}-r_{\hat{-}})-E_v(p_{\hat{-}})\right] {\hat\phi}_{-}(r_{\hat{-}},p_{\hat{-}})=\lambda\dashint\frac{dk_{\hat{-}}}{(p_{\hat{-}}-k_{\hat{-}})^2}\left[C(p_{\hat{-}},k_{\hat{-}},r_{\hat{-}}) {\hat\phi}_{-}(r_{\hat{-}},k_{\hat{-}})-S(p_{\hat{-}},k_{\hat{-}},r_{\hat{-}}) {\hat\phi}_{+}(r_{\hat{-}},k_{\hat{-}})\right] .\label{boundeq2}
	\end{align}
\end{subequations}
We again note here that Eqs.(\ref{boundeq1})-(\ref{boundeq2}) coincide with Eq.(4.12) of Ref.~\cite{BG} 
in the IFD as $\delta\to0$ because not only the energies of particle and anti-particle become 
$E_u(p_{\hat{-}})\overset{\delta\to 0}{\longrightarrow}E(p^1)$ and 
$E_v(p_{\hat{-}})\overset{\delta\to 0}{\longrightarrow}-E(p^1)$ but also  
the rest of the variables correspond to their IFD counterparts, e.g. $r_{\hat{+}}\overset{\delta\to 0}{\longrightarrow}r^0$, etc..

In the LFD limit $\mathbb{C} \to 0$ (or $\delta\to \pi/4$) on the other hand, 
as discussed earlier in 
Sec.~\ref{sub:behaviorLF}, 
$E_u(p_{\hat{-}})\overset{\delta\to \pi/4}{\longrightarrow}  B(p^+) + \frac{m^2}{2p^+}
= \frac{m^2-2\lambda}{2p^+}$
and $E_v(p_{\hat{-}}-r_{\hat{-}})\overset{\delta\to \pi/4}{\longrightarrow}  
- B(r^+ - p^+) - \frac{m^2}{2(r^+-p^+)} = - \frac{m^2-2\lambda}{2(r^+-p^+)}$ 
using Eq.(\ref{eqn:Bp-})
while $E_u(p_{\hat{-}}-r_{\hat{-}}) \overset{\delta\to \pi/4}{\longrightarrow} 
\frac{2(r^+-p^+)}{\mathbb{C}}+B(r^+-p^+)+\frac{m^2}{2(r^+-p^+)}$
and $E_v(p_{\hat{-}})\overset{\delta\to \pi/4}{\longrightarrow} -\frac{2p^+}{\mathbb{C}}$ for the bound-state kinematics $0 < p^+ < r^+$. 
In this limit then, noting 
$C(p^+,k^+,r^+) \to 1$ and $S(p^+,k^+,r^+) \to 0$ 
from Eq.(\ref{lfthetasol}), we note that Eq.(\ref{boundeq1}) 
gets reduced to 
\begin{equation}\label{boundeqlightfront}
\left[ -r^- +\frac{m^2-2\lambda}{2p^+}+\frac{m^2-2\lambda}{2(r^+-p^+)}\right]\phi(r^+,p^+)=\lambda\dashint\frac{dk^+}{(p^+-k^+)^2}\phi(r^+,k^+), 
\end{equation}
where $\phi(r^+,p^+)$ corresponds to ${\hat\phi}_{+}(r_{\hat{-}},p_{\hat{-}})$ in
the LFD limit. 
Also, the solution for ${\hat \phi}_{-} \to 0$ can be attained rather immediately from Eq.(\ref{boundeq2}) by dividing it by the energy denominator factor $\left[ r_{\hat{+}}+E_u(p_{\hat{-}}-r_{\hat{-}})-E_v(p_{\hat{-}})\right]$ and noting the correspondence $1/(r_{\hat{+}}+ E_u(p_{\hat{-}}-r_{\hat{-}}) - E_v(p_{\hat{-}})) \to 
\mathbb{C}/(2 r^+ + \mathbb{C}(r^- + B(r^+-p^+) + B(p^+) + \frac{m^2}{2(r^+-p^+)} +\frac{m^2}{2(p^+)} ) \to 0$ as $\mathbb{C} \to 0$.
Substituting the on-mass-shell condition $r^- = {\cal M}^2/(2 r^+)$ for the bound-state meson of mass 
${\cal M}$,
defining the manifestly boost-invariant light-front momentum fraction variables $x=\frac{p^+}{r^+} (0\leq x\leq 1)$ and $ y=\frac{k^+}{r^+} (0\leq y\leq 1)$ and multiplying both sides of the equation by $ (-2r^+) $, we obtain
\begin{equation}\label{boundeqlf}
\left[ {\cal M}^2 - \frac{m^2-2\lambda}{x} - \frac{m^2-2\lambda}{1-x}\right] \phi(x)= - 2\lambda\dashint_0^1\frac{dy}{(x-y)^2}\phi(y),
\end{equation}
where the $r^+$ independence of $\phi(r^+,p^+)$ and $\phi(r^+,k^+)$ is imposed in 
$\phi(x)$ and $\phi(y)$ due to the boost-invariance of the light-front bound-state equation 
correctly reproducing 't Hooft's bound-state equation, Eq.(25) in Ref.~\cite{tHooft}.

In general, relating $ E_u $'s and $ E_v $'s in Eq.(\ref{boundeq}) to the solutions of $ E $ in Eq.(\ref{EuEv}),
we summarize the interpolating coupled bound-state equations for ${\hat\phi}_{+}(r_{\hat{-}},p_{\hat{-}})$ and ${\hat\phi}_{-}(r_{\hat{-}},p_{\hat{-}})$ between IFD and LFD as follows: 

\begin{subequations}\label{finalboundeq}
	\begin{align}\label{finalboundeq1}
	&\left[ -r_{\hat{+}}+\frac{-\mathbb{S}p_{\hat{-}}+E(p_{\hat{-}})}{\mathbb{C}}+\frac{\mathbb{S}(p_{\hat{-}}-r_{\hat{-}})+E(p_{\hat{-}}-r_{\hat{-}})}{\mathbb{C}}\right] {\hat\phi}_{+}(r_{\hat{-}},p_{\hat{-}}) \notag\\
	&\qquad\qquad=\lambda\dashint\frac{dk_{\hat{-}}}{(p_{\hat{-}}-k_{\hat{-}})^2}\left[C(p_{\hat{-}},k_{\hat{-}},r_{\hat{-}}) {\hat\phi}_{+}(r_{\hat{-}},k_{\hat{-}})-S(p_{\hat{-}},k_{\hat{-}},r_{\hat{-}}) {\hat\phi}_{ -}(r_{\hat{-}},k_{\hat{-}})\right] , \\
	&\left[ r_{\hat{+}}+\frac{-\mathbb{S}(p_{\hat{-}}-r_{\hat{-}})+E(p_{\hat{-}}-r_{\hat{-}})}{\mathbb{C}}+\frac{\mathbb{S}p_{\hat{-}}+E(p_{\hat{-}})}{\mathbb{C}}\right] {\hat\phi}_{-}(r_{\hat{-}},p_{\hat{-}})\notag\\
	&\qquad\qquad=\lambda\dashint\frac{dk_{\hat{-}}}{(p_{\hat{-}}-k_{\hat{-}})^2}\left[C(p_{\hat{-}},k_{\hat{-}},r_{\hat{-}}) {\hat\phi}_{-}(r_{\hat{-}},k_{\hat{-}})-S(p_{\hat{-}},k_{\hat{-}},r_{\hat{-}}) {\hat\phi}_{+}(r_{\hat{-}},k_{\hat{-}})\right] . \label{finalboundeq2}
	\end{align}
\end{subequations}

In the next section, Sec.~\ref{sec:solbound}, we solve Eqs.(\ref{finalboundeq1})-(\ref{finalboundeq2}) using the solutions of Eqs.(\ref{gap_eq_inter_E}) and (\ref{gap_eq_inter_theta}) to determine the bound-state mass spectrum 
as well as the corresponding bound-state wavefunctions ${\hat\phi}_{+}(r_{\hat{-}},p_{\hat{-}})$ and ${\hat\phi}_{-}(r_{\hat{-}},p_{\hat{-}})$ interpolating between IFD and LFD.
\end{widetext}

\section{\label{sec:solbound} The Bound-State Solution}

Solving Eqs.(\ref{finalboundeq1})-(\ref{finalboundeq2}) in practice, we use 
$r^{\hat{+}}=\mathbb{C}r_{\hat{+}}+\mathbb{S}r_{\hat{-}}$ and scale 
the interpolating quark momentum variables 
$p_{\hat{-}}$ and $k_{\hat{-}}$ with respect to the 
interpolating bound-state momentum $r_{\hat{-}}$ introducing the 
interpolating momentum fraction variables $x=p_{\hat{-}}/r_{\hat{-}}$ and 
$y=k_{\hat{-}}/r_{\hat{-}}$, respectively~\footnote{For the computation with $r_{\hat{-}} = 0$ as in the IFD calculation at the meson rest frame~\cite{Li}, we do not scale the 
interpolating momentum variables but directly take $r_{\hat{-}} = 0$ in Eqs.(\ref{finalboundeq1})-(\ref{finalboundeq2}) as separately presented in Appendix~\ref{app:rest}. 
}.
\begin{widetext}
Noting the conversion between the upper and lower indices of the momentum variable, we rewrite Eqs.(\ref{finalboundeq1}) and (\ref{finalboundeq2}) respectively as follows, 
\begin{subequations}\label{eqn:boundeqx}
	\begin{align}\label{boundeqx1}
	&\left[ -r^{\hat{+}}+E(xr_{\hat{-}})+E(r_{\hat{-}}-xr_{\hat{-}})\right]r_{\hat{-}} {\hat\phi}_+(r_{\hat{-}},x)=\lambda\mathbb{C}\dashint\frac{dy}{(x-y)^2}\left[C(x,y,r_{\hat{-}}) {\hat\phi}_+(r_{\hat{-}},y)-S(x,y,r_{\hat{-}}) {\hat\phi}_-(r_{\hat{-}},y)\right] ,\\
	&\left[ r^{\hat{+}}+E(r_{\hat{-}}-xr_{\hat{-}})+E(xr_{\hat{-}})\right] r_{\hat{-}}{\hat\phi}_-(r_{\hat{-}},x)=\lambda\mathbb{C}\dashint\frac{dy}{(x-y)^2}\left[C(x,y,r_{\hat{-}}) {\hat\phi}_-(r_{\hat{-}},y)-S(x,y,r_{\hat{-}}) {\hat\phi}_+(r_{\hat{-}},y)\right] ,\label{boundeqx2}
	\end{align}
\end{subequations}
where 
\begin{equation}\label{Cxyr}
C(x,y,r_{\hat{-}})=\cos \left(\frac{\theta(xr_{\hat{-}})-\theta(yr_{\hat{-}})}{2}\right)\cos \left(\frac{\theta(r_{\hat{-}}-xr_{\hat{-}})-\theta(r_{\hat{-}}-yr_{\hat{-}})}{2}\right) ,
\end{equation}
and
\begin{equation}\label{Sxyr}
S(x,y,r_{\hat{-}})=\sin \left(\frac{\theta(xr_{\hat{-}})-\theta(yr_{\hat{-}})}{2}\right)\sin \left(\frac{\theta(r_{\hat{-}}-xr_{\hat{-}})-\theta(r_{\hat{-}}-yr_{\hat{-}})}{2}\right) .
\end{equation}
\end{widetext}
One may note here that the ordinary longitudinal momentum $r^1$ is related to the interpolating longitudinal momentum $r_{\hat{-}}$ as 
\begin{equation}
r^1=\frac{1}{\mathbb{C}}\left( r_{\hat{-}}\cos\delta-\sqrt{r_{\hat{-}}^2\cos^2\delta-\mathbb{C}(r_{\hat{-}}^2-{\cal M}^2\sin^2\delta)}\right)
\end{equation} 
due to the on-mass-shell condition for the meson mass ${\cal M}$ as $r^{\hat{\mu}} r_{\hat{\mu}} = {r^{\mu}}{r_{\mu}}={\cal M}^2$.
In this section, we solve the coupled equations Eqs.(\ref{boundeqx1}) and 
(\ref{boundeqx2}) to find the bound-state mass spectrum 
${\cal M}_{(n)}^2=\frac{(r^{\hat{+}}_{(n)})^2-(r_{\hat{-}})^2}{\mathbb{C}}$
and the corresponding wavefunctions ${\hat\phi}_{\pm}^{(n)}(r_{\hat{-}},x)$
for the $n$-th bound-state as we discuss in subsection~\ref{sub:spec} and subsection~\ref{sub:wavefunc}, respectively. Then, 
we also apply the $\hat\phi_{\pm}^{(n)}(r_{\hat{-}},x)$ solutions to obtain
the interpolating PDFs and discuss the comparison
with the quasi-PDFs in subsection~\ref{sub:quasipdf}.

\subsection{\label{sub:spec}Spectroscopy}

As discussed in Ref.~\cite{review}, the bound-state wavefunctions 
$\hat\phi_{\pm}^{(n)}(r_{\hat{-}},x)$ correspond to the coefficients of
the compound operators creating/annihilating color singlet quark and antiquark bound-states
in the generalized Bogoliubov transformation. 
In contrast to the Bogoliubov transformation for the fermion operators given by 
Eq.(\ref{transf}) with the coefficients satisfying the normalization
$\cos^2\zeta +\sin^2\zeta =1$, the coefficients of the generalized 
Bogoliubov transformation for the meson states satisfy the normalization condition~\cite{mov}
\begin{equation}\label{norm}
\int dx\left\lbrace  |{\hat\phi}_{+}^{(n)}(r_{\hat{-}},x)|^2-|{\hat\phi}_{-}^{(n)}(r_{\hat{-}},x)|^2\right\rbrace =1.
\end{equation}
In conjunction with the compound operators~\cite{review}, it is interesting to note 
that the scalar or fermionic matter fields transforming 
in the adjoint representation of $SU(N)$ have been discussed in 
the literature~\cite{DK}. 
 
In solving numerically the coupled integral equations, Eqs.~(\ref{boundeqx1}) and (\ref{boundeqx2}), we use the spectrum method illustrated in Refs.~\cite{Li} and \cite{mov}. 
While the the rest frame basis wavefunction~\cite{Li} was generalized
to the moving frame basis wavefunction~\cite{mov} in IFD, we further generalize it 
to the interpolating basis wavefunction applicable to any interpolation angle $\delta$ between IFD and LFD, i.e. 
\begin{align}\label{eqn:wfbasis}
&\Psi_m(\alpha,r_{\hat{-}},x)\notag\\
=&\sqrt{\frac{|r_{\hat{-}}|\alpha}{2^m m! \sqrt{\pi}}}{\rm e}^{-\frac{\alpha^2r_{\hat{-}}^2(1-2x)^2}{8}}H_m\left( \frac{\alpha r_{\hat{-}}}{2}(1-2x)\right),
\end{align}
where $ H_m $ is the $m$-th Hermite polynomial and $ \alpha $ is the variational parameter which can be tuned to minimize the mass of the ground state. 
The reason for the Hermite polynomial basis can be traced back to the similarity
between the $\lambda = 0$ free Hamiltonian expressed in terms of the quark-antiquark compound operators~\cite{review} and the simple harmonic oscillator Hamiltonian.  
Due to the charge conjugation symmetry of the bound-states, the wavefunctions of the mesons $\hat\phi_+ $ and $\hat\phi_- $ are then the superpositions of these basis functions
\begin{equation}\label{superposition}
\hat\phi_{\pm}^{(n)}(r_{\hat{-}},x)=	\begin{cases}
\sum_{m=0}^{N-1}a_m^{(n)\pm}\Psi_{2m}(\alpha,r_{\hat{-}},x),\ \ n\ {\rm even};\\
\sum_{m=0}^{N-1}b_m^{(n)\pm}\Psi_{2m+1}(\alpha,r_{\hat{-}},x),\ \ n\ {\rm odd}, 
\end{cases}
\end{equation}
where the momentum fraction $ x=p_{\hat{-}}/r_{\hat{-}} $ goes to the familiar light-front momentum fraction $ p^+/r^+ $ as $ \delta\to\pi/4 $.
While $x=p_{\hat{-}}/r_{\hat{-}}$ is unbounded, $x\in(-\infty,\infty)$, 
for $ 0\leq\delta<\pi/4 $ in solving the interpolating coupled bound-state equations 
given by Eqs. (\ref{boundeqx1}) - (\ref{boundeqx2}), $x$ gets bounded to be in $[0,1]$
consistently satisfying the bound-state kinematics $0 \leq p^+ \leq r^+$ in LFD at
$\delta = \pi/4$ (or $\mathbb{C} = 0$) and the two coupled equations, Eqs. (\ref{boundeqx1}) and (\ref{boundeqx2}), get reduced to the single light-front bound-state equation 
given by Eq.(\ref{boundeqlf}).
We check this LFD reduction by confirming the correspondence of the interpolating solutions 
${\hat\phi}_{+}(r_{\hat{-}},p_{\hat{-}}) \to \phi(x)$ and ${\hat\phi}_{-}(r_{\hat{-}},p_{\hat{-}}) \to 0$    
in the limit $\delta \to \pi/4$ (or $\mathbb{C} \to 0$).

In solving the coupled equations, Eqs. (\ref{boundeqx1}) - (\ref{boundeqx2}), for the ground state $n=0$ as well as the $n$ even excited states, we use the
orthonormal basis of $\Psi_{2m}(\alpha,r_{\hat{-}},x)$ with $m=0,1,2,...N-1$
due to the $x \leftrightarrow 1-x$ symmetry. 
In contrast, for the $n$ odd excited states, we use the
orthonormal basis of $\Psi_{2m+1}(\alpha,r_{\hat{-}},x)$ with $m=0,1,2,...N-1$
due to the  $x \leftrightarrow 1-x$ antisymmetry dictated by the charge conjugation symmetry.
By projecting Eqs.~(\ref{boundeqx1}) and ~(\ref{boundeqx2}) onto the given set of orthonormal basis functions and integrating over the momentum fraction $x$ on both sides, Eqs.~(\ref{boundeqx1})-(\ref{boundeqx2}) are then transformed into a matrix eigenvalue equation in the orthonormal basis given by Eq.~(\ref{superposition}) with $m=0,1,2,...N-1$, i.e.
$\Psi_{2m}$ or $\Psi_{2m+1}$ depending on
whether $n$ is even or odd. Due to the coupling between $\hat\phi_+ $ and $\hat\phi_- $, the size of the matrix to be diagonalized is then $2N\times 2N$ for either $n$ even or $n$ odd states.
While we get $2N$ of $r^{\hat{+}}_{(n)}$ eigenvalues by diagonalizing the $2N\times 2N$ matrix, what we actually find in the meson spectroscopy is just $N$ number of meson masses via ${\cal M}_{(n)}^2=\frac{(r^{\hat{+}}_{(n)})^2-(r_{\hat{-}})^2}{\mathbb{C}}$ as the half of the $2N$ eigenvalues of $r^{\hat{+}}_{(n)}$ are the same but opposite sign of the other half of the $r^{\hat{+}}_{(n)}$ eigenvalues. For the ground state $n=0$ and the $n$ even excited 
states, the $N$ number of ${\cal M}_{(n)}$ values are found from the lowest one as the ground 
state meson mass ${\cal M}_{(0)}$ to the exited state meson masses ${\cal M}_{(2)}$, ${\cal M}_{(4)}$, etc., all the way up to ${\cal M}_{2(N-1)}$. For the $n$ odd excited states, the $N$ number of 
${\cal M}_{(n)}$ values are found from the lowest one as the first excited state meson mass $M_{(1)}$ to the higher exited state meson masses ${\cal M}_{(3)}$, ${\cal M}_{(5)}$, etc.,
all the way up to ${\cal M}_{(2N-1)}$. 

In the numerical computation, we note that the numerical accuracy of the results depends 
not only on the number of basis functions but also the number of grid points for the numerical 
integrations. 
In practice, we take $N=20$ in Eq.(\ref{superposition}) to get 
the mass spectrum of the lowest 8 states, i.e. $n = 0,1,2,...,7$,
and the number of grid points as 200 for the computation of the integrations in Eqs.~(\ref{boundeqx1}) and (\ref{boundeqx2}). 
As our numerical method involves the values of the interpolation angle $\delta$ and the interpolating longitudinal momentum $r_{\hat{-}}$, we check the stability of our numerical results and their accuracy by varying
$\delta$ and $r_{\hat{-}}$. 
While our numerical program provides typically the numerical accuracy of up to 
4 or 5 digits for most $\delta$ and $r_{\hat{-}}$ values 
with the number of grid points being 200 for the computation of the integrations, we need to pay attention to the accuracy of the numerical values
for the mass spectrum when we take $\delta$ close to $\pi/4$ or $r_{\hat{-}}$ large. Especially, as the meson mass ${\cal M}_{(n)}=\sqrt{\frac{(r^{\hat{+}}_{(n)})^2-(r_{\hat{-}})^2}{\mathbb{C}}}$
gets smaller, the eigenvalues $r^{\hat{+}}_{(n)}$ that we obtain from Eqs.(\ref{boundeqx1}) and (\ref{boundeqx2}) gets closer to the value $r_{\hat{-}}$ that we take for our numerical computation so that it may lead to 
the cancellation of the two large values in getting the small value.
The demand of numerical accuracy gets even more enhanced when $\delta$ gets close to $\pi/4$ (or $\mathbb{C}$ gets close to zero) and $r_{\hat{-}}$ gets large as both the values in the numerator and the denomination get closer to zero.

In particular, as the bare quark mass $m \to 0$, we need to check 
our numerical accuracy paying attention to 
the GOR relation which was explicitly derived in the 't Hooft model~\cite{review} and identifying the pionic bound-state ${\cal M}_{(0)}^2 \sim m\sqrt{\lambda} \to 0$.
\begin{figure}
	\centering
	\includegraphics[width=1.0\linewidth]{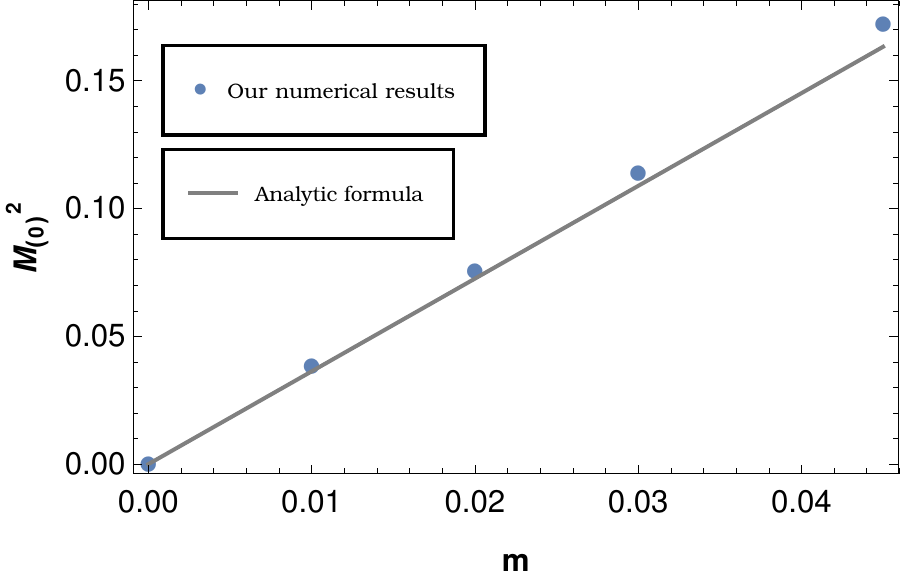}
	\caption{The bare quark mass $m$ dependence of ${{\cal M}_\pi}^2$. All quantities are in proper units of $ \sqrt{2\lambda} $. Our numerical results are compared with the analytic result given by Eq.(\ref{GOR}). }
	\label{fig:GOR}
\end{figure}
To confirm the validity of GOR relation numerically, we have computed ${\cal M}_{(0)}^2$ with 600 grid points (rather than 200 grid points normally used
in our computation) for the very small $m$ values ($m = 0, 0.01, 0.02, 0.03, 0.045$). The results are plotted in 
Fig.~\ref{fig:GOR}. As we can see in Fig.~\ref{fig:GOR}, we numerically confirm the behavior of ${\cal M}_{(0)}^2 \sim m\sqrt{\lambda}$ and indeed obtain the slope of the straight line
consistent with the analytic result given by~\footnote{There is the factor 2 difference between
Refs.~\cite{review} and ~\cite{Ellis,Brower,mov} in the coefficient of Eq.(\ref{GOR}). Our result is consistent 
with Ref.~\cite{Ellis,Brower,mov}.} 
\begin{equation}
\label{GOR}
{\cal M}_{\pi}^2 = - \frac{4 m <{\bar \psi} \psi>}{f_{\pi}^2} = \sqrt{\frac{8 \pi^2 m^2 \lambda}{3}},
\end{equation}
where we used Eq.(\ref{analytic-condensation}) for the vacuum condensation $<{\bar \psi} \psi>$ and the pion decay constant $f_\pi = \sqrt{N_c/\pi}$
derived~\cite{review} from the matrix element of the axial vector current between the pionic ground state and the non-trivial vacuum. Here, the pionic bound-state in the chiral limit
corresponds to the zero mass bound-state that occurs 
when both quarks have mass zero as pointed out by 't Hooft~\cite{tHooft}.
The corresponding bound-state wavefunction in LFD is given by $\phi(x) = 1$ for $x\in[0,1]$
as we will discuss in the next subsection, Sec.~\ref{sub:wavefunc}.  
These results are consistent with the discussions~\cite{review,mov,Zhicon,GSSW} on the SBCS in the 't Hooft model ($N_c \to \infty$) which does not contradict with the Coleman's theorem~\cite{Col} that would prohibit the SBCS in the case of a finite $N_c$.
As the result of ${\cal M}_{(0)}$ must be zero theoretically for the bare quark mass $m=0$, 
the scrutiny of the numerical sensitivity check depending on the values of $\delta$ and $r_{\hat{-}}$ as well as the number of computational grid points is maximally enhanced in the $m=0$ case. 
In Table~\ref{zero-mass-case}, the numerical values of ${\cal M}_{(0)}$ for the bare quark mass $m=0$ are listed depending on the values of $\delta$ and $r_{\hat{-}}$ as well as the number of computational grid points. 
We take the values of $r_{\hat{-}}$ as 
$r_{\hat{-}}= 0, 0.2 {\mathbf{M}_{0.18}}$, and $2 {\mathbf{M}_{0.18}}$, 
with the scale of ${\mathbf{M}_{0.18}}$ which is the lowest bound-state mass for $m=0.18$, i.e. ${\mathbf{M}_{0.18}}=0.88$, as the lowest bound-state mass for $m=0$ is zero, i.e. $\mathbf{M}_{0}=0$, and cannot be taken as any reference value of $r_{\hat{-}}$. 
While it is highly challenging to achieve the typical numerical accuracy mentioned earlier for the bare quark mass $m=0$, our numerical results appear to consistently approach the theoretical value ${\cal M}_{(0)} = 0$ as the number of grid points is increased from 200 to 600 for the ranges of the $\delta$ and $r_{\hat{-}}$ values in Table~\ref{zero-mass-case}. As $\delta$ gets close to $\pi/4$, 
our numerical computation demands much higher numerical accuracy. Although we haven't increased the number of grid points 
beyond 600, we anticipate that our numerical results would get closer and closer to zero
as we increase the number of grid points even if $\delta$ gets close to $\pi/4$. 
\begin{table}
	\caption{\label{tab:accucheck} Numerical results of the ground-state meson mass ${\cal M}_{(0)}$  depending on the values of $\delta$ and $r_{\hat{-}}$ as well as the number of computational grid points for the quark bare mass $m=0$ case. 
		All the mass spectra are given in six significant figures with the proper unit of $ \sqrt{2\lambda}$. The ground-state meson mass 
		$\mathbf{M}_{0.18}=0.88$ for $m=0.18$ is taken as the reference value of the interpolating longitudinal momentum $r_{\hat{-}}$. 
		The results for $r_{\hat{-}}=0$ were obtained with the method presented in 
		Appendix~\ref{app:rest}. 
	}
	\begin{ruledtabular}
		\label{zero-mass-case}
		\begin{tabular}{c|c|c|c}
			\centering
			$\delta$ & $r_{\hat{-}}$ & number of grid points & $ {\cal M}_{(0)}  $\\ \hline
			$	0$& $0$ &$ 200$&$ 0.0607101$\\
			&	& $600$ &$ 0.0361024$\\
			\hline
			$0	$& $0.2\mathbf{M}_{0.18}$ &$ 200$&$0.0632755$\\
			&	& $600$ &$ 0.0395562$\\
			\hline
			$0	$& $2\mathbf{M}_{0.18}$ &$ 200$&$0.107216$\\
			&	& $600$ &$ 0.0532456$\\
			
			\hline
			$	0.6$& $0$ &$ 200$&$ 0.0690707$\\
			&	& $600$ &$ 0.0407446$\\
			\hline
			$0.6	$& $0.2\mathbf{M}_{0.18}$ &$ 200$&$0.0678997$\\
			&	& $600$ &$0.0472921 $\\
			\hline
			$0.6	$& $2\mathbf{M}_{0.18}$ &$ 200$&$ 0.143686$\\
			&	& $600$ &$0.0599700 $\\
			
		\end{tabular}
	\end{ruledtabular}
\end{table}

In Table~\ref{tab:mesonmass}, we list the results of the meson mass spectra 
${\cal M}_{(n)}$ up to $n=7$ for the bare quark mass $m=0$ including ${\cal M}_{(0)}$ 
obtained with the 600 grid points for $\delta = 0, 0.6$ and 0.78. For the mass spectra of excited states, ${\cal M}_{(n)} (n=1,2,...,7)$, the results were obtained  with the number of
grid points just 200 for $\delta=0$ and 0.6 although the number of grid points for 
the $\delta = 0.78$ case was still kept as 600.
Similarly, we present the results of the meson mass spectra for the bare quark mass $m=0.18$ in Table~\ref{tab:mesonmass2}. Here, all the results ${\cal M}_{(n)} (n=0,1,2,...,7)$ including the ground-state were obtained with the 200 grid points
for $\delta=0$ and 0.6 while for $\delta=0.78$ with the 380 grid points. 
We note that our numerical results are consistent with each other up to the second digit 
after the decimal point for the $\delta$ values not close to $\pi/4$ such as $\delta=0$ and 0.6 and the $r_{\hat{-}}$ values not so large such as $r_{\hat{-}}=0$ and $0.2\mathbf{M}_{0.18}$
as listed in Tables~\ref{tab:mesonmass} and \ref{tab:mesonmass2}. 
Such stability persists in all the cases of the bare quark mass values 
($m=0, 0.045, 0.18, 1.00, 2.11$) that we considered for the computation of 
the meson mass spectra in this work. 

\begin{figure}
	\centering
	\includegraphics[width=1.0\linewidth]{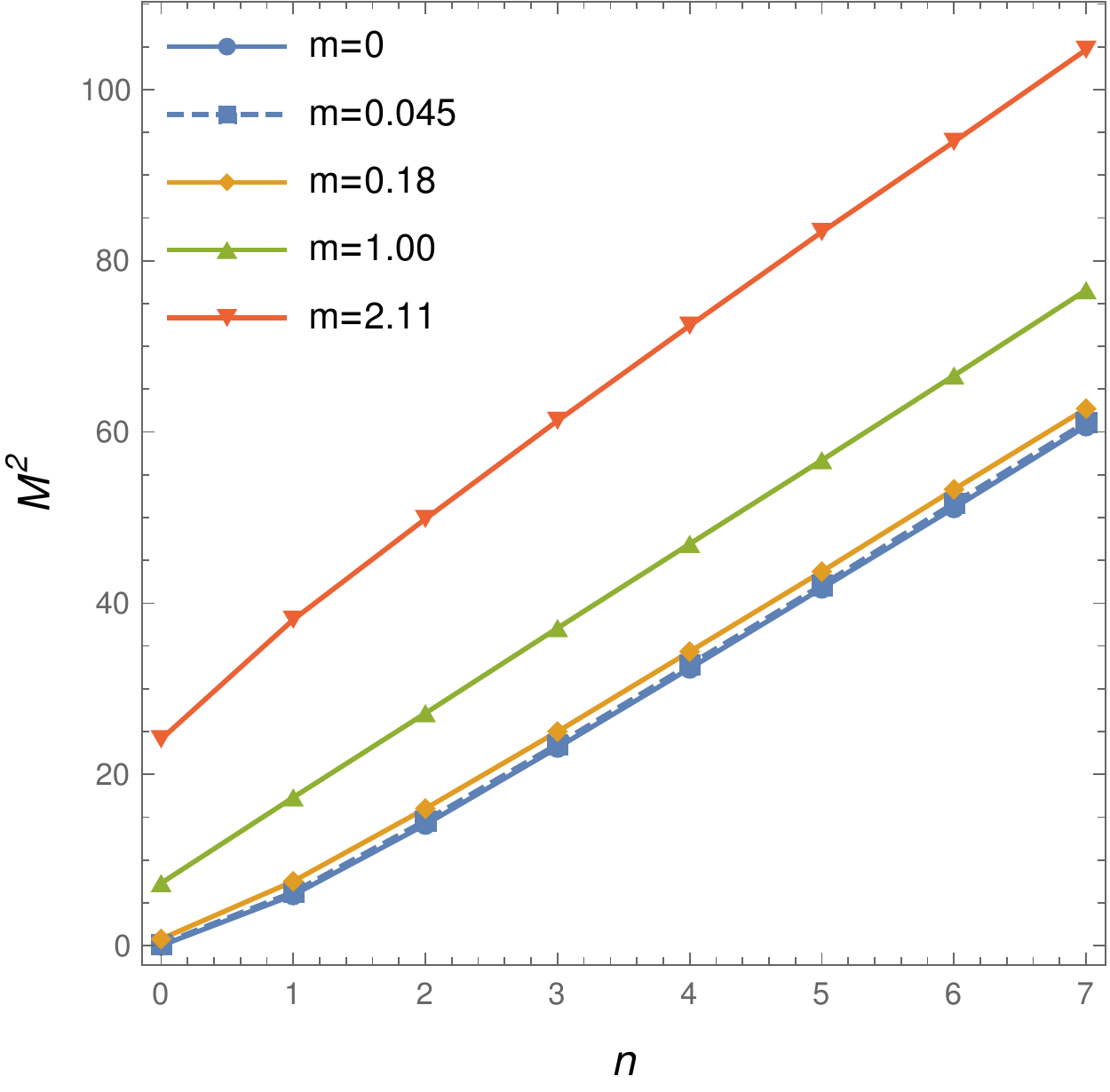}
	\caption{The feature of ``Regge trajectories" for  
the quark-antiquark bound-states each with the corresponding equal bare mass $m$. All quantities are in proper units of $ \sqrt{2\lambda} $. }
	\label{fig:mesonmassplot}
\end{figure}

In Table~\ref{tab:mass}, we summarize these stable numerical results up to the second
digit after the decimal point. Here, we take ${\cal M}_{(0)} = 0$ for the case $m=0$ from the theoretical SBCS ground. The values in Table~\ref{tab:mass} are also shown in Fig.~\ref{fig:mesonmassplot} depicting the feature of ``Regge trajectories" for  
the quark-antiquark bound-states each with the corresponding equal bare mass $m$
~\cite{tHooft,Li,mov,consistency,Brower}.
It is interesting to note that the Regge trajectory gets slightly modified
from just the linear trajectory behavior, developing a bit of curvature in the trajectory for the ground-state and the low-lying excited states. 
For the small mass, in particular $m=0$, the trajectory looks a little concave
shape while for the larger mass the curvature turns somewhat into   
a convex shape. This seems to reflect the fact that the GOR works in the chiral limit as shown in Fig.~\ref{fig:GOR} but the chiral symmetry gets broken as
the quark mass gets larger. The convex feature of the Regge trajectory
for the heavy quarkonia model was shown in Ref.~\cite{heavy-quarkonia-Regge}.

\begin{table}
	\caption{\label{tab:mesonmass}Meson mass spectra 
for the bare quark mass $m=0$ with the variation of $\delta$ and $r_{\hat{-}}$ values.
All the mass spectra are given in six significant figures with the proper unit of $ \sqrt{2\lambda} $. The ground-state meson mass $\mathbf{M}_{0.18}=0.88$ for $m=0.18$ is taken as the reference value of the interpolating longitudinal momentum $r_{\hat{-}}$.
The results for $r_{\hat{-}}=0$ were obtained with the method presented in 
Appendix~\ref{app:rest}. 
}
	\begin{ruledtabular}
		\begin{tabular}{c|c|c|c|c|c}
			\centering
				$\delta$ &  $r_{\hat{-}}$ & $ {\cal M}_{(0)} $ & $ {\cal M}_{(2)} $ & $ {\cal M}_{(4)} $ & $ {\cal M}_{(6)}$\\ \hline
				$	0$&$ 0$&$ 0.0361024 $ & $  3.76245 $ & $  5.68513 $ & $  7.15304$\\
				&   $0.2\mathbf{M}_{0.18}$ &$ 0.0395562 $ & $  3.76266 $ & $  5.68573 $ & $  7.15381$\\
				&   $2\mathbf{M}_{0.18}$ &$ 0.0532456 $ & $  3.76433 $ & $  5.68703 $ & $  7.16921$\\
			\hline
			 	$	0.6$&$ 0$&$ 0.0407446 $ & $  3.76302 $ & $  5.68552 $ & $  7.15329$\\
				&   $0.2\mathbf{M}_{0.18}$ &$0.0472921 $ & $  3.76285 $ & $  5.68542 $ & $  7.15361$\\
				&   $2\mathbf{M}_{0.18}$ &$ 0.0599700 $ & $  3.76441 $ & $  5.68690 $ & $  7.15444$\\
			\hline
				$	0.78$&$ 0$&$ 0.0884992 $ & $  3.76458 $ & $  5.68665 $ & $  7.15417$\\
			&   $0.2\mathbf{M}_{0.18}$ &$ 0.173979 $ & $  3.76292 $ & $  5.69002 $ & $  7.15655$\\
			&   $2\mathbf{M}_{0.18}$ &$ 0.106365 $ & $  3.77182 $ & $  5.69178 $ & $  7.16045$\\
			\hline\hline
			$\delta$ &  $r_{\hat{-}}$ & $ {\cal M}_{(1)} $ & $ {\cal M}_{(3)} $ & $ {\cal M}_{(5)} $ & $ {\cal M}_{(7)}$\\ \hline
			$	0$&$ 0$&$ 2.42728 $ & $  4.80619 $ & $  6.45726 $ & $  7.79124$\\
			&   $0.2\mathbf{M}_{0.18}$ &$ 2.42773 $ & $  4.80668 $ & $  6.45797 $ & $  7.79191$\\
			&   $2\mathbf{M}_{0.18}$ &$ 2.42861 $ & $  4.80603 $ & $  6.45736 $ & $  7.79204$\\
			\hline
			$	0.6$&$ 0$&$ 2.42772 $ & $  4.80668 $ & $  6.45756 $ & $  7.79145$\\
			&   $0.2\mathbf{M}_{0.18}$ &$ 2.42815 $ & $  4.80693 $ & $  6.45796 $ & $  7.79205$\\
			&   $2\mathbf{M}_{0.18}$ &$ 2.43215 $ & $  4.80696 $ & $  6.45845 $ & $  7.79172$\\
			\hline
			$	0.78$&$ 0$&$ 2.42906 $ & $  4.80807 $ & $  6.45857 $ & $  7.79239$\\
			&   $0.2\mathbf{M}_{0.18}$ &$ 2.43734 $ & $  4.80832 $ & $  6.45799 $ & $  7.79376$\\
			&   $2\mathbf{M}_{0.18}$ &$ 2.43520 $ & $  4.81589 $ & $  6.46355 $ & $  7.79905$\\
		\end{tabular}
	\end{ruledtabular}
\end{table}

\begin{table}
	\caption{\label{tab:mesonmass2}Meson mass spectra 
for the bare quark mass $m=0.18$ with the variation of $\delta$ and $r_{\hat{-}}$ values.
All the mass spectra are given in six significant figures with the proper unit of $ \sqrt{2\lambda} $. The ground-state meson mass $\mathbf{M}_{0.18}=0.88$ for $m=0.18$ is taken as the reference value of the interpolating longitudinal momentum $r_{\hat{-}}$.
The results for $r_{\hat{-}}=0$ were obtained with the method presented in 
Appendix~\ref{app:rest}. 
}
	\begin{ruledtabular}
		\begin{tabular}{c|c|c|c|c|c}
			\centering
			$\delta$ &  $r_{\hat{-}}$ & $ {\cal M}_{(0)} $ & $ {\cal M}_{(2)} $ & $ {\cal M}_{(4)} $ & $ {\cal M}_{(6)}$\\ \hline
			$	0$&$ 0$&$ 0.880457 $ & $  3.99902 $ & $  5.85781 $ & $  7.29550$\\
			&   $0.2\mathbf{M}_{0.18}$ &$0.880686 $ & $  3.99928 $ & $  5.85843 $ & $  7.29630$\\
			&   $2\mathbf{M}_{0.18}$ &$ 0.883080 $ & $  3.99921 $ & $  5.85774 $ & $  7.29676$\\
			\hline
			$	0.6$&$ 0$&$ 0.881753 $ & $  3.99943 $ & $  5.85807 $ & $  7.29566$\\
			&   $0.2\mathbf{M}_{0.18}$ &$0.881526 $ & $  3.99943 $ & $  5.85812 $ & $  7.29612$\\
			&   $2\mathbf{M}_{0.18}$ &$0.886997 $ & $  4.00048 $ & $  5.85938 $ & $  7.29641$\\
			\hline
			$	0.78$&$ 0$&$ 0.889730 $ & $  4.00233 $ & $  5.86016 $ & $  7.29718$\\
			&   $0.2\mathbf{M}_{0.18}$ &$ 0.856979 $ & $  3.99561 $ & $  5.85482 $ & $  7.29324$\\
			&   $2\mathbf{M}_{0.18}$ &$ 0.914992 $ & $  4.01176 $ & $  5.86721 $ & $  7.30555$\\
			\hline\hline
			$\delta$ &  $r_{\hat{-}}$ & $ {\cal M}_{(1)} $ & $ {\cal M}_{(3)} $ & $ {\cal M}_{(5)} $ & $ {\cal M}_{(7)}$\\ \hline
			$	0$&$ 0$&$ 2.73527 $ & $  5.00349 $ & $  6.61265 $ & $  7.92348$\\
			&   $0.2\mathbf{M}_{0.18}$ &$ 2.73565 $ & $  5.00401 $ & $  6.61336 $ & $  7.92401$\\
			&   $2\mathbf{M}_{0.18}$ &$2.73588 $ & $  5.00338 $ & $  6.61320 $ & $  7.92469$\\
			\hline
			$	0.6$&$ 0$&$ 2.73559 $ & $  5.00384 $ & $  6.61285 $ & $  7.92360$\\
			&   $0.2\mathbf{M}_{0.18}$ &$ 2.73595 $ & $  5.00406 $ & $  6.61329 $ & $  7.92419$\\
			&   $2\mathbf{M}_{0.18}$ &$ 2.73818 $ & $  5.00430 $ & $  6.61360 $ & $  7.92388$\\
			\hline
			$	0.78$&$ 0$&$ 2.73843 $ & $  5.00638 $ & $  6.61463 $ & $  7.92496$\\
			&   $0.2\mathbf{M}_{0.18}$ &$ 2.73213 $ & $  5.00048 $ & $  6.60966 $ & $  7.92062$\\
			&   $2\mathbf{M}_{0.18}$ &$ 2.75685 $ & $  5.01497 $ & $  6.62144 $ & $  7.93055$\\
		\end{tabular}
	\end{ruledtabular}
\end{table}

\begin{table}
	\caption{\label{tab:mass}Summary of meson mass spectra ${\cal M}_{(n)}$ 
($n=0,1,2,...,7$) for the bare quark mass values $m=0, 0.045, 0.18, 1.00, 2.11$ in 
the unit of $\sqrt{2\lambda} $.}
	\begin{ruledtabular}
		\begin{tabular}{l|cccccccc}
			\centering
			$n$ & 0 & 1 & 2 & 3 & 4 & 5 & 6 & 7\\ \hline
			$ m=0 $ & $ 0 $ & $ 2.43 $ & $ 3.76 $ & $ 4.81 $ & $ 5.69 $ & $ 6.46 $ & $ 7.15 $ & $ 7.79 $ \\ \hline
			$ m=0.045 $ & $ 0.42 $ & $2.50$ & $ 3.82 $ & $4.85$ & $ 5.73 $ & $6.49$ & $ 7.19 $ &  $ 7.82 $ \\ \hline
			$ m=0.18 $ & $ 0.88 $ & $ 2.74 $ & $ 4.00 $ & $ 5.00 $ & $ 5.86 $ & $ 6.61 $ & $ 7.30 $ & $ 7.92 $ \\ \hline
			$ m=1.00 $ & $ 2.70 $ & $ 4.16 $ & $ 5.21 $ & $ 6.09 $ & $ 6.85 $ & $ 7.53 $ & $ 8.16 $ & $ 8.75 $ \\ \hline
			$ m=2.11 $ & $ 4.91 $ & $ 6.17 $ & $ 7.06 $ & $ 7.83 $ & $ 8.51 $ & $ 9.13 $ & $ 9.69 $ & $ 10.23 $\\
		\end{tabular}
	\end{ruledtabular}
\end{table}

\subsection{\label{sub:wavefunc} Wavefunctions}
\begin{figure*}
	\centering
	\subfloat[]{
		\includegraphics[width=1.0\columnwidth]{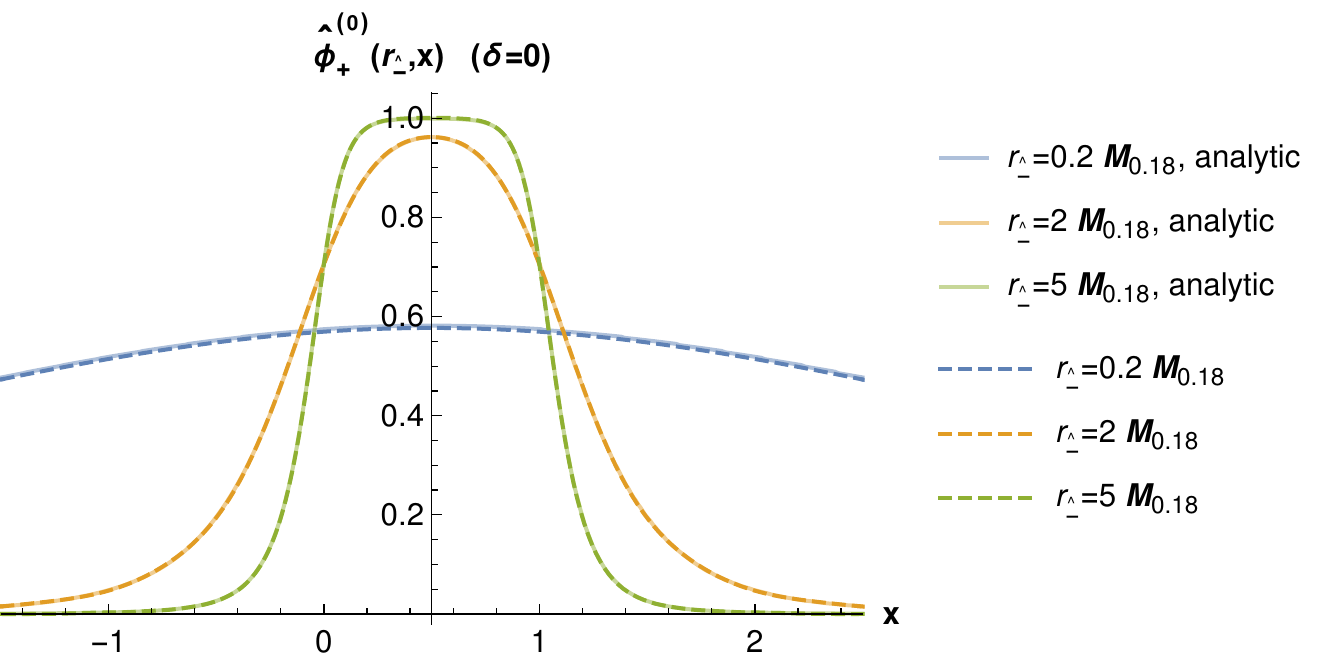}
		\label{fig:m0grzd0}
	}
	\centering
	\subfloat[]{
		\includegraphics[width=1.0\columnwidth]{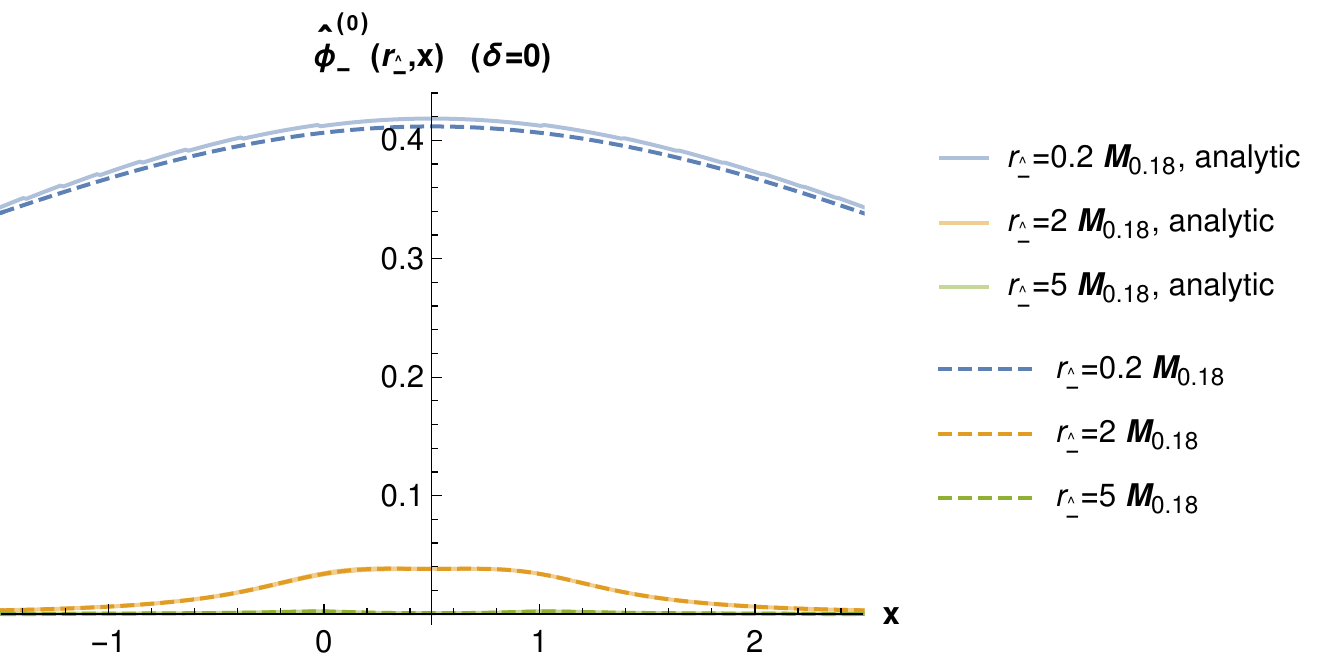}
		\label{fig:m0grfd0}
	}\\
	\centering
	\subfloat[]{
		\includegraphics[width=1.0\columnwidth]{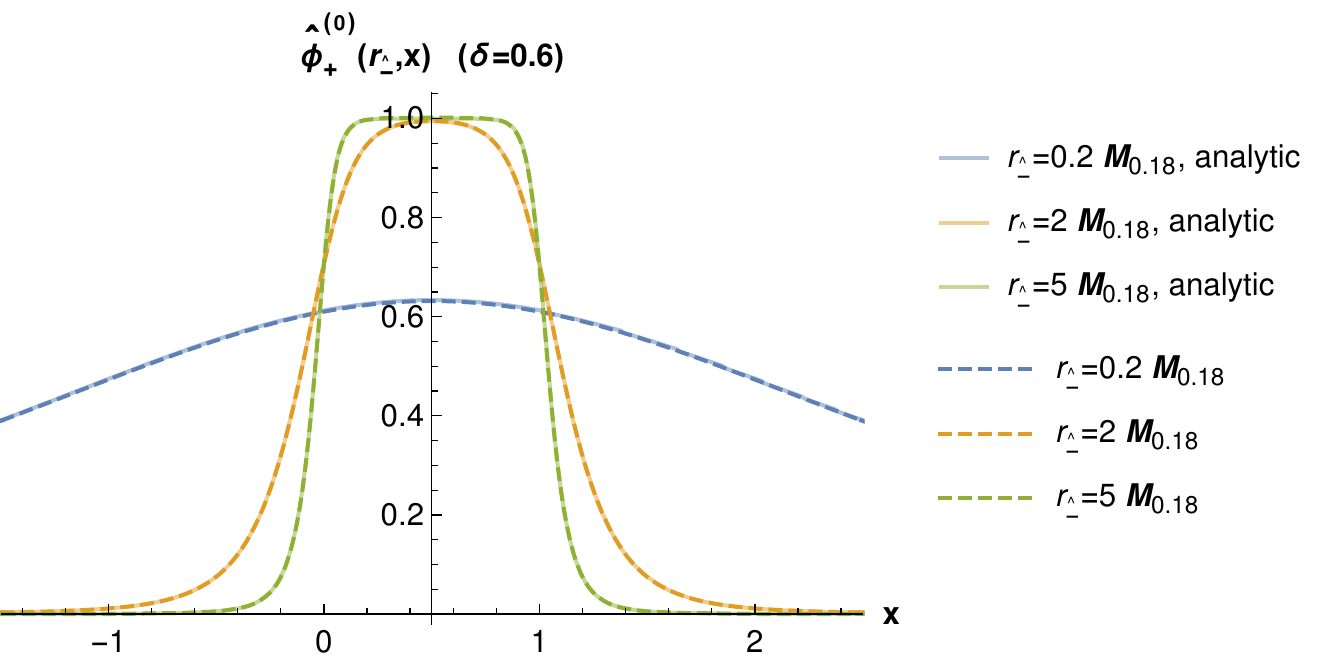}
		\label{fig:m0grzd06}
	}
	\centering
	\subfloat[]{
		\includegraphics[width=1.0\columnwidth]{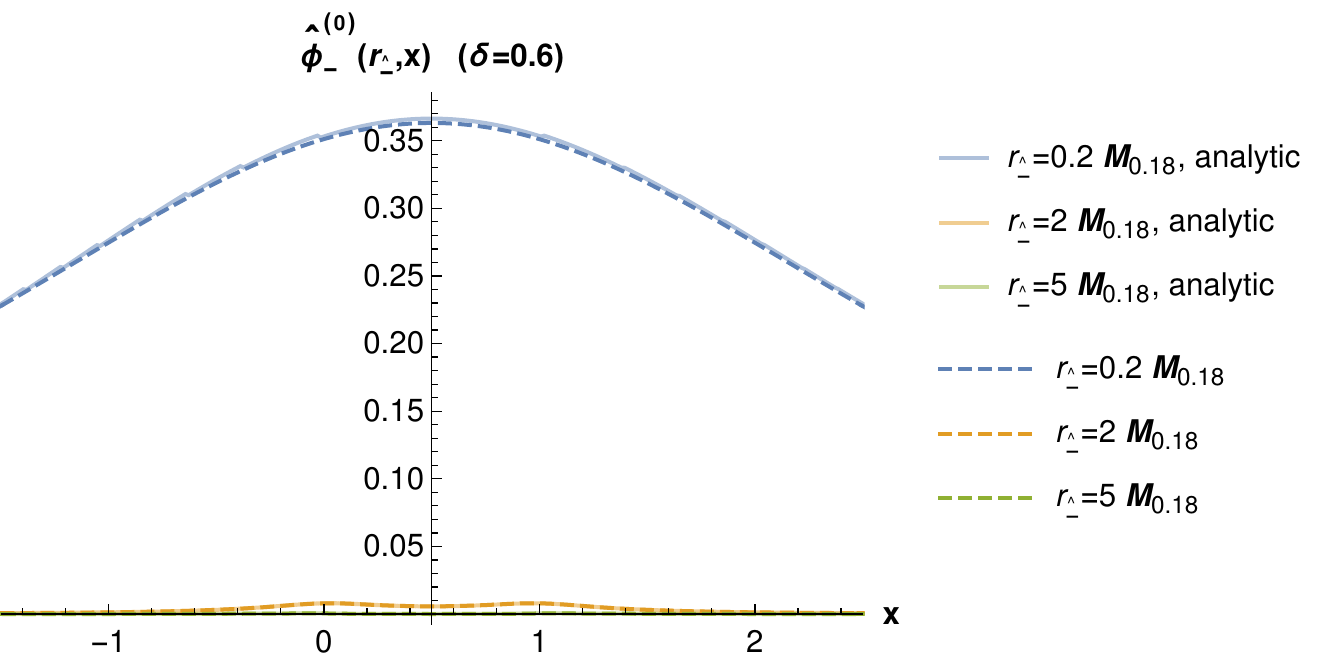}
		\label{fig:m0grfd06}
	}\\
	\centering
	\subfloat[]{
		\includegraphics[width=1.0\columnwidth]{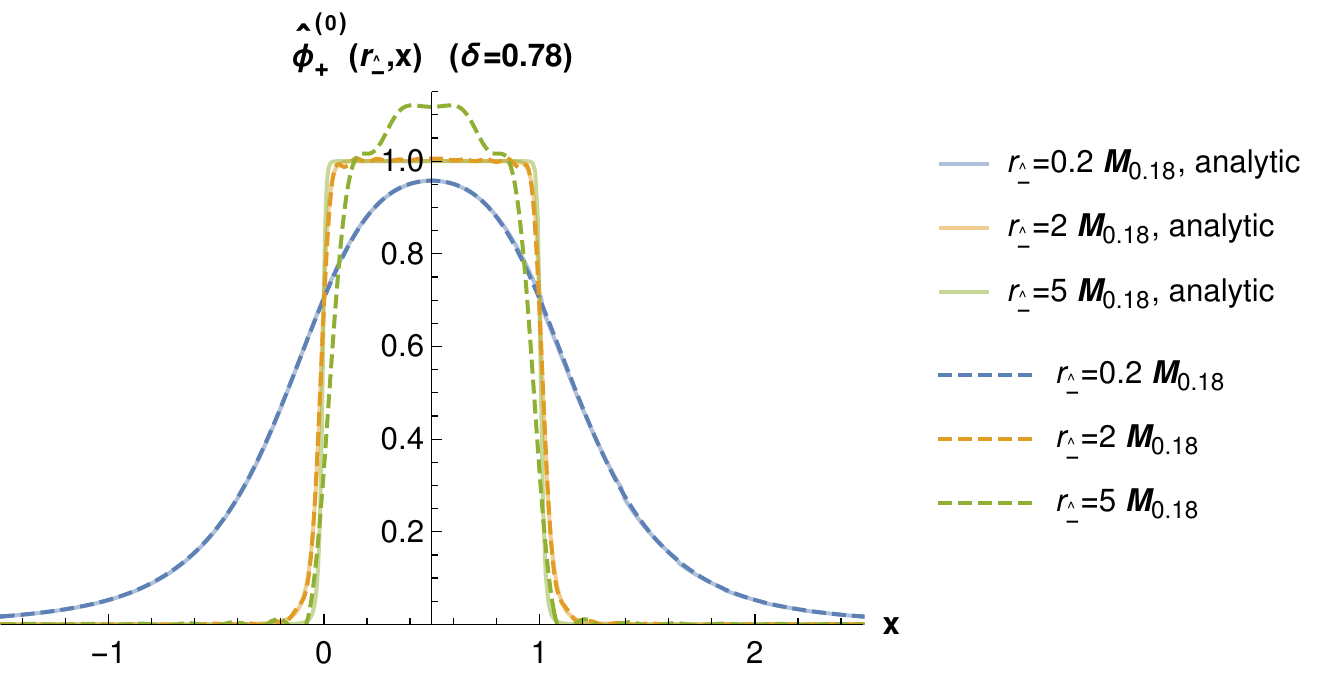}
		\label{fig:m0grzd078}
	}
	\centering
	\subfloat[]{
		\includegraphics[width=1.0\columnwidth]{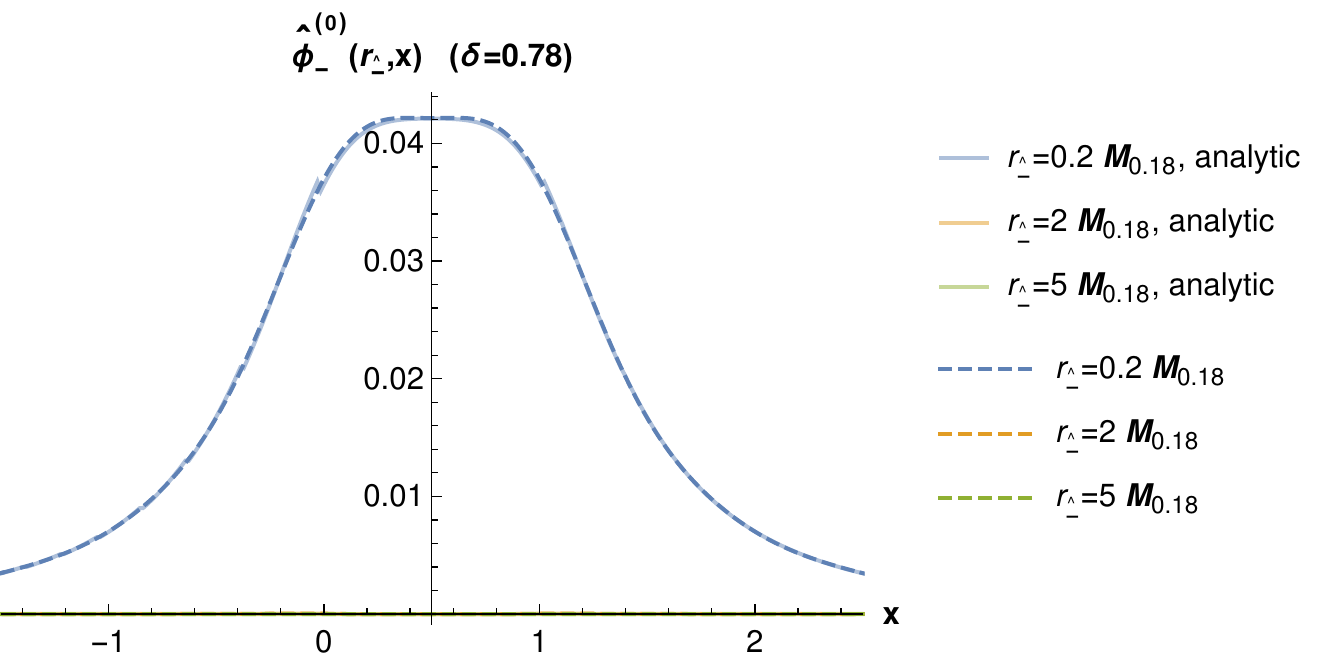}
		\label{fig:m0grfd078}
	}\\
	\caption{Ground state wave functions $\hat\phi_+^{(0)}(r_{\hat{-}},x)$ and $\hat\phi_-^{(0)}(r_{\hat{-}},x)$ for $ m=0 $. All quantities are in proper units of $ \sqrt{2\lambda} $.\label{fig:m0grz}}
\end{figure*}

\begin{figure*}
	\centering
	\subfloat[]{
		\includegraphics[width=1.0\columnwidth]{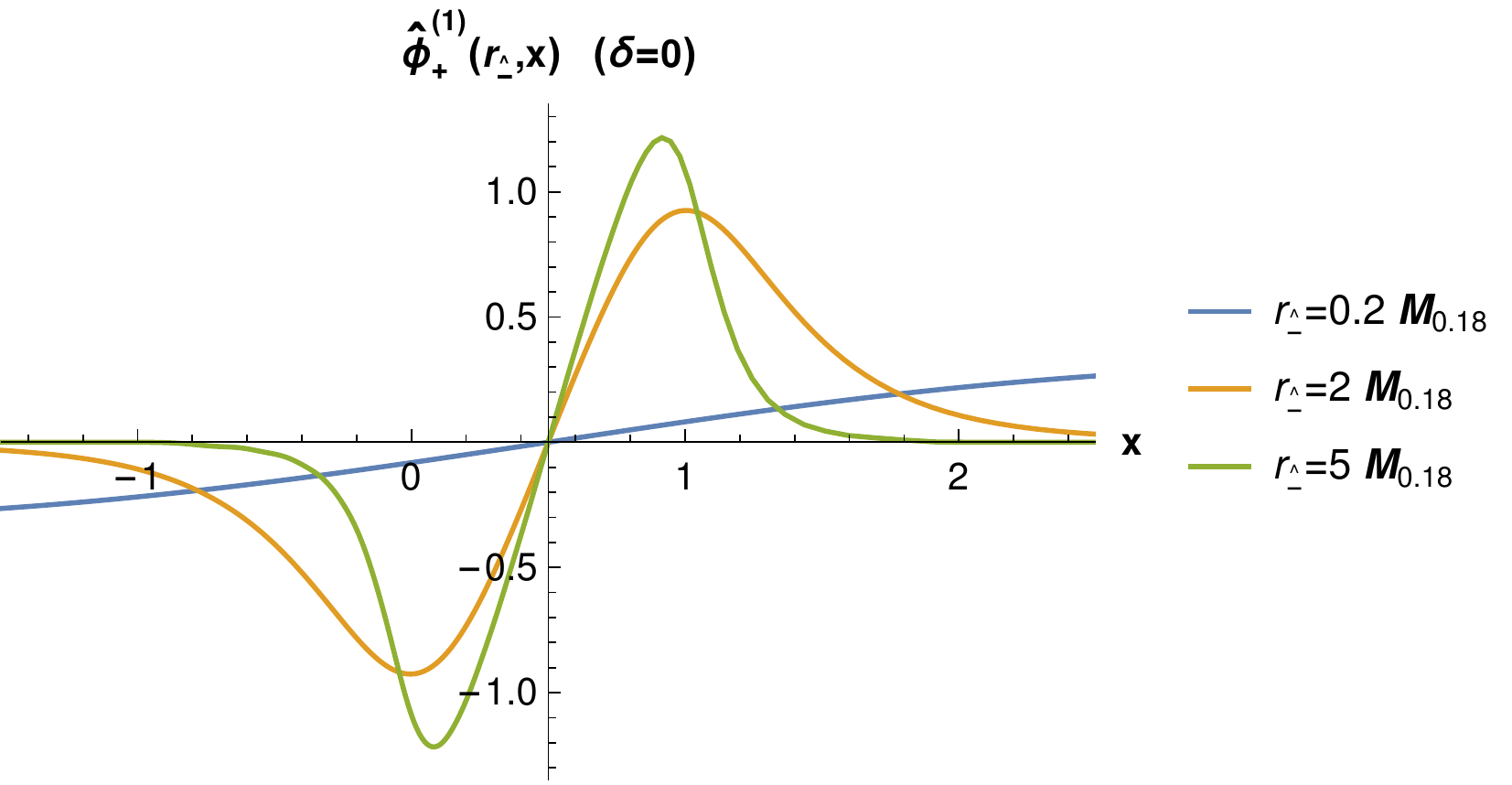}
		\label{fig:m0fezd0}
	}
	\centering
	\subfloat[]{
		\includegraphics[width=1.0\columnwidth]{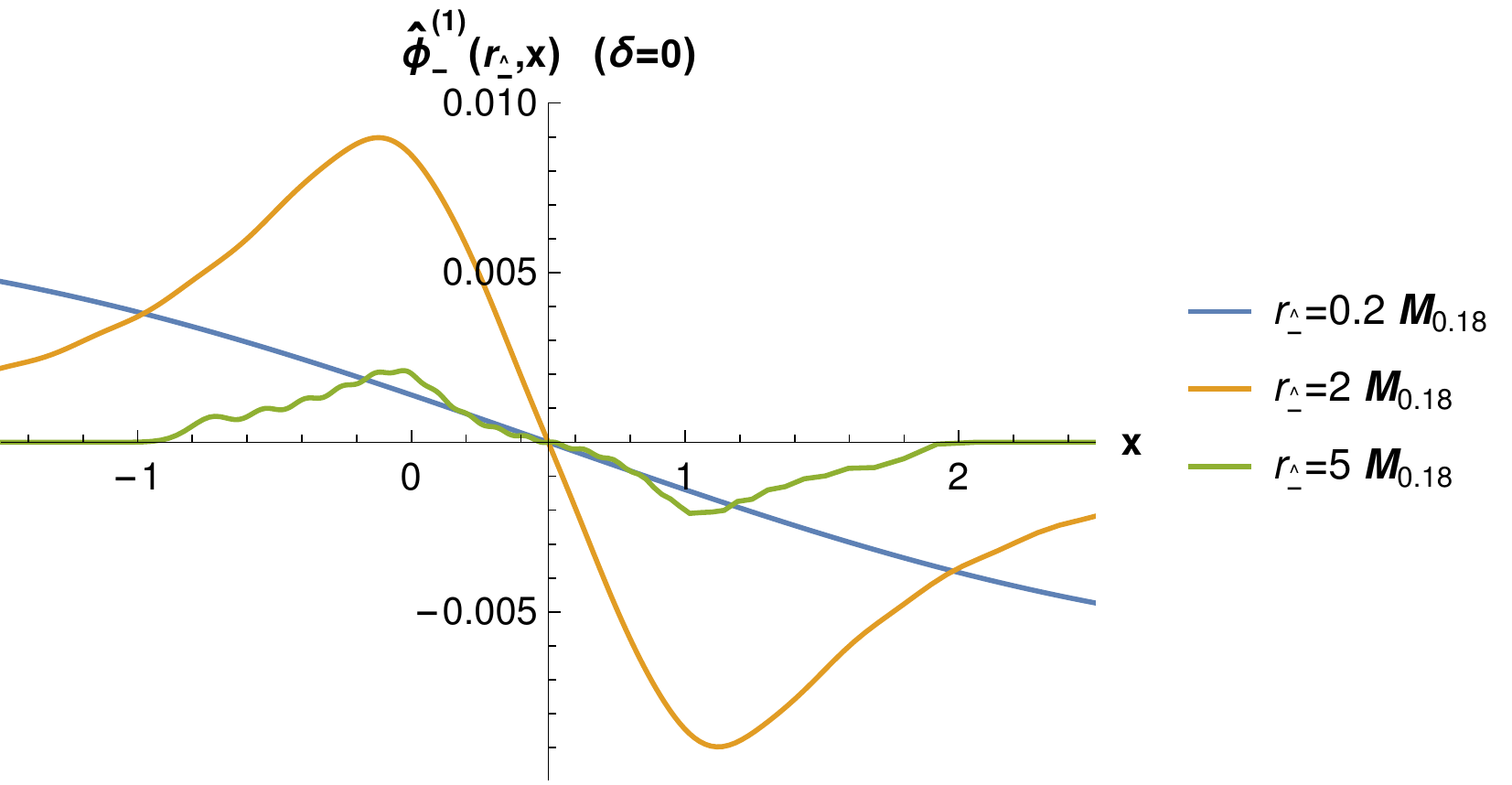}
		\label{fig:m0fefd0}
	}\\
	\centering
	\subfloat[]{
		\includegraphics[width=1.0\columnwidth]{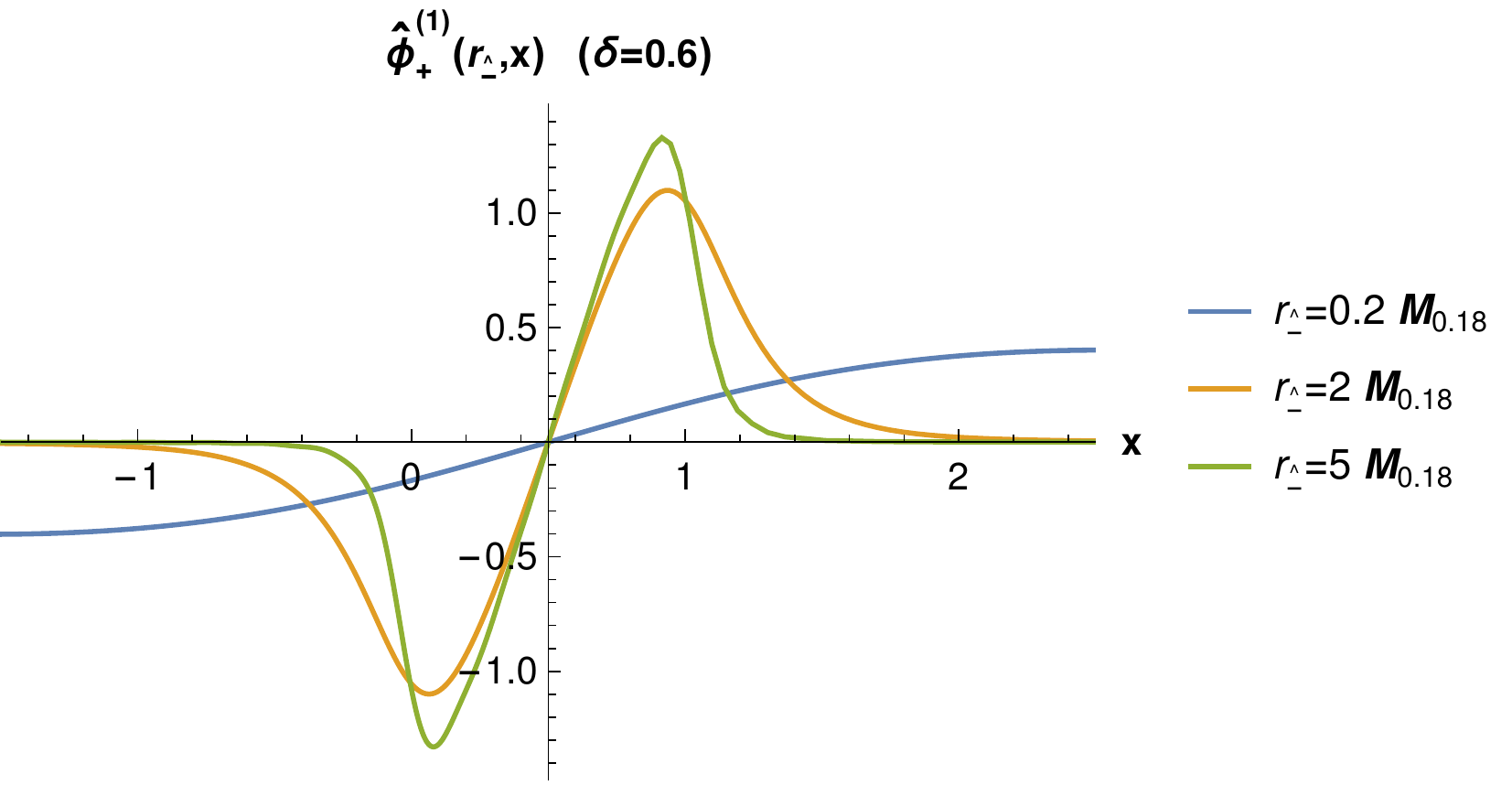}
		\label{fig:m0fezd06}
	}
	\centering
	\subfloat[]{
		\includegraphics[width=1.0\columnwidth]{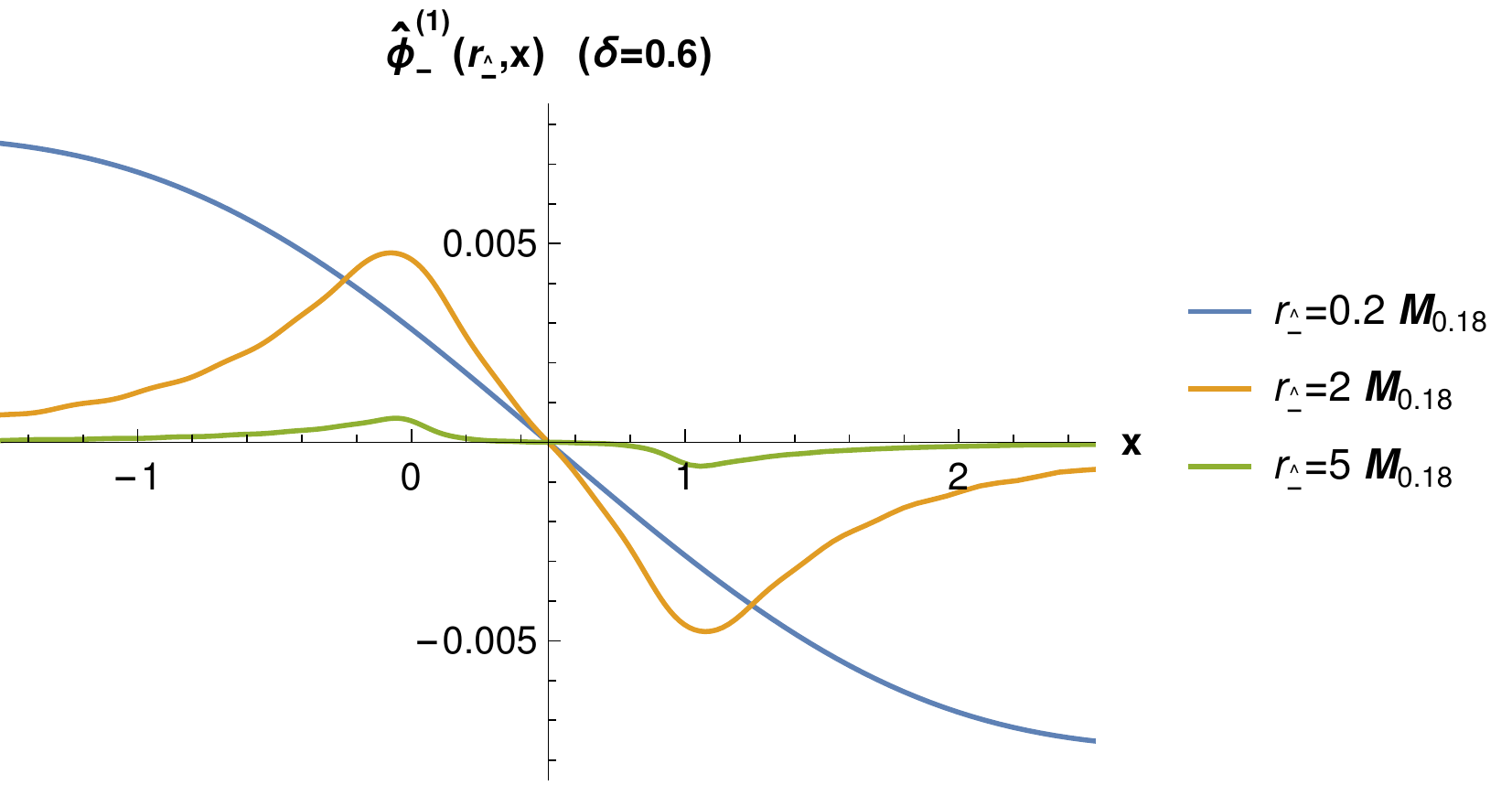}	\label{fig:m0fefd06}
	}\\
	\centering
	\subfloat[]{
		\includegraphics[width=1.0\columnwidth]{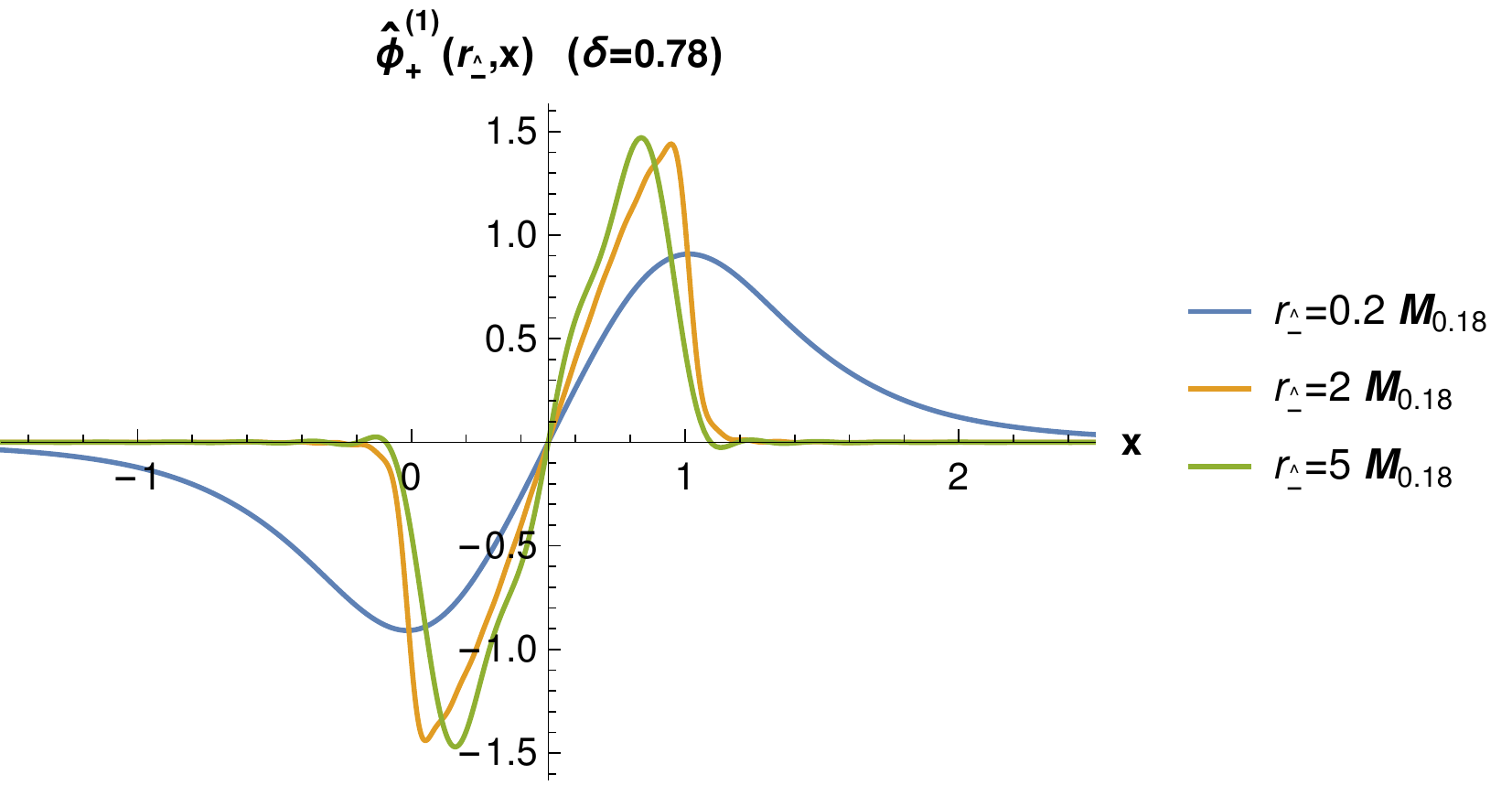}
		\label{fig:m0fezd078}
	}
	\centering
	\subfloat[]{
		\includegraphics[width=1.0\columnwidth]{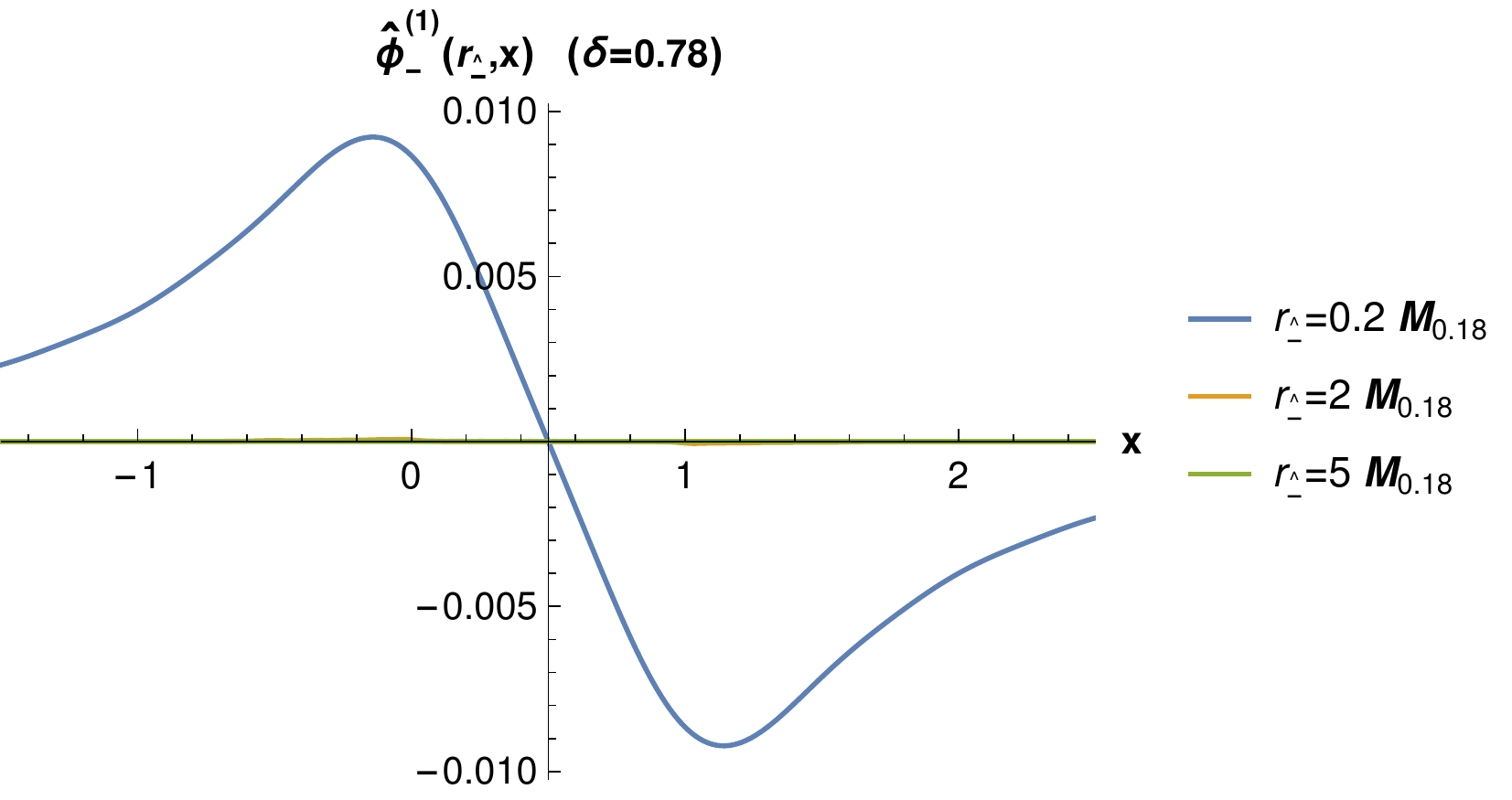}
		\label{fig:m0fefd078}
	}\\
	\caption{First excited state wave functions $\hat\phi_+^{(1)}(r_{\hat{-}},x)$ and  $\hat\phi_-^{(1)}(r_{\hat{-}},x)$ for $ m=0 $. All quantities are in proper units of $ \sqrt{2\lambda} $.\label{fig:m0fez}}
\end{figure*}

\begin{figure*}
	\centering
	\subfloat[]{
		\includegraphics[width=1.0\columnwidth]{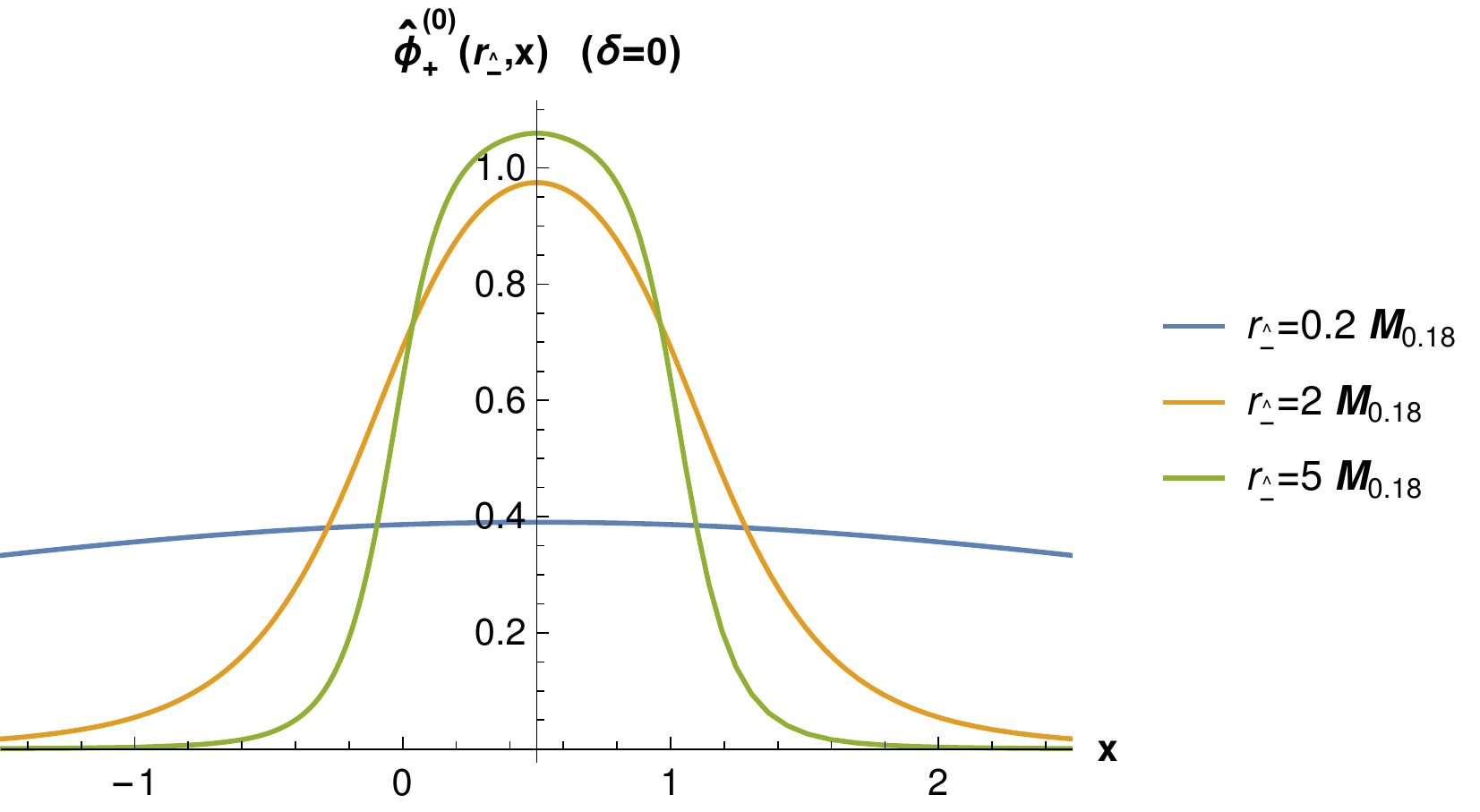}
		\label{fig:m018grzd0}
	}
	\centering
	\subfloat[]{
		\includegraphics[width=1.0\columnwidth]{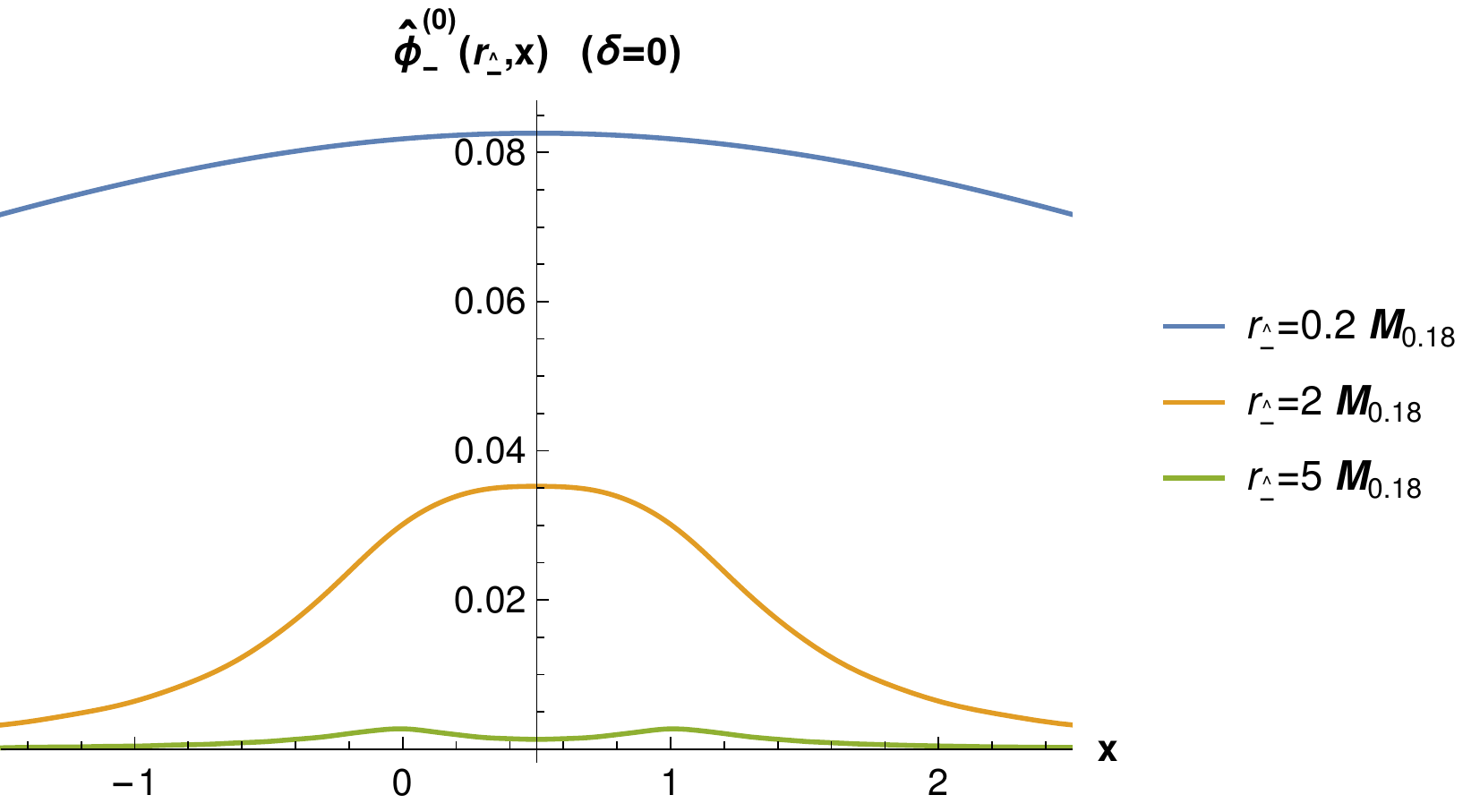}
		\label{fig:m018grfd0}
	}\\
	\centering
	\subfloat[]{
		\includegraphics[width=1.0\columnwidth]{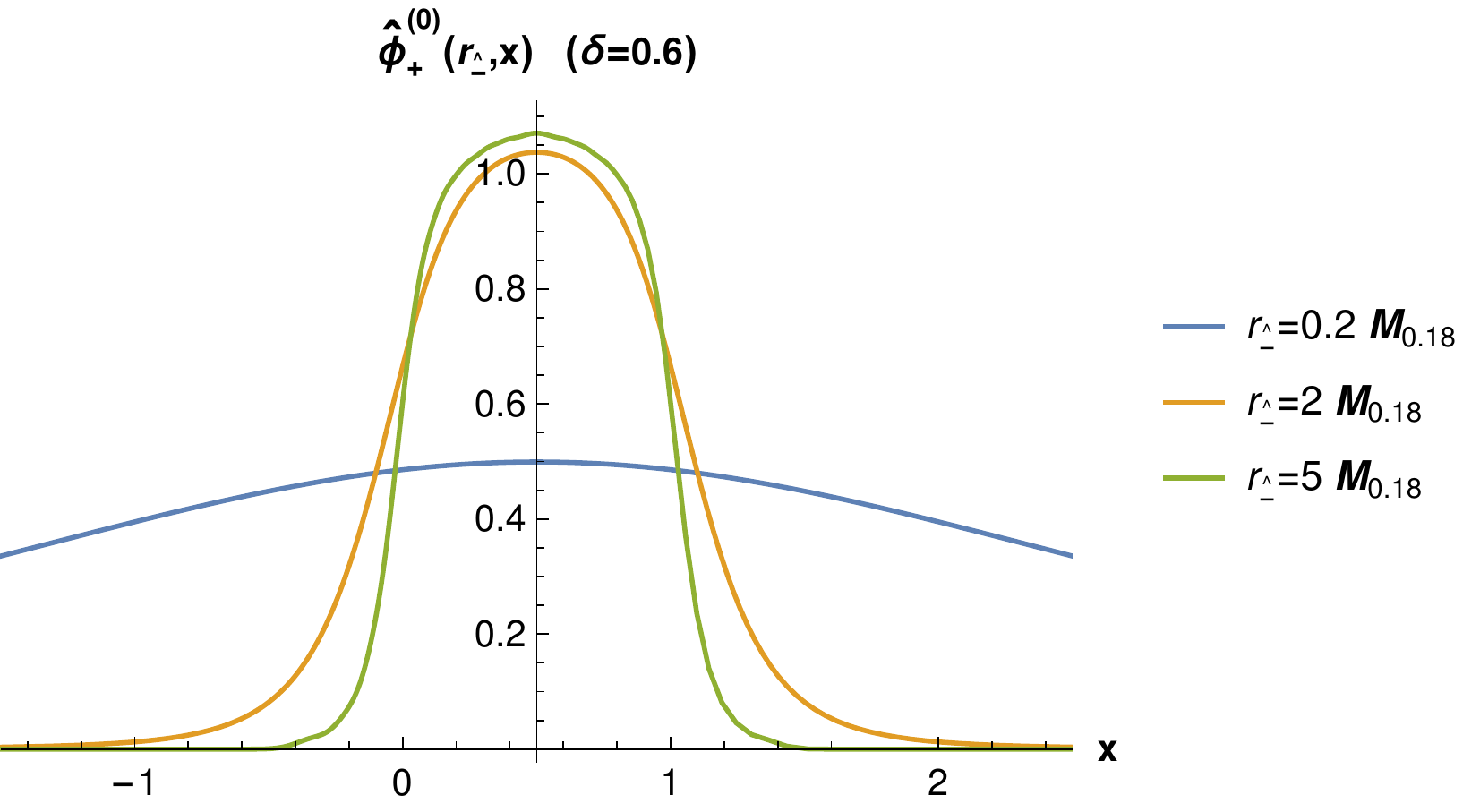}
		\label{fig:m018grzd06}
	}
	\centering
	\subfloat[]{
		\includegraphics[width=1.0\columnwidth]{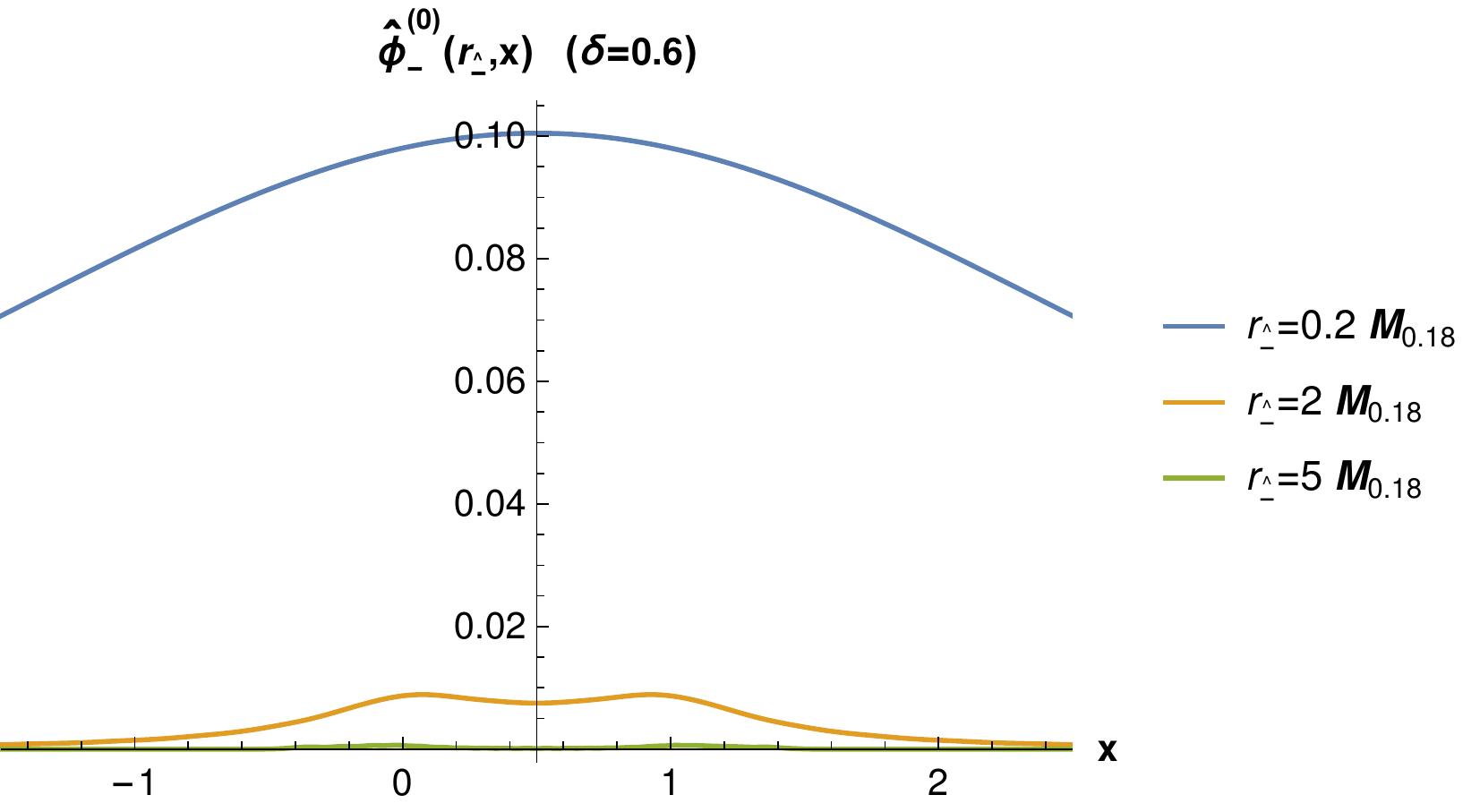}
		\label{fig:m018grfd06}
	}\\
	\centering
	\subfloat[]{
		\includegraphics[width=1.0\columnwidth]{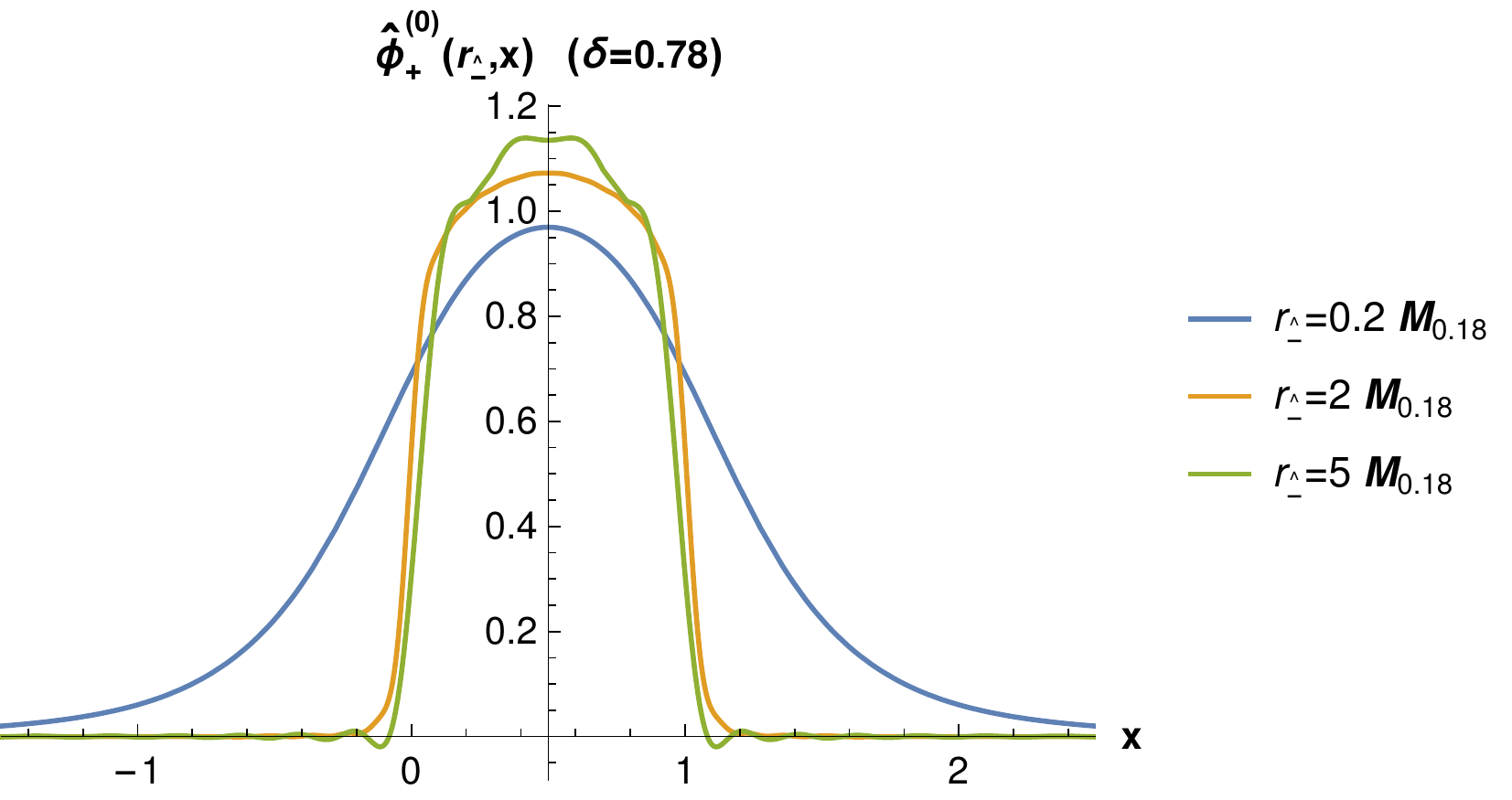}
		\label{fig:m018grzd078}
	}
	\centering
	\subfloat[]{
		\includegraphics[width=1.0\columnwidth]{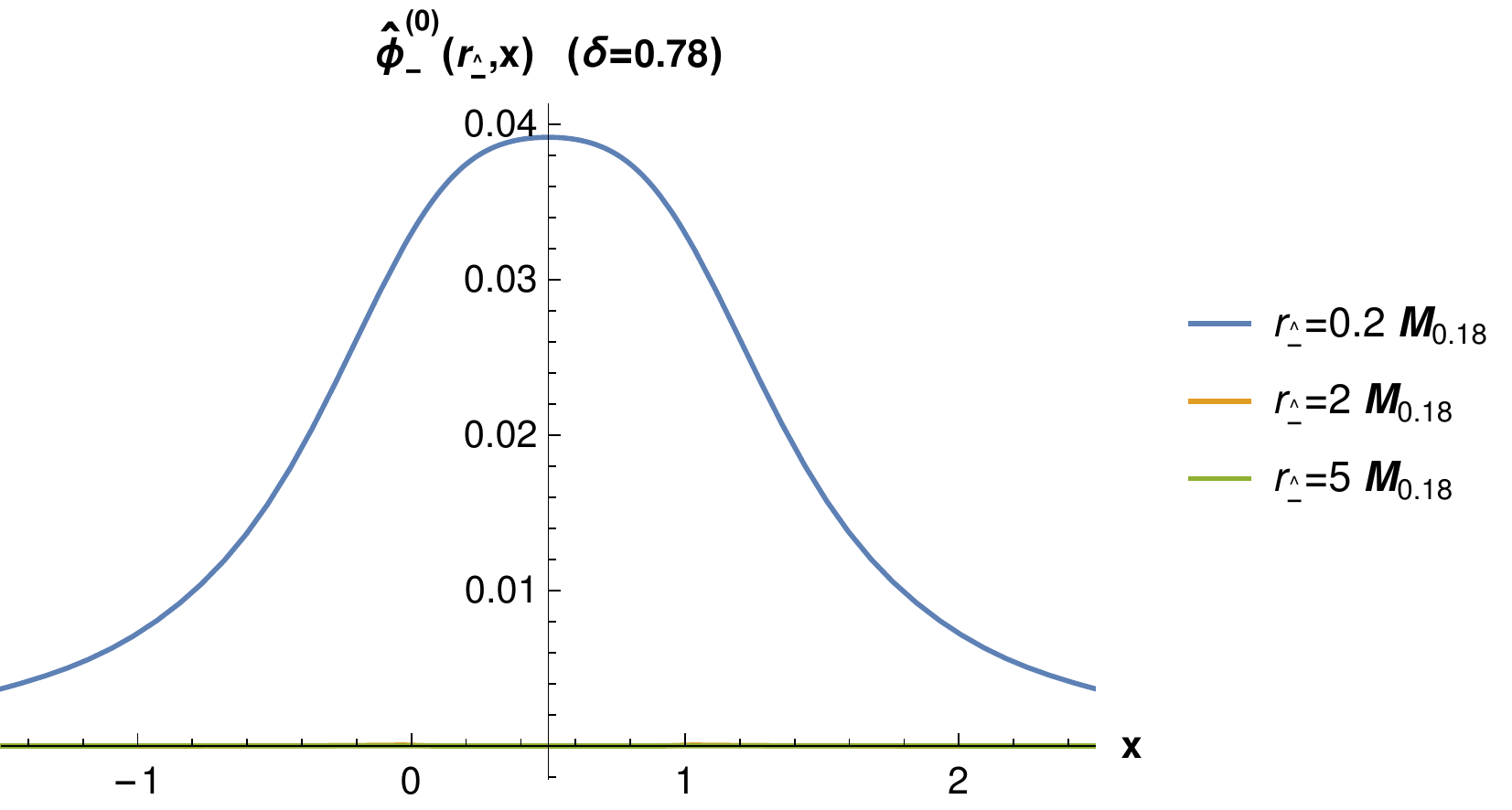}
		\label{fig:m018grfd078}
	}\\
	\caption{Ground state wave functions $\hat\phi_+^{(0)}(r_{\hat{-}},x)$ and $\hat\phi_-^{(0)}(r_{\hat{-}},x)$ for $ m=0.18 $. All quantities are in proper units of $ \sqrt{2\lambda} $.\label{fig:m018grz}}
\end{figure*}

\begin{figure*}
	\centering
	\subfloat[]{
		\includegraphics[width=1.0\columnwidth]{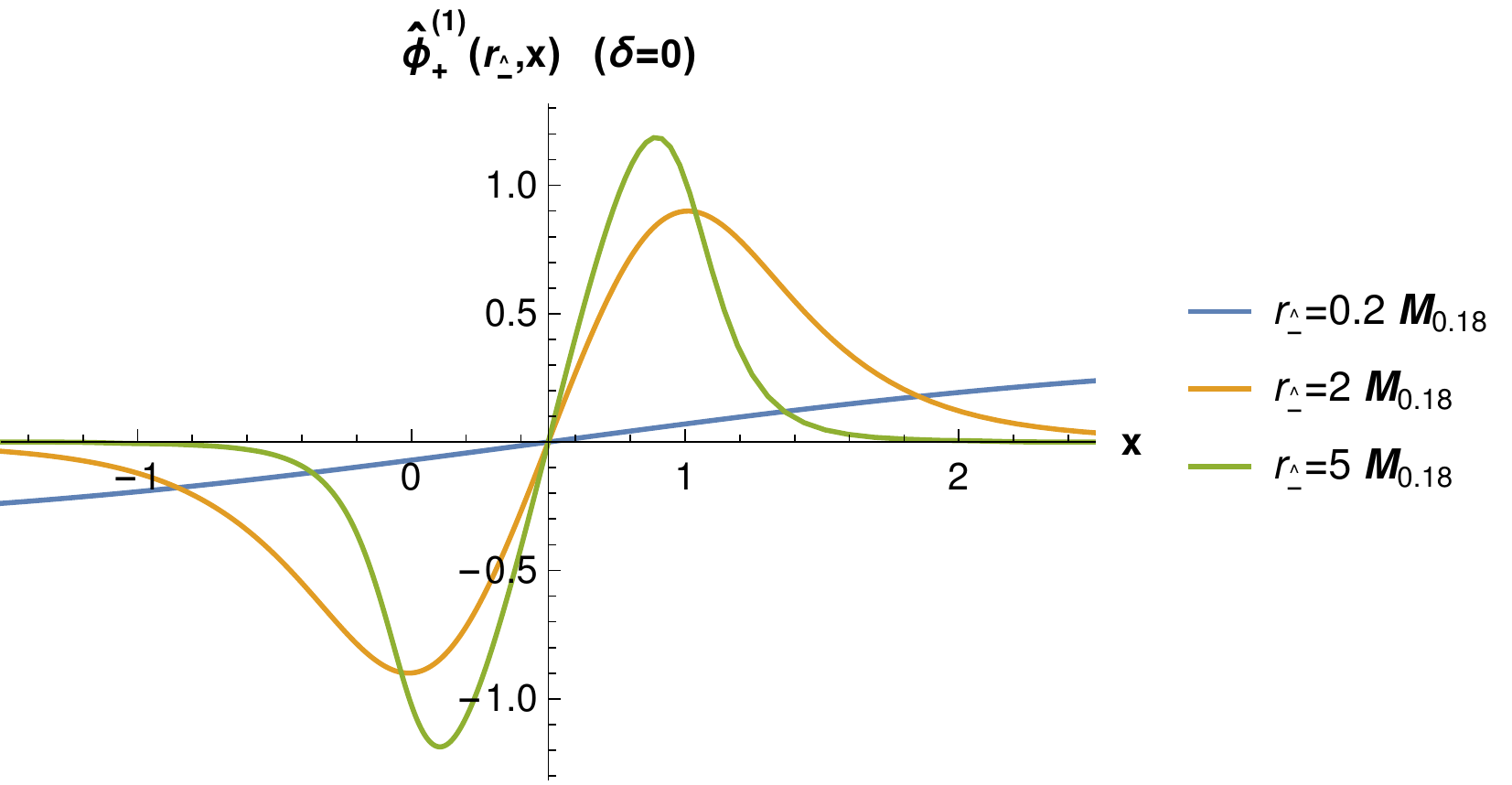}
		\label{fig:m018fezd0}
	}
	\centering
	\subfloat[]{
		\includegraphics[width=1.0\columnwidth]{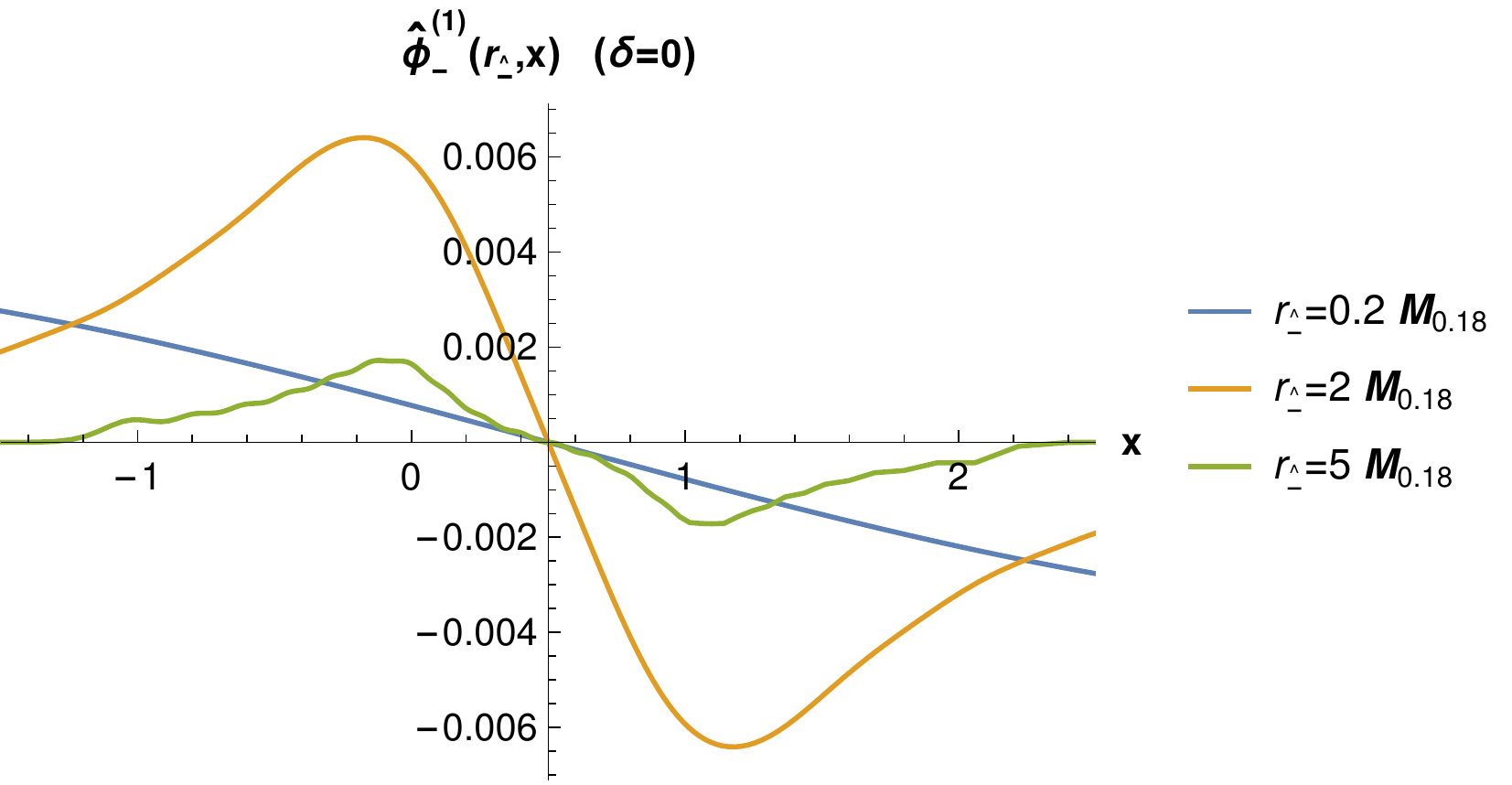}
		\label{fig:m018fefd0}
	}\\
	\centering
	\subfloat[]{
		\includegraphics[width=1.0\columnwidth]{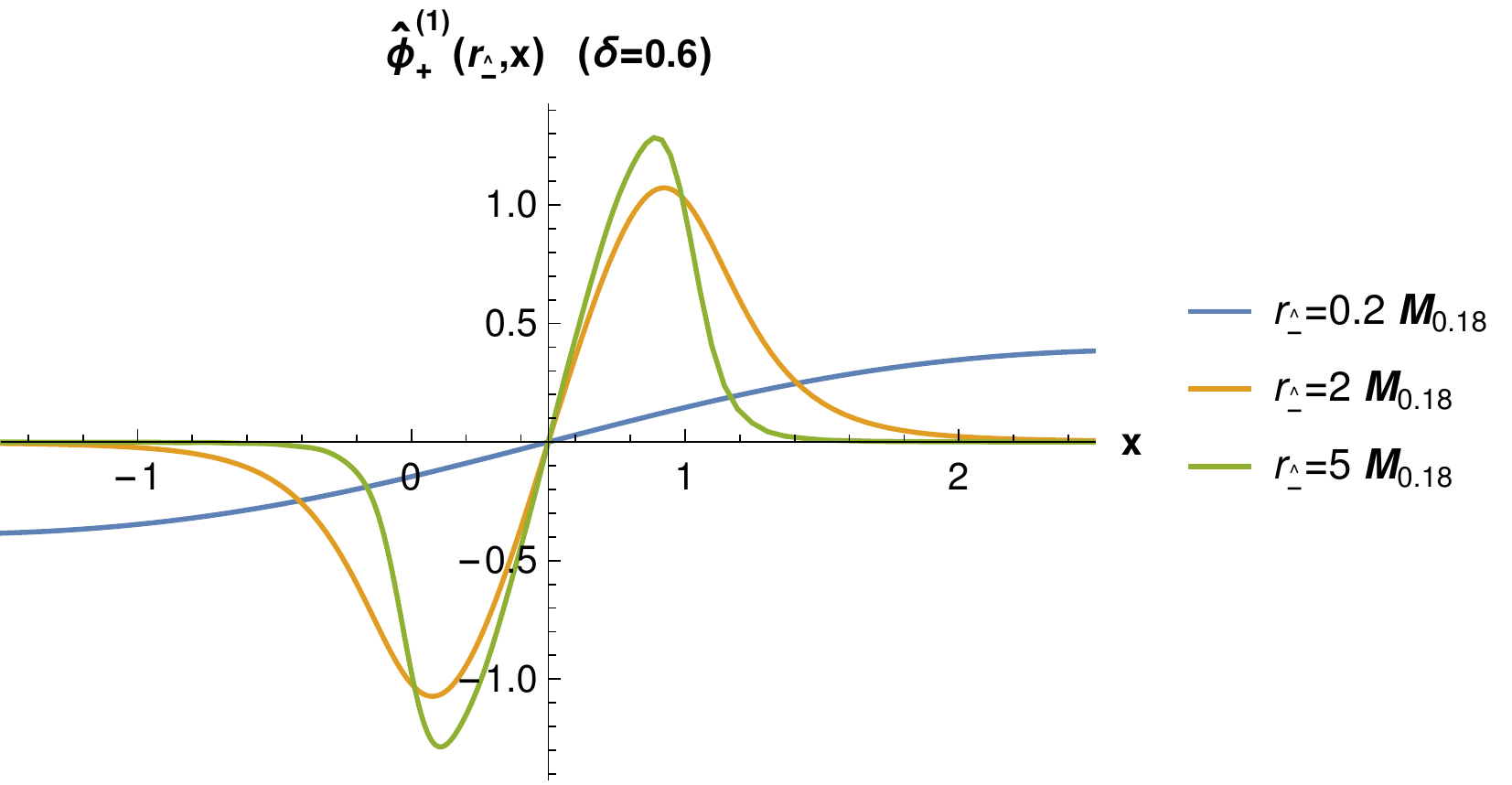}
		\label{fig:m018fezd06}
	}
	\centering
	\subfloat[]{
		\includegraphics[width=1.0\columnwidth]{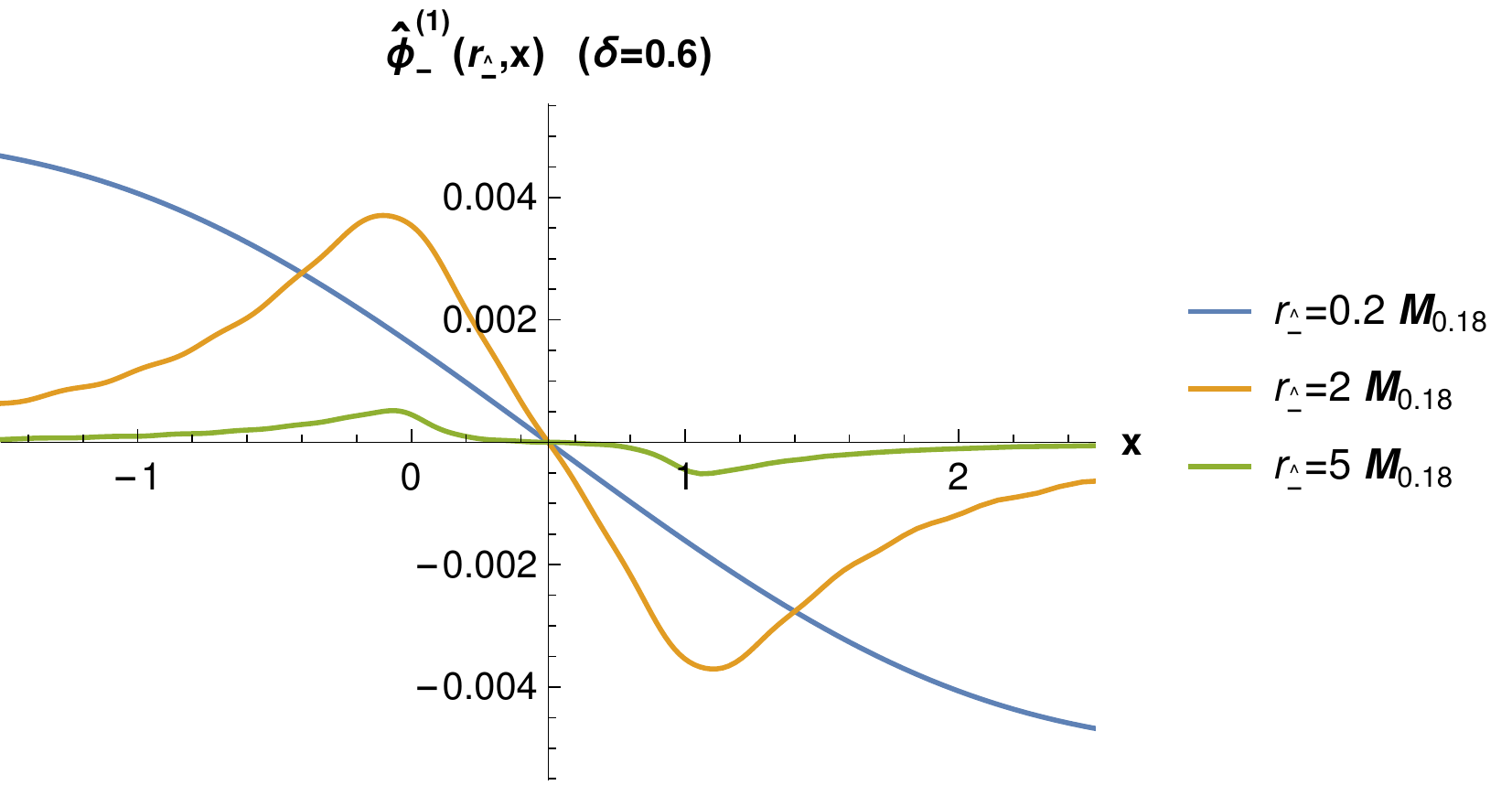}
		\label{fig:m018fefd06}
	}\\
	\centering
	\subfloat[]{
		\includegraphics[width=1.0\columnwidth]{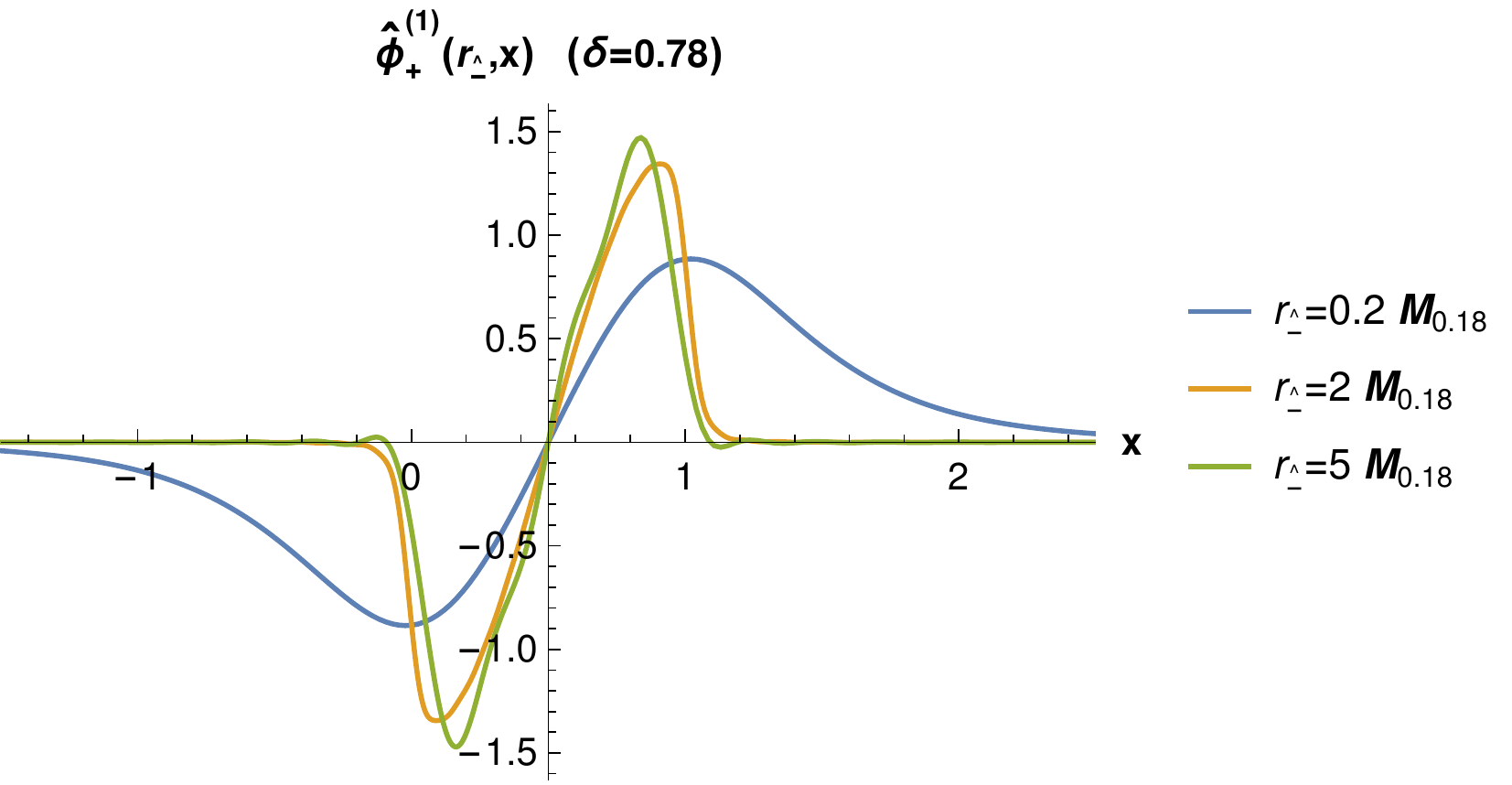}
		\label{fig:m018fezd078}
	}
	\centering
	\subfloat[]{
		\includegraphics[width=1.0\columnwidth]{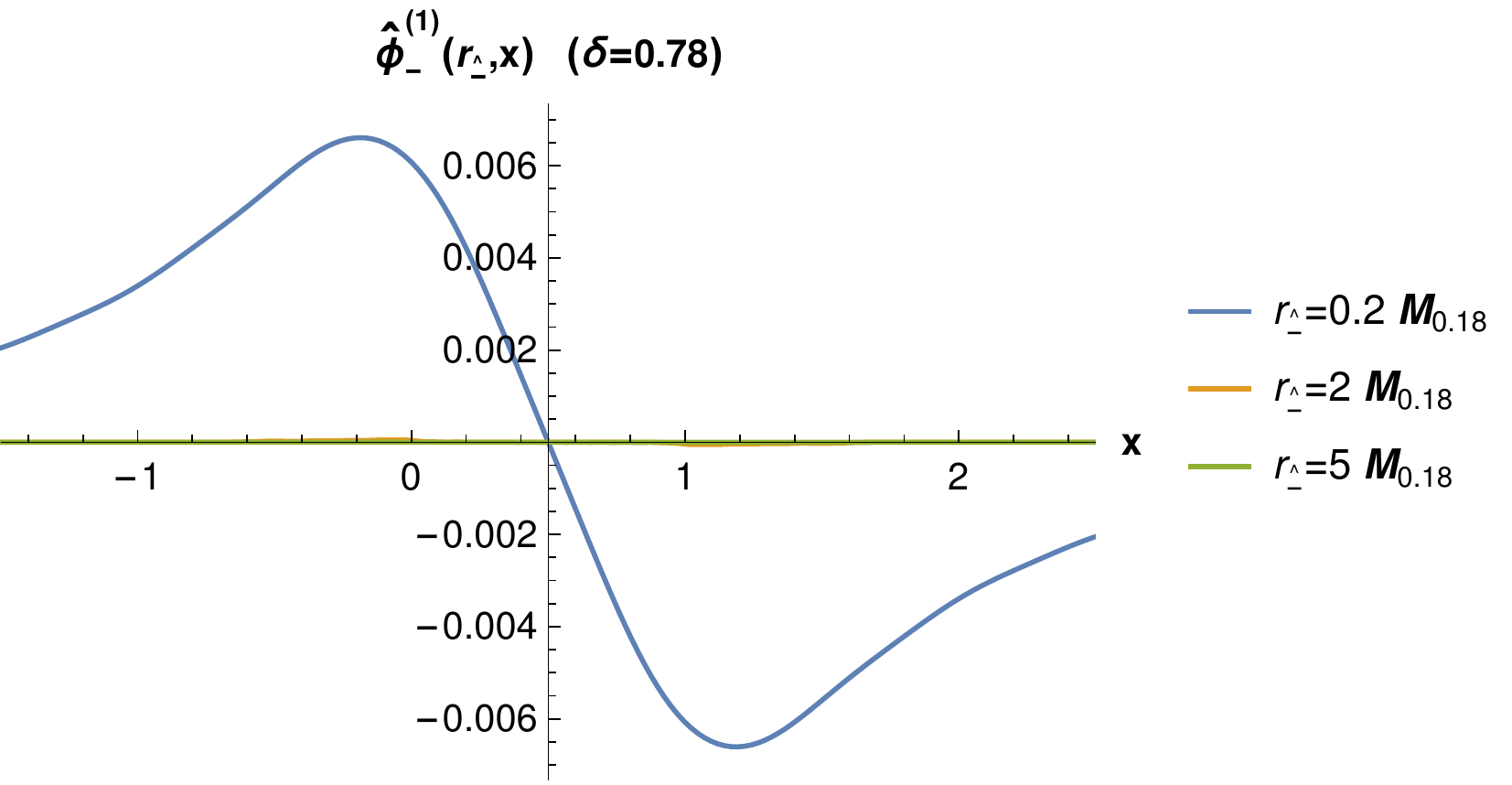}
		\label{fig:m018fefd078}
	}\\
	\caption{First excited state wave functions $\hat\phi_+^{(1)}(r_{\hat{-}},x)$ and $\hat\phi_-^{(1)}(r_{\hat{-}},x)$ for $ m=0.18 $. All quantities are in proper units of $ \sqrt{2\lambda} $.\label{fig:m018fez}}
\end{figure*}
We present here our numerical solutions of the bound-state wavefunctions 
$\hat\phi_{\pm}^{(n)}(r_{\hat{-}},x)$ interpolating between IFD and LFD for 
the ground state $n=0$ and the first excited state $n=1$ with $r_{\hat{-}} \neq 0$
in terms of the interpolating longitudinal momentum fraction variable $x=p_{\hat{-}}/r_{\hat{-}}$. The results for $r_{\hat{-}} = 0$ are presented separately in Appendix~\ref{app:rest} in terms of 
the variable $\xi = \tan^{-1}(p_{\hat{-}})$ without scaling the interpolating momentum variable $p_{\hat{-}}$ with respect to $r_{\hat{-}}$. In the chiral limit, where
the GOR relation ${\cal M}_{(0)}^2 \sim m\sqrt{\lambda} \to 0$ is satisfied, 
the analytic solution of the pionic ground-state wavefunction $\phi_{\pm}$ in IFD, i.e.
$\hat\phi_{\pm}^{(0)}$ for $\delta =0$, is known~\cite{review} in terms of the IFD longitudinal momentum variables $p^1$ and $r^1$ of the quark and the pionic meson, respectively.
Corresponding the IFD longitudinal momentum $p^1$ and $r^1$
to the interpolating longitudinal momentum $p_{\hat{-}}$ and $r_{\hat{-}}$
and confirming the consistency with the LFD analytic mass gap solution
discussed in Sec.~\ref{sub:behaviorLF}, we note that the corresponding analytic solution in the interpolating formulation is given by
\begin{align}
\label{pionic-wavefunction}
{\hat{\phi}}^{(0)}_{\pm}(r_{\hat{-}}, p_{\hat{-}}) &= \frac{1}{2}\left( \cos\frac{\theta(r_{\hat{-}}-p_{\hat{-}}) -\theta(p_{\hat{-}})}{2} \right.\notag\\
&\left.  \ \ \ \pm \sin\frac{\theta(r_{\hat{-}}-p_{\hat{-}}) +\theta(p_{\hat{-}})}{2} \right) ,
\end{align} 
where the normalization is taken to satisfy Eq.(\ref{norm})~\cite{mov}.
In Fig.~\ref{fig:m0grz}, our numerical results of the plus 
and minus components of the bound-state wavefunction for the ground state,
i.e. $\hat\phi_{\pm}^{(0)}(r_{\hat{-}},x)$, are shown for the bare quark mass value $m=0$ in comparison with the interpolating analytic solution given by Eq.(\ref{pionic-wavefunction}). 
The results of $\delta=0, 0.6$ and 0.78 are shown in the top, middle and bottom panels, respectively.
In each panel, the results of $r_{\hat{-}}=0.2 \mathbf{M}_{0.18}, 2 \mathbf{M}_{0.18}$ and $5 \mathbf{M}_{0.18}$ 
with the scale of the ground-state meson mass $\mathbf{M}_{0.18}=0.88$, i.e. 
$r_{\hat{-}}=0.176, 1.76$ and $4.4$ (all in units of $\sqrt{2\lambda}$), are depicted by the solid lines for the analytic results and the dashed lines for the numerical results in blue, yellow, and green, respectively.
Our numerical results coincide with the analytic results given by Eq.(\ref{pionic-wavefunction}) as shown in Fig.~\ref{fig:m0grz} except for some wiggle and bulge in the numerical result of $\delta = 0.78$ and $r_{\hat{-}}=5\mathbf{M}_{0.18}$ in Fig.~\ref{fig:m0grzd078} due to the numerical sensitivity
near the LFD ($\mathbb{C}\to 0$) and large longitudinal momentum $r_{\hat{-}}$. 
Our results in Fig.~\ref{fig:m0grz} appear to confirm the validity of our numerical results as well as the analytic results. 
As the longitudinal momentum 
$r_{\hat{-}}$, i.e. $r^1$ for $\delta=0$, gets large, the numerical results of  
$\hat\phi_{+}^{(0)}(r_{\hat{-}},x)$ approach to $\phi(x) = 1$ for $x\in [0,1]$, 
which is the solution of the 't Hooft equation given by Eq.~(\ref{boundeqlf}) in LFD~\cite{tHooft,Brower},
while $\hat\phi_{-}^{(0)}(r_{\hat{-}},x)$ results tend to diminish although for very small momentum, e.g., $r^1=0.2 \mathbf{M}_{0.18}$, it is still of comparable order of magnitude to $\hat\phi_{+}^{(0)}(r_{\hat{-}},x)$ as noted also in Ref.~\cite{mov}. 
Ref.~\cite{pdf} also noted that for light mesons the large-momentum IFD numerical results approach the exact light-front solution very slowly. While 
the LFD solution for $m=0$ exhibits an infinite slope at the endpoints $x=0,1$,
such feature is not achieved in the IFD large momentum method ~\cite{mov,pdf}. 
As $\delta$ gets closer to $\pi/4$, however, the resemblance to the LFD solutions is attained 
even in the smaller longitudinal momentum (e.g. $r_{\hat{-}}=2 \mathbf{M}_{0.18}$)
and thus the large $r_{\hat{-}}$ (e.g. $r_{\hat{-}}=5 \mathbf{M}_{0.18}$)
does not need to be taken for the confirmation of the LFD solutions.
The similar behavior of resemblance to the LFD results
depending on the values of $\delta$ and $r_{\hat{-}}$
is also found in the first excited states $\hat\phi_{+}^{(1)}(r_{\hat{-}},x)$ and $\hat\phi_{-}^{(1)}(r_{\hat{-}},x)$ shown in Fig.~\ref{fig:m0fez}. 
As dictated by the charge conjugation symmetry, under the exchange of $x \leftrightarrow 1-x$, $\hat\phi_{+}^{(1)}(r_{\hat{-}},x)$ and $\hat\phi_{-}^{(1)}(r_{\hat{-}},x)$ are antisymmetric while $\hat\phi_{+}^{(0)}(r_{\hat{-}},x)$ and $\hat\phi_{-}^{(0)}(r_{\hat{-}},x)$ are symmetric.
We note also that our results for $\delta=0$ shown
in the top panels of Figs.~\ref{fig:m0grz}-\ref{fig:m0fez}, i.e. the plots of Figs.~\ref{fig:m0grzd0},\ref{fig:m0grfd0},\ref{fig:m0fezd0},\ref{fig:m0fefd0} appear to be consistent with the results in Ref.~\cite{mov} although only a qualitative comparison can be made as different momentum values are taken for the moving frames in Ref.~\cite{mov} compared to what we present here. 

In contrast to the case of $m=0$, there are no known analytic solutions
of the bound-state wavefunctions for $m\neq 0$. 
While we present the numerical results of $\hat\phi_{\pm}^{(0)}(r_{\hat{-}},x)$ and $\hat\phi_{\pm}^{(1)}(r_{\hat{-}},x)$ for the cases of $m= 0.045, 1.0$ and 2.11 in Appendix~\ref{app:figures}, we take here  $m=0.18$ to correspond with the spectroscopy discussion in the last subsection (Sec.~\ref{sub:spec}) and show its numerical results of $\hat\phi_{\pm}^{(n)}(r_{\hat{-}},x)$ for the ground
state $n=0$ and the first excited state $n=1$.  
In Fig.~\ref{fig:m018grz}, the numerical results of 
$\hat\phi_{+}^{(0)}(r_{\hat{-}},x)$ and $\hat\phi_{-}^{(0)}(r_{\hat{-}},x)$  for $\delta = 0, 0.6$ and $0.78$ are shown in the 
top, middle and bottom panels, respectively. 
Similarly, in Fig.~\ref{fig:m018fez}, the numerical results of $\hat\phi_{+}^{(1)}(r_{\hat{-}},x)$ and $\hat\phi_{-}^{(1)}(r_{\hat{-}},x)$  for $\delta = 0, 0.6$ and $0.78$ are shown in the 
top, middle and bottom panels, respectively. 
In each panel, the results of $r_{\hat{-}}=0.2 \mathbf{M}_{0.18}, 2 \mathbf{M}_{0.18}$ and $5 \mathbf{M}_{0.18}$ 
with the scale of $\mathbf{M}_{0.18}=0.88$, i.e. 
$r_{\hat{-}}=0.176, 1.76$ and $4.4$ (all in units of $\sqrt{2\lambda}$), are depicted by the solid lines in blue, yellow, and green, respectively.
As noted for the case of $m=0$, the large-momentum IFD ($\delta=0$) numerical results approach the LFD results very slowly also for the case of $m=0.18$. While the LFD results ($\hat\phi_{+}^{(n)}(r_{\hat{-}},x)=\phi^{(n)}(x)$) should be constrained in the $x$-region $[0,1]$, the IFD results in the top left panel of
Fig.~\ref{fig:m018grz}, i.e. Fig.~\ref{fig:m018grzd0}, exhibit rather long tails outside the $[0,1]$ region even for the pretty large longitudinal momentum $r^1 = 5 \mathbf{M}_{0.18} = 4.4$.   
For $\delta=0.78$, however, i.e., very close to the LFD ($\pi/4\approx 0.785398$),
shown in Fig.~\ref{fig:m018grzd078},
the wavefunctions for the relatively larger momenta
$r_{\hat{-}}=2\mathbf{M}_{0.18}=1.76$ and $5\mathbf{M}_{0.18} =4.4$ almost coincide with each other while closely fitting in the region $[0,1]$ although the result with very small longitudinal momentum ($r_{\hat{-}}=0.2\mathbf{M}_{0.18}=0.176$) has a long tail
outside the $[0,1]$ region similar to the IFD ($\delta=0$) result. 
While we notice some wiggle and bulge in $\hat\phi_{+}^{(0)}(r_{\hat{-}},x) (\delta=0.78)$ for $r_{\hat{-}} = 5 \mathbf{M}_{0.18}$, we didn't pursue any further
numerical accuracy as we understand that it is due to the computational sensitivity arising in the interpolation region where $\mathbb{C}$ gets close to $0$ in particular
as $r_{\hat{-}}$ gets very large.
Since the LFD results of $\hat\phi_{-}^{(n)}(r_{\hat{-}},x)$ must vanish as discussed
in the derivation of the 't Hooft's bound-state equation given by Eq.~(\ref{boundeqlf}),
it is manifest in Fig.~\ref{fig:m018grfd0} that 
the large-momentum IFD ($\delta=0$) results approach to the LFD ($\delta=\pi/4$) result again very slow for $\hat\phi_{-}^{(0)}(r_{\hat{-}}=r^1,x)$ while the 
$\delta=0.78$ results in Fig.~\ref{fig:m018grfd078} are rather immediately 
close to the LFD result.
Fig.~\ref{fig:m018fez} shows the first excited state wavefunctions $\hat\phi_{+}^{(1)}(r_{\hat{-}},x)$ and $\hat\phi_{-}^{(1)}(r_{\hat{-}},x)$ for the input bare quark mass value $m=0.18$. The results for $m=0.18$ in Fig.~\ref{fig:m018fez} look quite similar to the results for $m=0$ in Fig.~\ref{fig:m0fez}. 
They share the same feature of the charge conjugation symmetry, under the exchange of $x \leftrightarrow 1-x$, i.e. $\hat\phi_{+}^{(1)}(r_{\hat{-}},x)$ and $\hat\phi_{-}^{(1)}(r_{\hat{-}},x)$ are antisymmetric while $\hat\phi_{+}^{(0)}(r_{\hat{-}},x)$ and $\hat\phi_{-}^{(0)}(r_{\hat{-}},x)$ are symmetric. They also share the similar behavior of resemblance to the LFD results depending on the values of $\delta$ and $r_{\hat{-}}$, which we have discussed for the ground state previously.

\subsection{\label{sub:quasipdf}Quasi-PDFs}
\begin{figure*}
	\centering
	\subfloat[]{
		\includegraphics[width=0.5\linewidth]{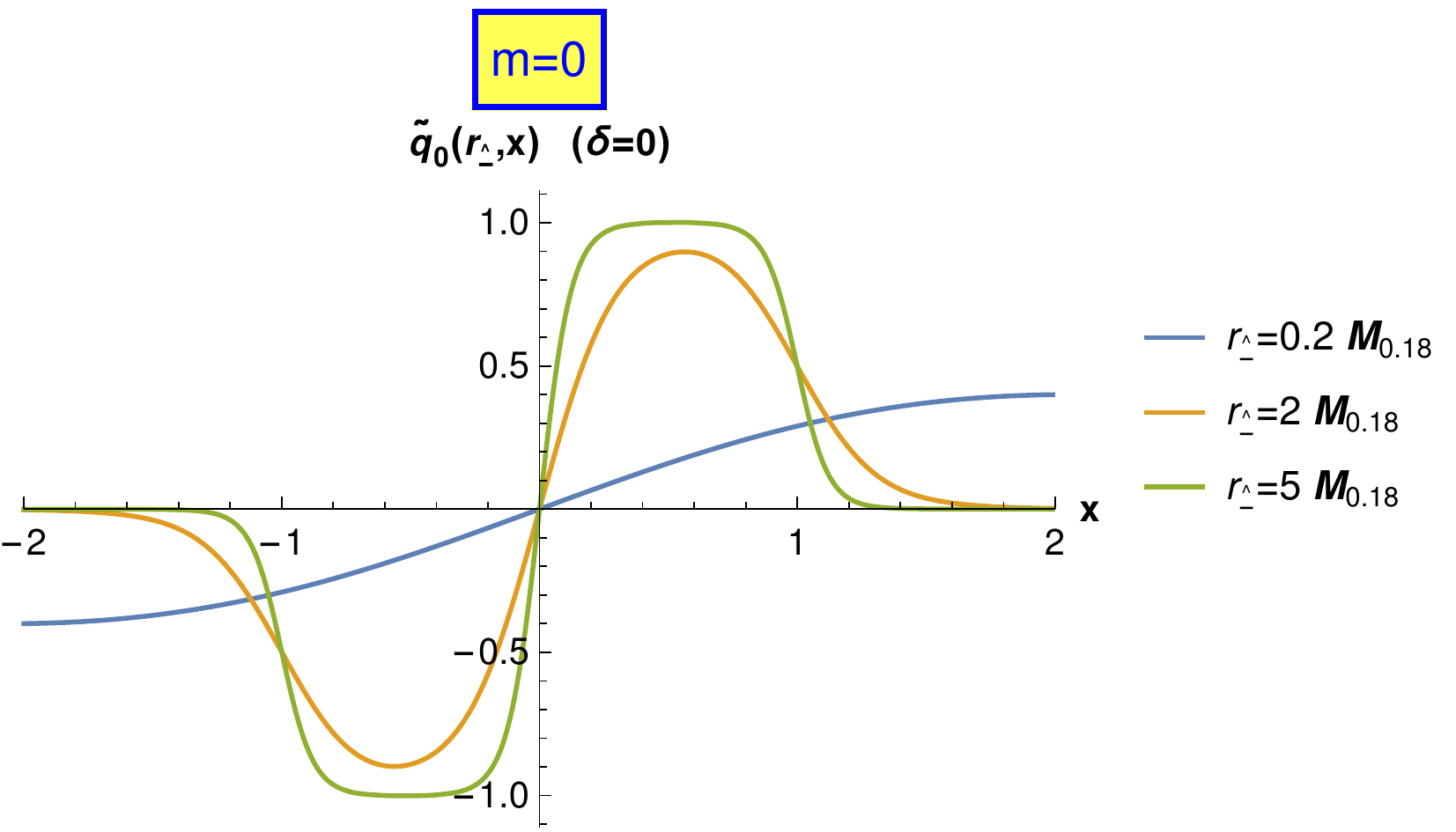}
		\label{fig:m0_delta0_grquasipdf}
	}
	\centering
	\subfloat[]{
		\includegraphics[width=0.5\linewidth]{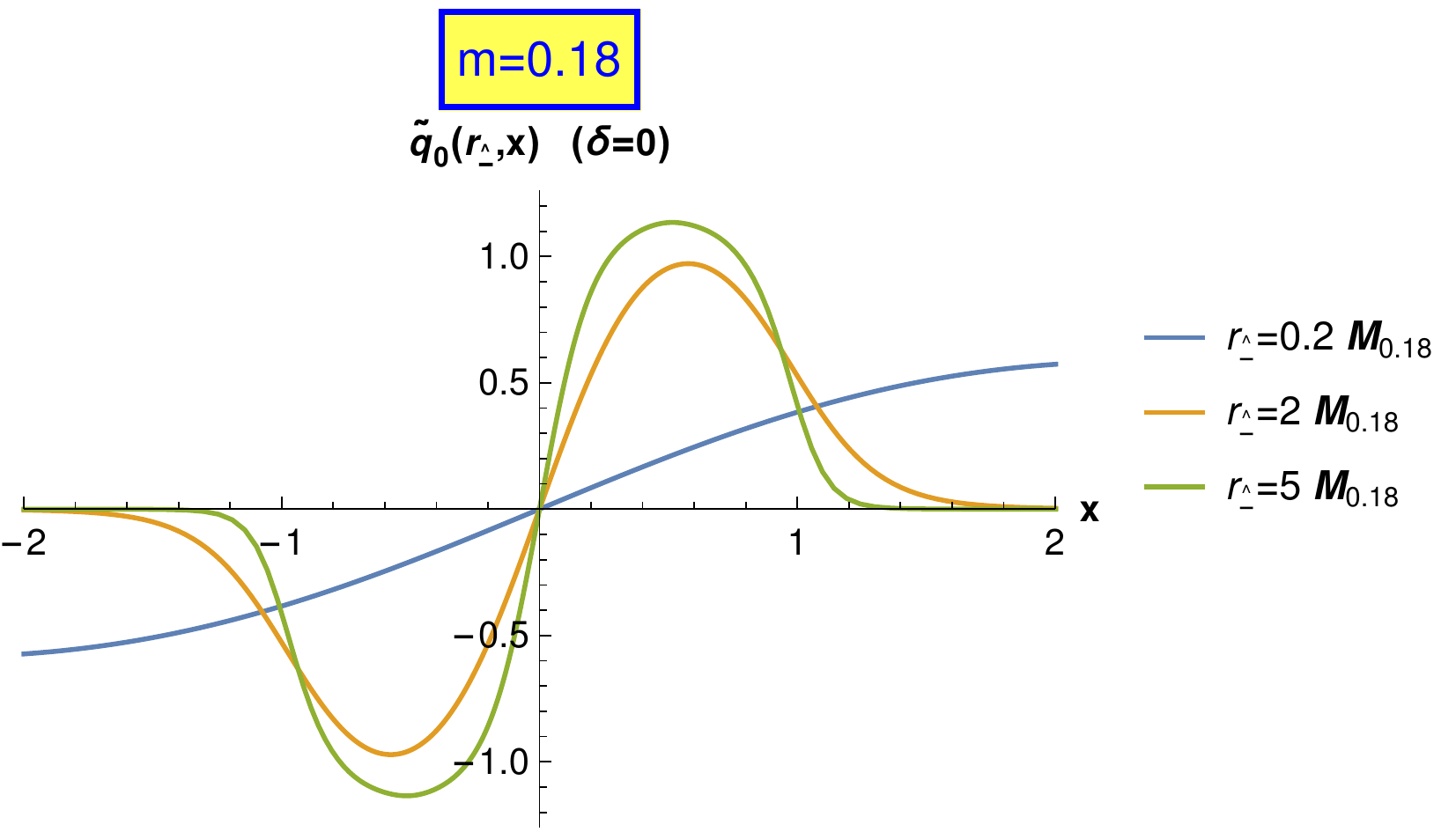}
		\label{fig:m018_delta0_grquasipdf}
	}
	\caption{Quasi-PDFs in IFD ($\delta =0$) for the ground state ($n=0$) wave functions of (a) $m=0$, (b) $m=0.18$
		. All quantities are in proper units of $ \sqrt{2\lambda} $.\label{fig:delta0_grquasipdf}}
\end{figure*}

\begin{figure*}
	\centering
	\subfloat[]{
		\includegraphics[width=0.5\linewidth]{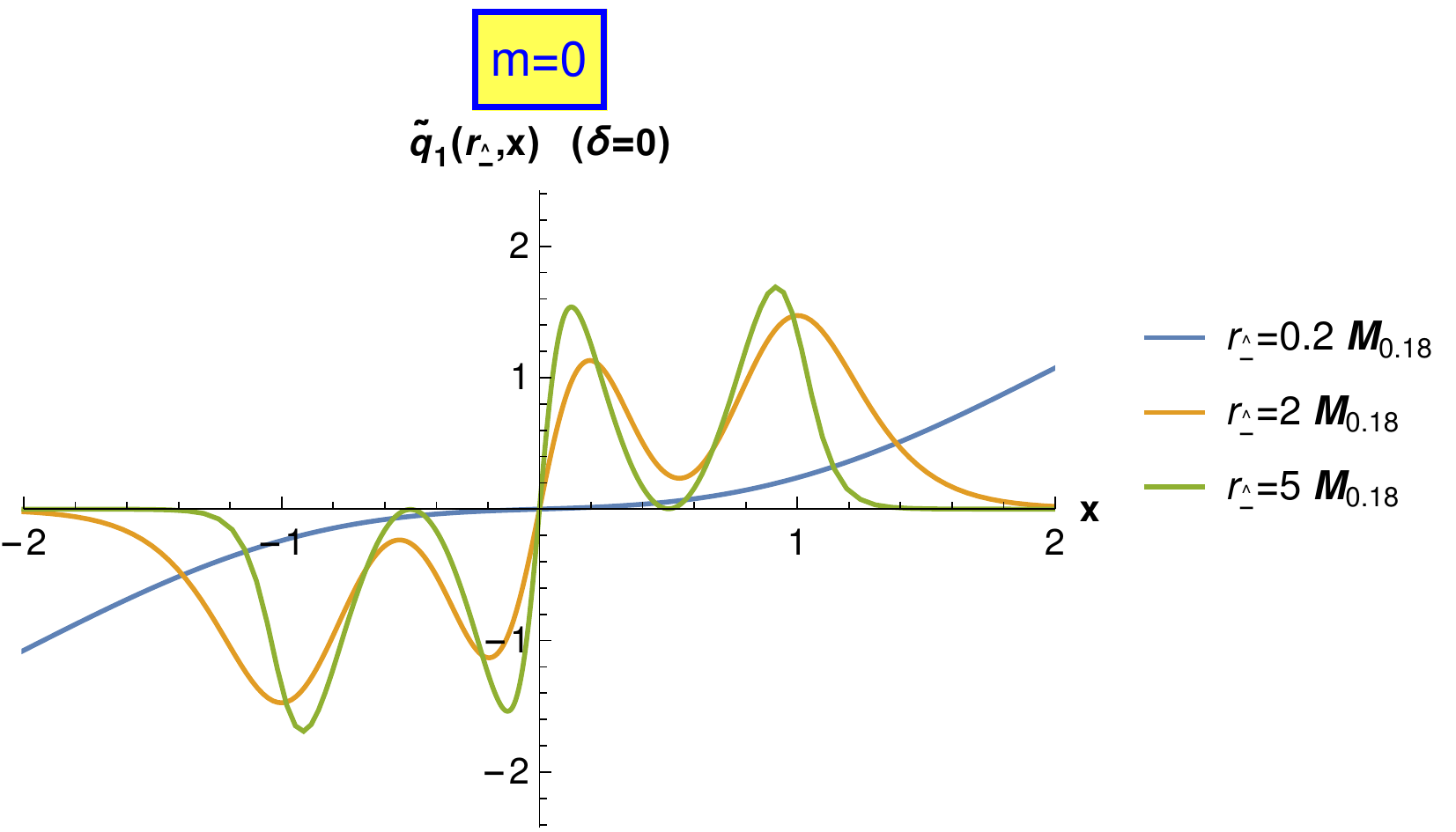}
		\label{fig:m0_delta0_fequasipdf}
	}
	\centering
	\subfloat[]{
		\includegraphics[width=0.5\linewidth]{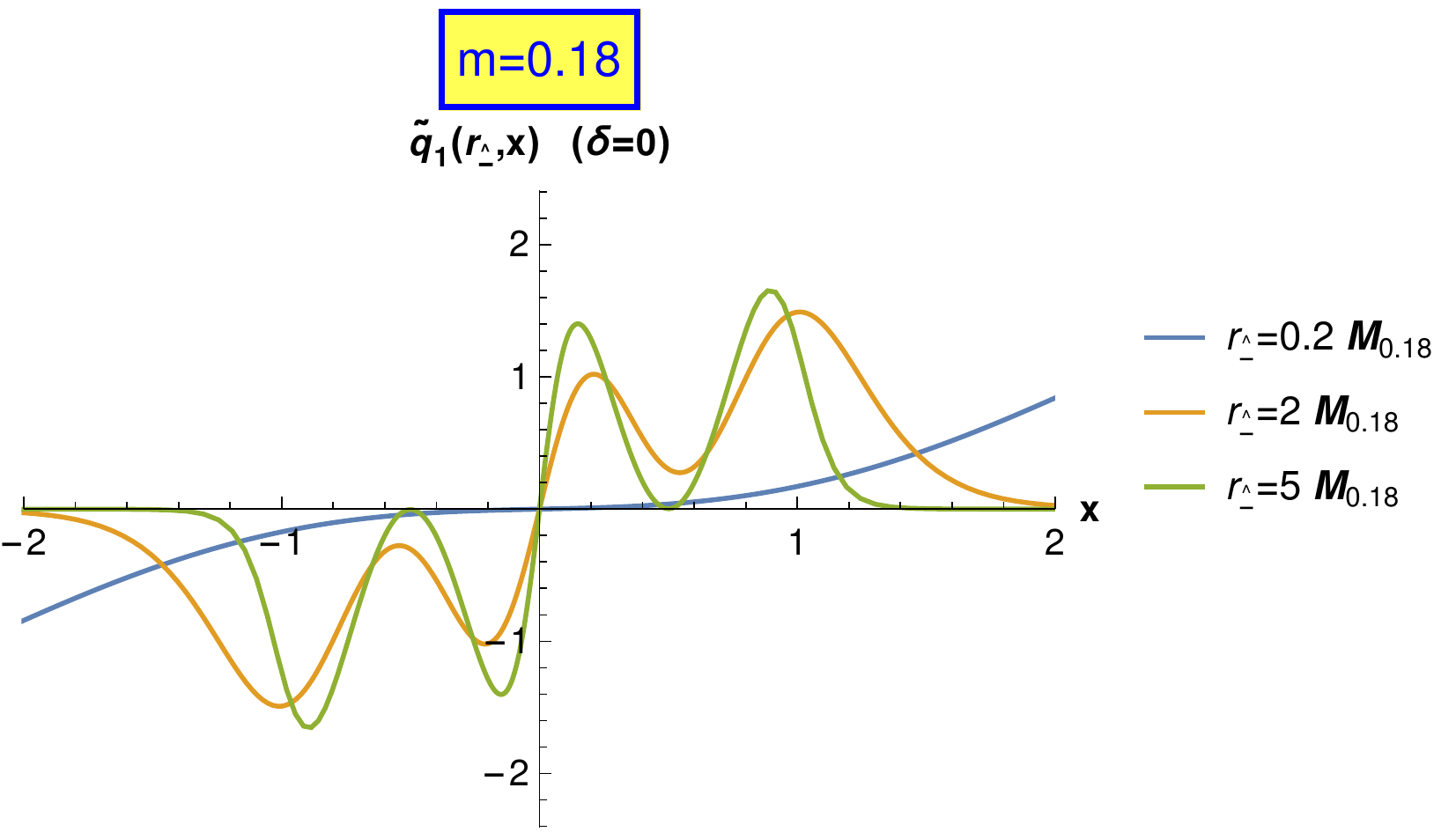}
		\label{fig:m018_delta0_fequasipdf}
	}
	\caption{Quasi-PDFs in IFD ($\delta=0$) for the first excited state ($n=1$) wave functions of (a) $m=0$, (b) $m=0.18$
		. All quantities are in proper units of $ \sqrt{2\lambda} $.\label{fig:delta0_fequasipdf}}
\end{figure*}

\begin{figure*}
	\centering
	\subfloat[]{
		\includegraphics[width=0.5\linewidth]{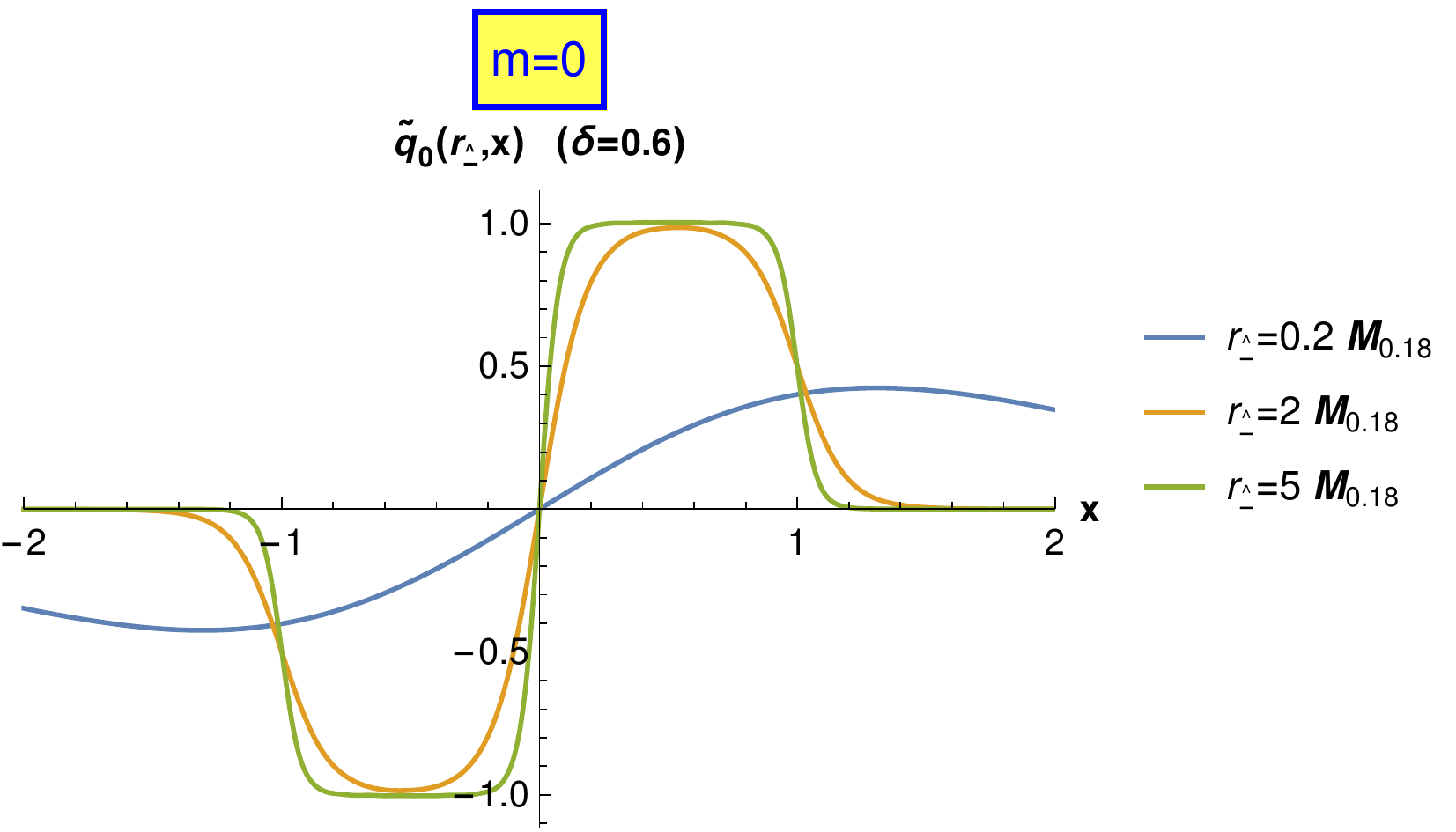}
		\label{fig:m0_delta06_grquasipdf}
	}
	\centering
	\subfloat[]{
		\includegraphics[width=0.5\linewidth]{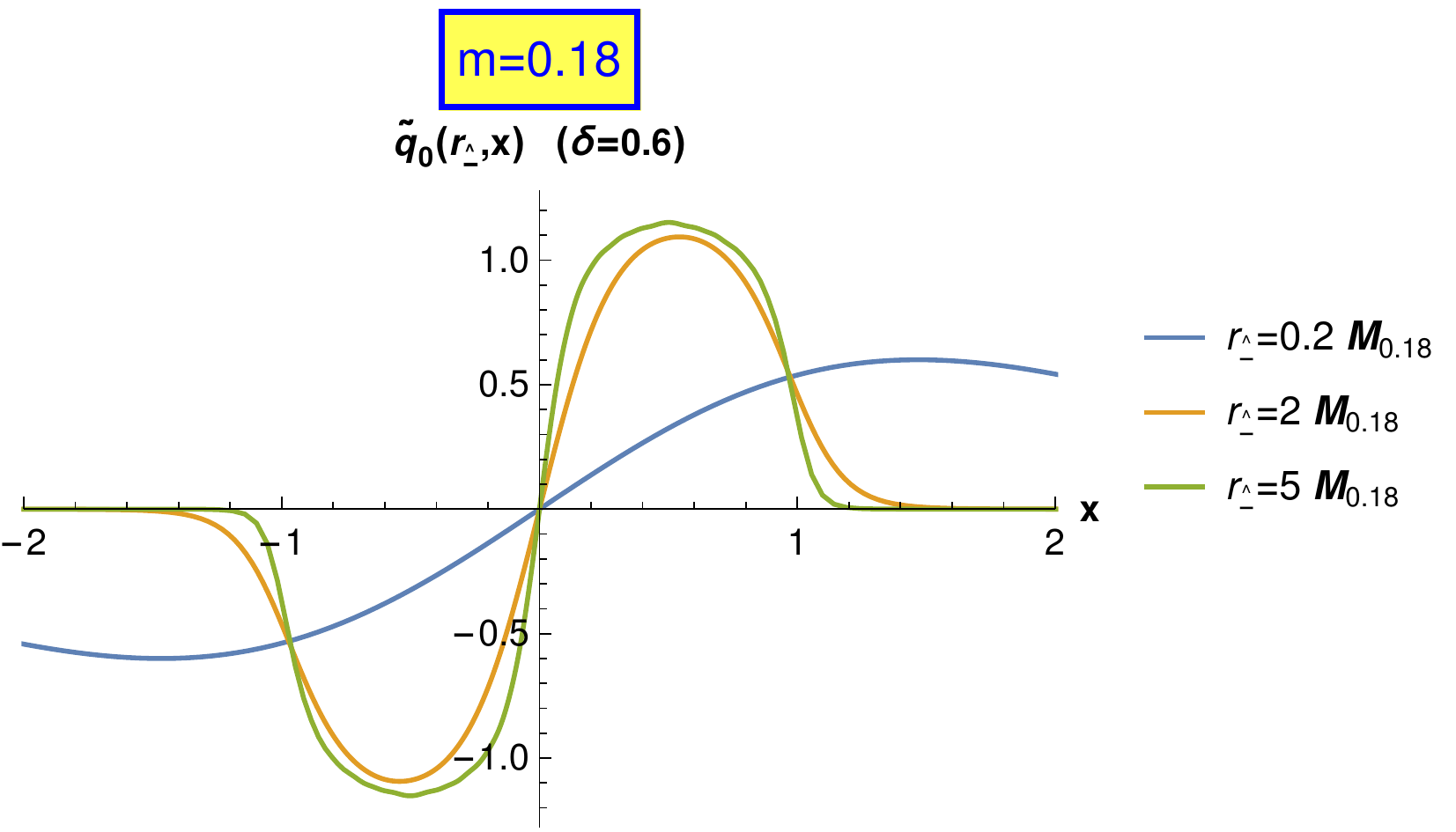}
		\label{fig:m018_delta06_grquasipdf}
	}
	\caption{$ \delta=0.6$ interpolating ``quasi-PDFs" for the ground state ($n=0$) wave functions of (a) $m=0$, (b) $m=0.18$
		. All quantities are in proper units of $ \sqrt{2\lambda} $.\label{fig:delta06_grquasipdf}}
\end{figure*}

\begin{figure*}
	\centering
	\subfloat[]{
		\includegraphics[width=0.5\linewidth]{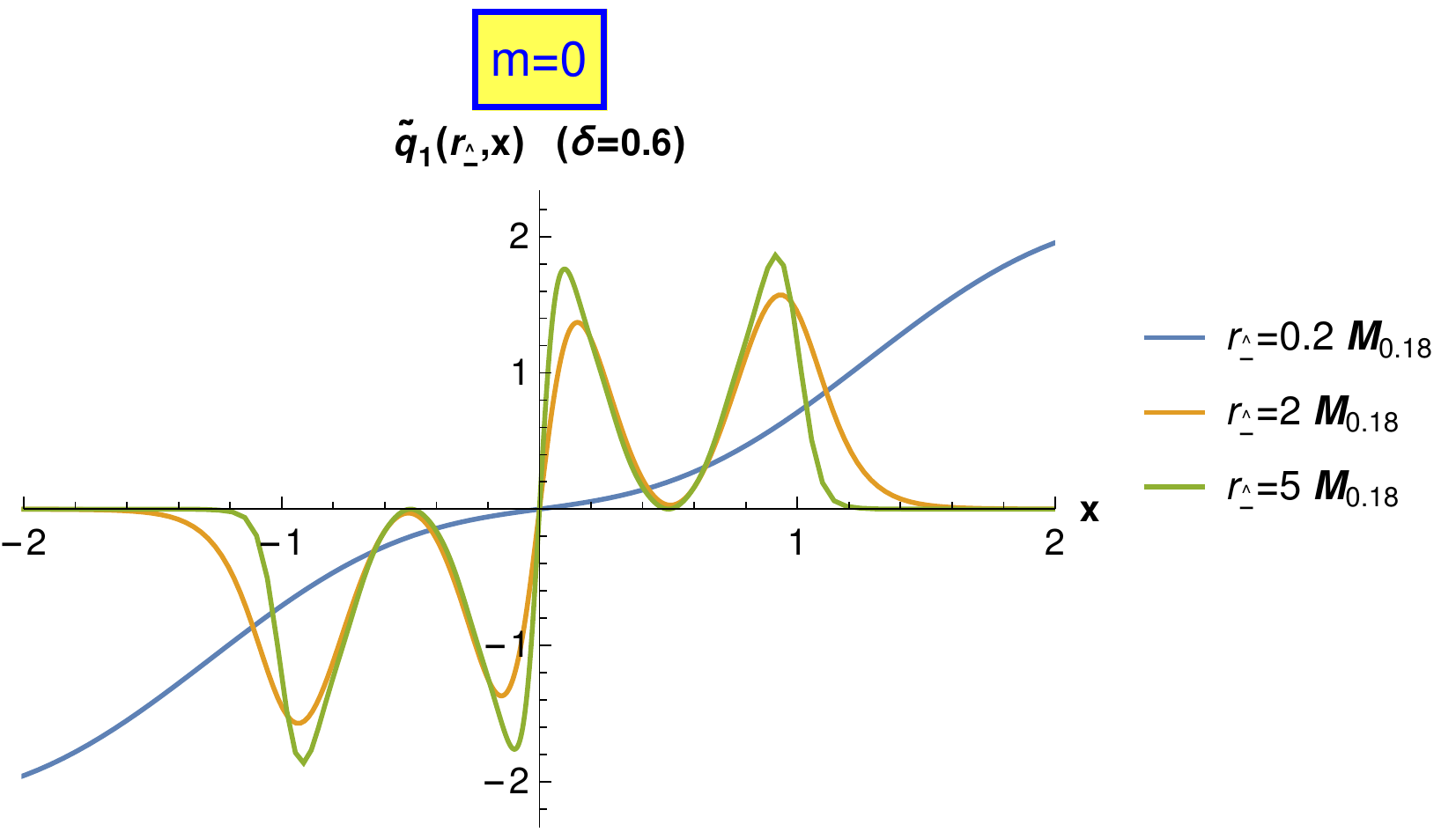}
		\label{fig:m0_delta06_fequasipdf}
	}
	\centering
	\subfloat[]{
		\includegraphics[width=0.5\linewidth]{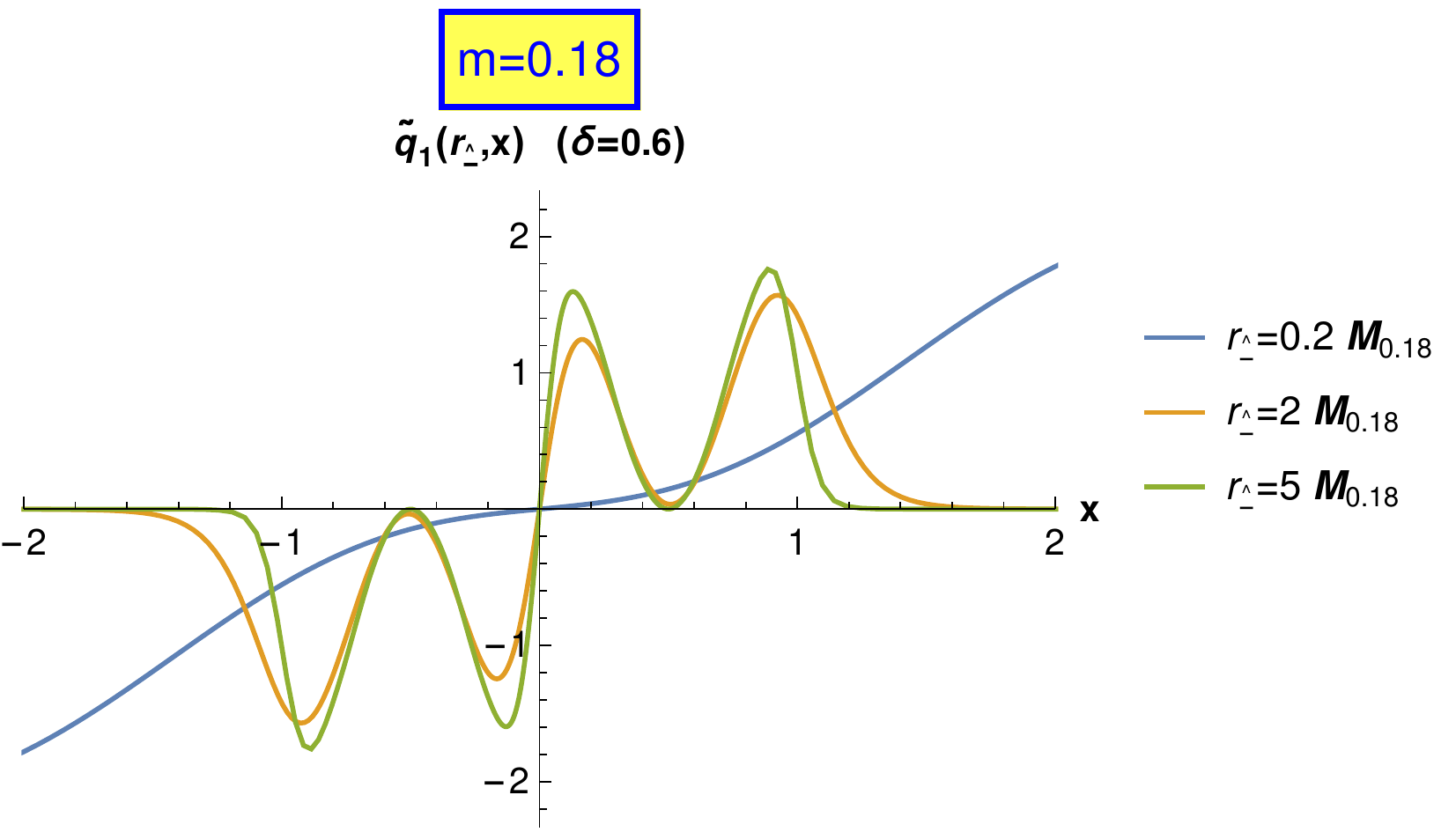}
		\label{fig:m018_delta06_fequasipdf}
	}
	\caption{$ \delta=0.6$ interpolating ``quasi-PDFs" for the first excited state ($n=1$) wave functions of (a) $m=0$, (b) $m=0.18$
		. All quantities are in proper units of $ \sqrt{2\lambda} $.\label{fig:delta06_fequasipdf}}
\end{figure*}

\begin{figure*}
	\centering
	\subfloat[]{
		\includegraphics[width=0.5\linewidth]{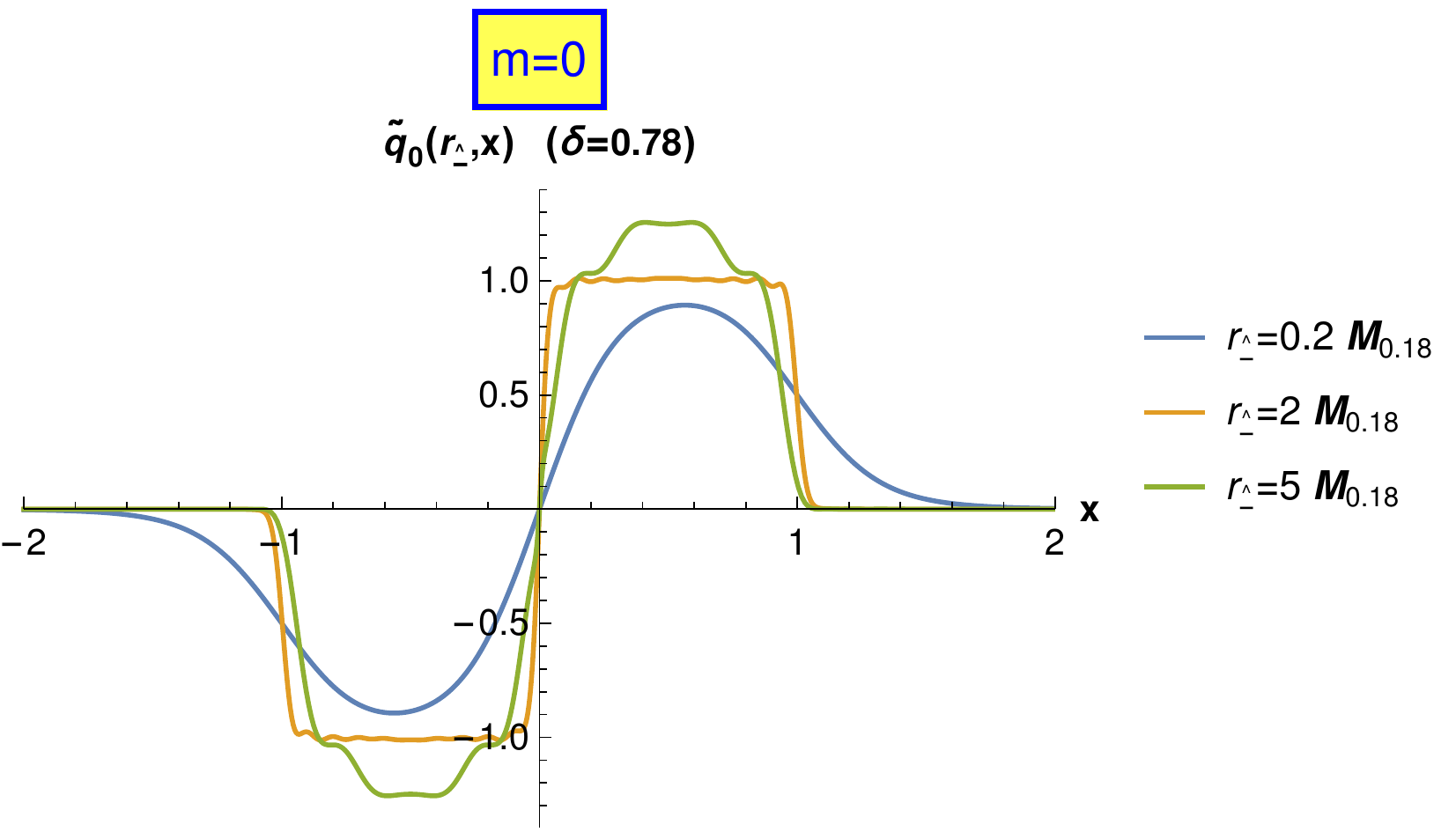}
		\label{fig:m0_delta078_grquasipdf}
	}
	\centering
	\subfloat[]{
		\includegraphics[width=0.5\linewidth]{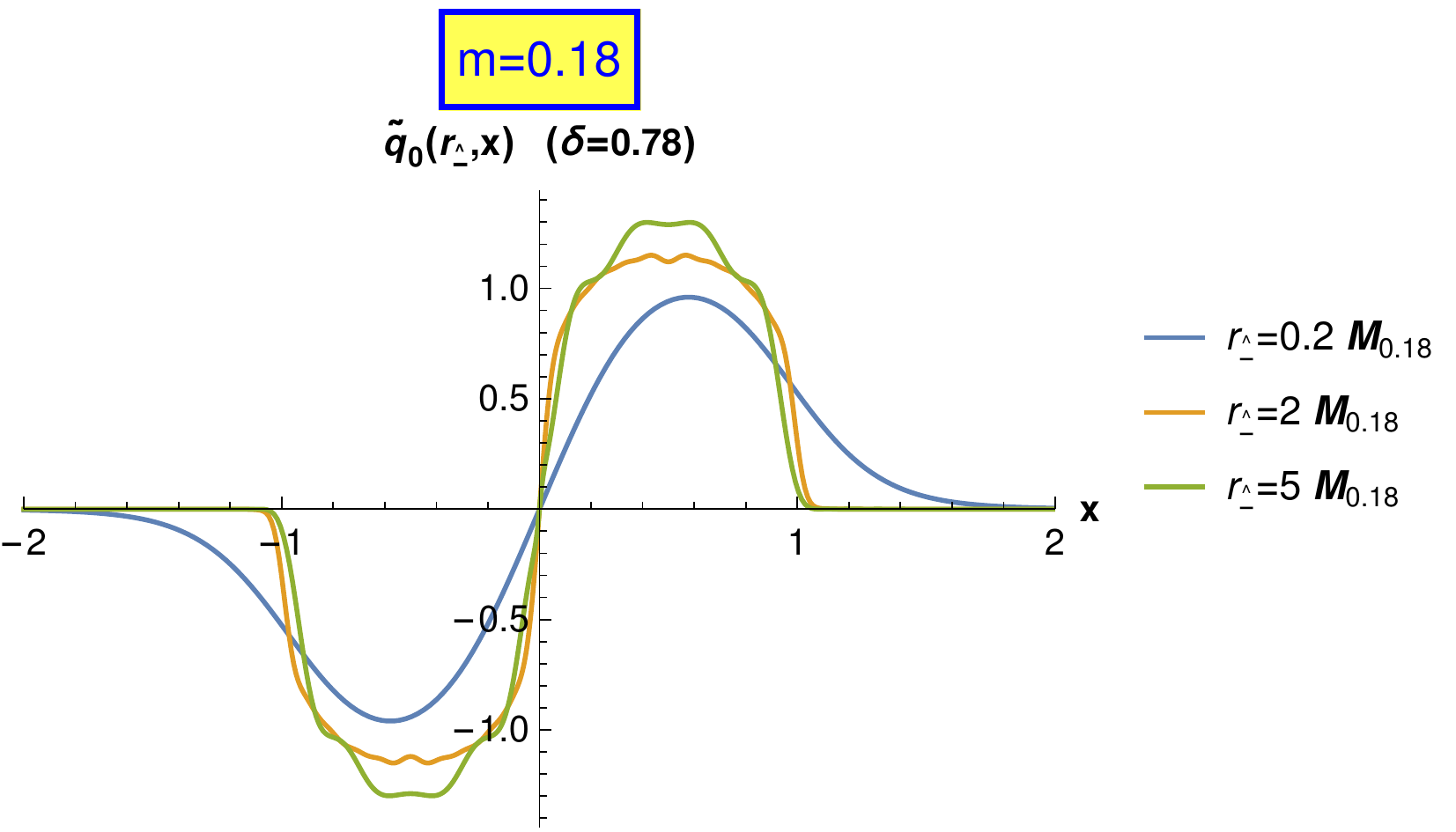}
		\label{fig:m018_delta078_grquasipdf}
	}
	\caption{$ \delta=0.78$ interpolating ``quasi-PDFs" for the ground state ($n=0$) wave functions of (a) $m=0$, (b) $m=0.18$. 
	All quantities are in proper units of $ \sqrt{2\lambda} $.\label{fig:delta078_grquasipdf}}
\end{figure*}

\begin{figure*}
	\centering
	\subfloat[]{
		\includegraphics[width=0.5\linewidth]{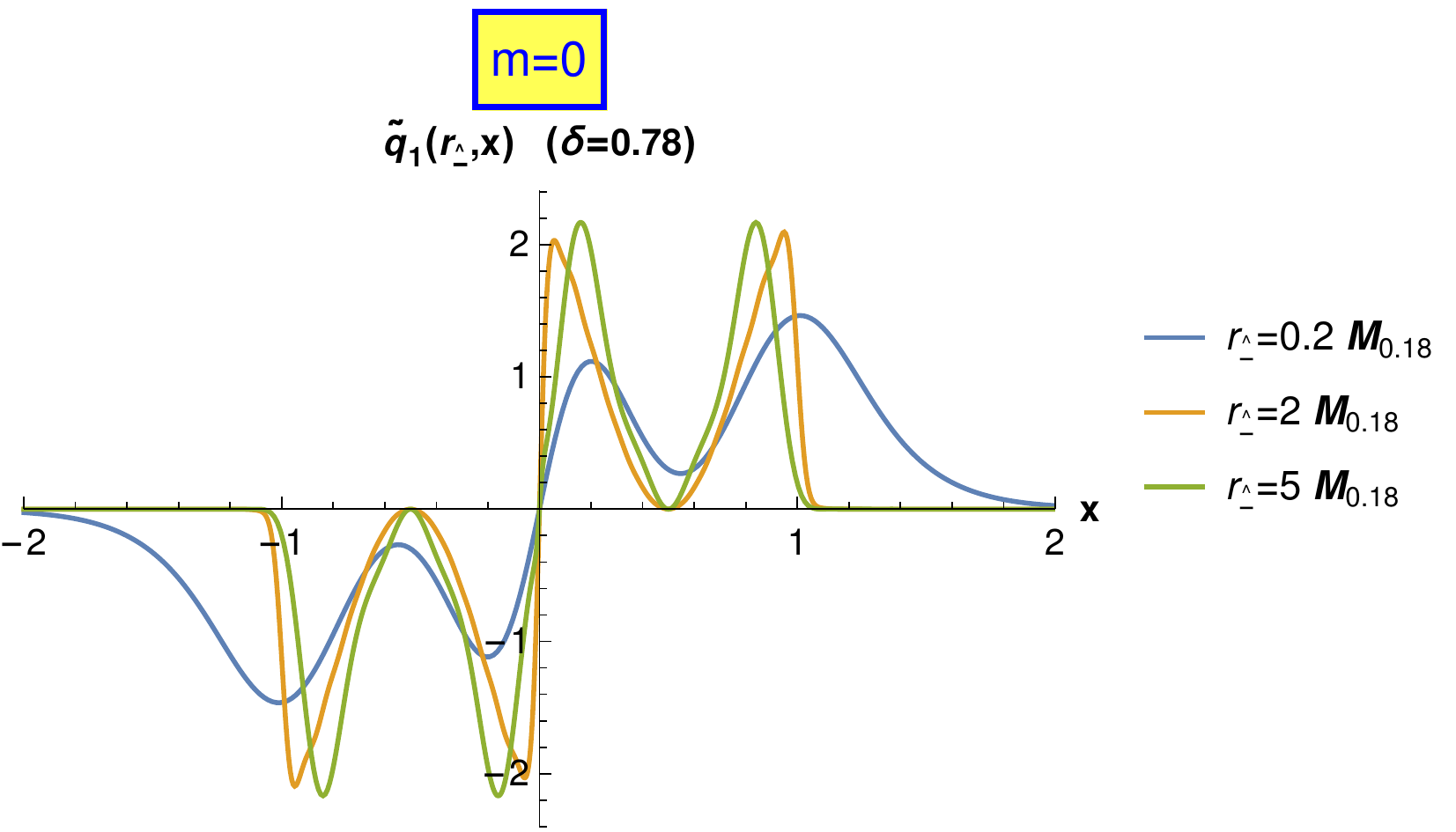}
		\label{fig:m0_delta078_fequasipdf}
	}
	\centering
	\subfloat[]{
		\includegraphics[width=0.5\linewidth]{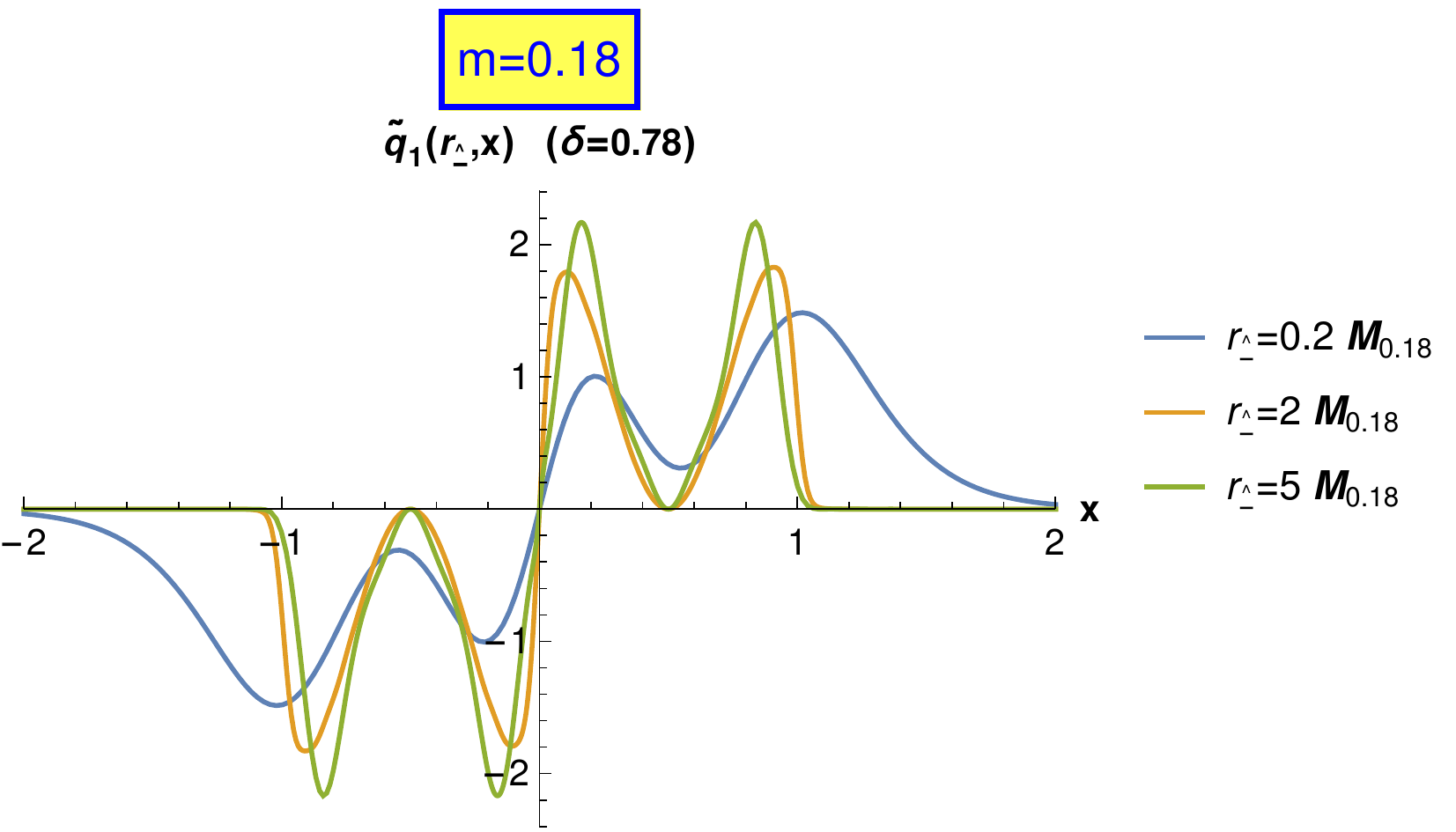}
		\label{fig:m018_delta078_fequasipdf}
	}
	\caption{$ \delta=0.78$ interpolating ``quasi-PDFs" for the first excited state ($n=1$) wave functions of (a) $m=0$, (b) $m=0.18$. All quantities are in proper units of $ \sqrt{2\lambda} $.\label{fig:delta078_fequasipdf}}
\end{figure*}

Since we obtained the bound-state wavefunctions, we now apply them to compute 
the so-called quasi-PDFs which have been discussed extensively even in the 't Hooft model application~\cite{pdf} due to the possibility of computing directly the longitudinal momentum fraction $x$-dependence of the parton distributions in Euclidean lattice approach using the large momentum effective field theory (LaMET) program~\cite{XJ-LaMET}. In our interpolating 't Hooft model computation, the ``quasi-PDFs" may be defined as the following matrix element for the $n$-th state of the meson with the interpolating longitudinal momentum $r_{\hat{-}}$:
\begin{align}\label{intepolating-pdf-gauge-link} 
\tilde{q}_{(n)}(r_{\hat{-}},x)=&\int_{-\infty}^{+\infty}\frac{dx^{\hat{-}}}{4\pi}\ {\rm e}^{ix^{\hat{-}}r_{\hat{-}}}<r^{\hat{+}}_{(n)},r_{\hat{-}}\ |\ \bar{\psi}(x^{\hat{-}})\ \cdot\notag\\
&\ \ \ \gamma_{\hat{-}}\ {\cal W}[x^{\hat{-}},0]\ \psi(0)\ |\ r^{\hat{+}}_{(n)},r_{\hat{-}}>_C,
\end{align}
where $r^{\hat{+}}_{(n)}=\sqrt{r_{\hat{-}}^2+\mathbb{C}{\cal M}_{(n)}^2}$ as obtained from Eq.(\ref{eqn:energy_momentum}), and the range of the longitudinal momentum fraction
$x=p_{\hat{-}}/r_{\hat{-}}$ is unconstrained, $-\infty<x<+\infty$, for $ 0\leq\delta<\pi/4 $,
while bounded, $0 \leq x\leq 1$, for $\delta=\pi/4$.
One should note that this definition of the ``quasi-PDFs" is not unique, 
e.g. taking $\gamma^{\hat{+}}$ instead of $\gamma_{\hat{-}}$ in front of the 
interpolating gauge link ${\cal W}[x^{\hat{-}},0]$ in Eq.(\ref{intepolating-pdf-gauge-link}), but still uniquely approach the PDF defined in the LFD as $\delta \to \pi/4$, whichever definition is taken. While one may consider the so-called universality class ~\cite{Hatta} of the interpolating ``quasi-PDFs", we note that the definition given by Eq.(\ref{intepolating-pdf-gauge-link}) coincides with the canonical definition of 
the quasi-PDFs in IFD ($\delta =0$)~\cite{pdf}. 
While it has been discussed which definition approaches the PDFs in LFD
faster for the perspectives of the LaMET program~\cite{XJ-LaMET,pdf}, 
we will take the definition given by Eq.(\ref{intepolating-pdf-gauge-link}) in this work and discuss our numerical results corresponding to this definition.
The interpolating gauge link 
\begin{equation}\label{gauge-link}
{\cal W}[x^{\hat{-}},0]={\cal P}\left[{\rm exp}\left( -ig\int_{0}^{x^{\hat{-}}}dx^{'\hat{-}}A_{\hat{-}}(x^{'\hat{-}})\right)  \right] 
\end{equation}
inserted in Eq.~(\ref{intepolating-pdf-gauge-link}) assures the gauge invariance of the interpolating ``quasi-PDFs". The subscript $C$ in Eq.~(\ref{intepolating-pdf-gauge-link}) indicates the removal of the disconnected contribution discussed~\cite{collins} for the forward matrix element computation:
\begin{align}\label{disconnected-intepolating-pdf} 
& <\ r^{\hat{+}}_{(n)},r_{\hat{-}}|\ r^{\hat{+}}_{(n)},r_{\hat{-}}>\int_{-\infty}^{+\infty}\frac{dx^{\hat{-}}}{4\pi}\ {\rm e}^{ix^{\hat{-}}r_{\hat{-}}} < \Omega |\ \bar{\psi}(x^{\hat{-}})\cdot \notag \\
& \ \ \ \ \gamma_{\hat{-}}\ {\cal W}[x^{\hat{-}},0]\ \psi(0)\ |\Omega>.
\end{align} 

As we adopted the axial gauge in the interpolation form, i.e. $ A_{\hat{-}}^a=0 $, 
the gauge link becomes an identity and the quantization procedure illustrated in
Sec.~\ref{sub:Ham} yields 
\begin{align}
\label{interpol-pdf}
\tilde{q}_{(n)}(r_{\hat{-}},x)&=\frac{r_{(n)}^{\hat{+}}}{r_{\hat{-}}}\sin\theta(xr_{\hat{-}})\left[\hat\phi_+^{(n)}(r_{\hat{-}},x)^2+\hat\phi_-^{(n)}(r_{\hat{-}},x)^2\right.\notag\\
&+\left.\hat\phi_+^{(n)}(r_{\hat{-}},-x)^2+\hat\phi_-^{(n)}(r_{\hat{-}},-x)^2 \right]. 
\end{align}
For $\delta=0$, i.e. IFD, Eq.(\ref{interpol-pdf}) coincides with Eq.(74) of Ref.~\cite{pdf}. 

Based on this formula, we compute the interpolating ``quasi-PDFs" 
for the cases of $\delta = 0, 0.6$ and 0.78. 
In Figs.~\ref{fig:delta0_grquasipdf} and \ref{fig:delta0_fequasipdf},
the interpolating ``quasi-PDFs" of the ground state ($n=0$) and the first excited state ($n=1$) are shown for the case of $\delta=0$, respectively. The two panels in
each of these figures exhibit our numerical results for the bare quark mass 
$m=0$ and $m=0.18$, respectively. 
Numerical results for other mass cases ($m=$0.045, 1.0 and 2.11) are summarized in Appendix~\ref{quasi-PDF-other-masses}. 
In each panel, the
results for the meson longitudinal momentum $r_{\hat{-}}=r^1=0.2\mathbf{M}_{0.18}, 2\mathbf{M}_{0.18}, 5\mathbf{M}_{0.18}$ are depicted by blue, yellow and green solid lines, respectively. 
The general behaviors of our numerical results with respect to the 
variation of $r^1$ values from small ($0.2\mathbf{M}_{0.18}$) to large 
($5\mathbf{M}_{0.18}$) agree with the results presented in Ref.~\cite{pdf}, although different longitudinal 
momentum values were taken between ours and Ref.~\cite{pdf}.
As $r^1$ gets larger, the numerical results of the quasi-PDFs resemble 
the PDFs in LFD more closely fitting in $x\in[0,1]$ and $[-1,0]$.
It is our interest to take a look at the rate of achieving
the resemblance to the PDFs in LFD as $\delta$ gets away from the IFD ($\delta=0$) and $r_{\hat{-}}$ gets larger.
The numerical results of the ground state ($n=0$) and the first excited state ($n=1$) are shown for the case of $\delta = 0.6$ in Figs.~\ref{fig:delta06_grquasipdf}
and ~\ref{fig:delta06_fequasipdf} and  for the case of $\delta = 0.78$ in Figs.~\ref{fig:delta078_grquasipdf} and ~\ref{fig:delta078_fequasipdf}, respectively.  
As noticed previously in Figs.~\ref{fig:m0grz} and ~\ref{fig:m018grz}, 
the wiggle and bulge in $\hat\phi_{+}^{(0)}(r_{\hat{-}},x) (\delta=0.78)$ for $r_{\hat{-}} = 5 \mathbf{M}_{0.18}$ is due to the computational sensitivity arising in the interpolation region where $\mathbb{C}$ gets close to $0$ in particular
as $r_{\hat{-}}$ gets very large. The corresponding wiggle and bulge in 
$\tilde{q}_{(0)}(r_{\hat{-}},x)$ is noticed also in Fig.~\ref{fig:delta078_grquasipdf}.
Besides such numerical sensitivity for the very large value of $r_{\hat{-}}$,
it is apparent that the $\delta=0.78$ results are rather immediately 
close to the LFD result.

Interestingly, our numerical results of the interpolating ``quasi-PDFs" in the moving 
frames indicate a possibility to utilize both variation of 
$\delta$ and $r_{\hat{-}}$ to attain the LFD result more effectively. Namely, 
one may not need to boost the longitudinal momentum $r_{\hat{-}}$ too large 
but search for a ``sweet spot" by varying both $\delta$ and $r_{\hat{-}}$ together to obtain the ``LFD-like" result. In IFD, $\delta=0$ is fixed and thus the boost
to the large longitudinal momentum is necessary for a successful approach
to the LFD result. However, in the interpolating formulation between the IFD and
the LFD, the LFD result can be approached even at rather small $r_{\hat{-}}$.
Moreover, the application to the lattice formulation may be also possible 
with the existing technique of Wick rotation replacing the ordinary 
instant form time $x^0$ by the interpolating time $x^{\hat{+}}$ in the process
of taking the ``imaginary time" in the lattice
 as far as $\delta$ remains in the region $0 \le \delta < \pi/4$
avoiding the light-like surface $\delta = \pi/4$. 
As discussed in the later part of Sec.~\ref{FPandCM}, one can 
match the Minkowsky space and the Euclidean space
confirming the correspondence given by
Eq.(\ref{eqn:covpsquared}).

\begin{figure*}
	\centering
	\subfloat[]{
		\includegraphics[width=0.5\linewidth]{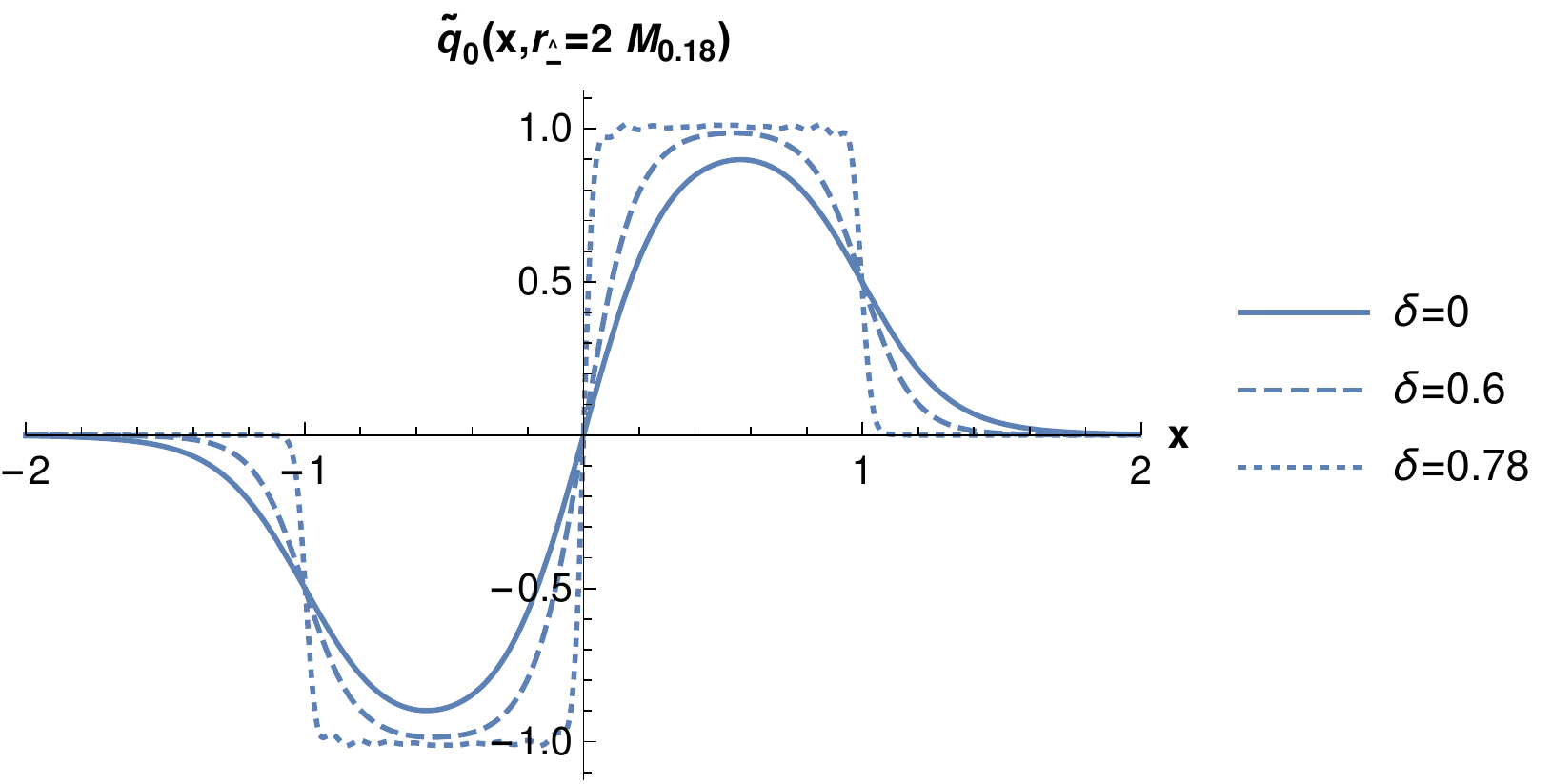}
		\label{fig:m0mom2grquasipdfmomplot}
	}
	\centering
	\subfloat[]{
		\includegraphics[width=0.5\linewidth]{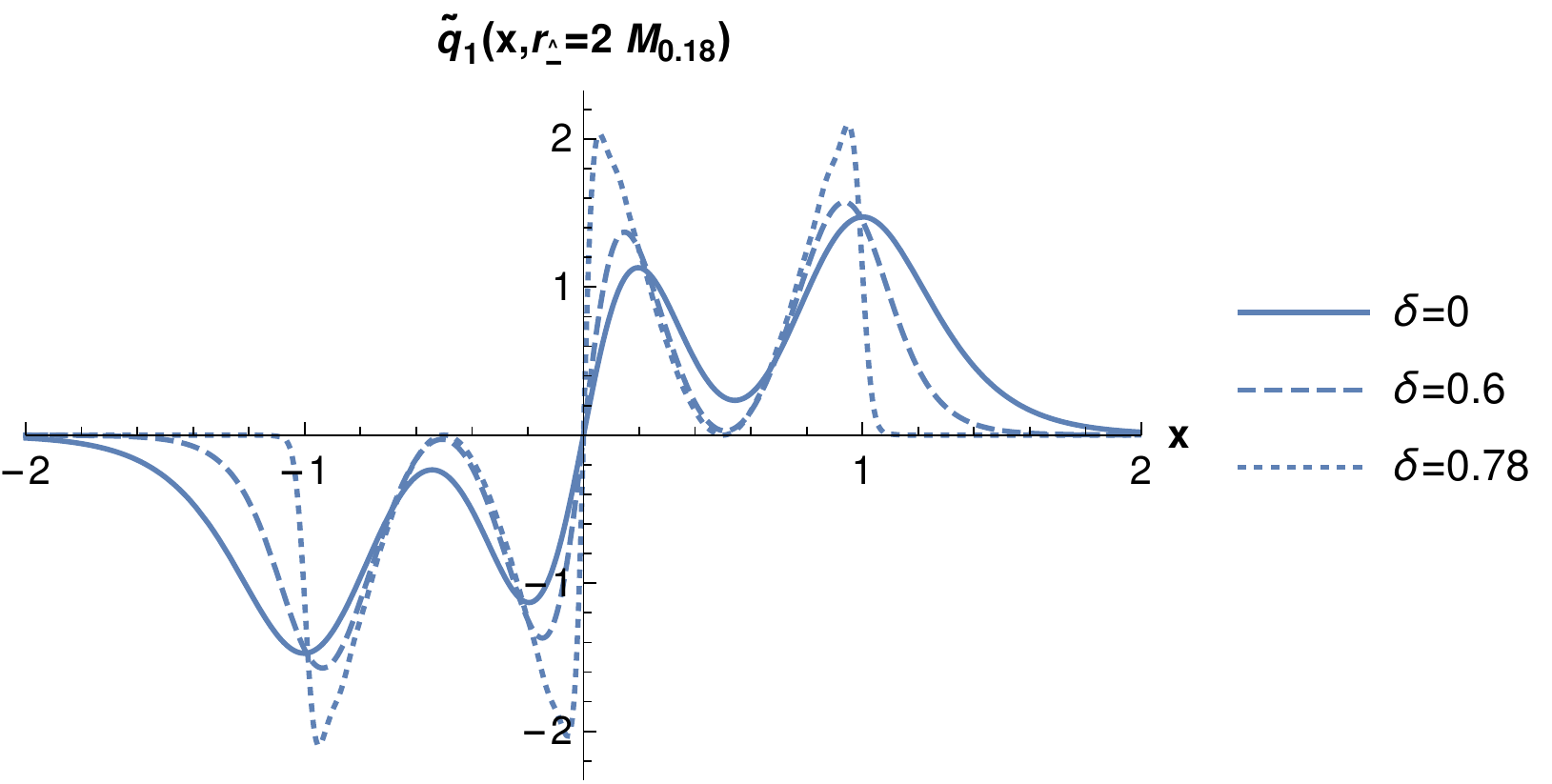}
		\label{fig:m0mom2fequasipdfmomplot}
	}
	\caption{Interpolating ``quasi-PDFs" at the fixed 
value of  $r_{\hat{-}}=2\mathbf{M}_{0.18}$	
for the (a) ground state and (b) first excited state for $m=0$ 
in three cases of $\delta=0$, $\delta=0.6$ and $\delta=0.78$. All quantities are in proper units of $ \sqrt{2\lambda} $.
	\label{fig:m0mom2quasipdfmomplot}}
\end{figure*}

For an illustration of the $\delta$ variation for a given finite $r_{\hat{-}}$, we take $r_{\hat{-}}=2\mathbf{M}_{0.18}$ for the case of $m=0$ and show the ``quasi-PDFs" of the ground-state and the first excited-state for the variation of $\delta$ parameter as $\delta=0$, $\delta=0.6$ and $\delta=0.78$ in Fig.~\ref{fig:m0mom2quasipdfmomplot}. It indicates that a pretty slow approach to the LFD result in the large-momentum IFD can be fairly well expedited by taking $\delta$ away from the IFD ($\delta=0$) and getting closer to the LFD ($\delta=\pi/4$) while the same value of the longitudinal momentum $r_{\hat{-}}=2\mathbf{M}_{0.18}$ is taken. The numerical sensitivity arising near $\delta=\pi/4$ for the large $r_{\hat{-}}$, e.g. the ``wiggle and bulge" mentioned for $r_{\hat{-}}=5\mathbf{M}_{0.18}$ in Figs.~\ref{fig:delta078_grquasipdf} and ~\ref{fig:delta078_fequasipdf} for 
$\delta=0.78$, is also dodged by taking the smaller value of $r_{\hat{-}}$, e.g. $r_{\hat{-}}=2\mathbf{M}_{0.18}$ for this illustration. We think it would be 
worthwhile to explore this idea of utilizing the interpolating formulation between IFD and LFD for the application to the lattice computation.

For further application of the interpolating bound-state wavefunctions, 
one can also consider the interpolating ``quasi-distribution amplitude (quasi-DA)"
which may be written as
\begin{align}
\label{quasi-DA}
&\tilde{\Phi}_{(n)}(r_{\hat{-}},x)=\frac{1}{f^{(n)}}\int_{-\infty}^{+\infty}\frac{dx^{\hat{-}}}{2\pi}{\rm e}^{i(x-\frac{1}{2})r_{\hat{-}}x^{\hat{-}}}\cdot\notag\\
&<\ r^{\hat{+}}_{(n)},r_{\hat{-}}|\bar{\psi}(\frac{x^{\hat{-}}}{2}){\cal W}[\frac{x^{\hat{-}}}{2},-\frac{x^{\hat{-}}}{2}]\gamma_{\hat{-}}\gamma_5\psi(-\frac{x^{\hat{-}}}{2})|\Omega>,
\end{align}
where $\cal W$ is the gauge link introduced in Eq.~(\ref{gauge-link}), and $f^{(n)}$ is the decay constant mentioned in Sec.~\ref{sub:spec}. 
As mentioned in the definition of the interpolating ``quasi-PDFs" given by 
Eq.(\ref{intepolating-pdf-gauge-link}), the definition of the interpolating ``quasi-DAs" is 
also not unique, e.g. $\gamma^{\hat{+}}$ can be taken instead of $\gamma_{\hat{-}}$ in front of the interpolating gauge link ${\cal W}[\frac{x^{\hat{-}}}{2},-\frac{x^{\hat{-}}}{2}]$ in Eq.(\ref{quasi-DA}). Whichever definition is taken, they all uniquely approach the DA defined in the LFD as $\delta \to \pi/4$, belonging to the same universality class~\cite{Hatta} as mentioned for the case of interpolating ``quasi-PDFs". Using the definition given by Eq.(\ref{quasi-DA}) which coincides with the canonical definition of 
the quasi-DAs in IFD ($\delta =0$)~\cite{pdf}, we note that the interpolating ``quasi-DAs" of the even-$n$ mesonic states can be formulated as
\begin{align}\label{quasi-DA-explicit}
\tilde{\Phi}_{(n)}(r_{\hat{-}},x)=&\frac{1}{f^{(n)}}\sqrt{\frac{N_c r^{\hat{+}}}{\pi r_{\hat{-}}}}\sin\frac{\theta\left(xr_{\hat{-}}\right)+\theta\left((1-x)r_{\hat{-}}\right)}{2}\notag\\
\times&\left[\hat\phi_{+}^{(n)}(r_{\hat{-}},x)+\hat\phi_{-}^{(n)}(r_{\hat{-}},x)\right],
\end{align}
where $\hat\phi_{\pm}^{(n)}(r_{\hat{-}},x)$ denote the interpolating mesonic wave functions associated with the $n$-th excited mesonic state.
The normalization condition of the interpolating ``quasi-DAs" given by
\begin{equation}\label{quasi-DA-norm}
\int_{-\infty}^{+\infty}dx \tilde{\Phi}_{(2n)}(r_{\hat{-}},x)=1
\end{equation}
is consistent with the explicit form of the decay constant $f^{(n)}$ given by
\begin{equation}\label{decay-constant}
f^{(n)}=\begin{cases}
&\sqrt{\frac{N_c r^{\hat{+}}}{\pi r_{\hat{-}}}}\int_{-\infty}^{+\infty}dx \sin\frac{\theta\left(xr_{\hat{-}}\right)+\theta\left((1-x)r_{\hat{-}}\right)}{2}\\
&\ \ \ \ \ \ \ \times\left[\hat\phi_{+}^{(n)}(r_{\hat{-}},x)+\hat\phi_{-}^{(n)}(r_{\hat{-}},x)\right]\ \ \ \ {\rm even}\ n;\\
&0\ \ \ \ \ \ \ \ \ \ \ \ \ \ \ \ \ \ \ \ \ \ \ \ \ \ \ \ \ \ \ \ \ \ \ \ \ \ \ \ \ \ \ \ \ \ \ \ {\rm odd}\ n.
\end{cases}
\end{equation}
While the corresponding results for the IFD ($\delta =0$) have been 
worked out in Refs.~\cite{mov,pdf}, it is interesting to note that 
the analytic result for the pion decay constant $f_\pi = \sqrt{N_c/\pi}$~\cite{review}
can be immediately obtained by taking the LFD solution in the chiral limit,
i.e. $\hat\phi_{+}^{(0)}(r_{\hat{-}},x)=\phi(x) = 1$ for $x\in [0,1]$ and 
$\hat\phi_{-}^{(0)}(r_{\hat{-}},x)= 0$, as well as  
$\theta(x r_{\hat{-}})= \theta(x r^+)= \theta(p^+) = \frac{\pi}{2}$
and $\theta((1-x) r_{\hat{-}})= \theta((1-x) r^+)= \theta(r^+ - p^+) = \frac{\pi}{2}$,
noting $\frac{r^{\hat{+}}}{r_{\hat{-}}} = \frac{r^{+}}{r_{-}} =1$ in LFD. 
As the DAs in LFD are directly involved with the QCD factorization theorem for the hard exclusive reactions involving hadrons, it would be useful to explore  
the utility of the interpolating ``quasi-DAs" further in the future works.

\section{\label{sec:conclusion}Conclusion and Outlook}
In this work, we interpolated the 't Hooft model (i.e. $\text{QCD}_2$ in the large $N_c$ limit) with the interpolation angle $\delta$ between IFD ($\delta=0$) and LFD ($\delta=\pi/4$) and analyzed its nontrivial vacuum effects
on the quark mass and wavefunction renormalization as well as the corresponding meson mass and wavefunction properties taking the meson 
as the quark-antiquark bound-state. We derived the interpolating mass gap equation between IFD and LFD using not only the algebraic method based on the Bogoliubov transformation between the trivial and nontrivial vacuum as well as the bare and dressed quark but also the diagrammatic method based on the self-consistent embodiment of the quark self-energy. Our mass gap solutions agree not only with the LFD result in Ref.~\cite{tHooft} for $\delta=\pi/4$
but also with the IFD results in Refs.~\cite{Li, mov} for $\delta=0$. 
The renormalized chiral condensate was computed and the agreement of the result in the chiral limit was verified with the exact result in Ref.~\cite{Zhicon}. Its invariance regardless of the $\delta$ values between IFD and LFD was also confirmed.
Taking into account the wavefunction renormalization factor $F(p_{\hat{-}})$ as well as the mass function $M(p_{\hat{-}})$ and expressing the dressed quark propagator $S(p)$ in terms of $F(p_{\hat{-}})$ and $M(p_{\hat{-}})$ as given by Eq.~(\ref{eqn:Sponeway}), we resolved the issue of $E(p_{\hat{-}})$ not being always positive for $m \lesssim 0.56$ discussed in Ref.~\cite{BG}. Extending the interpolating energy-momentum dispersion relation of the on-mass-shell particle given by Eq.~(\ref{eqn:energy_momentum}) 
to the case of the dressed quark with the rescaled variable given by Eq.~(\ref{eqn:energy_momentum_dressed}), we obtained the interpolation angle independent energy function ${\tilde E}(p'_{\hat{-}})$. Typical profiles of $ {\tilde E}(p'_{\hat{-}})$ were exemplified in Fig.~\ref{fig:m0018fepppplot}. 

Utilizing the dressed fermion propagators, 
we then derived the quark-antiquark bound-state equation 
interpolating between IFD and LFD for the equal bare quark and antiquark mass $m$ and solved numerically the corresponding bound-state equations. 
From the numerical solutions of the spectroscopy, we find that the meson mass spectrum is independent of interpolation angle between the IFD and LFD as expected for physical observables. 
In particular, for the bare quark mass $m \to 0$, we confirmed 
the GOR behavior of the pionic ground-state mass square 
${\cal M}_{(0)}^2 \sim m\sqrt{\lambda} \to 0$ as shown in Fig.~\ref{fig:GOR}.
Our result is consistent with the discussions~\cite{review,mov,Zhicon,GSSW} on the SBCS in the 't Hooft model ($N_c \to \infty$). 
Plotting the meson mass spectra ${\cal M}_{(n)}$ 
($n=0,1,2,...,7$) for various $m$ values as summarized in
in Table~\ref{tab:mass}, we also observe the Regge trajectory feature 
as shown in Fig.~\ref{fig:mesonmassplot}. 
The corresponding bound-state wavefunctions 
$\hat\phi_{\pm}^{(n)}(r_{\hat{-}},x)$ were obtained, in particular, for the low-lying states, 
i.e., $n=$ 0 and 1 states, and were applied to the 
interpolating formulation of ``quasi-PDFs".
The results of the wavefunctions clearly dictate the charge conjugation symmetry, exhibiting the symmetric and antisymmetric behaviors of 
$\hat\phi_{\pm}^{(0)}(r_{\hat{-}},x)$ and 
$\hat\phi_{\pm}^{(1)}(r_{\hat{-}},x)$, respectively, 
under the exchange of $x \leftrightarrow 1-x$. 
It is also interesting to note that the massless Goldstone boson
cannot exist in the rest frame due to the null normalization 
from the equivalence between $\hat\phi_{+}^{(0)}(p_{\hat{-}})$ and 
$\hat\phi_{-}^{(0)}(p_{\hat{-}})$ for the massless ground-state
in the rest frame as shown in Fig.~\ref{fig:m0rfn0ana}. 

Applying the bound-state wavefunctions $\hat\phi_{\pm}^{(n)}(r_{\hat{-}},x)$ 
for the computation of the interpolating ``quasi-PDFs" given by Eq.~(\ref{interpol-pdf}), we note the consistency with the observation made in Ref.~\cite{pdf}
for the quasi-PDFs at $\delta=0$ (IFD) that there exists considerable difference between the shapes of the LFD result and the IFD quasi-PDF result for 
the light mesons. Our results indicate that the slow approach to the LFD-like results may be remedied by varying the interpolation parameter $\delta$ as well 
as the interpolating longitudinal meson momentum $r_{\hat{-}}$. 
For the future work, one may explore such idea to search for the ``sweet spot'' of $\delta$ and $r_{\hat{-}}$ to attain most effective computation with the least sensitive numerical errors in getting the LFD result. Extending
the Wick rotation technique to the interpolating time $x^{\hat{+}}$, the idea
of searching for the ``sweet spot'' may be applicable to the usual lattice formulation in the Eunclidean space. This would be in good contrast to
the recent application of the present interpolating formulation to 
the two-dimensional $\phi^4$ theory using the discretization technique in Minkowsky space consistent with the discrete light-cone quantization (DLCQ) approach~\cite{Hiller,Kent-DLCQ}. It will be interesting to explore both ``Euclidean" and ``Minkowsky" numerical approaches implementing the interpolating formulation between IFD and LFD.

\begin{acknowledgments}
	This work was supported by the U.S. Department of
	Energy Grant No. DE-FG02-03ER41260. This research used resources of the National
	Energy Research Scientific Computing Center,
	which is supported by the Office of Science of
	the U.S. Department of Energy under Contract
	No. DE-AC02-05CH11231.
\end{acknowledgments}

\appendix

\section{\label{app:Bogo}Bogoliubov transformation for the interpolating spinors between IFD and LFD}

In this Appendix, we summarize the interpolating spinors and $\gamma$ matrices in the $1+1$ dimensional chiral representation and the Bogoliubov transformation between the free and interacting spinors as well as the corresponding creation/annihilation operators. 

For the representation of the spinors~\cite{Spi}, we adopt the chiral representation (CR), under which the usual $ \gamma $ matrices in IFD for the $1+1$ dimensions are
given by the Pauli matrices:
\begin{align}\label{eqn:gamma}
\gamma^0=\sigma_1=\left(\begin{array}{cc}
0 & 1\\
1 & 0
\end{array}\right) , \gamma^1=i\sigma_2=\left(\begin{array}{cc}
0 & 1\\
-1 & 0
\end{array} \right) ,\notag\\
\gamma^5=\gamma_5=-\gamma_0\gamma_1=\gamma^0\gamma^1=-\sigma_3=\left( \begin{array}{cc}
-1 & 0\\
0 & 1
\end{array}\right).
\end{align}
These can be transformed into standard representation (SR) easily by a transformation matrix
\begin{equation}\label{eqn:CRtoSR}
S=S^{\dagger}=\frac{1}{\sqrt{2}}\left(\begin{array}{cc}
1&1\\
1&-1
\end{array} \right) 
\end{equation}
through
\begin{equation}\label{eqn:gammaCRtoSR}
\gamma^{\mu}_{\mathrm{SR}}=S\gamma^{\mu}_{\mathrm{CR}}S^{\dagger}.
\end{equation}
In the SR representation, the free spinors in the rest frame in IFD are typically
given by 
\begin{equation}\label{eqn:uvSR}
u^{(0)}_{\mathrm{SR}}(p^1=0)=\sqrt{2 m}\left( \begin{array}{c}
1\\
0
\end{array}\right),\  v^{(0)}_{\mathrm{SR}}(p^1=0)=\sqrt{2 m}\left( \begin{array}{c}
0\\
1
\end{array}\right) ,
\end{equation}
where we take the normalization factor ${\bar u}^{(0)}_{\mathrm{SR}}u^{(0)}_{\mathrm{SR}} = 2m$ in conformity with the standard textbooks~\cite{text}.
The corresponding free spinors in the chiral representation are used in this work  without denoting the ``CR" specification: 
\begin{align}\label{eqn:uvCR}
u^{(0)}(p^1=0)=S\cdot u^{(0)}_{\mathrm{SR}}(p^1=0)=\sqrt{m}\left( \begin{array}{c}
1\\
1
\end{array}\right),\notag\\ 
v^{(0)}(p^1=0)=S\cdot v^{(0)}_{\mathrm{SR}}(p^1=0)=\sqrt{m}\left( \begin{array}{c}
1\\
-1
\end{array}\right) .
\end{align}
The IFD spinors in the moving frame with the energy 
$E_p = \sqrt{(p^1)^2 + m^2}$ are then obtained as
\begin{align}\label{uIFD}
u^{(0)}(p^1)&= B(\eta) u^{(0)}(0) = \sqrt{m}
\left( \begin{array}{cc}
e^{-\frac{\eta}{2}} & 0 \\
0 & e^{\frac{\eta}{2}}
\end{array}\right) \left( \begin{array}{cc}
1\\
1
\end{array}\right) \notag\\
&=
\left( \begin{array}{c}
\sqrt{E_p-p^1}\\
\sqrt{E_p+p^1}
\end{array}\right),
\end{align}
and
\begin{equation}\label{vIFD}
v^{(0)}(-p^1)= B(-\eta) v^{(0)}(0) =\left( \begin{array}{c}
\sqrt{E_p+p^1}\\
-\sqrt{E_p-p^1}
\end{array}\right),
\end{equation}
where the usual boost operator $B(\eta)$ with the rapidity
$\eta= {\rm tanh}^{-1} \frac{p^1}{E_p}$  
and the longitudinal boost generator $ K^1 $ 
is given by
\begin{equation}\label{boost}
B(\eta)=\exp\left( -i\eta\cdot K^1\right) = \exp\left( -\frac{1}{2}\eta\sigma_3\right)
= \exp\left( \frac{1}{2}\eta\gamma_5\right)  .
\end{equation}
In terms of the interpolating momentum variables, the rapidity $\eta$
can be written~\cite{Spi} as 
\begin{equation}\label{rapidity}
\eta = \log \left( \frac{p^{\hat{+}} + p_{\hat{-}}}{m\left( \cos\delta + \sin\delta\right)} \right), 
\end{equation}
where one can note the following equality as well
\begin{equation}\label{rapiditym}
- \eta = \log \left( \frac{p^{\hat{+}} - p_{\hat{-}}}{m\left( \cos\delta - \sin\delta\right)} \right) .
\end{equation}
The boost operator $B(\eta)$ can then be written in terms of the interpolating momentum variables as
\begin{align}\label{interpol-boost}
B(\eta) = 
\left( \begin{array}{cc}
\sqrt{\frac{p^{\hat{+}} - p_{\hat{-}}}{ m(\cos\delta - \sin\delta) }}
& 0 \\
0 &
\sqrt{\frac{p^{\hat{+}} + p_{\hat{-}}}{ m(\cos\delta + \sin\delta) }}
\end{array}\right) ,
\end{align}
so that the boosted interpolating spinors are given by
\begin{equation}
u^{(0)}(p_{\hat{-}})=\left( \begin{array}{c}
\sqrt{\frac{p^{\hat{+}}-p_{\hat{-}}}{\cos\delta-\sin\delta}}\\
\sqrt{\frac{p^{\hat{+}}+p_{\hat{-}}}{\cos\delta+\sin\delta}}
\end{array}\right)\label{u_inter_free},
\end{equation}
and
\begin{equation}
v^{(0)}(-p_{\hat{-}})=\left( \begin{array}{c}
\sqrt{\frac{p^{\hat{+}}+p_{\hat{-}}}{\cos\delta-\sin\delta}}\\
-\sqrt{\frac{p^{\hat{+}}-p_{\hat{-}}}{\cos\delta+\sin\delta}}
\end{array}\right)\label{v_inter_free}.
\end{equation}
Although Eqs.(\ref{u_inter_free})-(\ref{v_inter_free}) are expressed 
in terms of the interpolating momentum variables while 
Eqs.(\ref{uIFD})-(\ref{vIFD}) are written in terms of the IFD momentum variables, 
one should note that they are intrinsically the same spinors with respect to each other
as we have shown in the above derivation. 
As $ \delta\to\frac{\pi}{4} $, i.e., in the LFD limit, Eqs.(\ref{u_inter_free})-(\ref{v_inter_free}) coincide with the LFD spinors ~\cite{Spi}
\begin{equation}
\label{uLFD}
u^{(0)}(p^+)=\frac{1}{\sqrt{\sqrt{2}p^+}}\left( \begin{array}{c}
m\\
\sqrt{2}p^+
\end{array}\right),
\end{equation}
and
\begin{equation}
\label{vLFD}
v^{(0)}(p^+)=\frac{1}{\sqrt{\sqrt{2}p^+}}\left( \begin{array}{c}
m\\
-\sqrt{2}p^+
\end{array}\right) ,
\end{equation}
where one may note the correspondence
\begin{equation}\label{limit_relation}
\frac{p^{\hat{+}}-p_{\hat{-}}}{\mathbb{C}}\overset{\mathbb C \to 0}{\longrightarrow}\frac{m^2}{2p^+},
\end{equation}
with $ p^{\hat{+}}\to p^+ $, and $ p_{\hat{-}}\to p^+ $.
Thus, the interpolating spinors given by Eqs.(\ref{u_inter_free})-(\ref{v_inter_free})
are nothing but the same spinors as given by Eqs.(\ref{uIFD})-(\ref{vIFD}) in IFD 
and Eqs.(\ref{uLFD})-(\ref{vLFD}) in LFD, respectively. There are no differences
in the spinors except the expression difference in terms of the momentum variables
taken in each different form of the dynamics.

Dropping the ``CR" specification again for the $\gamma$ matrices, we follow the link given by Eq.(\ref{eqn:interpolangle}) to get the following interpolating $ \gamma $ matrices in the chiral representation as
\begin{align}\label{eqn:inter_gamma}
\gamma^{\hat{+}}&=\left( \begin{array}{cc}
0 & \sqrt{1+\mathbb{S}}\\
\sqrt{1-\mathbb{S}} & 0
\end{array}\right)
,\notag\\ \gamma^{\hat{-}}&=\left( \begin{array}{cc}
0  & -\sqrt{1-\mathbb{S}}\\
\sqrt{1+\mathbb{S}} & 0
\end{array}\right) 
.
\end{align}
Lowering the indices with the interpolating metric given by Eq.(\ref{eqn:metric}), 
we also get
\begin{align}\label{eqn:inter_gamma_sub}
\gamma_{\hat{+}}&=\mathbb{C}\gamma^{\hat{+}}+\mathbb{S}\gamma^{\hat{-}}=\left( \begin{array}{cc}
0  & \sqrt{1-\mathbb{S}}\\
\sqrt{1+\mathbb{S}} & 0
\end{array}\right) 
,\notag\\
\gamma_{\hat{-}}&=\mathbb{S}\gamma^{\hat{+}}-\mathbb{C}\gamma^{\hat{-}}=\left( \begin{array}{cc}
0  & \sqrt{1+\mathbb{S}}\\
-\sqrt{1-\mathbb{S}} & 0
\end{array}\right) , 
\end{align}
where $\sqrt{1\pm\mathbb{S}}$ can be identically given by $\cos\delta\pm\sin\delta $ with $ \cos\delta \geq\sin\delta $ always due to $0\le \delta \le \frac{\pi}{4}$.
One can explicitly check that the interpolating $ \gamma $ matrices satisfy $ \left\lbrace \gamma^{\hat{\mu}},\gamma^{\hat{\nu}}\right\rbrace =2g^{\hat{\mu}\hat{\nu}}\cdot\mathbf{I}_{2\times 2} $, in particular,
\begin{equation}\label{eqn:inter_gamma_relation}
(\gamma^{\hat{+}})^2=(\gamma_{\hat{+}})^2=\mathbb{C}\cdot\mathbf{I}_{2\times 2},\ 
(\gamma^{\hat{-}})^2=(\gamma_{\hat{-}})^2=-\mathbb{C}\cdot\mathbf{I}_{2\times 2}.
\end{equation}
As $ \delta\to\pi/4 $, one gets the usual LFD $ \gamma $ matrices given by
\begin{align}\label{eqn:lf_gamma}
\gamma^+=\left( \gamma^0+\gamma^1\right) /\sqrt{2}=\left( \begin{array}{cc}
0 & \sqrt{2}\\
0 & 0
\end{array}\right) =\gamma_-,\notag\\
\gamma^-=\left( \gamma^0-\gamma^1\right) /\sqrt{2}=\left( \begin{array}{cc}
0 & 0\\
\sqrt{2} & 0
\end{array}\right)=\gamma_+ ,
\end{align}
where $ \{\gamma^{\mu},\gamma^{\nu}\}=2g^{\mu\nu}\cdot\mathbf{I}_{2\times 2} $
and $ (\gamma^+)^2=(\gamma^-)^2=0 $. 

Now, using the Bogoliubov transformation given by Eq.(\ref{transf})
for the creation/annihilation operators of the quark/anti-quark fields
as well as the boost operation for the free quark/anti-quark fields, 
we get the following relationship between the interacting spinors and 
the free spinors at rest:
\begin{equation}\label{eqn:Bogo_u}
u(p_{\hat{-}})=T(p_{\hat{-}})u^{(0)}(0),\ \  v(-p_{\hat{-}})=T(p_{\hat{-}})\ v^{(0)}(0) ,
\end{equation}
where
\begin{align}\label{eqn:TpItp}
T(p_{\hat{-}})&=\sqrt{\frac{p^{\hat{+}}}{\sqrt{\mathbb{C}}m}}\exp[-\frac{\gamma_{\hat{-}}}{\sqrt{\mathbb{C}}}\frac{\theta(p_{\hat{-}})}{2}]\notag\\
&=
\sqrt{\frac{p^{\hat{+}}}{\sqrt{\mathbb{C}}m}}\exp[-\frac{\gamma_{\hat{-}}}{\sqrt{\mathbb{C}}}\frac{\theta_f(p_{\hat{-}})+2\zeta(p_{\hat{-}}) }{2}]\notag\\
&=
T_f(p_{\hat{-}})
\exp[-\frac{\gamma_{\hat{-}}}{\sqrt{\mathbb{C}}}\zeta(p_{\hat{-}})]\
\end{align}
with 
\begin{align}
{\rm e}^{-\frac{\gamma_{\hat{-}}}{\sqrt{\mathbb{C}}}\zeta(p_{\hat{-}})}
&= \cos\zeta(p_{\hat{-}})\cdot\mathbf{I}_{2\times 2}-\sin\zeta(p_{\hat{-}})\frac{\gamma_{\hat{-}}}{\sqrt{\mathbb{C}}}  \notag  \\
&=
\left( \begin{array}{cc}
\cos\zeta(p_{\hat{-}}) & -\sin\zeta(p_{\hat{-}})\sqrt{\frac{1+\mathbb{S}}{\mathbb{C}}}\\
\sin\zeta(p_{\hat{-}})\sqrt{\frac{1-\mathbb{S}}{\mathbb{C}}} & \cos\zeta(p_{\hat{-}})
\end{array}\right) ,
\end{align}
and
\begin{widetext}
\begin{align}\label{eqn:TpfItp}
T_f(p_{\hat{-}})&=\sqrt{\frac{p^{\hat{+}}}{\sqrt{\mathbb{C}}m}}\exp[-\frac{\gamma_{\hat{-}}}{\sqrt{\mathbb{C}}}\frac{\theta_f(p_{\hat{-}})}{2}]=\sqrt{\frac{p^{\hat{+}}}{\sqrt{\mathbb{C}}m}}\left( \cos\frac{\theta_f(p_{\hat{-}})}{2}\cdot\mathbf{I}_{2\times 2}-\sin\frac{\theta_f(p_{\hat{-}})}{2}\cdot\frac{\gamma_{\hat{-}}}{\sqrt{\mathbb{C}}}\right) .
\end{align}
\end{widetext}
Note here that the free spinors in the moving frame are related to each other as $\frac{\gamma_{\hat{-}}}{\sqrt{\mathbb{C}}}\cdot u^{(0)}(p_{\hat{-}})=v^{(0)}(-p_{\hat{-}})$ and $\frac{\gamma_{\hat{-}}}{\sqrt{\mathbb{C}}}\cdot v^{(0)}(-p_{\hat{-}})=-u^{(0)}(p_{\hat{-}})$, and that 
the free spinors in the moving frame and the rest frame are related to each other 
\begin{align}
u^{(0)}(p_{\hat{-}})&=\sqrt{\frac{p^{\hat{+}}}{\sqrt{\mathbb{C}}m}}\exp[-\frac{\gamma_{\hat{-}}}{\sqrt{\mathbb{C}}}\frac{\theta_f(p_{\hat{-}})}{2}]u^{(0)}(0)\notag\\
v^{(0)}(-p_{\hat{-}})&=\sqrt{\frac{p^{\hat{+}}}{\sqrt{\mathbb{C}}m}}\exp[-\frac{\gamma_{\hat{-}}}{\sqrt{\mathbb{C}}}\frac{\theta_f(p_{\hat{-}})}{2}]v^{(0)}(0),
\end{align}
where $\theta_f(p_{\hat{-}})$ is given by Eq.(\ref{eqn:thetaf})
and is related in the IFD case to the rapidity $\eta$ as 
$\sin \theta_f = \tanh \eta$, while in general they are related by $\tanh\eta=\frac{\sin\theta_f -\tan\delta}{1-\sin\theta_f \tan\delta}$, 
or in other words, 
$\sin\theta_f=\frac{\tanh\eta+\tan\delta}{1+\tan\delta\tanh\eta}$.
From Eq.(\ref{eqn:Bogo_u}), one can get the interacting quark/anti-quark spinors 
given by Eqs.(\ref{eqn:uspinor}) and (\ref{eqn:vspinor}) in Sec.\ref{sub:Ham}
as well as Eqs.(\ref{Gamma^+`}) and (\ref{Gamma^-`}). \\


\section{\label{app:Var}Minimization of the vacuum energy with respect to the Bogoliubov angle}

In this Appendix, we show that the mass gap equation given by Eq.(\ref{gap_eq_inter_theta}) can also be obtained by minimizing the vacuum energy 
$\mathcal{E}_v$ in Eq.~(\ref{eqn:vacuum_energy}) with respect to the Bogoliubov angle, $\theta(p_{\hat{-}})$ in addition to the methods presented in Sec.~\ref{sec:self}.
Recall that 
\begin{widetext}
	\begin{align}
	\mathcal{E}_v=\int\frac{dp_{\hat{-}}}{2\pi}\text{Tr}\left[ \left(-\gamma^0\gamma^{\hat{-}}p_{\hat{-}}+m\gamma^0 \right)\Gamma^-(p_{\hat{-}})\right] +\frac{\lambda}{2}\int \frac{dp_{\hat{-}}}{2\pi}\int \frac{dk_{\hat{-}}}{(p_{\hat{-}}-k_{\hat{-}})^2}\text{Tr}\left[\gamma^0\gamma^{\hat{+}}\Gamma^+(k_{\hat{-}})\gamma^0\gamma^{\hat{+}} \Gamma^-(p_{\hat{-}})\right] ,\label{eqn:vacuum_energy_re}
	\end{align}
where $\Gamma^{\pm}$ is defined by Eqs.(\ref{Gamma^+`}) and (\ref{Gamma^-`}).
Let us now compute the small variation of $\mathcal{E}_v$ as
\begin{align}
\delta\mathcal{E}_v&=\int\frac{dp_{\hat{-}}}{2\pi}\text{Tr}\left[ \left(-\gamma^0\gamma^{\hat{-}}p_{\hat{-}}+m\gamma^0 \right)\delta\Gamma^-(p_{\hat{-}})\right] 
+\frac{\lambda}{2}\int \frac{dp_{\hat{-}}}{2\pi}\int \frac{dk_{\hat{-}}}{(p_{\hat{-}}-k_{\hat{-}})^2}\text{Tr}\left[\gamma^0\gamma^{\hat{+}}\delta\Gamma^+(k_{\hat{-}})\gamma^0\gamma^{\hat{+}} \Gamma^-(p_{\hat{-}})\right]\notag\\
&+\frac{\lambda}{2}\int \frac{dp_{\hat{-}}}{2\pi}\int \frac{dk_{\hat{-}}}{(p_{\hat{-}}-k_{\hat{-}})^2}\text{Tr}\left[\gamma^0\gamma^{\hat{+}}\Gamma^+(k_{\hat{-}})\gamma^0\gamma^{\hat{+}} \delta\Gamma^-(p_{\hat{-}})\right] .\label{eqn:var_vacuum_energy}
\end{align}
In the second term of the above equation, we are able to swap variables $p_{\hat{-}}$ and $k_{\hat{-}}$, i.e.
\begin{align}
&\int\frac{dp_{\hat{-}}dk_{\hat{-}}}{(p_{\hat{-}}-k_{\hat{-}})^2}\text{Tr}\left[\gamma^0\gamma^{\hat{+}}\delta\Gamma^+(k_{\hat{-}})\gamma^0\gamma^{\hat{+}} \Gamma^-(p_{\hat{-}})\right]
=\int\frac{dp_{\hat{-}}dk_{\hat{-}}}{(p_{\hat{-}}-k_{\hat{-}})^2}\text{Tr}\left[\gamma^0\gamma^{\hat{+}}\delta\Gamma^+(p_{\hat{-}})\gamma^0\gamma^{\hat{+}} \Gamma^-(k_{\hat{-}})\right]\notag\\
=&-\int\frac{dp_{\hat{-}}dk_{\hat{-}}}{(p_{\hat{-}}-k_{\hat{-}})^2}\text{Tr}\left[\gamma^0\gamma^{\hat{+}}\delta\Gamma^-(p_{\hat{-}})\gamma^0\gamma^{\hat{+}} \Gamma^-(k_{\hat{-}})\right],
\end{align}
where we used the fact that the sum of $\Gamma^+(p_{\hat{-}})$ and 
$\Gamma^-(p_{\hat{-}})$ is independent of $p_{\hat{-}}$ from the second line
to the third line. 
Thus, we obtain
\begin{align}
\delta\mathcal{E}_v&=\int\frac{dp_{\hat{-}}}{2\pi}\text{Tr}\left[ \left(-\gamma^0\gamma^{\hat{-}}p_{\hat{-}}+m\gamma^0 \right)\delta\Gamma^-(p_{\hat{-}})\right] \notag\\
&+\frac{\lambda}{2}\int \frac{dp_{\hat{-}}}{2\pi}\int \frac{dk_{\hat{-}}}{(p_{\hat{-}}-k_{\hat{-}})^2}\text{Tr}\left[\gamma^0\gamma^{\hat{+}}\Gamma^+(k_{\hat{-}})\gamma^0\gamma^{\hat{+}} \delta\Gamma^-(p_{\hat{-}})-\gamma^0\gamma^{\hat{+}}\delta\Gamma^-(p_{\hat{-}})\gamma^0\gamma^{\hat{+}} \Gamma^-(k_{\hat{-}})\right] .\label{eqn:var_vacuum_energy_2}
\end{align}
The functional differentiation of $\mathcal{E}_v$ relative to $\theta(p_{\hat{-}})$
for a given $p_{\hat{-}}$ is then given by
\begin{align}
\frac{\delta\mathcal{E}_v}{\delta\theta(p_{\hat{-}})}&=\text{Tr}\left[ \left(-\gamma^0\gamma^{\hat{-}}p_{\hat{-}}+m\gamma^0\right) \frac{\delta\Gamma^-(p_{\hat{-}})}{\delta\theta(p_{\hat{-}})}\right] \notag\\
&+ \frac{\lambda}{2}
\int \frac{dk_{\hat{-}}}{(p_{\hat{-}}-k_{\hat{-}})^2}\text{Tr}\left[
\left( 
\gamma^0\gamma^{\hat{+}} \Gamma^+(k_{\hat{-}})
\gamma^0\gamma^{\hat{+}} \frac{\delta\Gamma^-(p_{\hat{-}})}{\delta\theta(p_{\hat{-}})}
-\gamma^0\gamma^{\hat{+}}
\frac{\delta\Gamma^-(p_{\hat{-}})}{\delta\theta(p_{\hat{-}})}
\gamma^0\gamma^{\hat{+}}
\Gamma^-(k_{\hat{-}})\right) \right] =0,
\label{diff_vacuum_energy}
\end{align}
i.e.  
\begin{align}
&\text{Tr}\left\lbrace 
\left( \begin{array}{cc}
-(\cos\delta+\sin\delta)p_{\hat{-}} & m \\
m&(\cos\delta-\sin\delta)p_{\hat{-}}
\end{array}\right)
\left( \begin{array}{cc}
\frac{\cos\theta(p_{\hat{-}})}{2(\cos\delta-\sin\delta)} & \frac{\sin\theta(p_{\hat{-}})}{2\sqrt{\mathbb{C}}} \\
\frac{\sin\theta(p_{\hat{-}})}{2\sqrt{\mathbb{C}}} & \frac{-\cos\theta(p_{\hat{-}})}{2(\cos\delta+\sin\delta)}
\end{array}\right)
+ \frac{\lambda}{2}\int \frac{dk_{\hat{-}}}{(p_{\hat{-}}-k_{\hat{-}})^2}\right.\notag\\
&\left.\times\left[ 
\left( \begin{array}{cc}
\frac{1-\sin\theta(k_{\hat{-}})}{2} & \frac{\cos\delta-\sin\delta}{2\sqrt{\mathbb{C}}}\cos\theta(k_{\hat{-}})\\
\frac{\cos\delta+\sin\delta}{2\sqrt{\mathbb{C}}}\cos\theta(k_{\hat{-}}) &\frac{1+\sin\theta(k_{\hat{-}})}{2}
\end{array}\right)
\left( \begin{array}{cc}
\frac{\cos\theta(p_{\hat{-}})}{2} & \frac{(\cos\delta-\sin\delta)}{2\sqrt{\mathbb{C}}}\sin\theta(p_{\hat{-}}) \\
\frac{(\cos\delta+\sin\delta)}{2\sqrt{\mathbb{C}}}\sin\theta(p_{\hat{-}})  & \frac{-\cos\theta(p_{\hat{-}})}{2}
\end{array}\right)\right.\right.\notag\\
&-\left.\left.
\left( \begin{array}{cc}
\frac{\cos\theta(p_{\hat{-}})}{2} & \frac{(\cos\delta-\sin\delta)}{2\sqrt{\mathbb{C}}}\sin\theta(p_{\hat{-}}) \\
\frac{(\cos\delta+\sin\delta)}{2\sqrt{\mathbb{C}}}\sin\theta(p_{\hat{-}})  & \frac{-\cos\theta(p_{\hat{-}})}{2}
\end{array}\right)
\left( \begin{array}{cc}
\frac{1+\sin\theta(k_{\hat{-}})}{2} & -\frac{\cos\delta-\sin\delta}{2\sqrt{\mathbb{C}}}\cos\theta(k_{\hat{-}})\\
-\frac{\cos\delta+\sin\delta}{2\sqrt{\mathbb{C}}}\cos\theta(k_{\hat{-}}) &\frac{1-\sin\theta(k_{\hat{-}})}{2}
\end{array}\right)
\right] \right\rbrace  =0 .
\label{geteq}
\end{align}
The computation of the trace leads to the gap equation given by 
Eq.(\ref{gap_eq_inter_theta}). 

\end{widetext}

\section{\label{app:FreeComparison} 
Treatment of the $\lambda = 0$ (Free) case vs.  
the $\lambda \neq 0$ (Interacting) case with respect to
the mass dimension $\sqrt{2\lambda}$ }
\begin{figure*}
	\centering
	\subfloat[]{
		\includegraphics[width=0.5\linewidth]{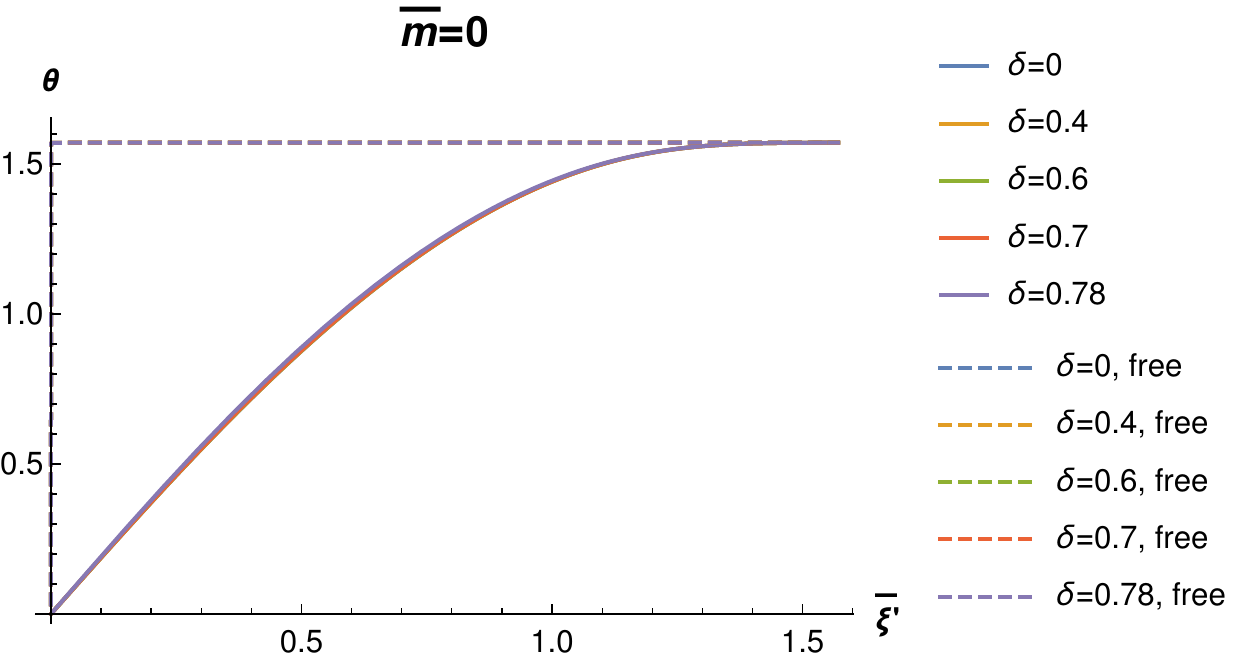}
		\label{fig:m0thetaintervsfreere}}
	\centering
	\subfloat[]{
		\includegraphics[width=0.5\linewidth]{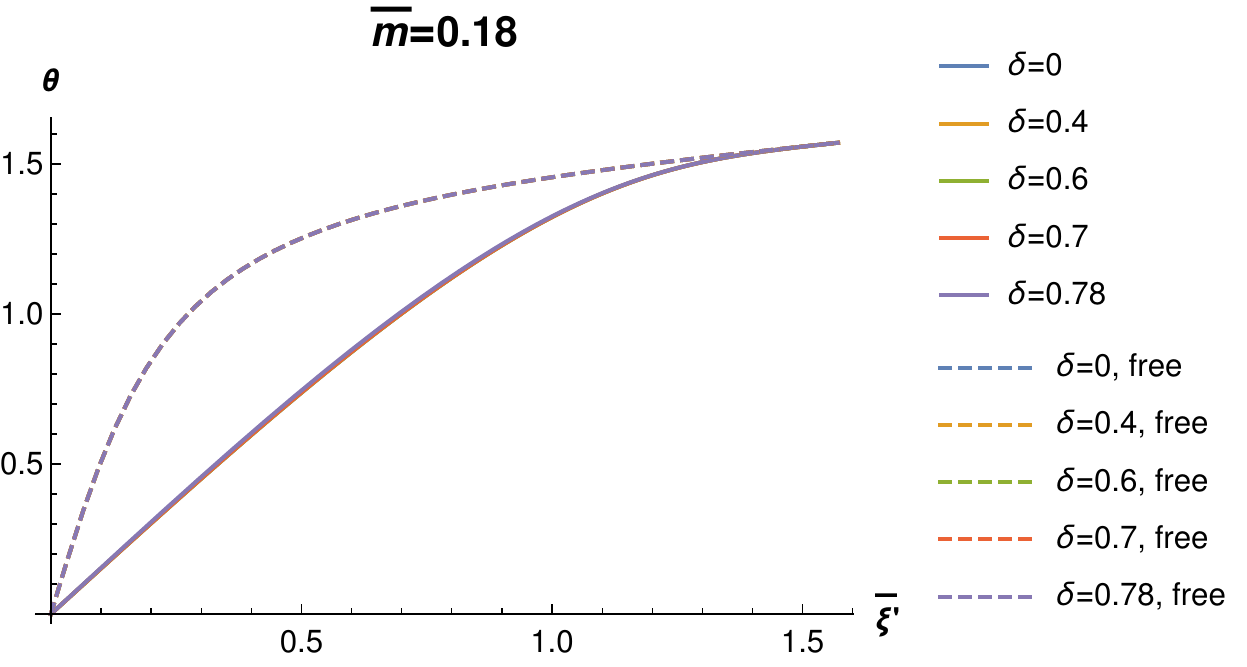}
		\label{fig:m018thetaintervsfreere}}
	\\
	\centering
	\subfloat[]{
		\includegraphics[width=0.5\linewidth]{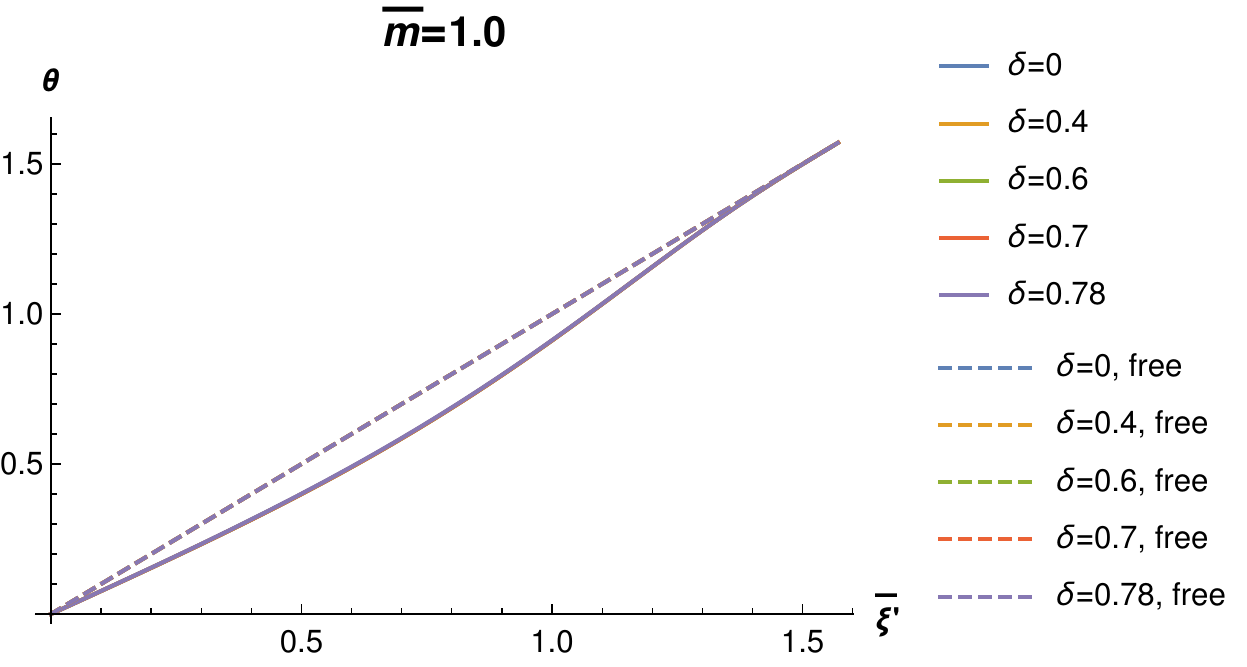}
		\label{fig:m100thetaintervsfreere}}
	\centering
	\subfloat[]{
		\includegraphics[width=0.5\linewidth]{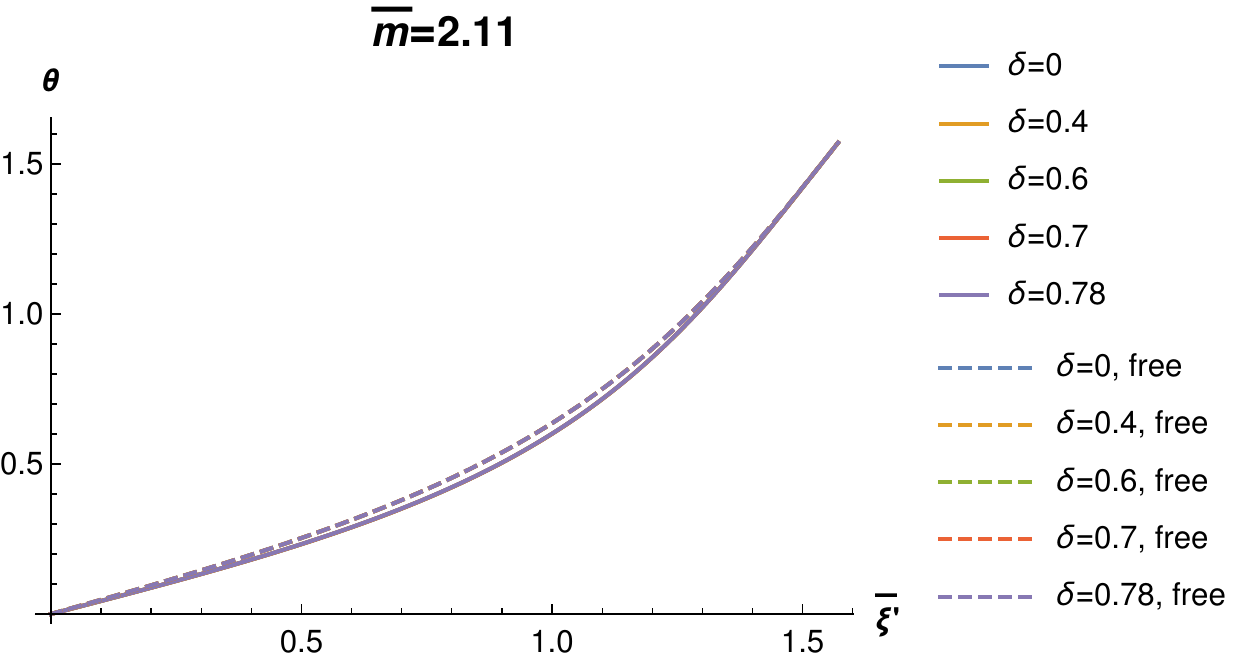}
		\label{fig:m211thetaintervsfreere}}
	\caption{\label{fig:solutionthetaintervsfreere}The numerical solutions of $ \theta(\bar{p}'_{\hat{-}}) $ versus the free solutions $\theta_f(\bar{p}'_{\hat{-}})$ for a few choices of quark mass (a) $\bar{m}=0$ (b) $\bar{m}=0.18$ (c) $\bar{m}=1.0$ and (d) $\bar{m}=2.11$, for several different interpolation angles.}
\end{figure*}

As the 't Hooft coupling $\lambda$ given by Eq.(\ref{lambda-def}) has the 
mass-square dimension, we scaled out the mass dimension $ \sqrt{2\lambda} $
and used the dimensionless mass $m$, longitudinal momentum 
$p_{\hat{-}}$, etc., in presenting all the figures and tables of our work. 
Of course, we could have explicitly defined
the dimensionless variables, e.g., denoted by
\begin{equation}\label{primem}
\bar{p}_{\hat{-}}=\frac{p_{\hat{-}}}{\sqrt{2\lambda}},\ \bar{E}=\frac{E}{\sqrt{2\lambda}}, \bar{m}=\frac{m}{\sqrt{2\lambda}},
\end{equation}
and rewrite, for example, Eqs.~(\ref{gap_eq_inter_E}) and ~(\ref{gap_eq_inter_theta}) 
as
\begin{align}\label{gap_eq_inter_E_rescaled_m}
\bar{E}(\bar{p}_{\hat{-}})&=\bar{p}_{\hat{-}}\sin\theta(\bar{p}_{\hat{-}})+\sqrt{\mathbb{C}}\bar{m}\cos\theta(\bar{p}_{\hat{-}})\notag\\
&+\mathbb{C}\cdot\frac{1}{4}\dashint\frac{d\bar{k}_{\hat{-}}}{(\bar{p}_{\hat{-}}-\bar{k}_{\hat{-}})^2}\cos\left( \theta(\bar{p}_{\hat{-}})-\theta(\bar{k}_{\hat{-}})\right),
\end{align}
and 
\begin{align}\label{gap_eq_inter_theta_rescaled_m}
\bar{p}_{\hat{-}}\cos\theta(\bar{p}_{\hat{-}})-\sqrt{\mathbb{C}}\bar{m}
\sin\theta(\bar{p}_{\hat{-}})&=\mathbb{C}\cdot\frac{1}{4}\dashint\frac{d\bar{k}_{\hat{-}}}{(\bar{p}_{\hat{-}}-\bar{k}_{\hat{-}})^2}\notag\\
&\times\sin\left( \theta(\bar{p}_{\hat{-}})-\theta(\bar{k}_{\hat{-}})\right),
\end{align}
respectively. Similarly, we can also scale out the interpolation angle dependence by defining the following rescaled variables 
\begin{equation}\label{prime}
\bar{p}'_{\hat{-}}=\frac{\bar{p}_{\hat{-}}}{\sqrt{\mathbb{C}}},\ \bar{E}'=\frac{\bar{E}}{\sqrt{\mathbb{C}}},
\end{equation}
to reduce Eqs.~(\ref{gap_eq_inter_E_rescaled_m}) and ~(\ref{gap_eq_inter_theta_rescaled_m}) as  
\begin{align}\label{gap_eq_inter_E_rescaled}
\bar{E}'(\bar{p}'_{\hat{-}})&=\bar{p}'_{\hat{-}}\sin\theta(\bar{p}'_{\hat{-}})+\bar{m}\cos\theta(\bar{p}'_{\hat{-}})\notag\\
&+\frac{1}{4}\dashint\frac{d\bar{k}'_{\hat{-}}}{(\bar{p}'_{\hat{-}}-\bar{k}'_{\hat{-}})^2}\cos\left( \theta(\bar{p}'_{\hat{-}})-\theta(\bar{k}'_{\hat{-}})\right) ,
\end{align}
and
\begin{align}\label{gap_eq_inter_theta_rescaled}
\bar{p}'_{\hat{-}}\cos\theta(\bar{p}'_{\hat{-}})-\bar{m}
\sin\theta(\bar{p}'_{\hat{-}})&=\frac{1}{4}\dashint\frac{d\bar{k}'_{\hat{-}}}{(\bar{p}'_{\hat{-}}-\bar{k}'_{\hat{-}})^2}\notag\\
&\times\sin\left( \theta(\bar{p}'_{\hat{-}})-\theta(\bar{k}'_{\hat{-}})\right) ,
\end{align}
respectively. Indeed, we used such rescaled variables in Eqs.(\ref{con_rescale}),
(\ref{reduced-energy-momentum-DR}) and (\ref{eqn:energy_momentum_dressed}) 
to present the corresponding results without any dependence of $\delta\in[0,\pi/4]$, confirming that the physical results are indeed invariant regardless of the interpolation angle $\delta$ as they must be. However, one should note the contrast between
the scaling by the dimensionful parameter $\sqrt{2\lambda}$
and the scaling by the dimensionless parameter 
$\sqrt{\mathbb{C}}$.
While the rescaling over the dimensionless variable $\mathbb{C}$ includes the limit of $\mathbb{C}=0$, i.e. LFD, 
the rescaling over the dimensionful variable $ \sqrt{2\lambda} $ cannot include
the limit to $\lambda = 0$. 
Namely, the free theory without any interaction must be
distinguished from the interacting theory and should be 
discussed separately.
For $\lambda=0$, in fact, the mass gap solution 
$\theta(p_{\hat{-}})=\theta_f(p_{\hat{-}})$ can be 
immediately found even analytically by taking the 
right-hand-side of Eq.(\ref{gap_eq_inter_theta}) to be zero, i.e.
\begin{equation}\label{eqn:freethetasolution}
\theta_f(p_{\hat{-}})=\arctan\left( \frac{p_{\hat{-}}}{\sqrt{\mathbb{C}}m}\right),
\end{equation} 
where the dimensionless ratio $p_{\hat{-}}/m$ can still be written as 
${\bar p}_{\hat{-}}/{\bar m}$ with the cancellation of the $\lambda =0$ factor in the ratio.
In terms of the rescaled variables in Eqs.(\ref{primem}) and (\ref{prime}), this analytic free solution becomes 
\begin{equation}\label{eqn:freethetasolutionre}
\theta_f(\bar{p}'_{\hat{-}})=\arctan\left( \frac{\bar{p}'_{\hat{-}}}{\bar{m}}\right) .
\end{equation}

In Fig.~\ref{fig:solutionthetaintervsfreere}, we plot the interacting mass gap solution 
$\theta(\bar{p}'_{\hat{-}})$ in comparison with the free mass gap solution
$\theta_f(\bar{p}'_{\hat{-}})$ as functions of rescaled variable $\bar{\xi}' =  \tan^{-1} \bar{p}'_{\hat{-}}$. Note here that the interacting mass gap solution $\theta(\bar{p}'_{\hat{-}})$ includes the LFD solution analytically given by 
Eq.(\ref{lfthetasol}) while the free analytic solution $\theta_f(\bar{p}'_{\hat{-}})$ is clearly distinguished from the interacting solution $\theta(\bar{p}'_{\hat{-}})$, although the difference between the free solution and the interacting solution gets reduced as $\bar{m}$ gets larger.
It confirms that the entire nontrivial contributions from the interaction
to the LFD solution are provided by the zero-mode $p^+ = 0$ as 
the finite $\bar{p}'_{\hat{-}}$ can be attained only if  
$\bar{p}_{\hat{-}} = \frac{p^+}{\sqrt{2\lambda}} = 0$
in the limit $\sqrt{\mathbb{C}} \rightarrow 0$ (LFD).

\section{\label{app:figures}Mesonic wavefunctions for 
$m=$ 0.045, 1.0 and 2.11 in the unit of $\sqrt{2\lambda}$}

\begin{figure*}
	\centering
	\subfloat[]{
		\includegraphics[width=0.5\linewidth]{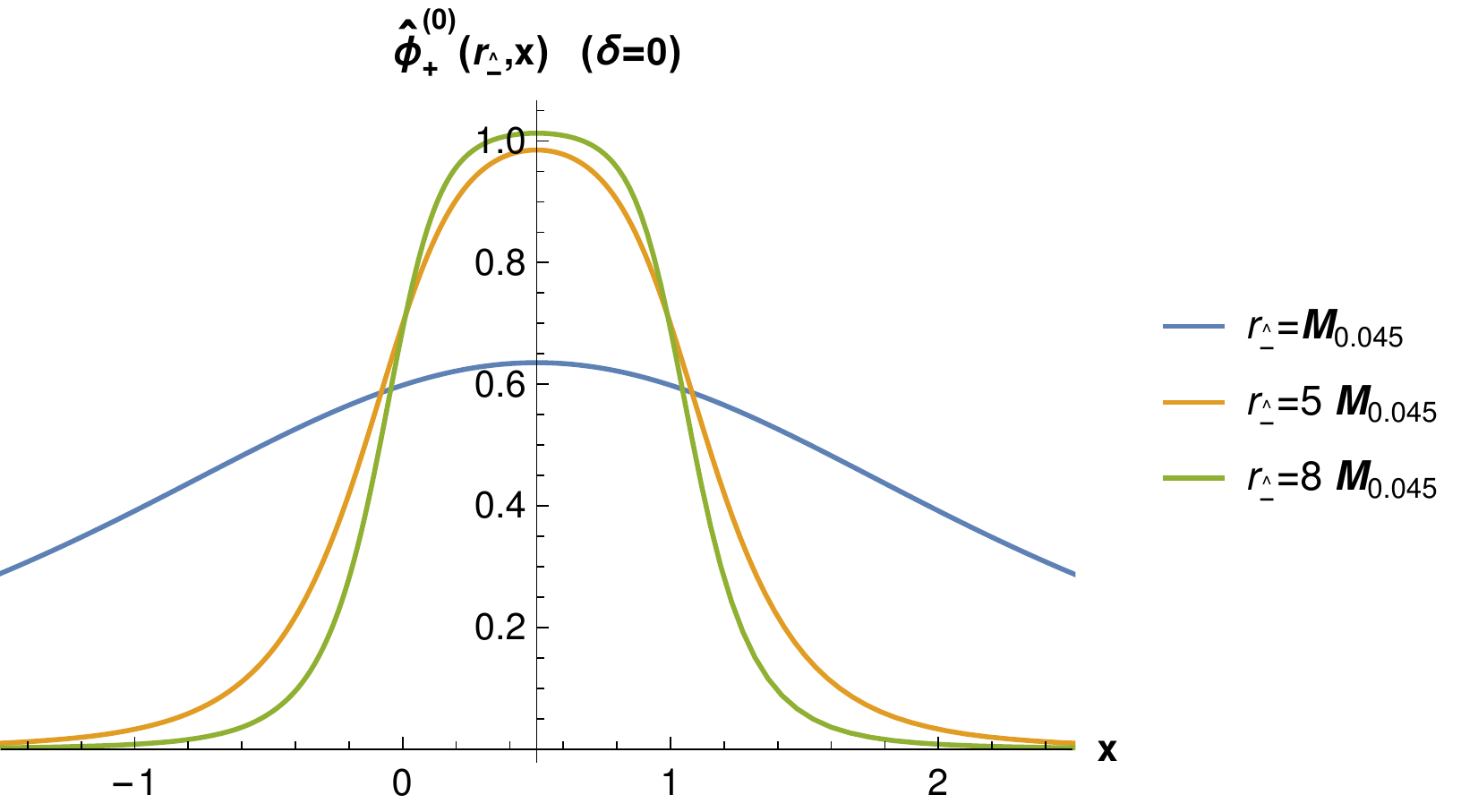}
		\label{fig:m0045grzd0}
	}
	\centering
	\subfloat[]{
		\includegraphics[width=0.5\linewidth]{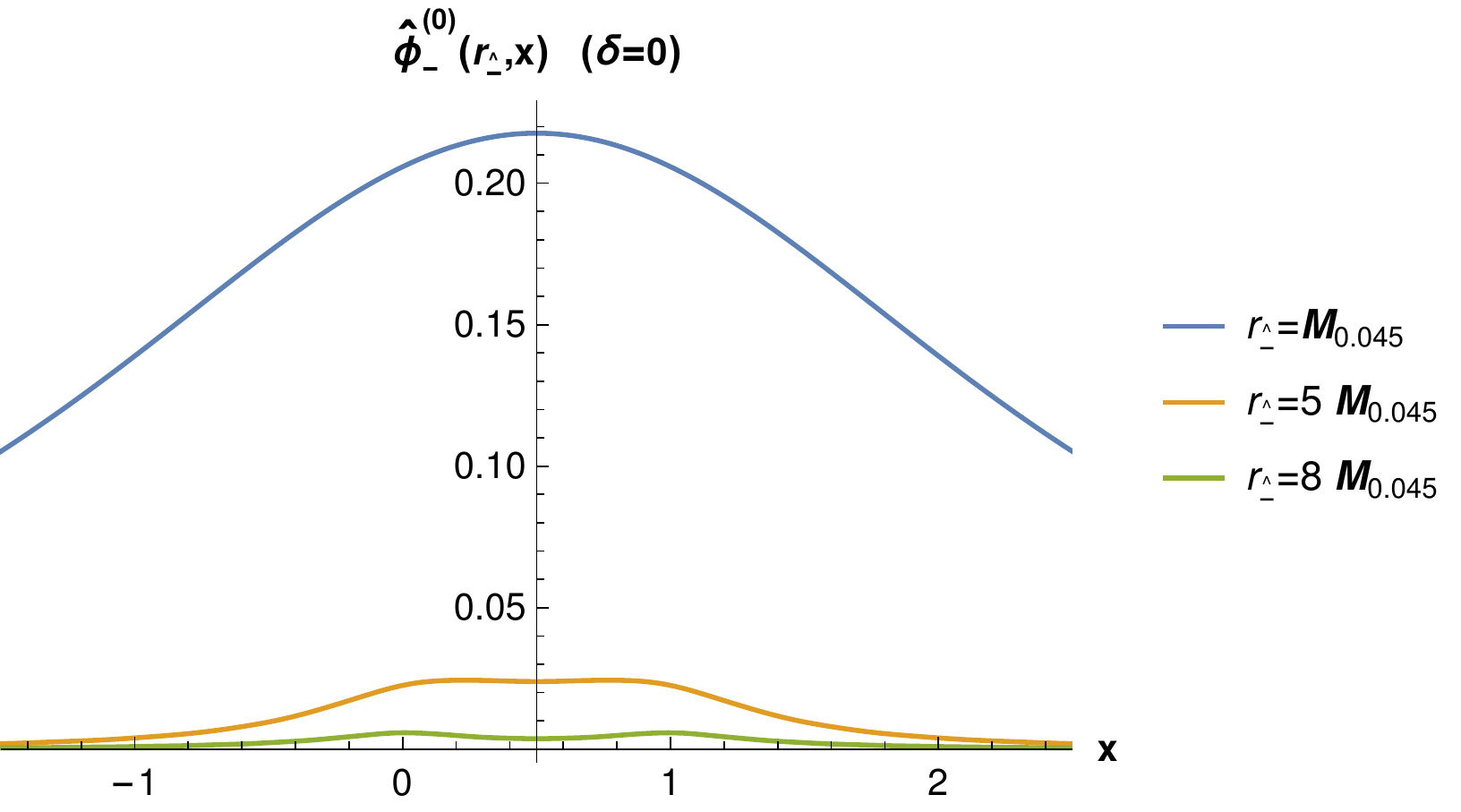}
		\label{fig:m0045grfd0}
	}
	\\
	\centering
	\subfloat[]{
		\includegraphics[width=0.5\linewidth]{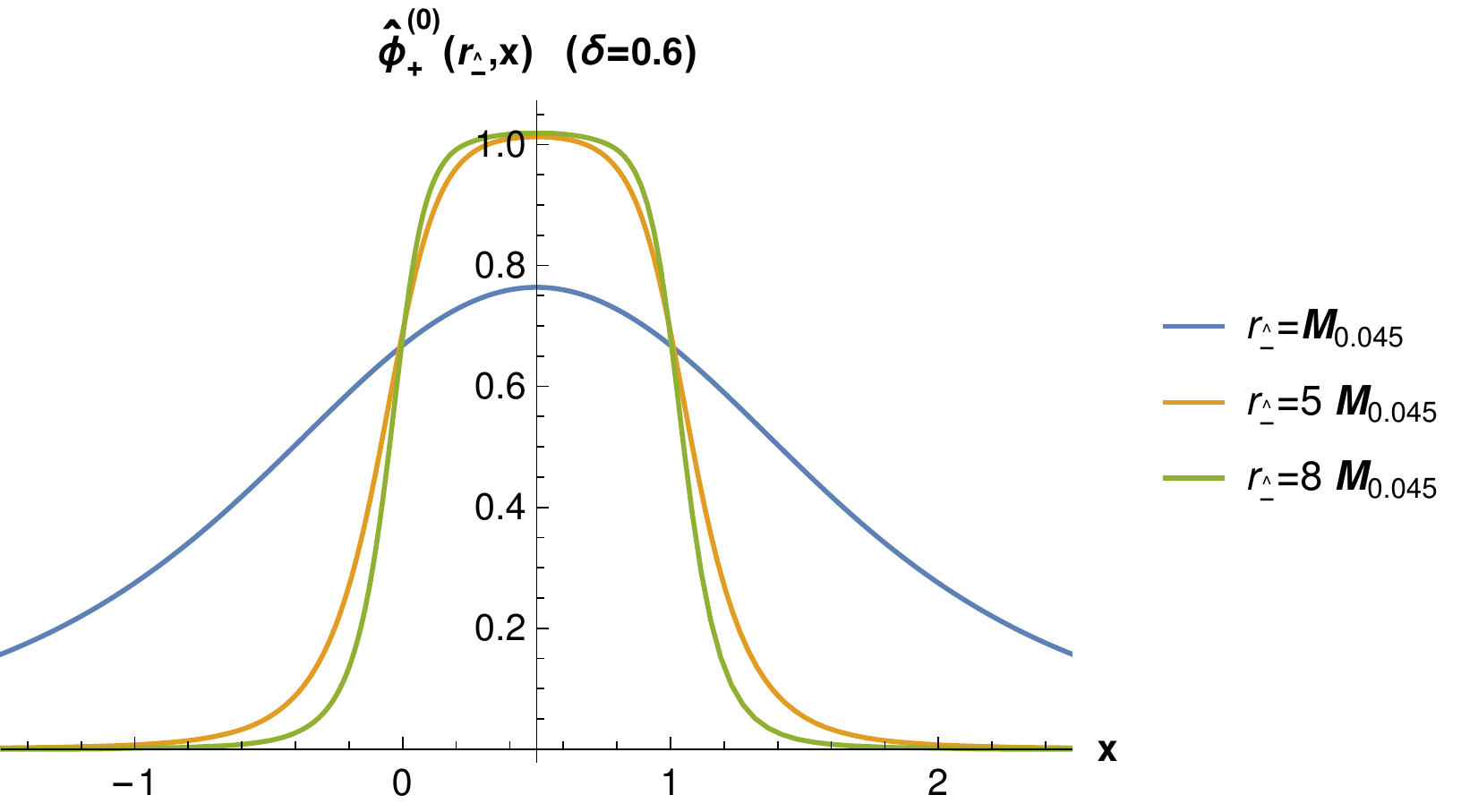}
		\label{fig:m0045grzd06}
	}
	\centering
	\subfloat[]{
		\includegraphics[width=0.5\linewidth]{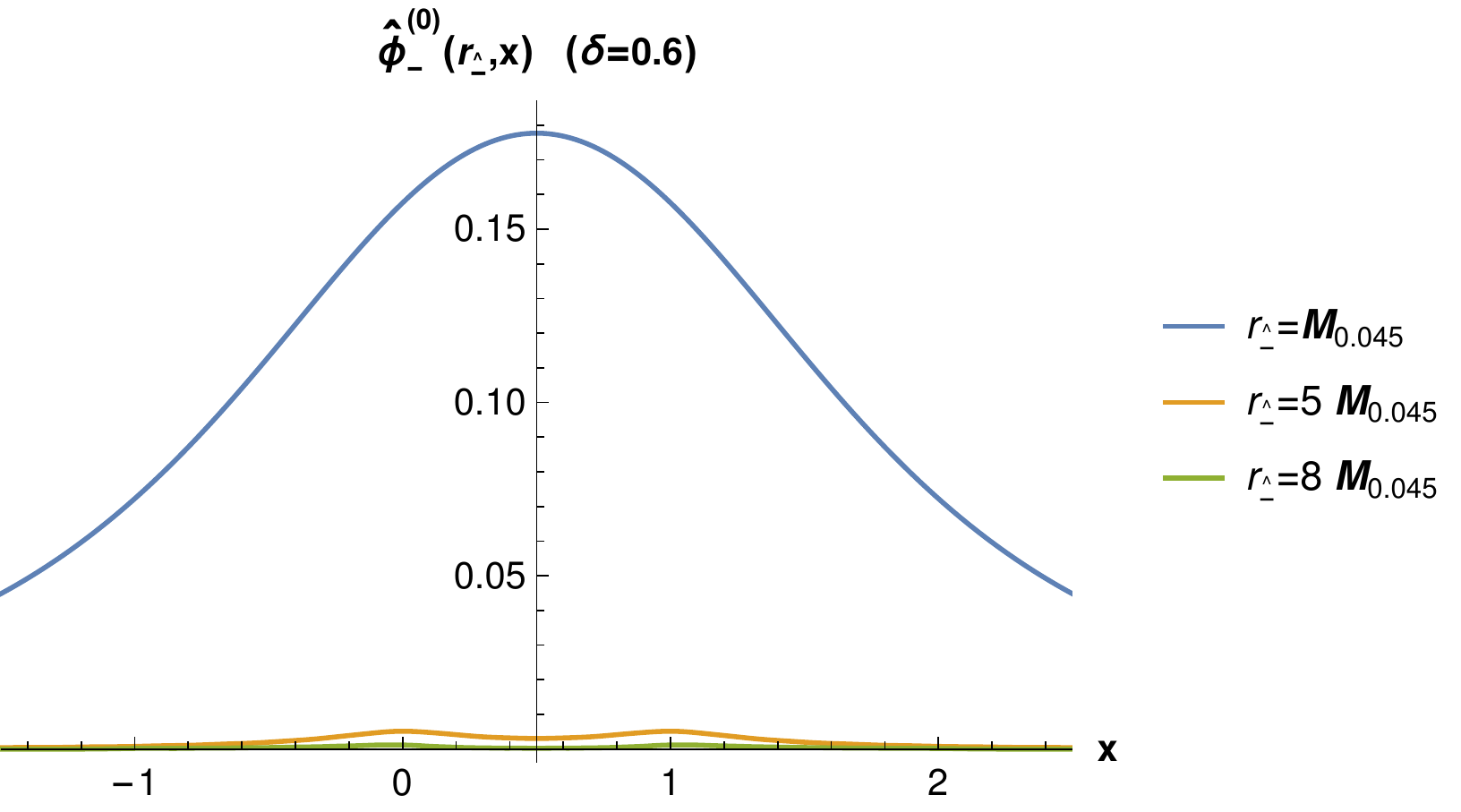}
		\label{fig:m0045grfd06}
	}
	\\
	\centering
	\subfloat[]{
		\includegraphics[width=0.5\linewidth]{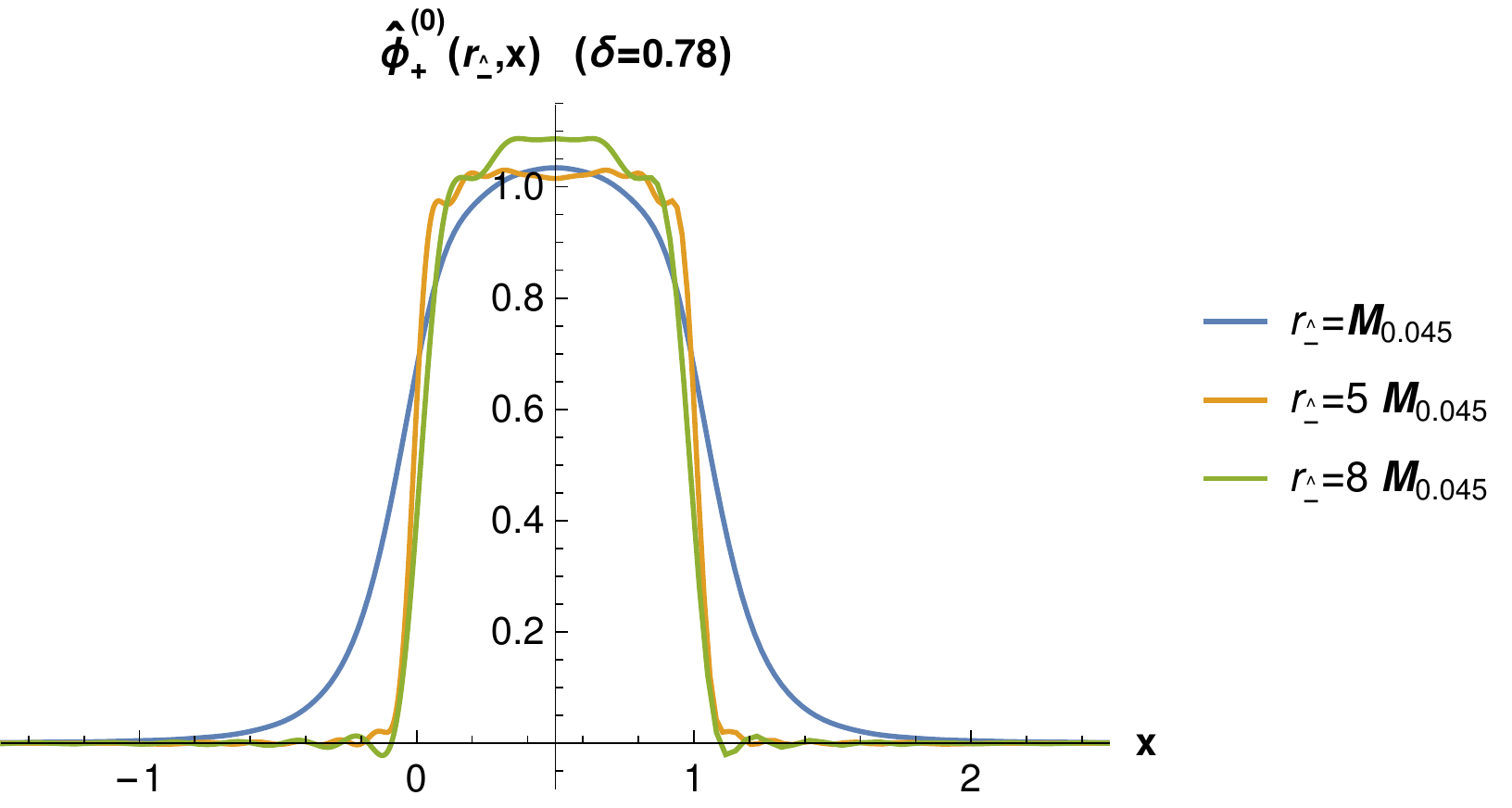}
		\label{fig:m0045grzd078}
	}
	\centering
	\subfloat[]{
		\includegraphics[width=0.5\linewidth]{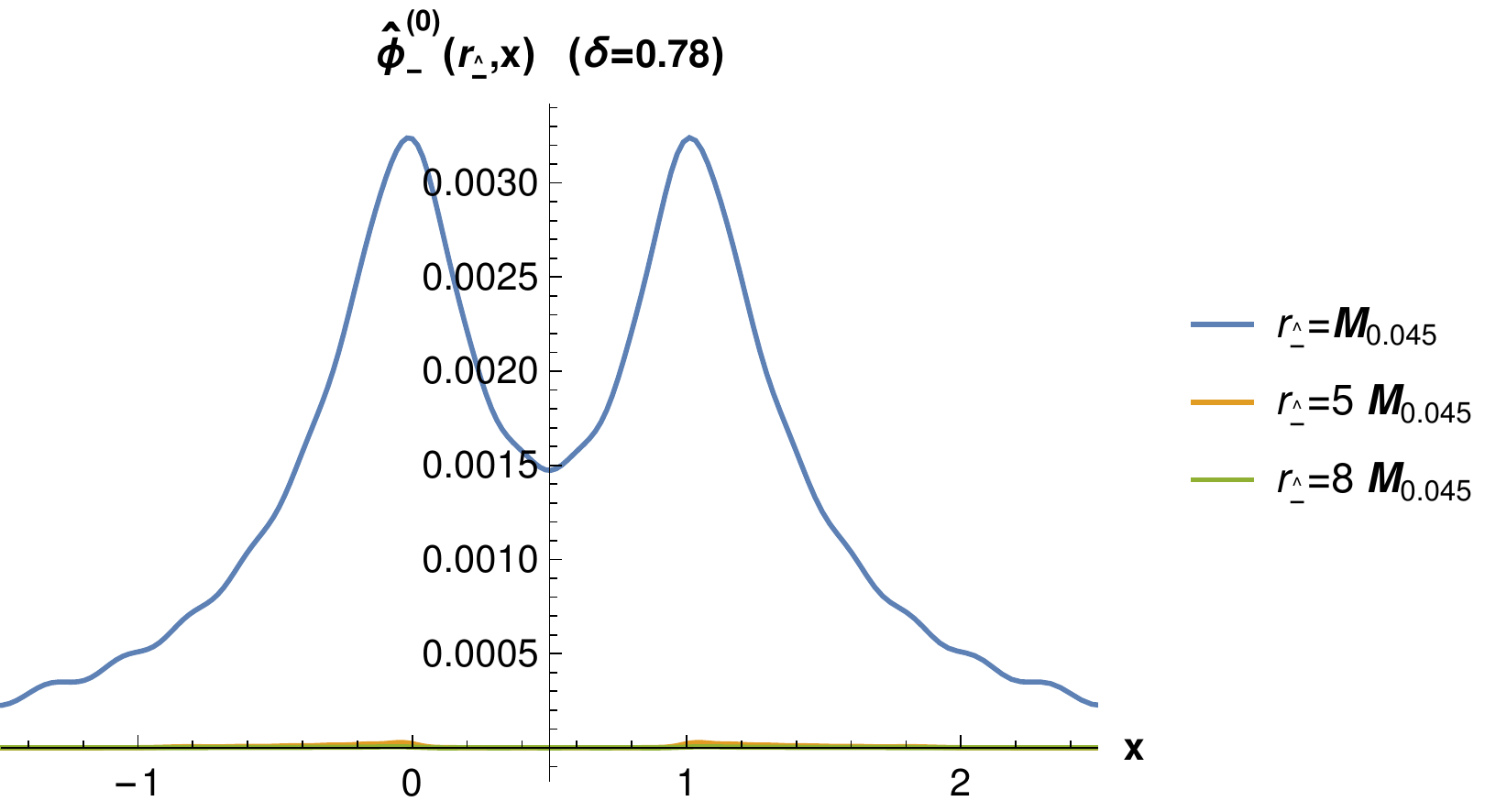}
		\label{fig:m0045grfd078}
	}
	\caption{Ground state wave functions $\hat\phi_+^{(0)}(r_{\hat{-}},x)$ and $\hat\phi_-^{(0)}(r_{\hat{-}},x)$ for $ m=0.045 $. All quantities are in proper units of $ \sqrt{2\lambda} $.\label{fig:m0045grz}}
\end{figure*}

\begin{figure*}
	\centering
	\subfloat[]{
		\includegraphics[width=0.5\linewidth]{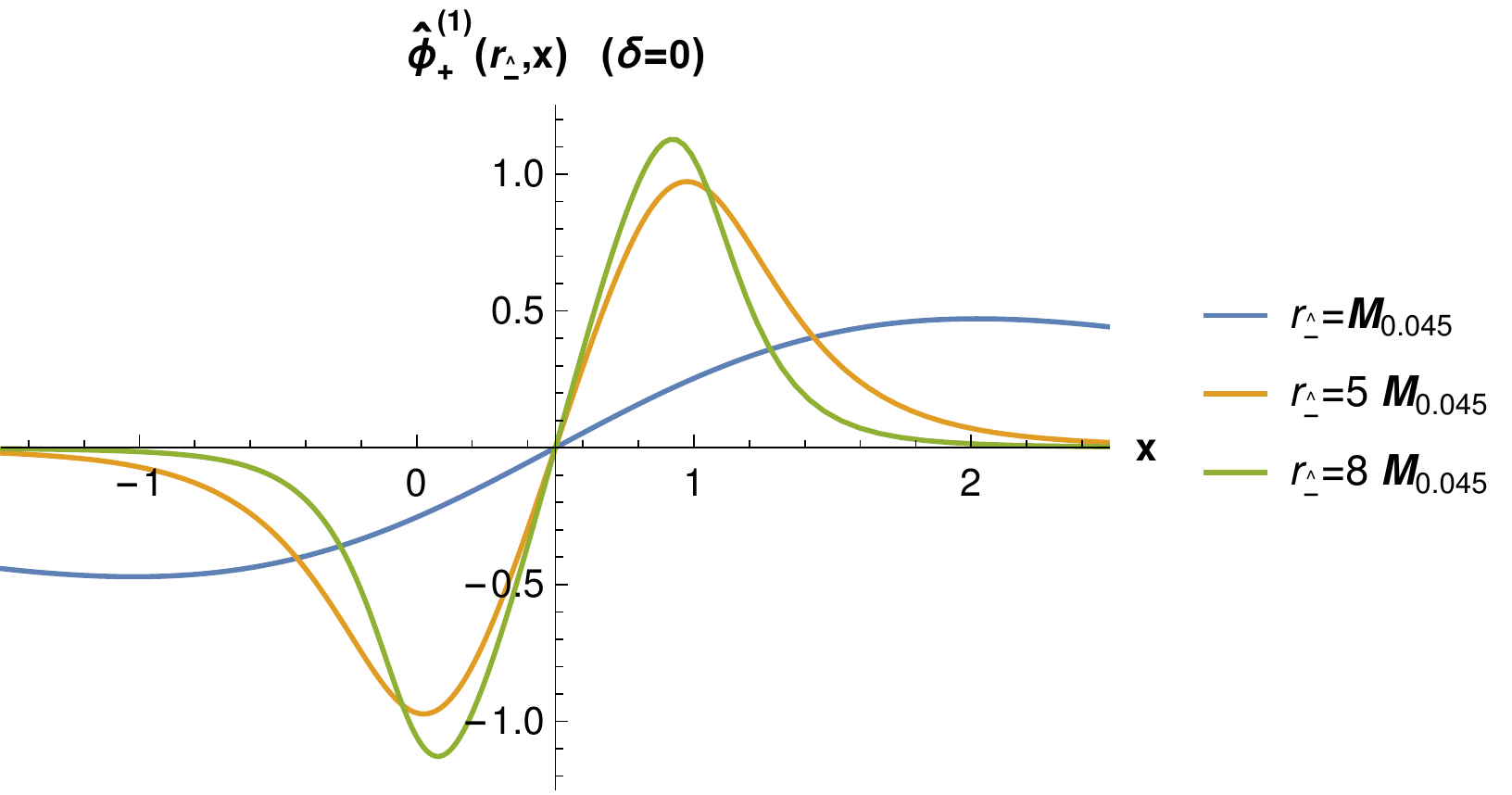}
		\label{fig:m0045fezd0}
	}
	\centering
	\subfloat[]{
		\includegraphics[width=0.5\linewidth]{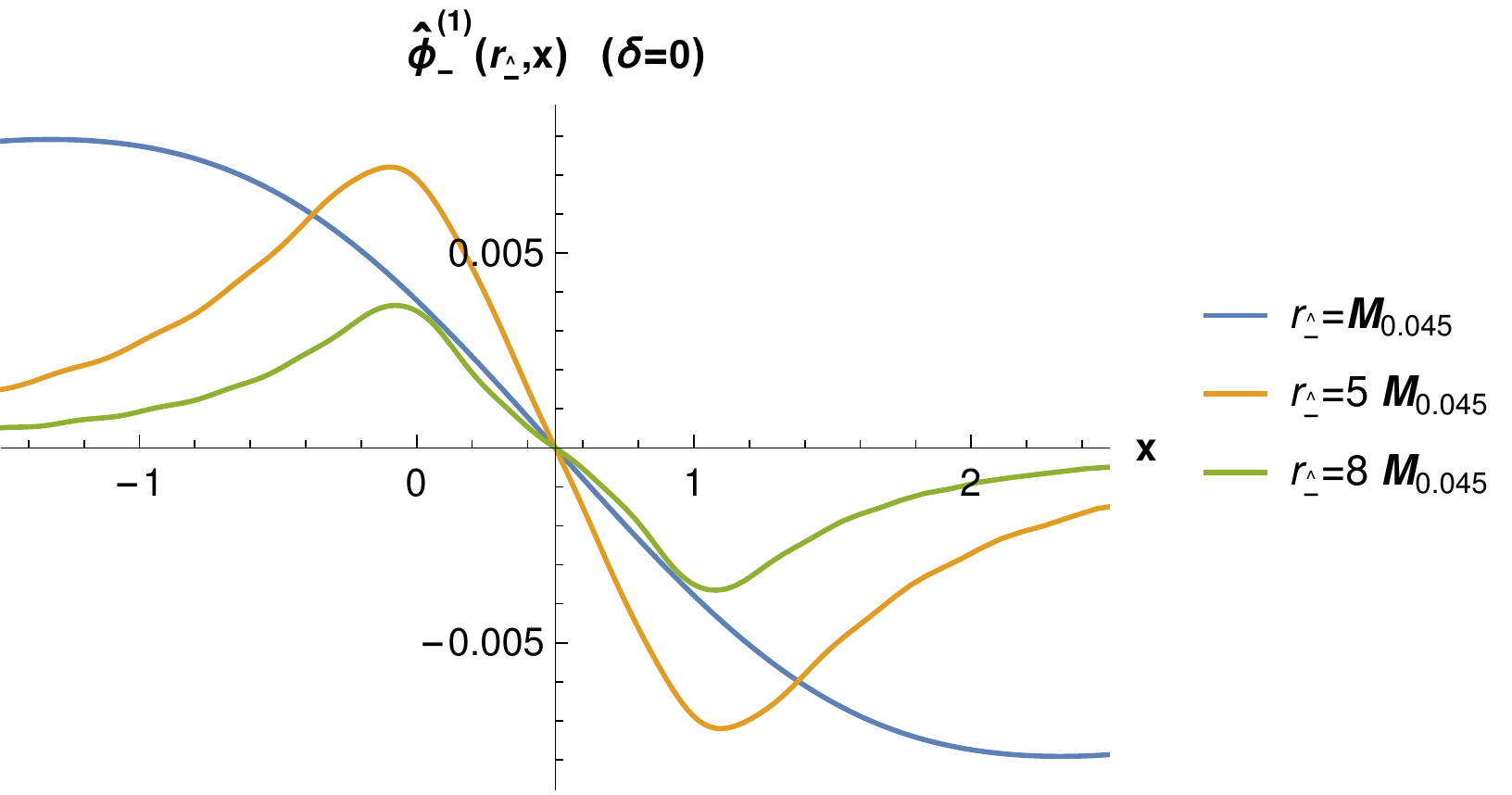}
		\label{fig:m0045fefd0}
	}\\
	\centering
	\subfloat[]{
		\includegraphics[width=0.5\linewidth]{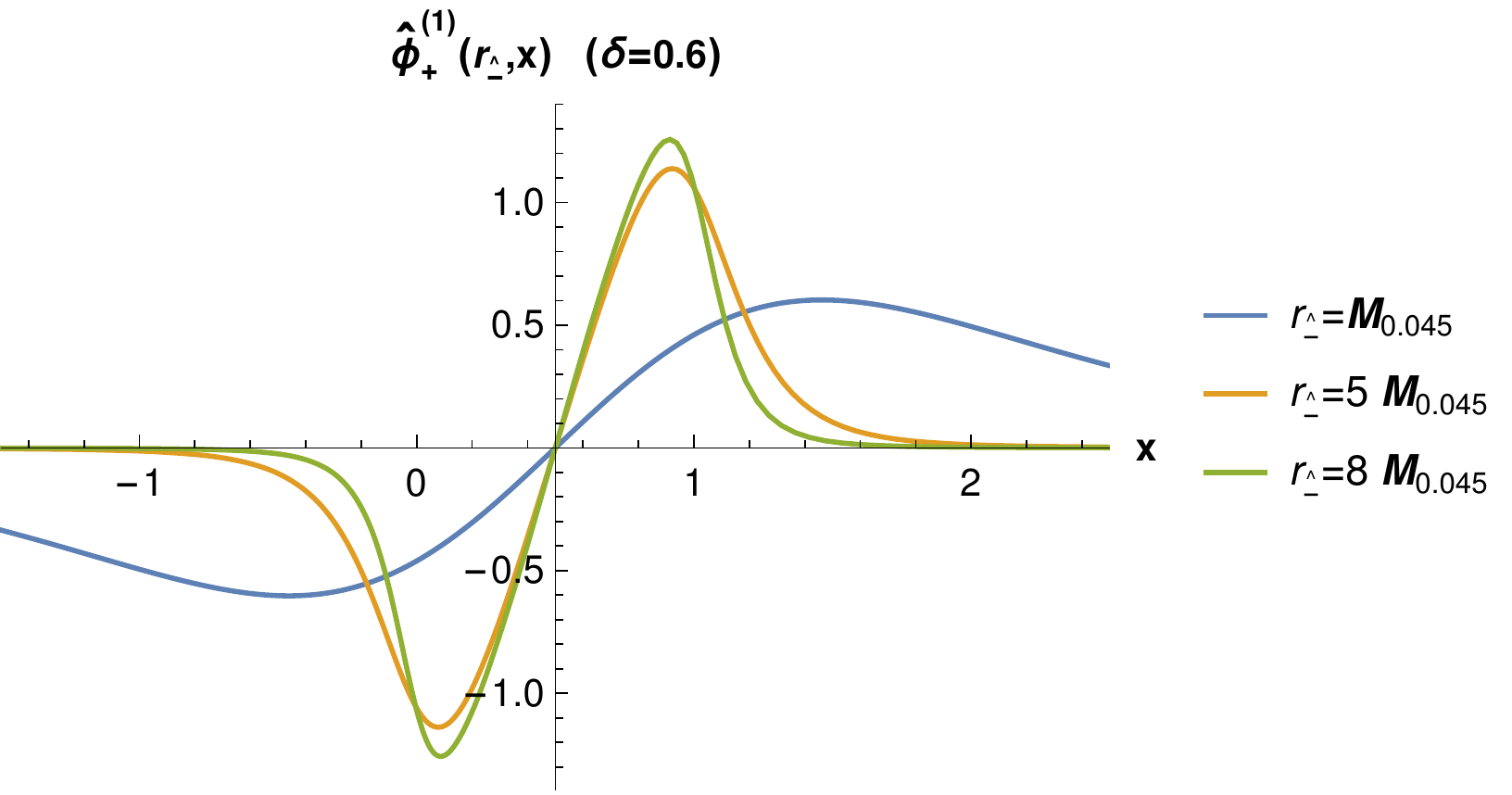}
		\label{fig:m0045fezd06}
	}
	\centering
	\subfloat[]{
		\includegraphics[width=0.5\linewidth]{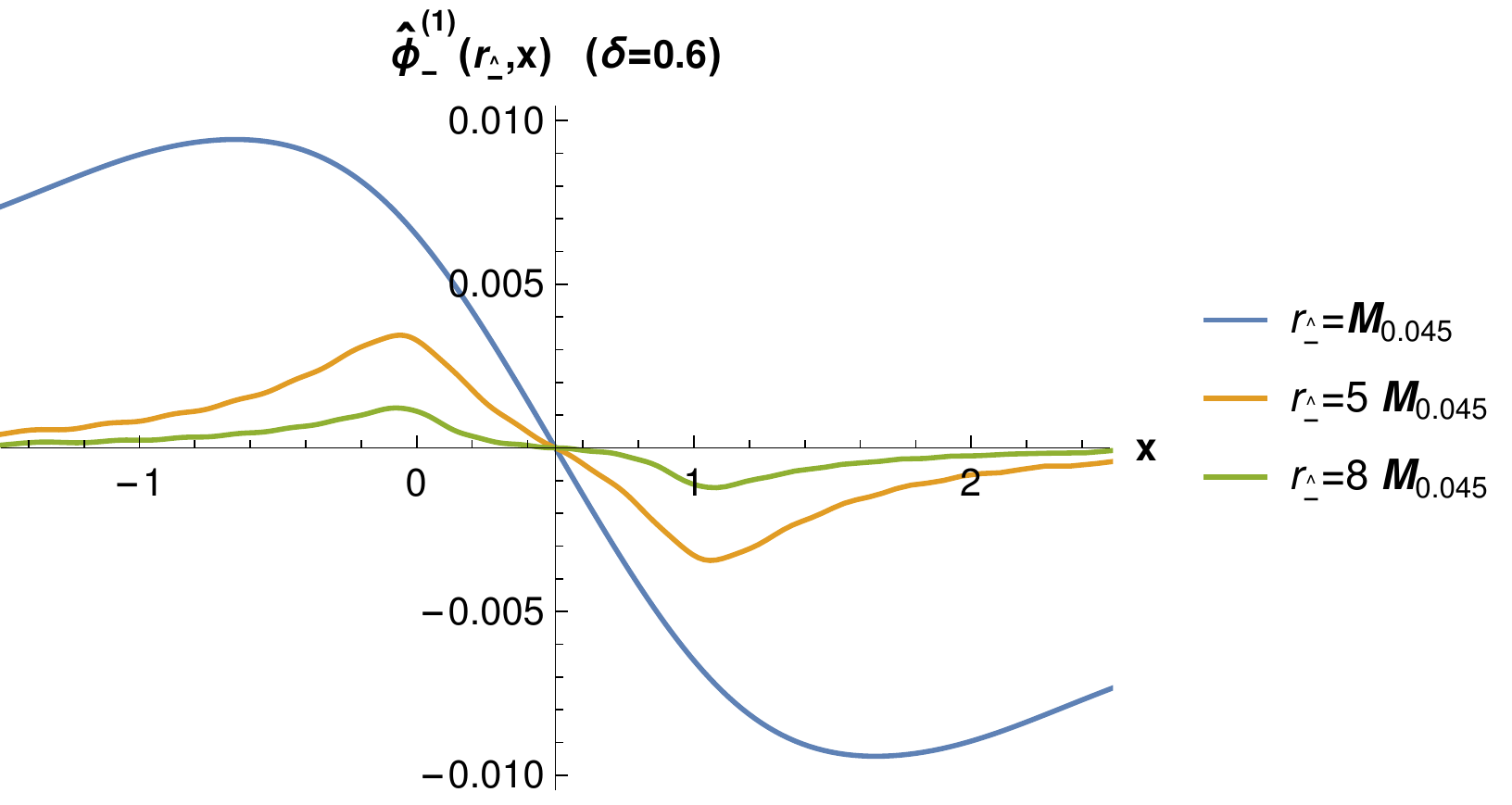}	
		\label{fig:m0045fefd06}
	}\\
	\centering
	\subfloat[]{
		\includegraphics[width=0.5\linewidth]{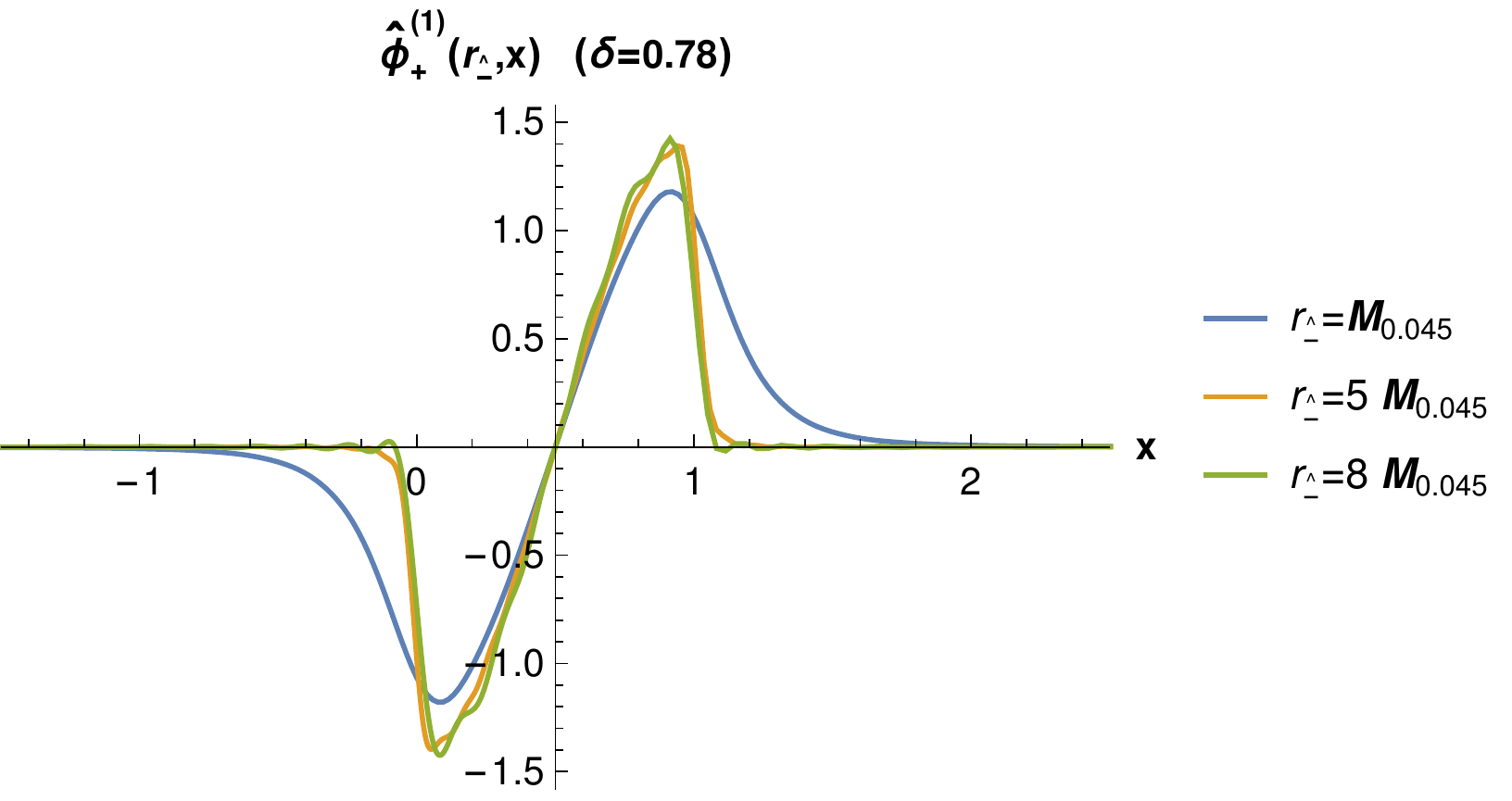}
		\label{fig:m0045fezd078}
	}
	\centering
	\subfloat[]{
		\includegraphics[width=0.5\linewidth]{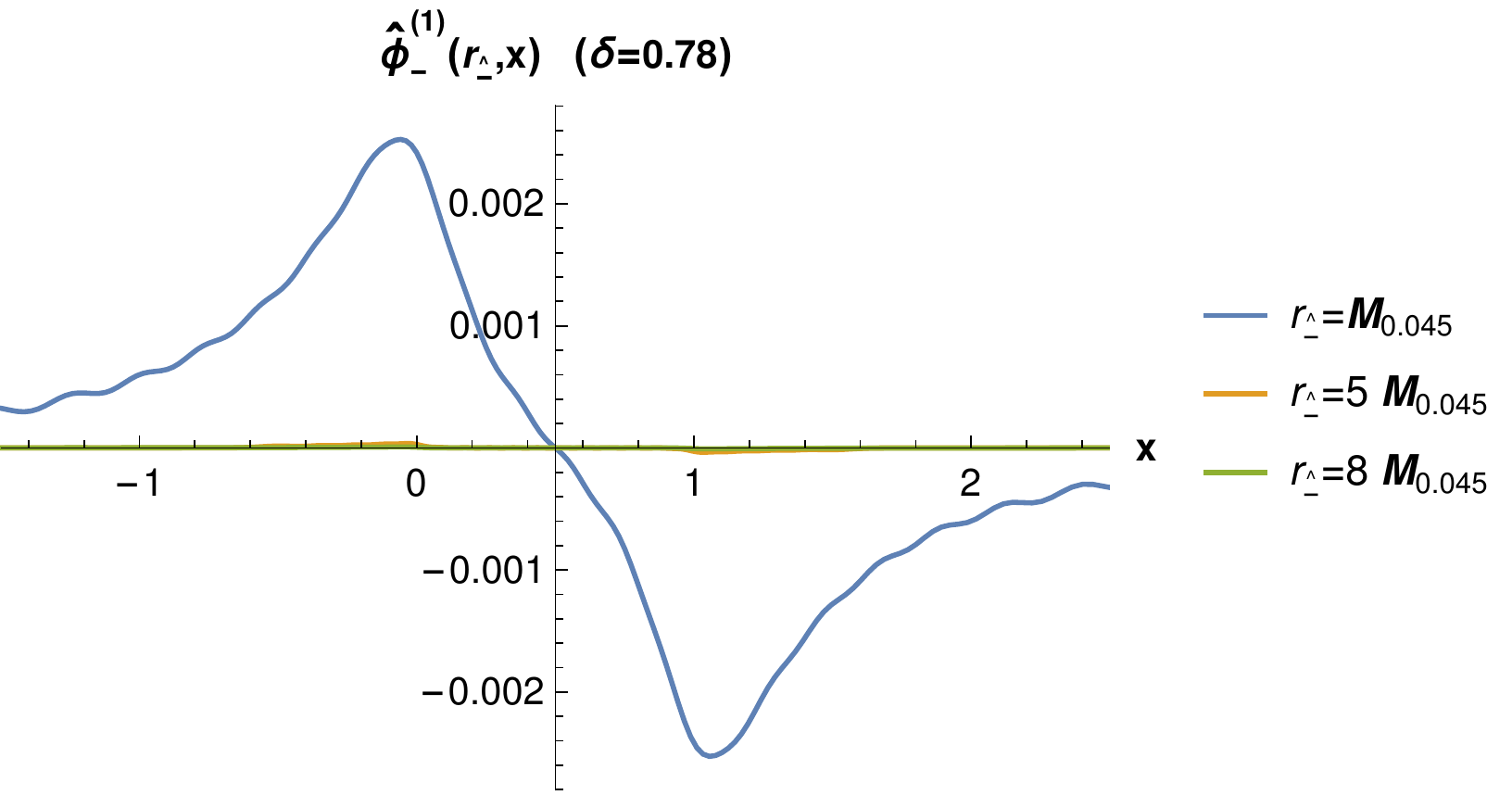}
		\label{fig:m0045fefd078}
	}\\
	\caption{First excited state wave functions $\hat\phi_+^{(1)}(r_{\hat{-}},x)$ and $\hat\phi_-^{(1)}(r_{\hat{-}},x)$ for $ m=0.045 $. All quantities are in proper units of $ \sqrt{2\lambda} $.\label{fig:m0045fez}}
\end{figure*}

\begin{figure*} 
	\centering
	\subfloat[]{
		\includegraphics[width=0.5\linewidth]{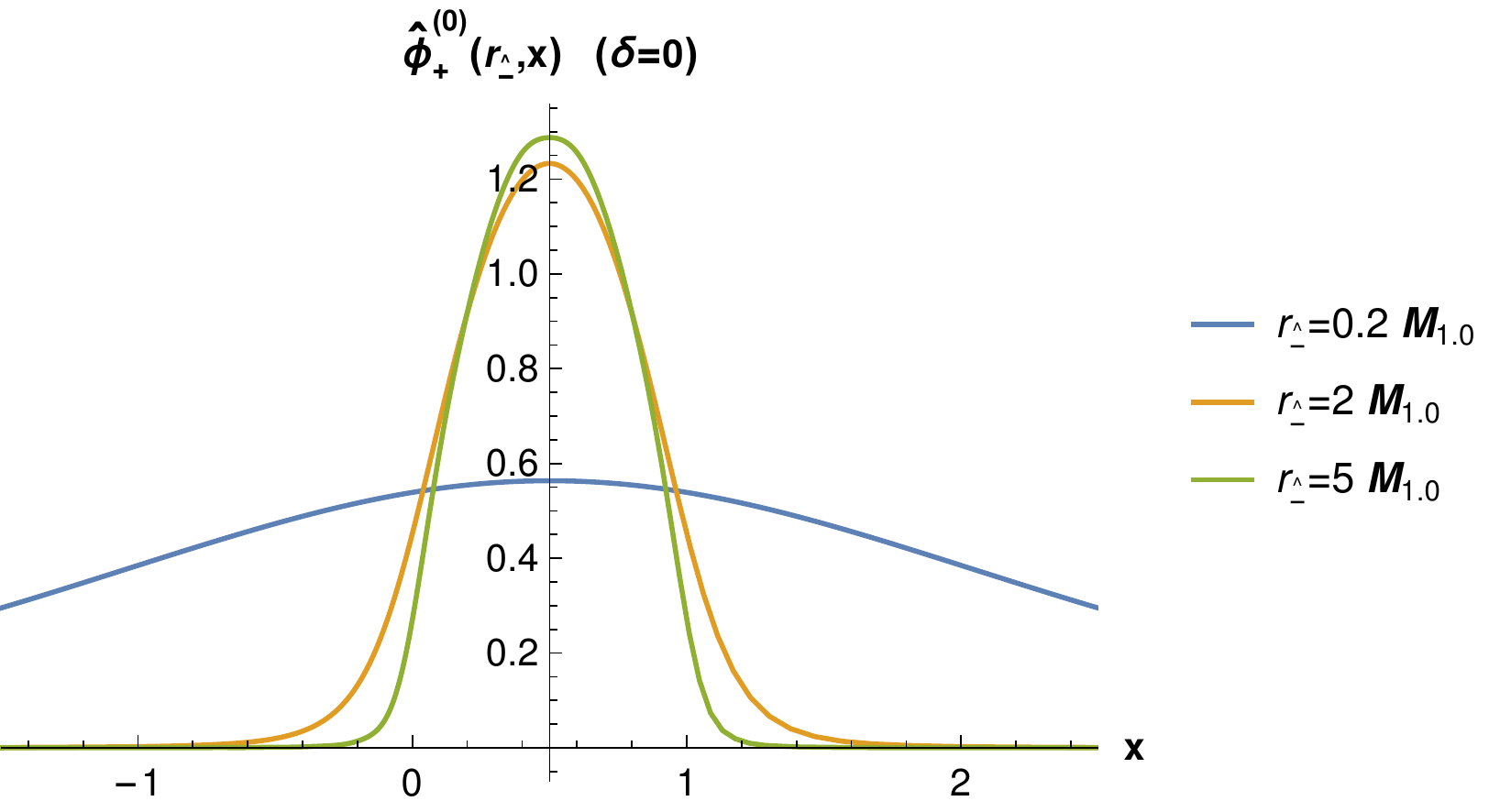}
		\label{fig:m100grzd0}
	}
	\centering
	\subfloat[]{
		\includegraphics[width=0.5\linewidth]{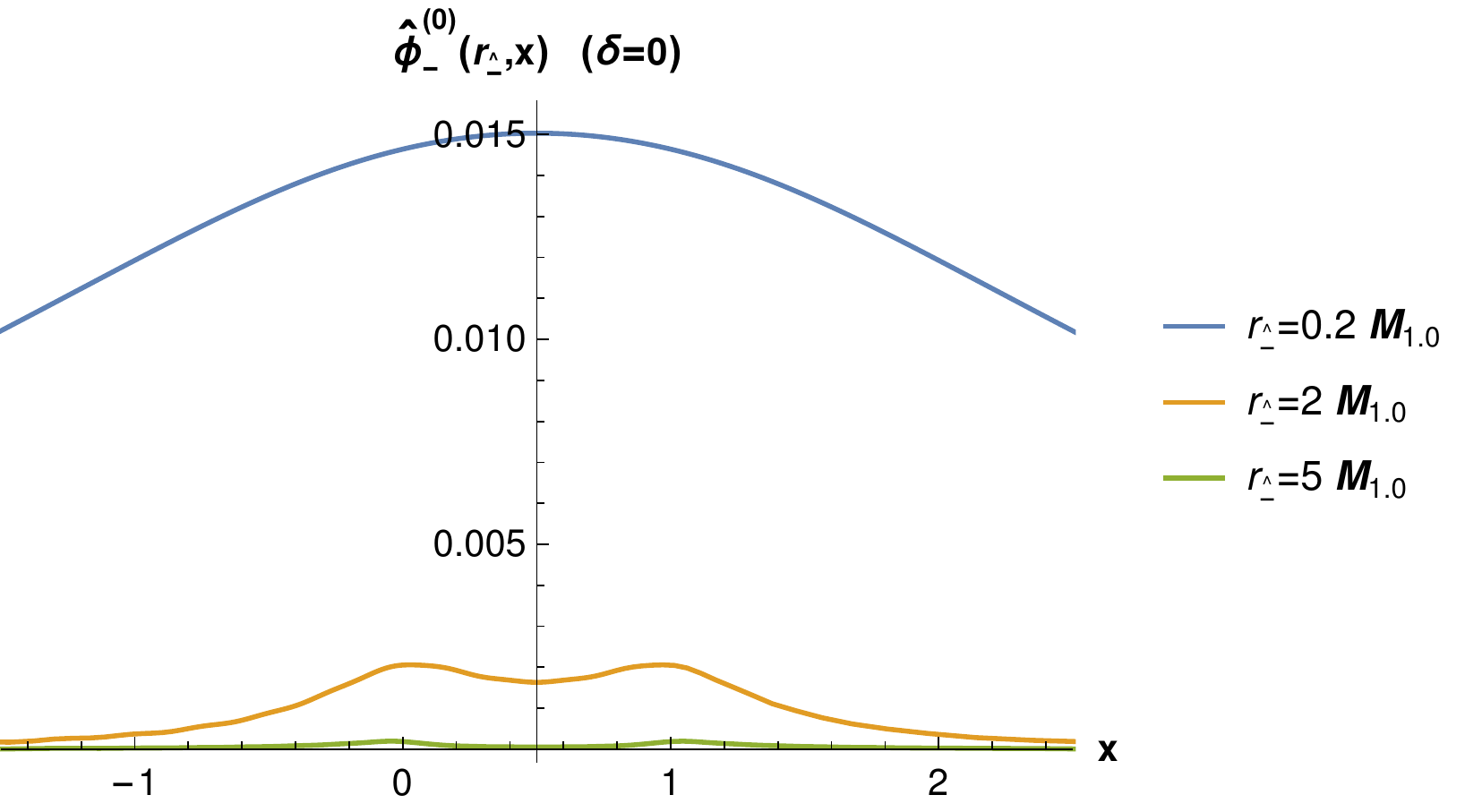}
		\label{fig:m100grfd0}
	}
	\\
	\centering
	\subfloat[]{
		\includegraphics[width=0.5\linewidth]{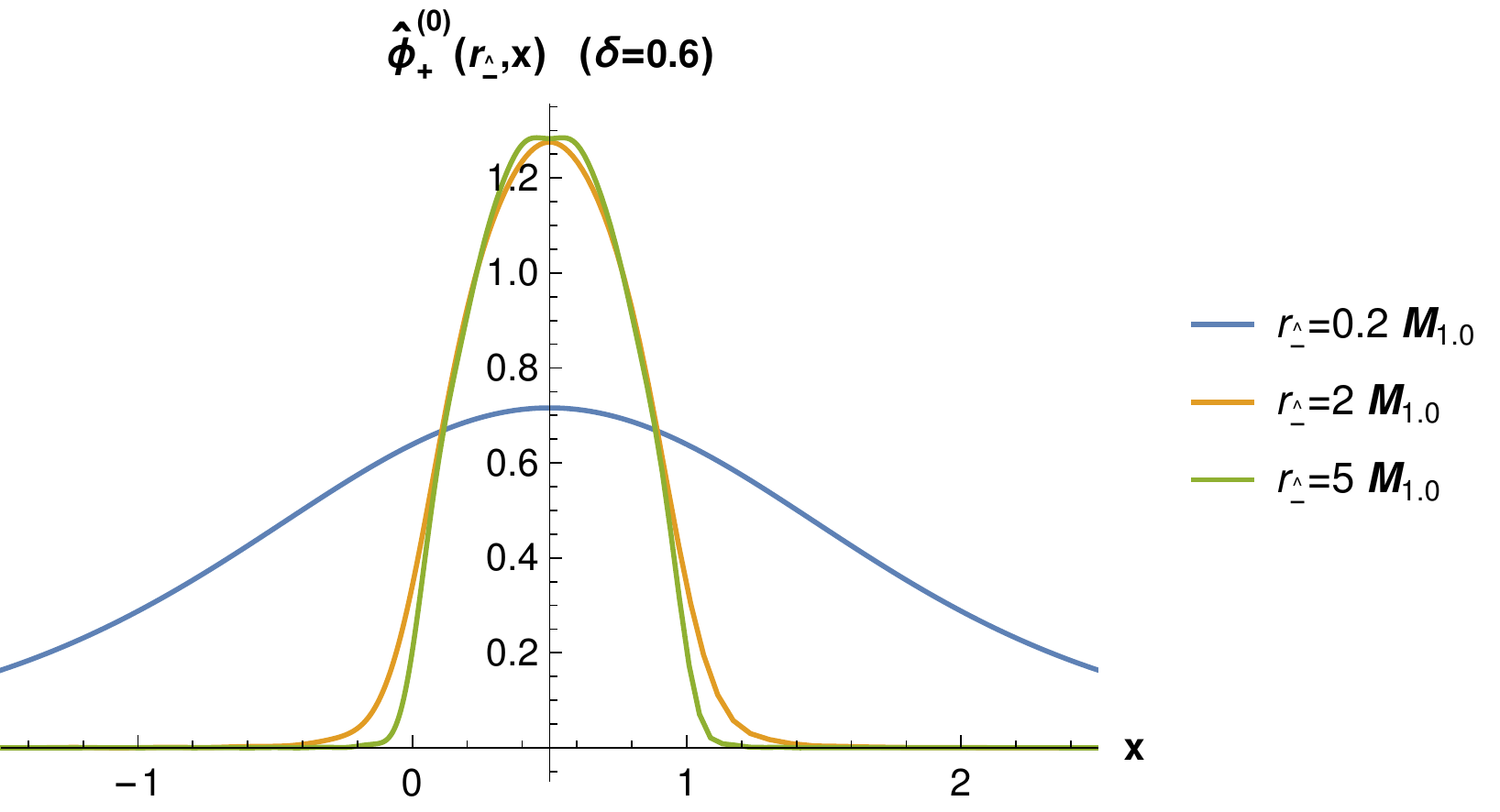}
		\label{fig:m100grzd06}
	}
	\centering
	\subfloat[]{
		\includegraphics[width=0.5\linewidth]{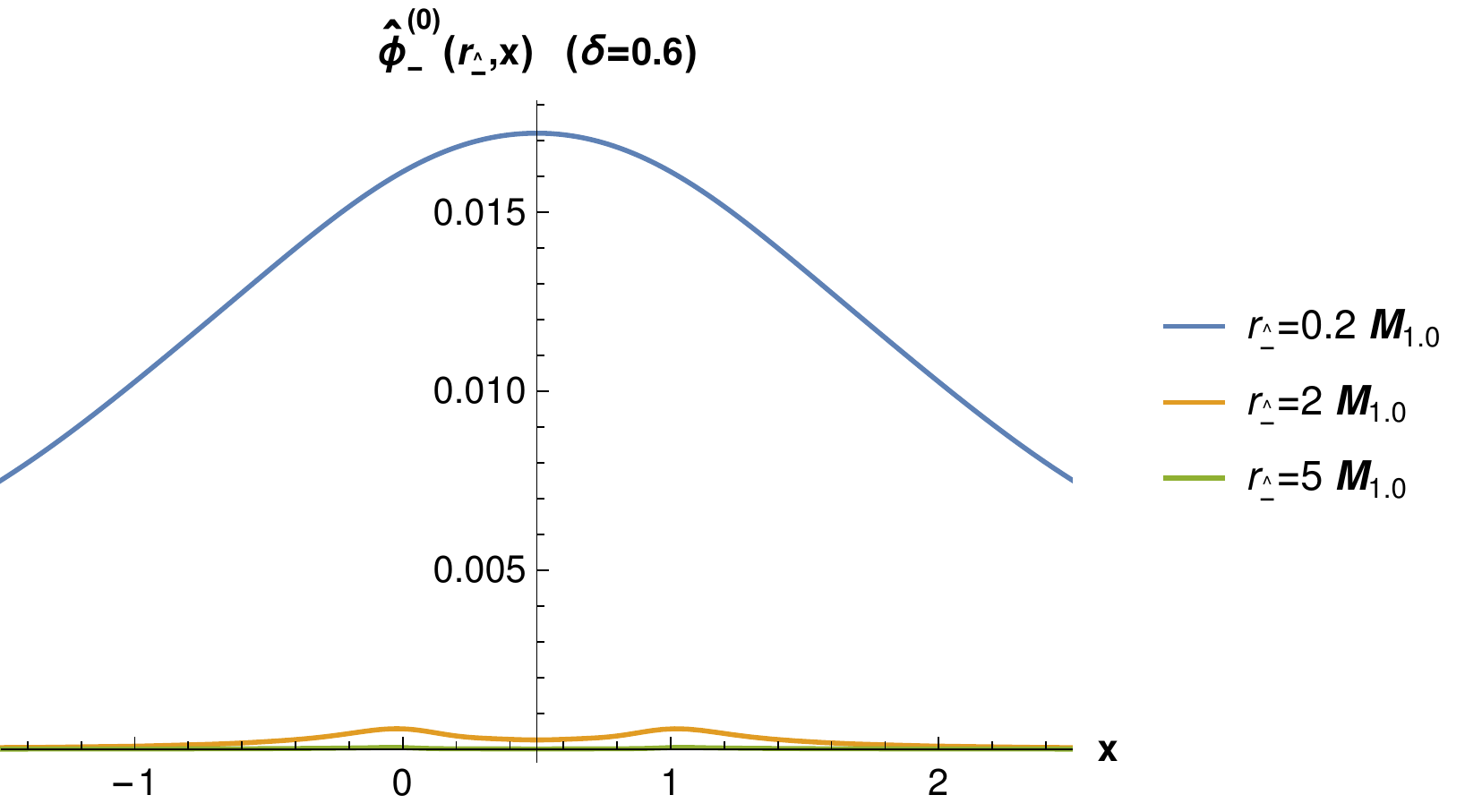}
		\label{fig:m100grfd06}
	}
	\\
	\centering
	\subfloat[]{
		\includegraphics[width=0.5\linewidth]{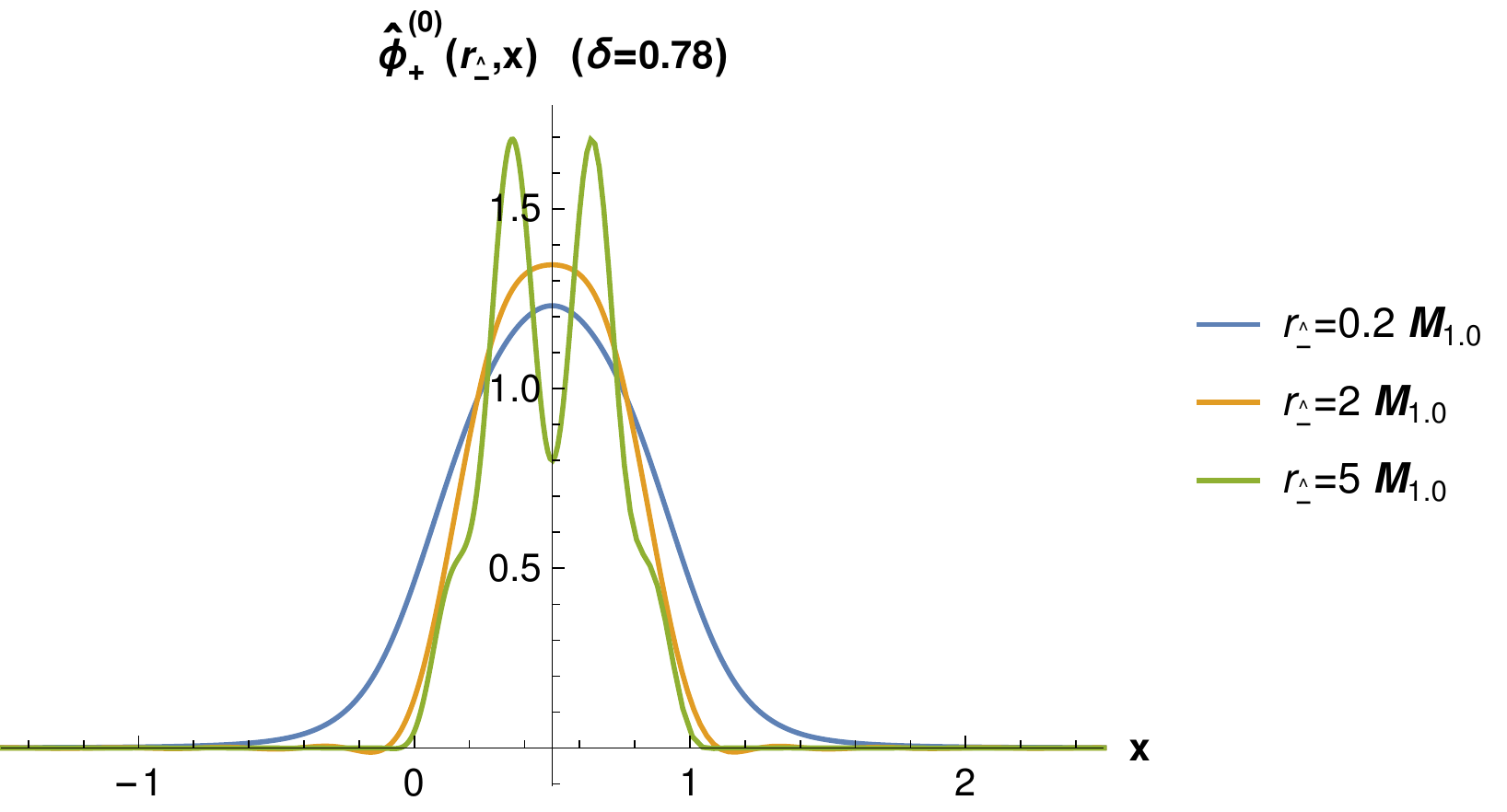}
		\label{fig:m100grzd078}
	}
	\centering
	\subfloat[]{
		\includegraphics[width=0.5\linewidth]{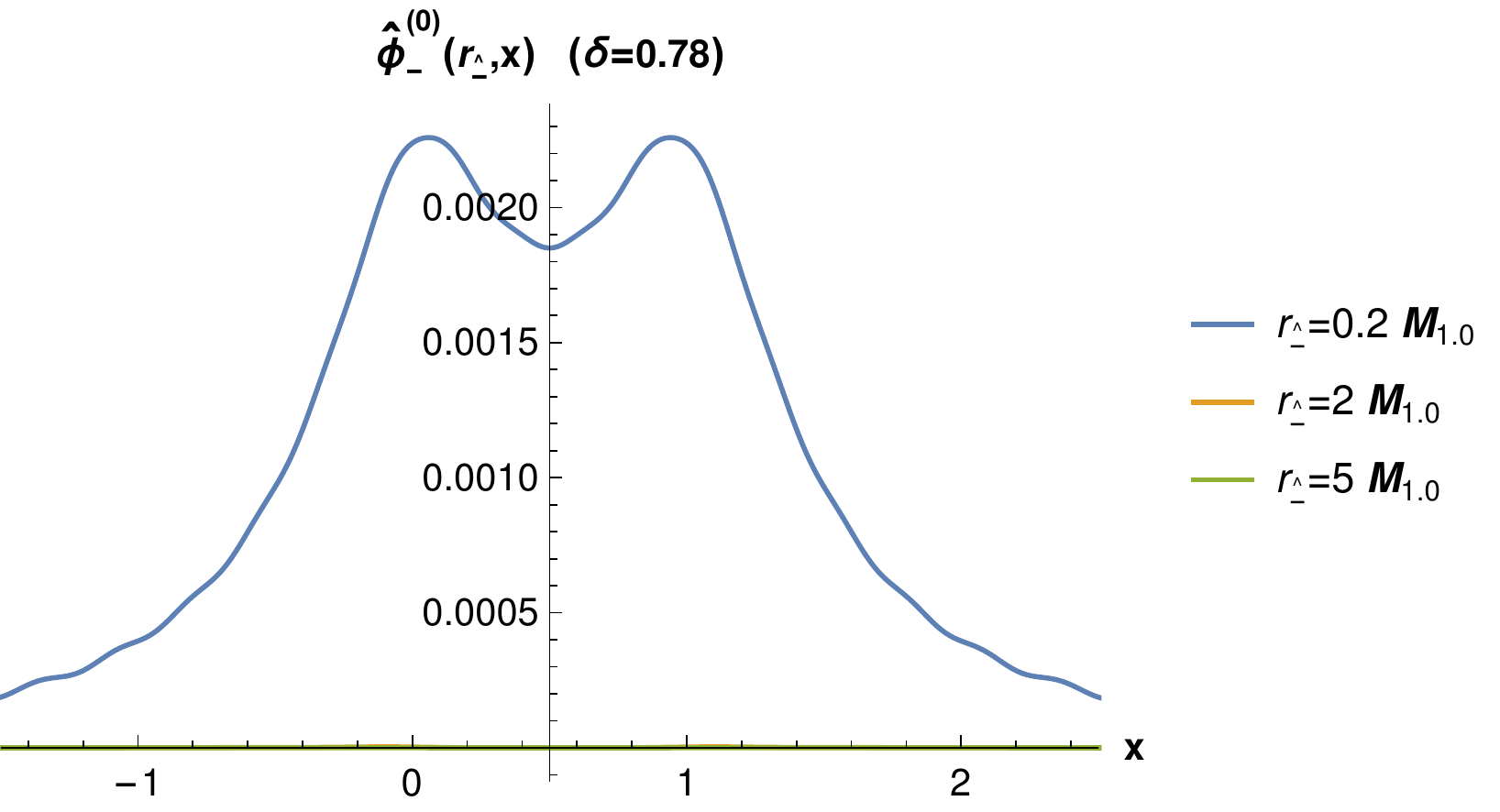}
		\label{fig:m100grfd078}
	}
	\\
	\caption{Ground state wave functions $\hat\phi_+^{(0)}(r_{\hat{-}},x)$ and $\hat\phi_-^{(0)}(r_{\hat{-}},x)$ for $ m=1.0 $. All quantities are in proper units of $ \sqrt{2\lambda} $.\label{fig:m100grz}}
\end{figure*}

\begin{figure*}
	\centering
	\subfloat[]{
		\includegraphics[width=0.5\linewidth]{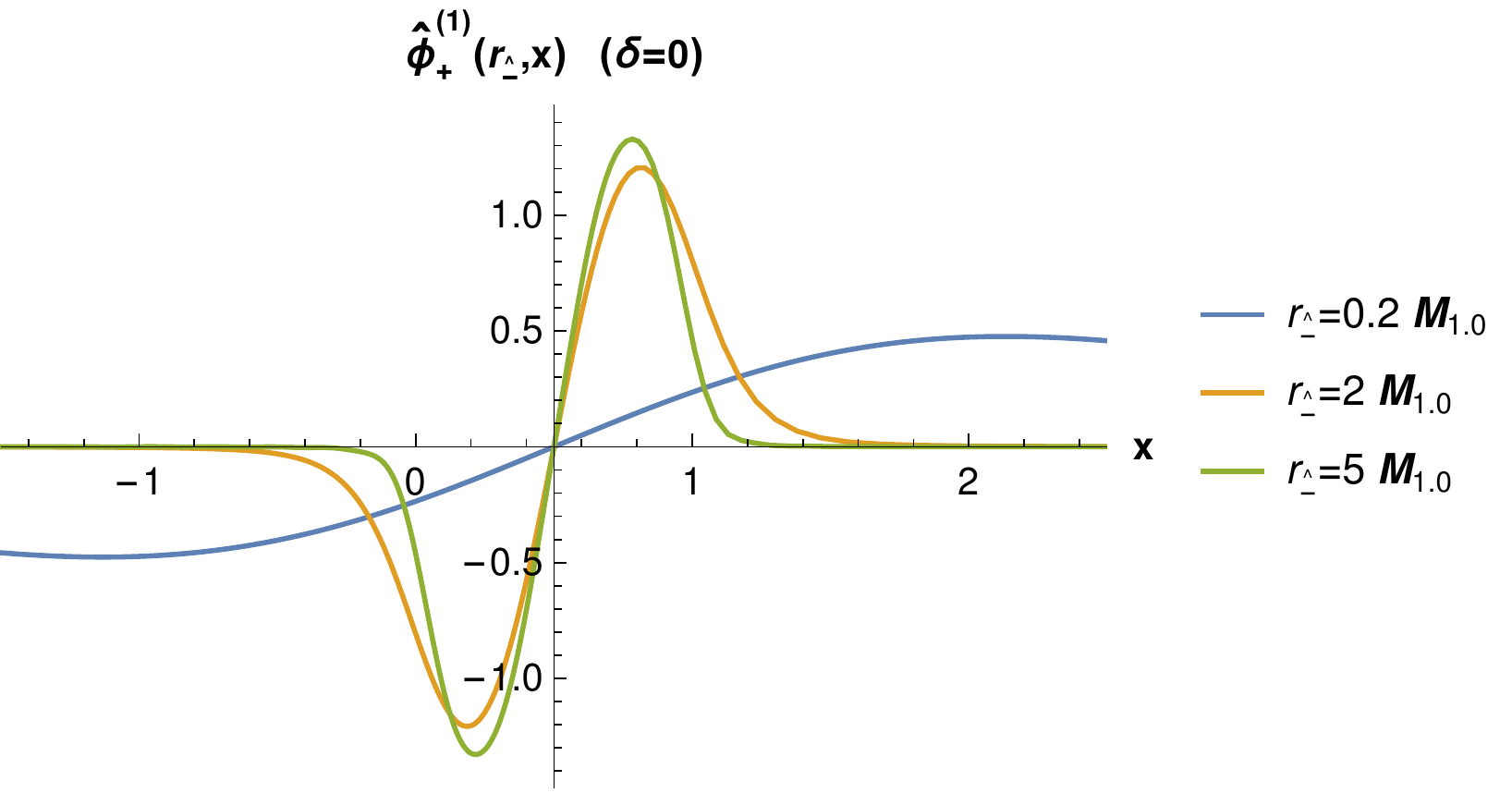}
		\label{fig:m100fezd0}
	}
	\centering
	\subfloat[]{
		\includegraphics[width=0.5\linewidth]{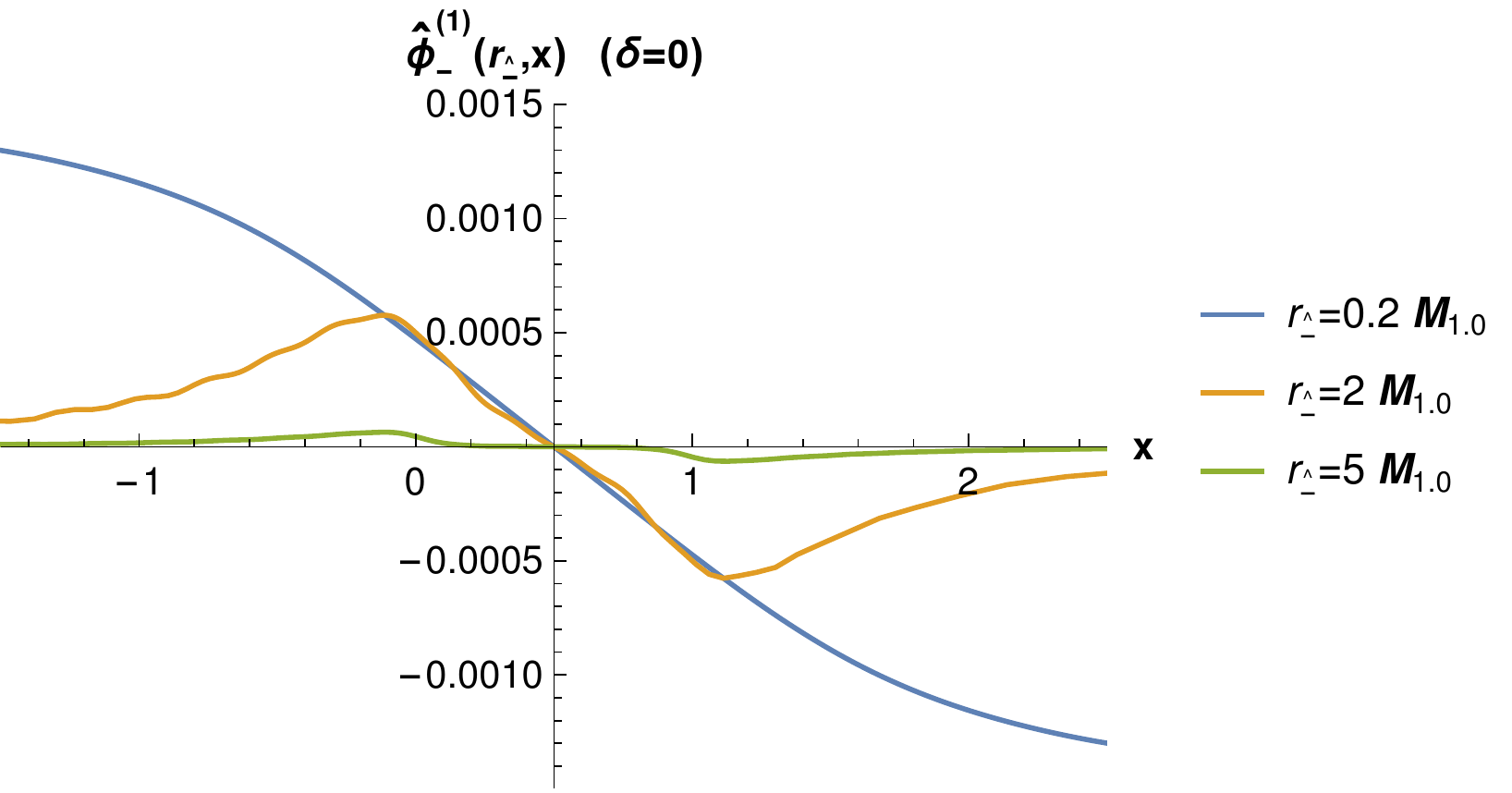}
		\label{fig:m100fefd0}
	}
	\\
	\centering
	\subfloat[]{
		\includegraphics[width=0.5\linewidth]{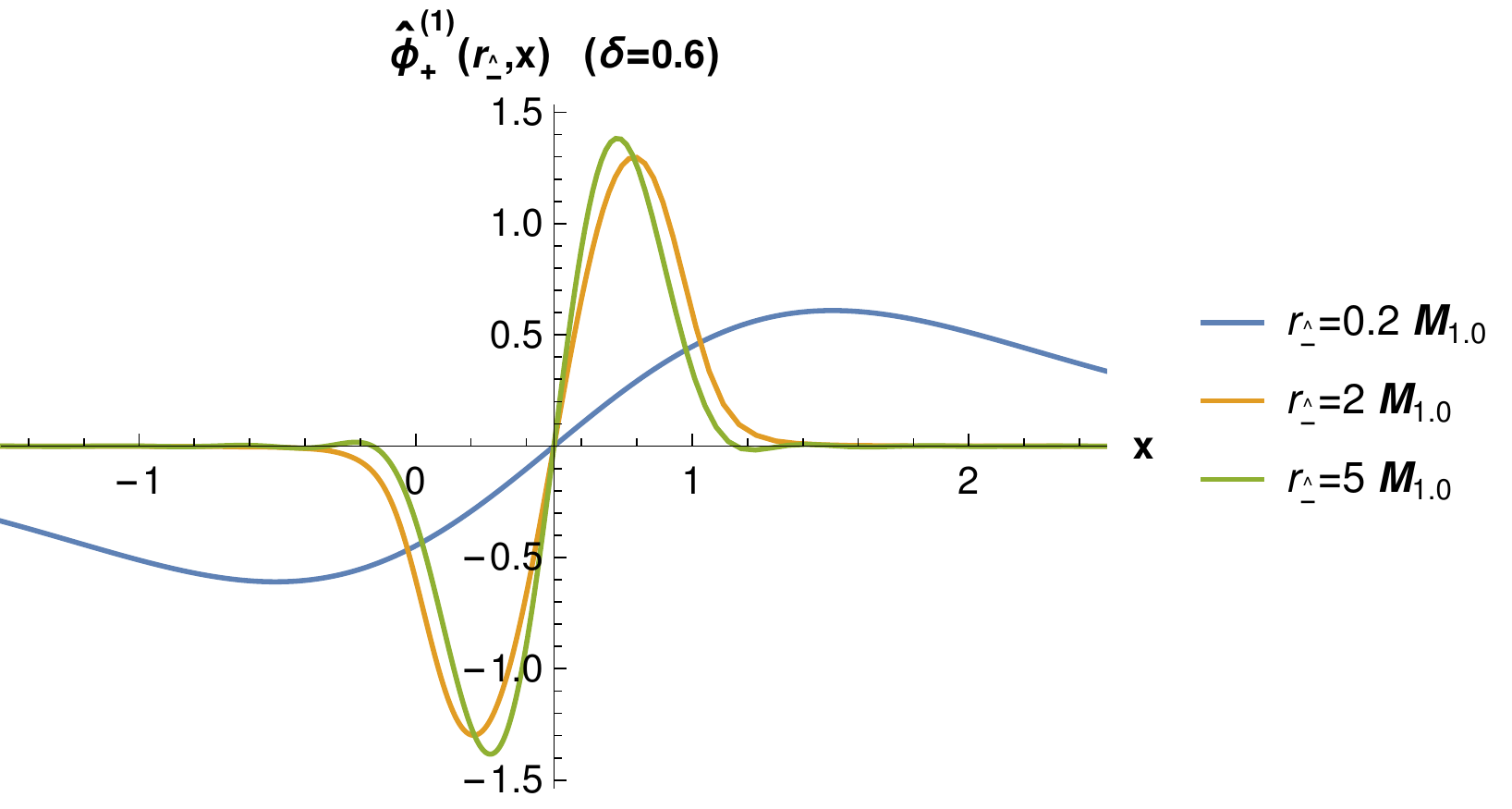}
		\label{fig:m100fezd06}
	}
	\centering
	\subfloat[]{
		\includegraphics[width=0.5\linewidth]{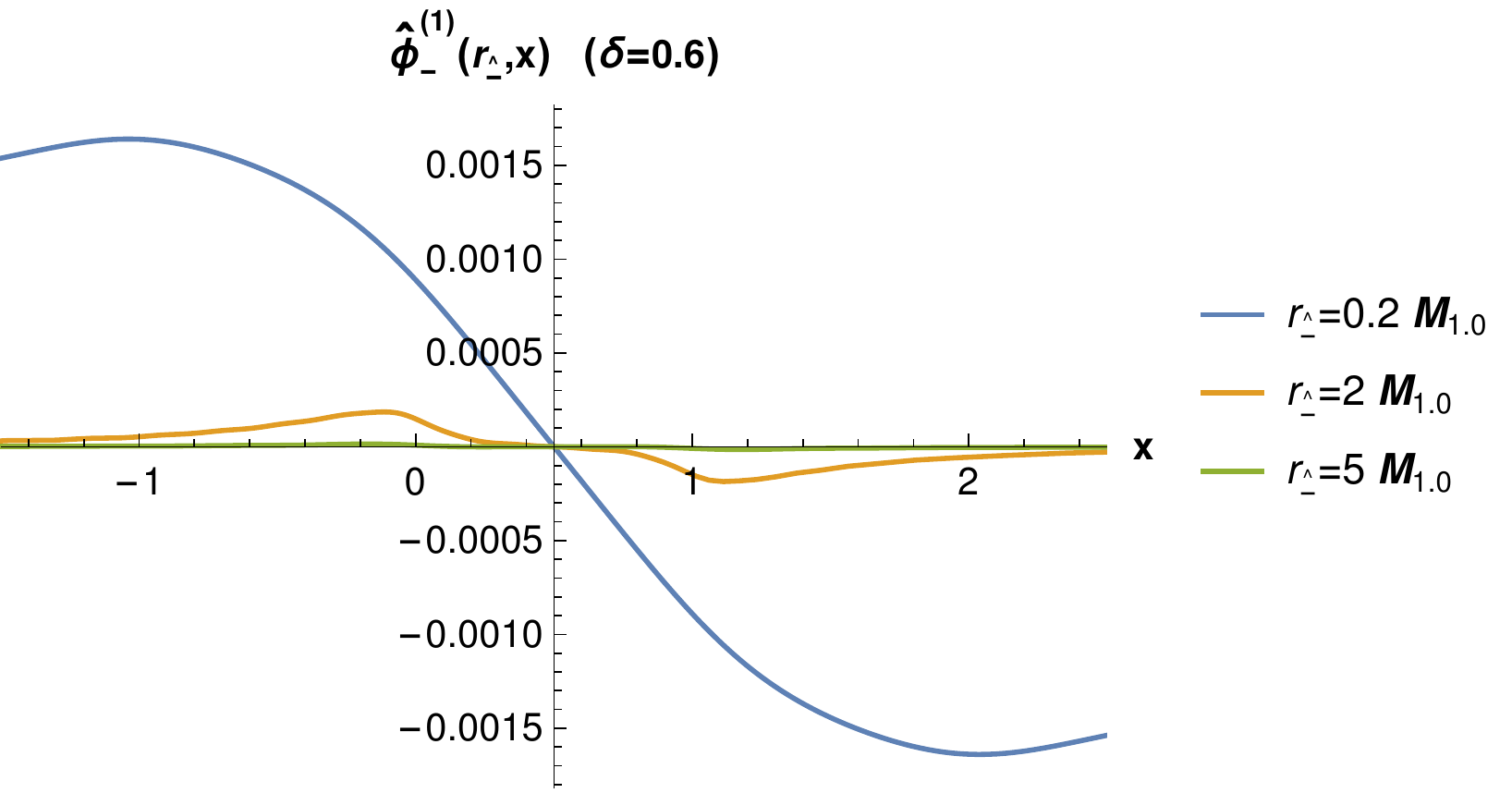}	
		\label{fig:m100fefd06}
	}\\
	\centering
	\subfloat[]{
		\includegraphics[width=0.5\linewidth]{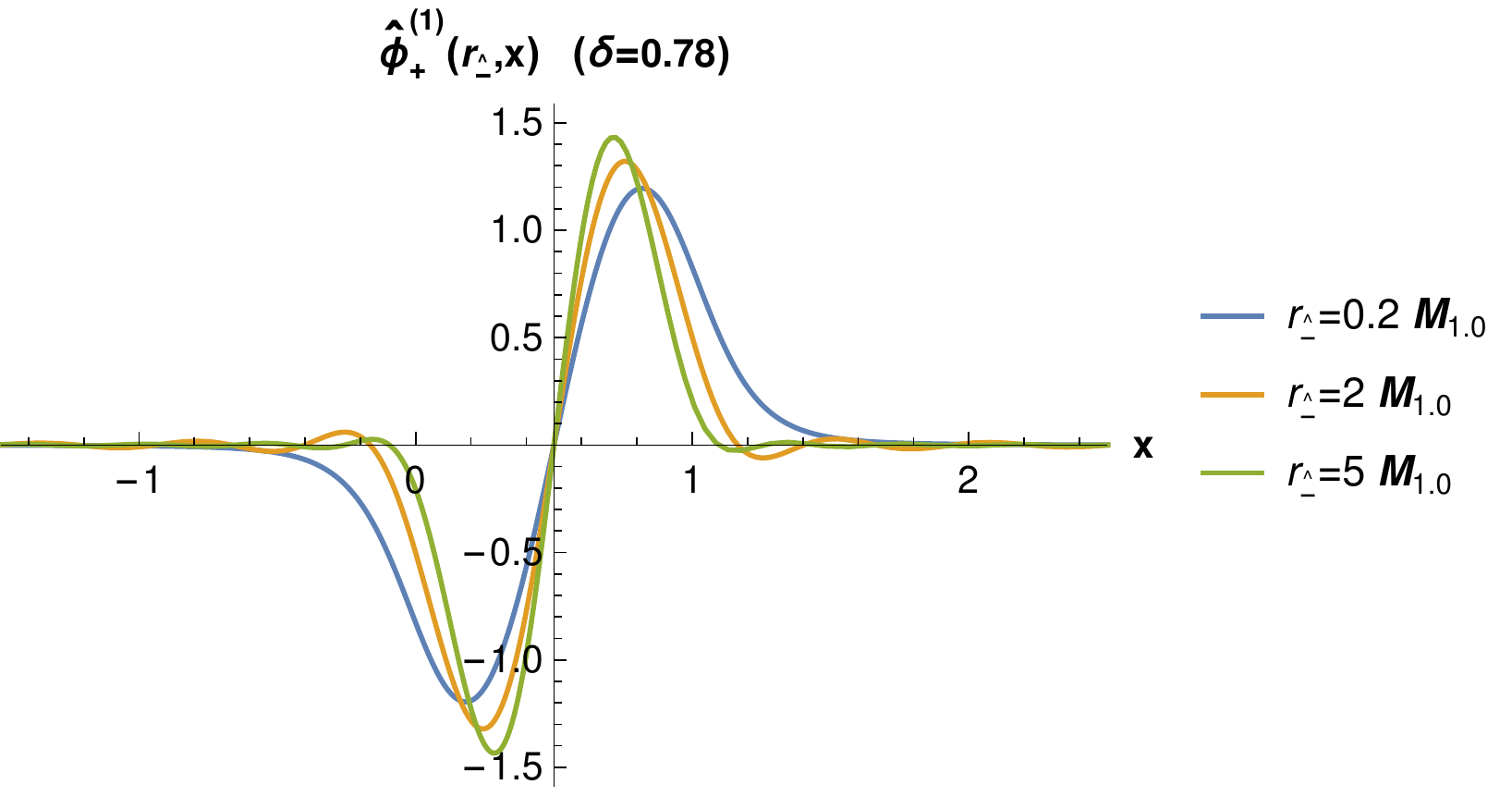}
		\label{fig:m100fezd078}
	}
	\centering
	\subfloat[]{
		\includegraphics[width=0.5\linewidth]{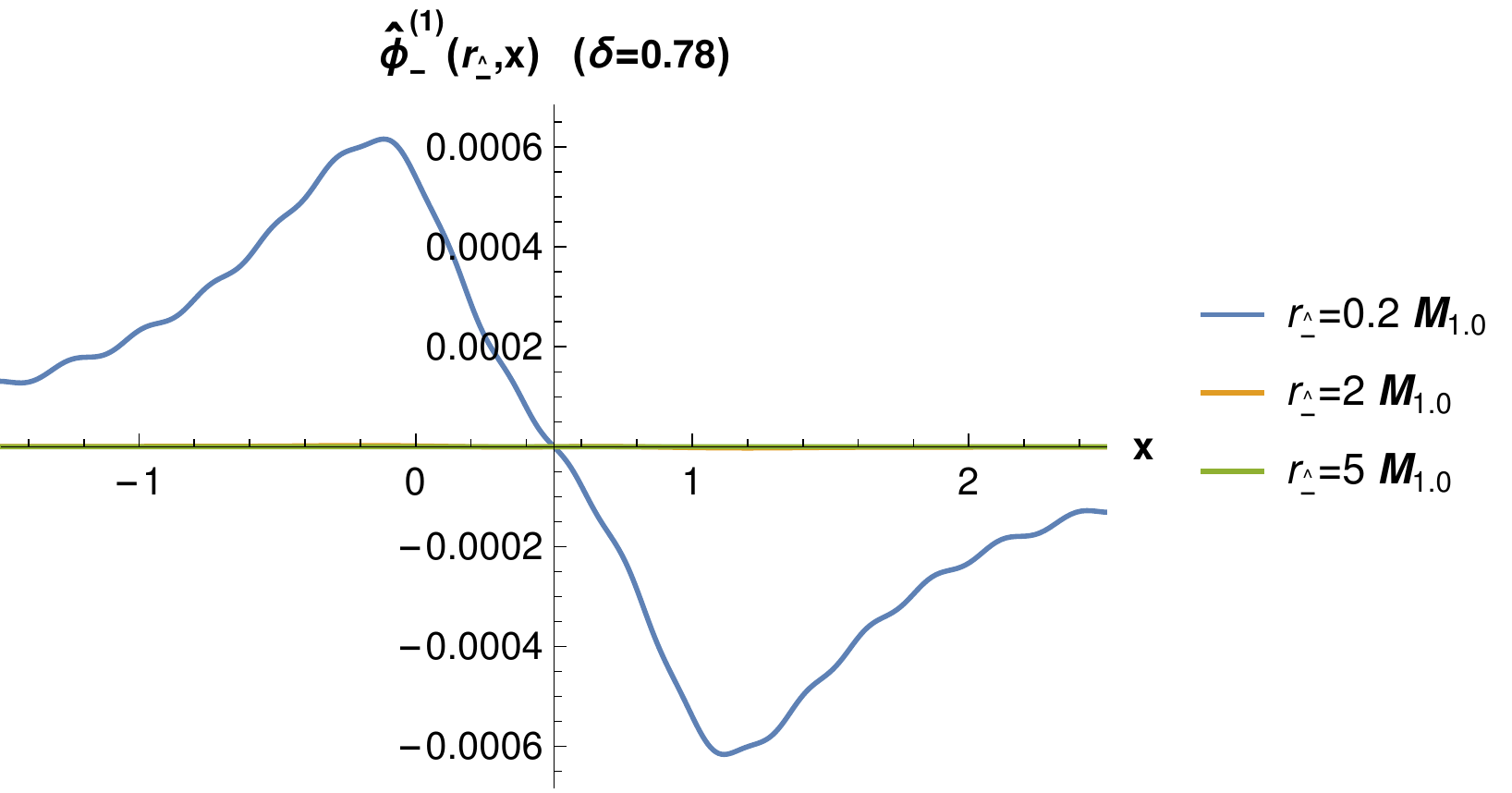}
		\label{fig:m100fefd078}
	}\\
	\caption{First excited state wave functions $\hat\phi_+^{(1)}(r_{\hat{-}},x)$ and $\hat\phi_-^{(1)}(r_{\hat{-}},x)$ for $ m=1.0 $. All quantities are in proper units of $ \sqrt{2\lambda} $.\label{fig:m100fez}}
\end{figure*}

\begin{figure*}
	\centering
	\subfloat[]{
		\includegraphics[width=0.5\linewidth]{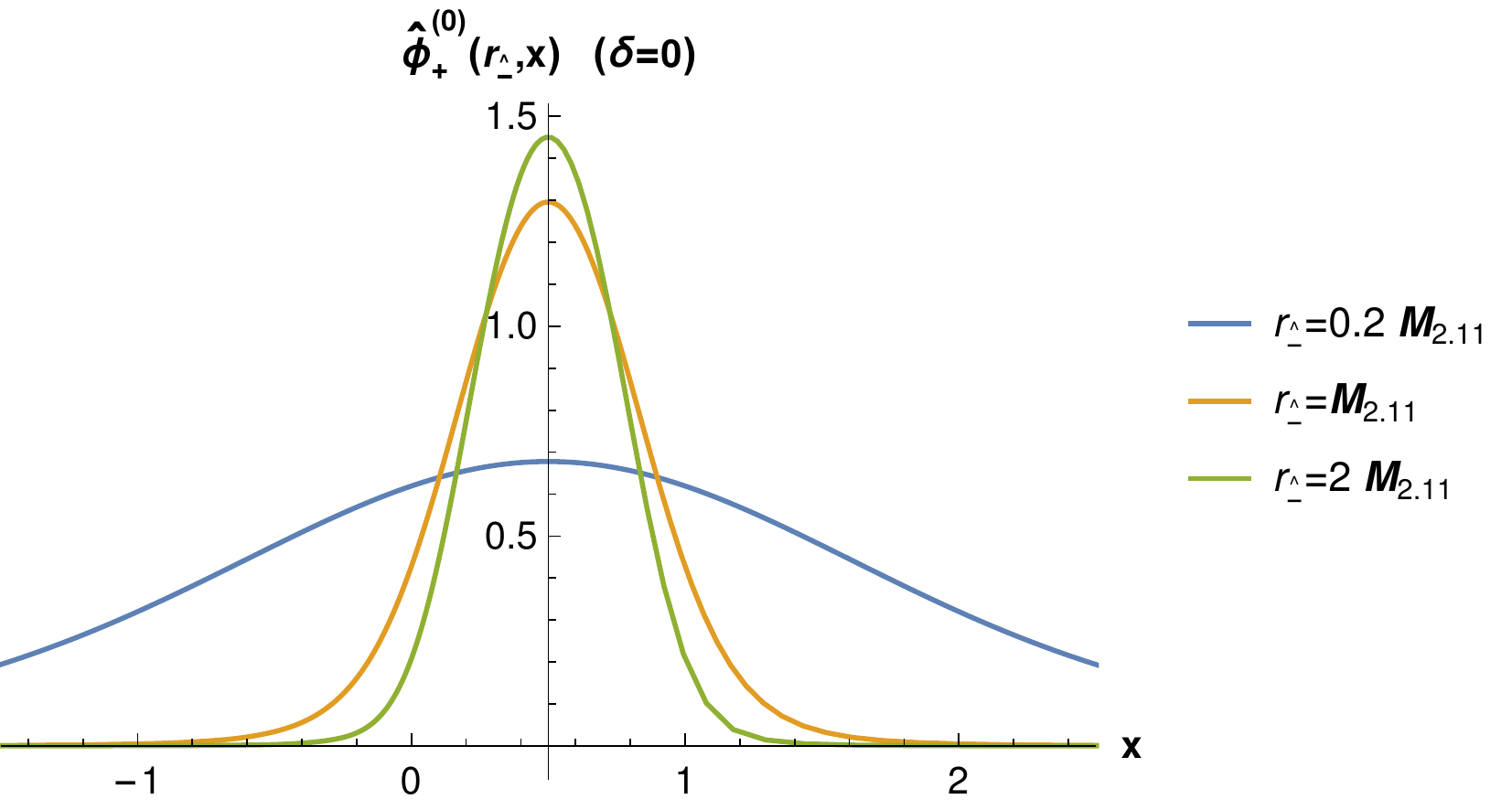}
		\label{fig:m211grzd0}
	}
	\centering
	\subfloat[]{
		\includegraphics[width=0.5\linewidth]{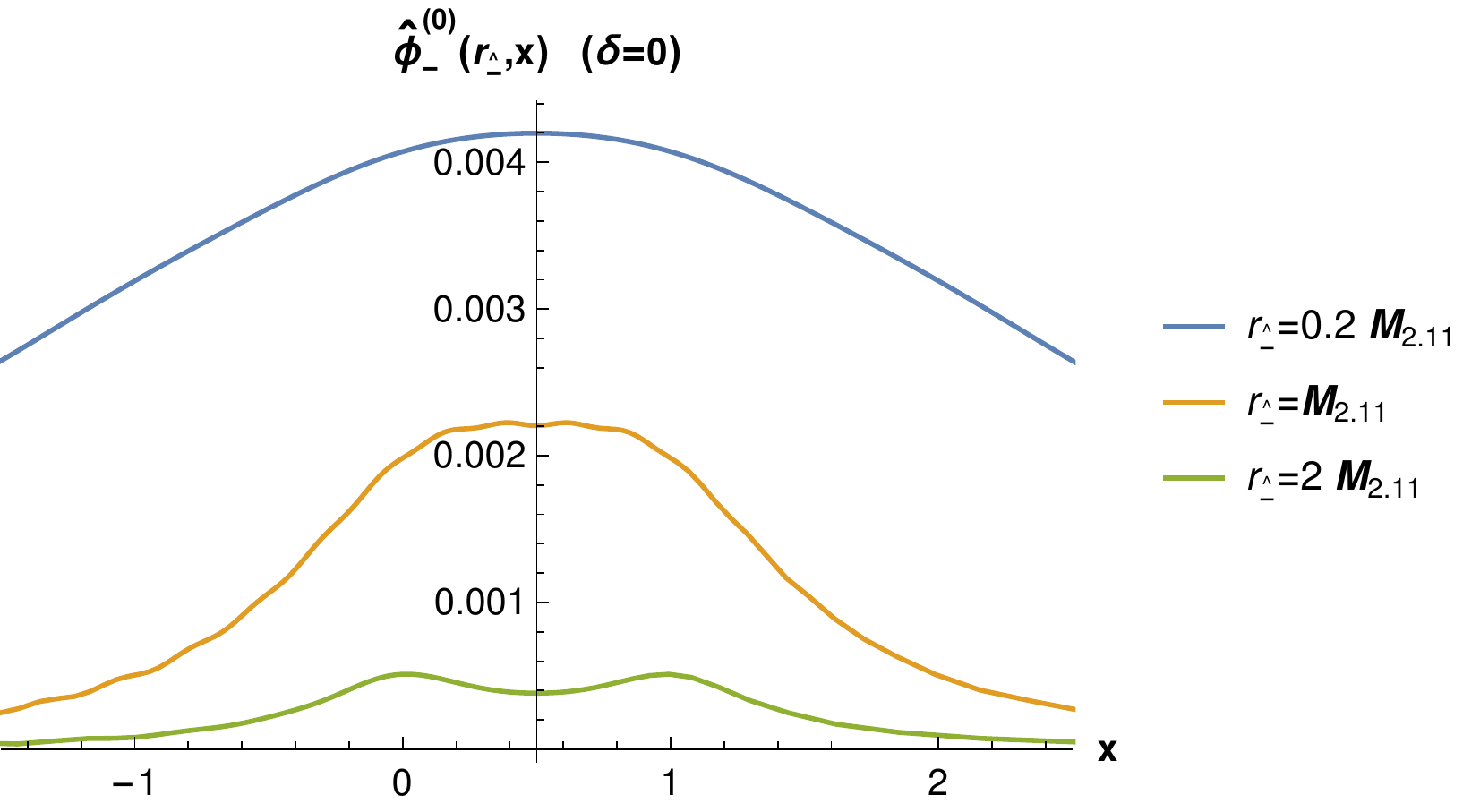}
		\label{fig:m211grfd0}
	}\\
	\centering
	\subfloat[]{
		\includegraphics[width=0.5\linewidth]{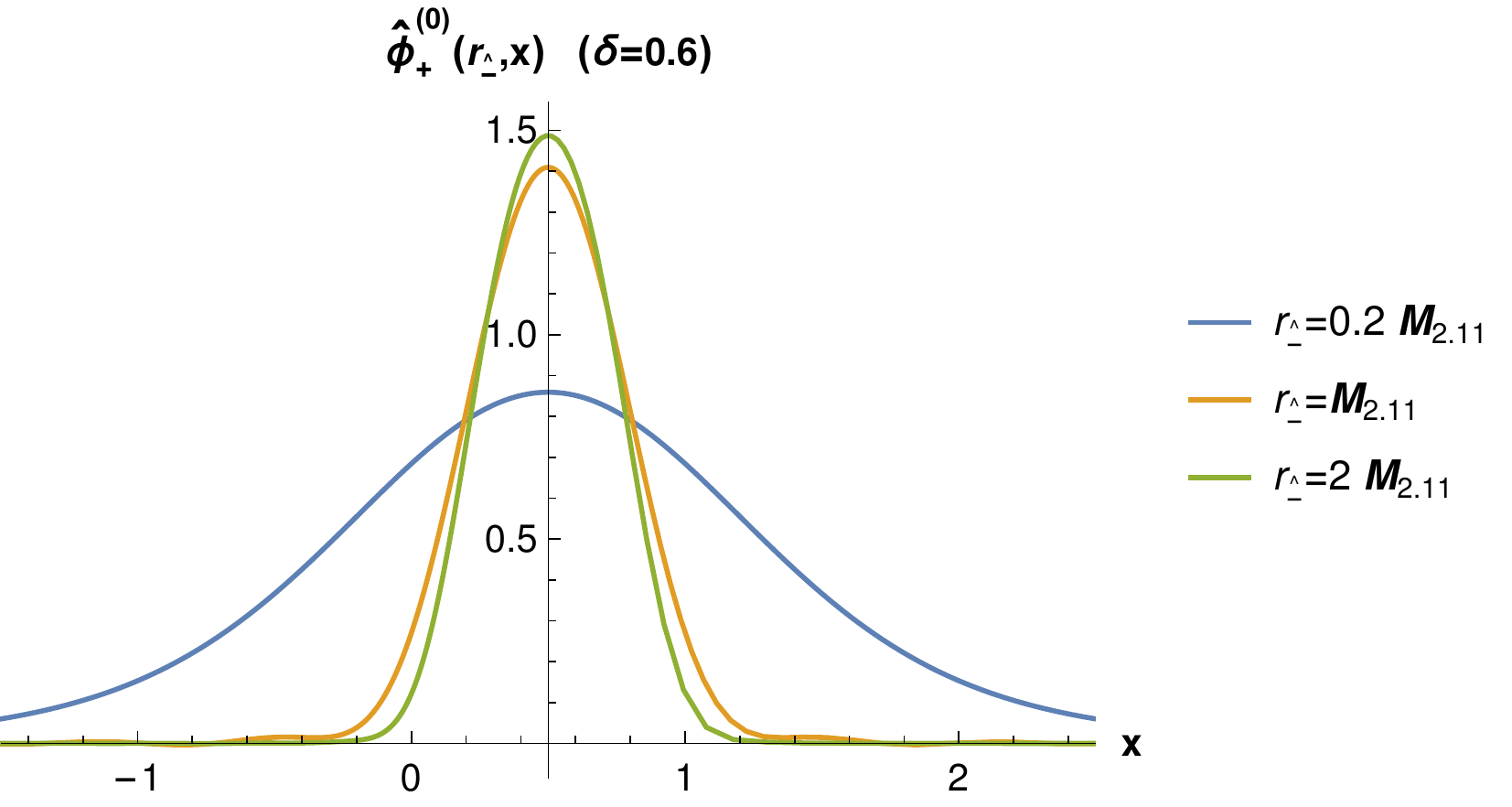}
		\label{fig:m211grzd06}
	}
	\centering
	\subfloat[]{
		\includegraphics[width=0.5\linewidth]{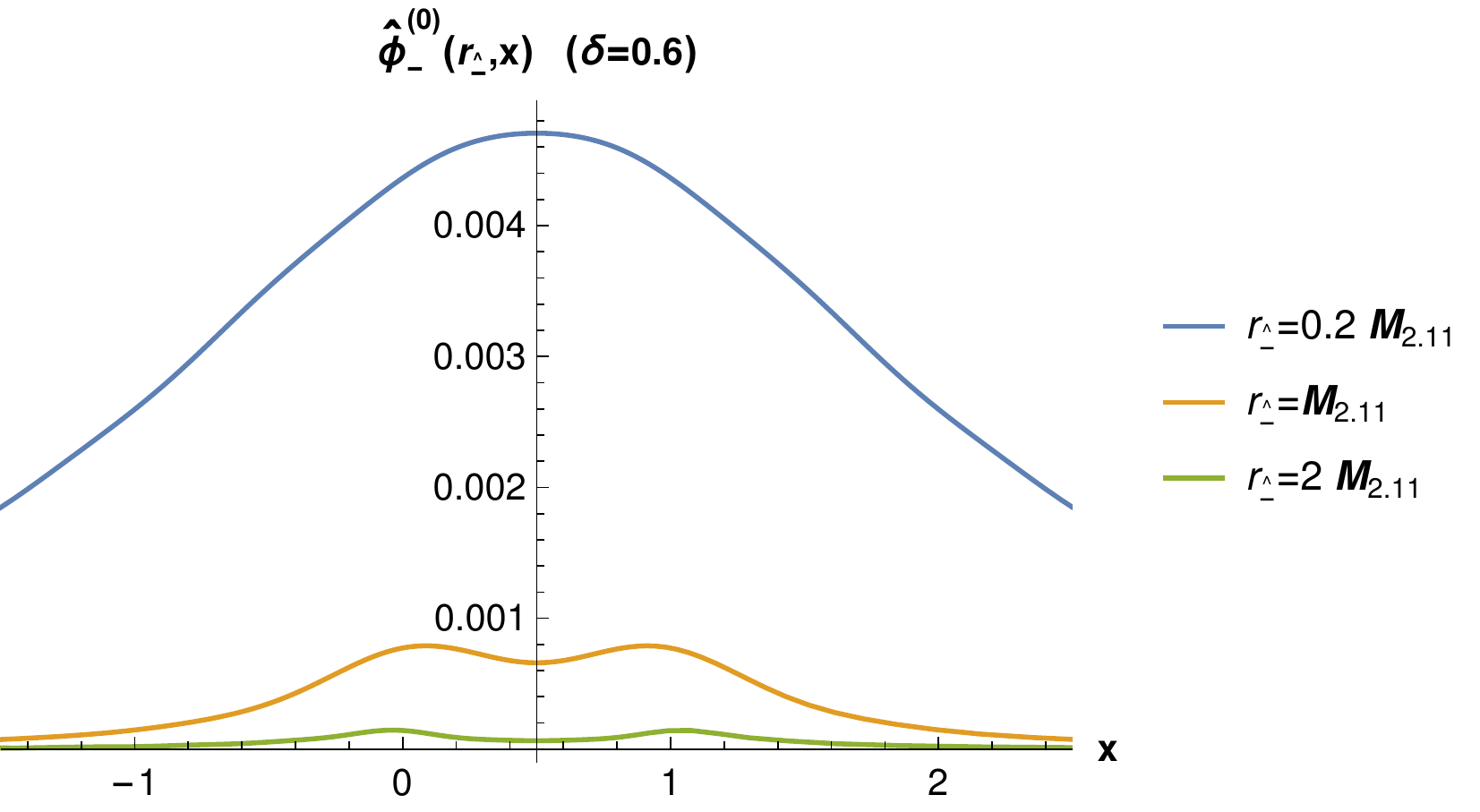}
		\label{fig:m211grfd06}
	}\\
	\centering
	\subfloat[]{
		\includegraphics[width=0.5\linewidth]{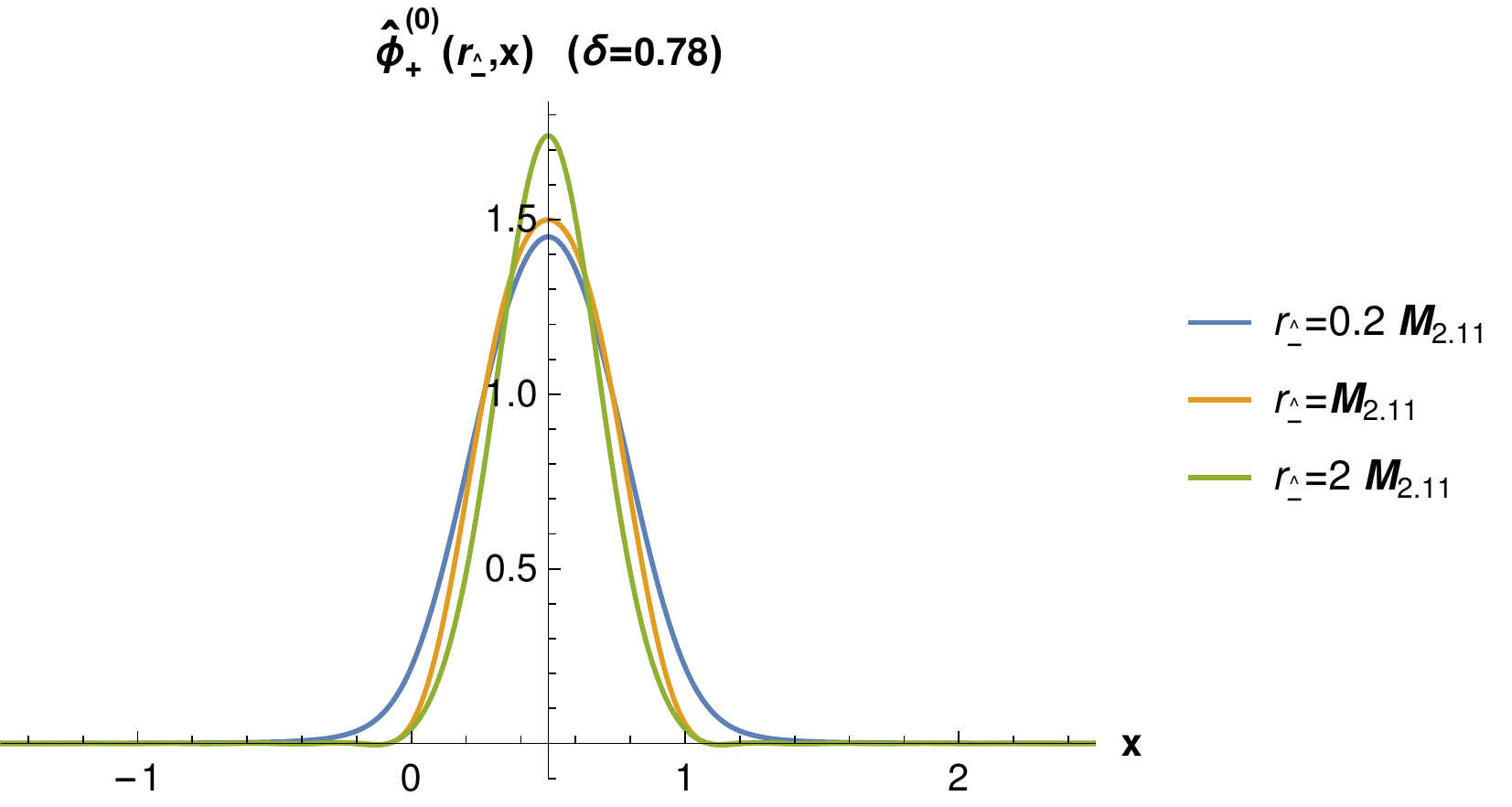}
		\label{fig:m211grzd078}
	}
	\centering
	\subfloat[]{
		\includegraphics[width=0.5\linewidth]{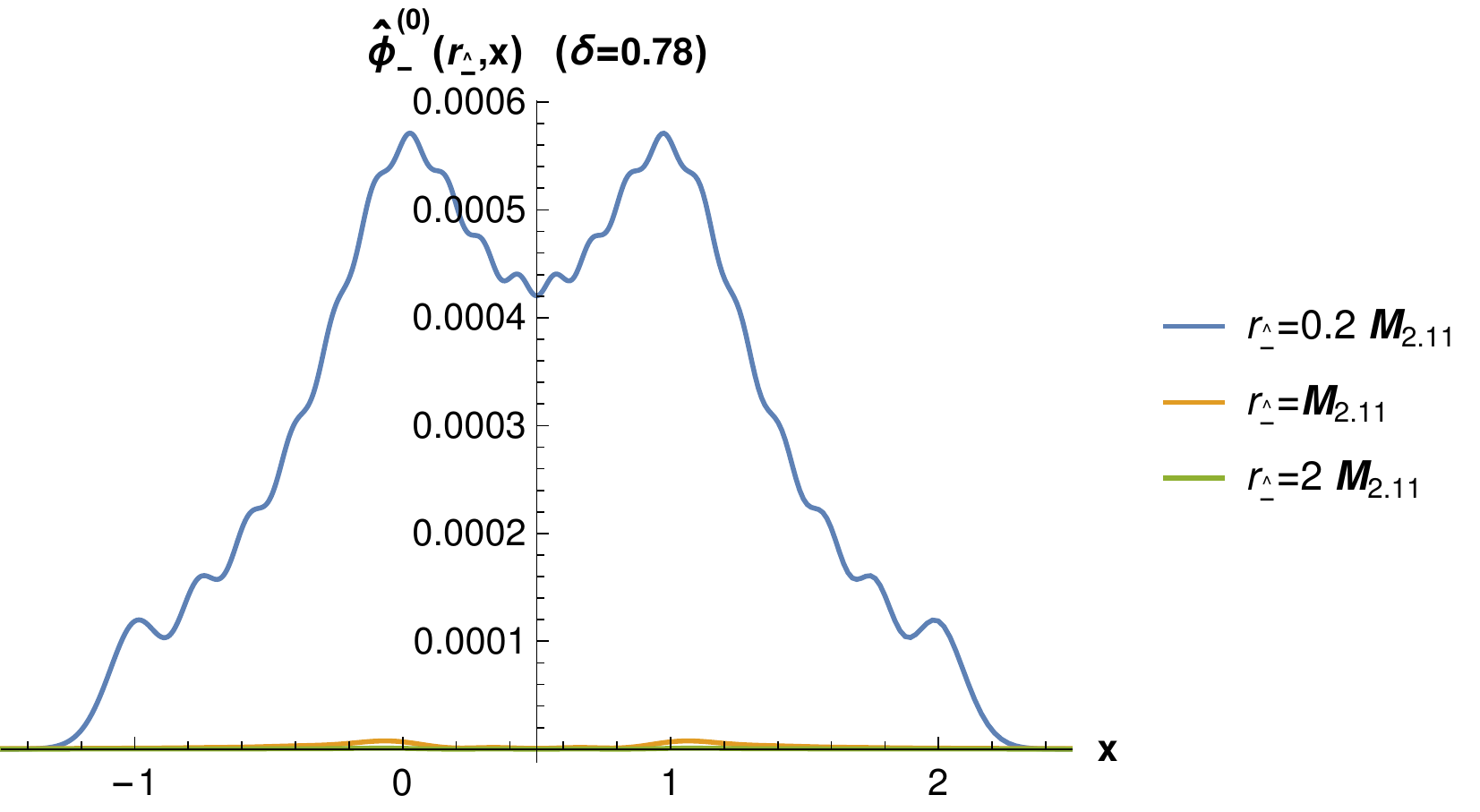}
		\label{fig:m211grfd078}
	}\\
	\caption{Ground state wave functions $\hat\phi_+^{(0)}(r_{\hat{-}},x)$ and $\hat\phi_-^{(0)}(r_{\hat{-}},x)$ for $ m=2.11 $. All quantities are in proper units of $ \sqrt{2\lambda} $.\label{fig:m211grz}}
\end{figure*}

\begin{figure*}
	\centering
	\subfloat[]{
		\includegraphics[width=0.5\linewidth]{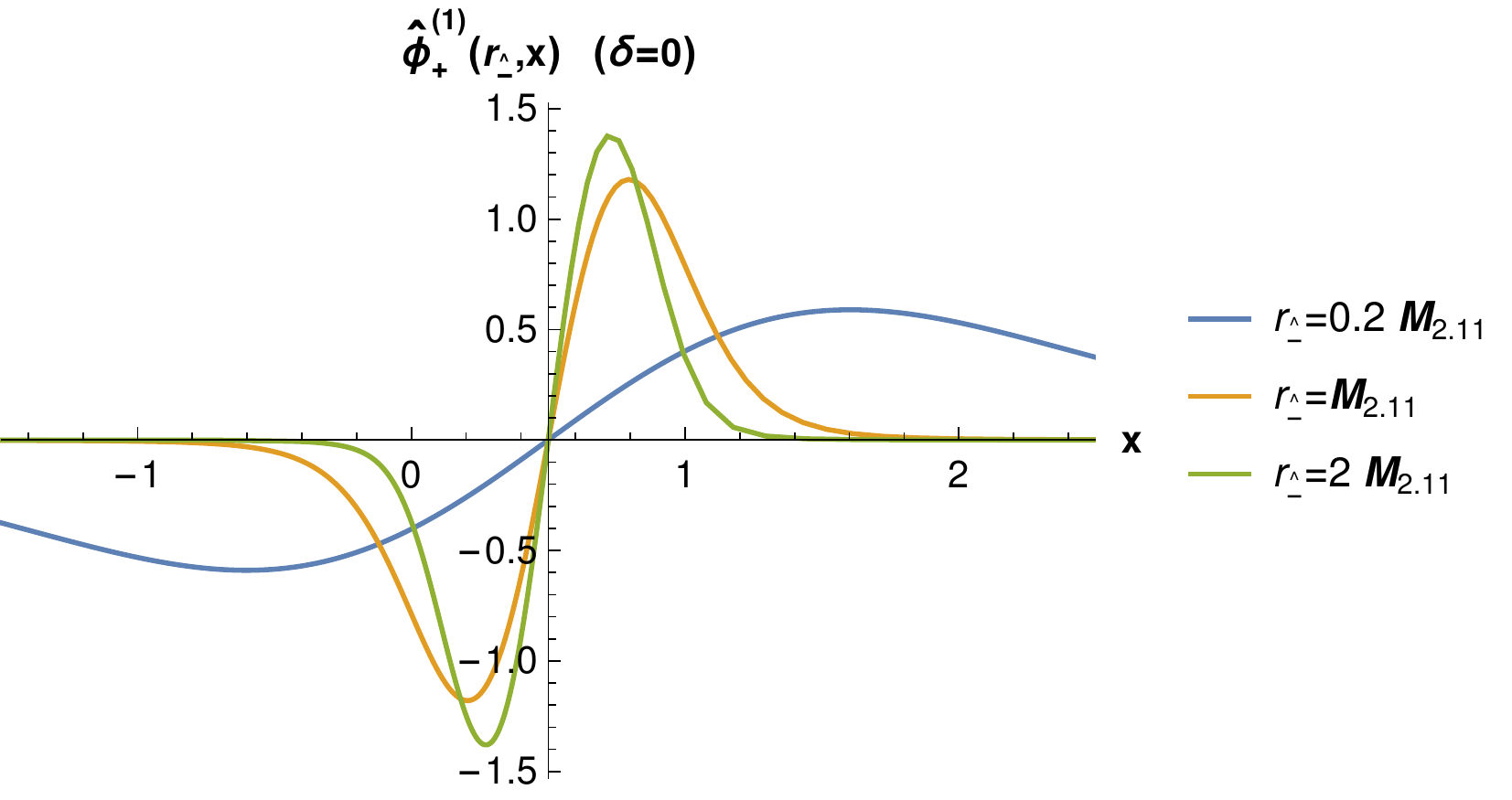}
		\label{fig:m211fezd0}
	}
	\centering
	\subfloat[]{
		\includegraphics[width=0.5\linewidth]{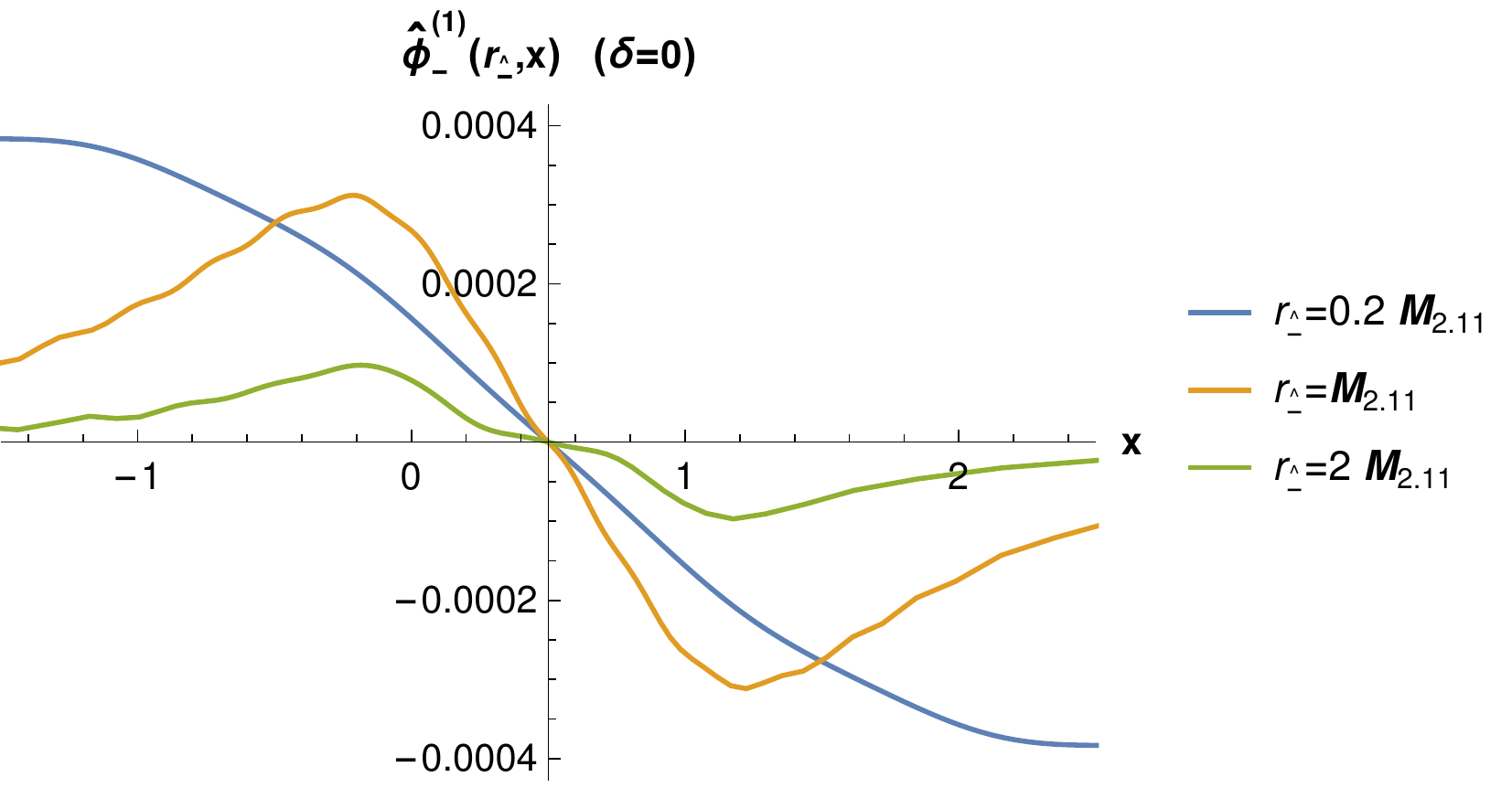}
		\label{fig:m211fefd0}
	}\\
	\centering
	\subfloat[]{
		\includegraphics[width=0.5\linewidth]{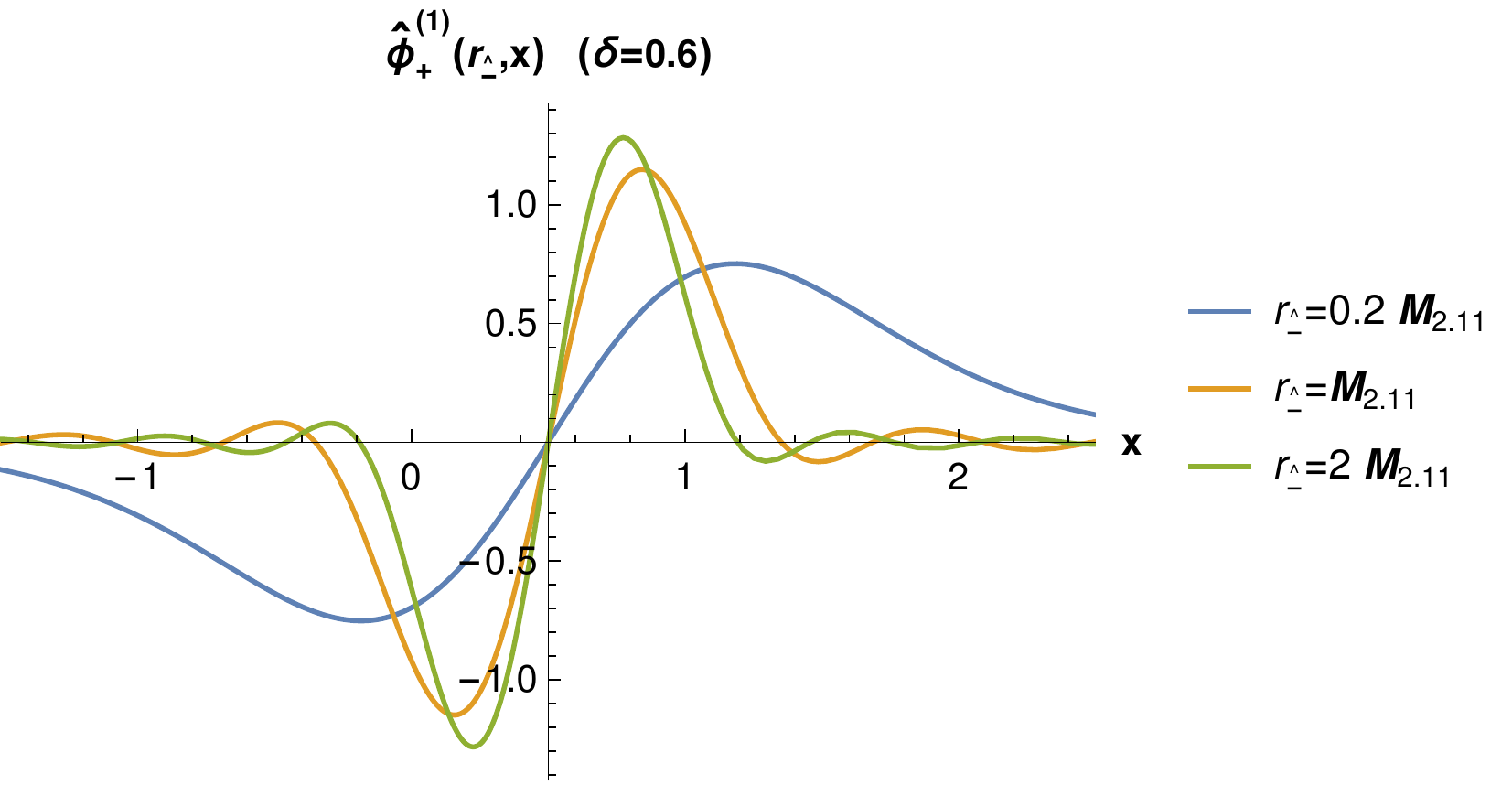}
		\label{fig:m211fezd06}
	}
	\centering
	\subfloat[]{
		\includegraphics[width=0.5\linewidth]{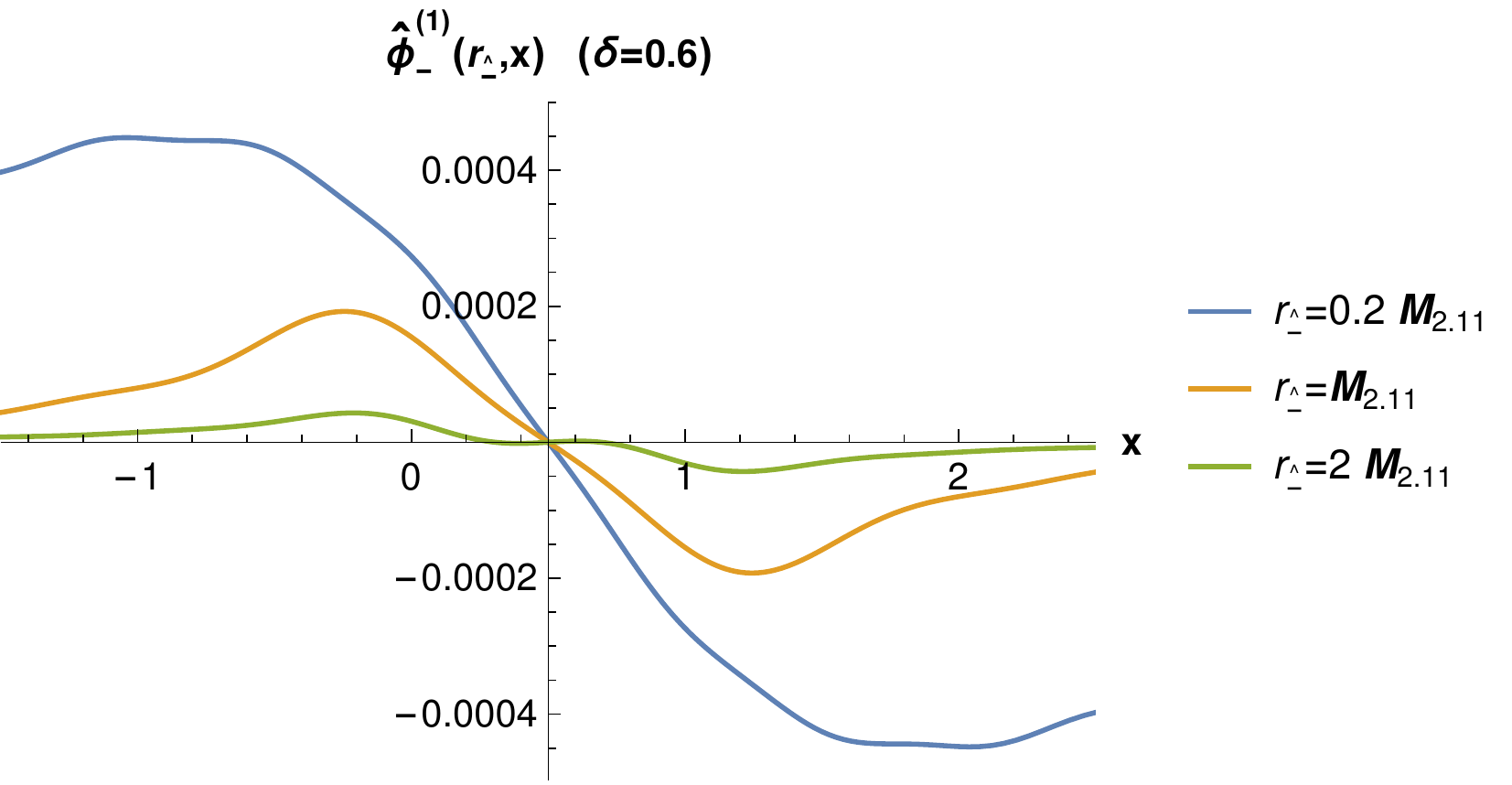}	
		\label{fig:m211fefd06}
	}\\
	\centering
	\subfloat[]{
		\includegraphics[width=0.5\linewidth]{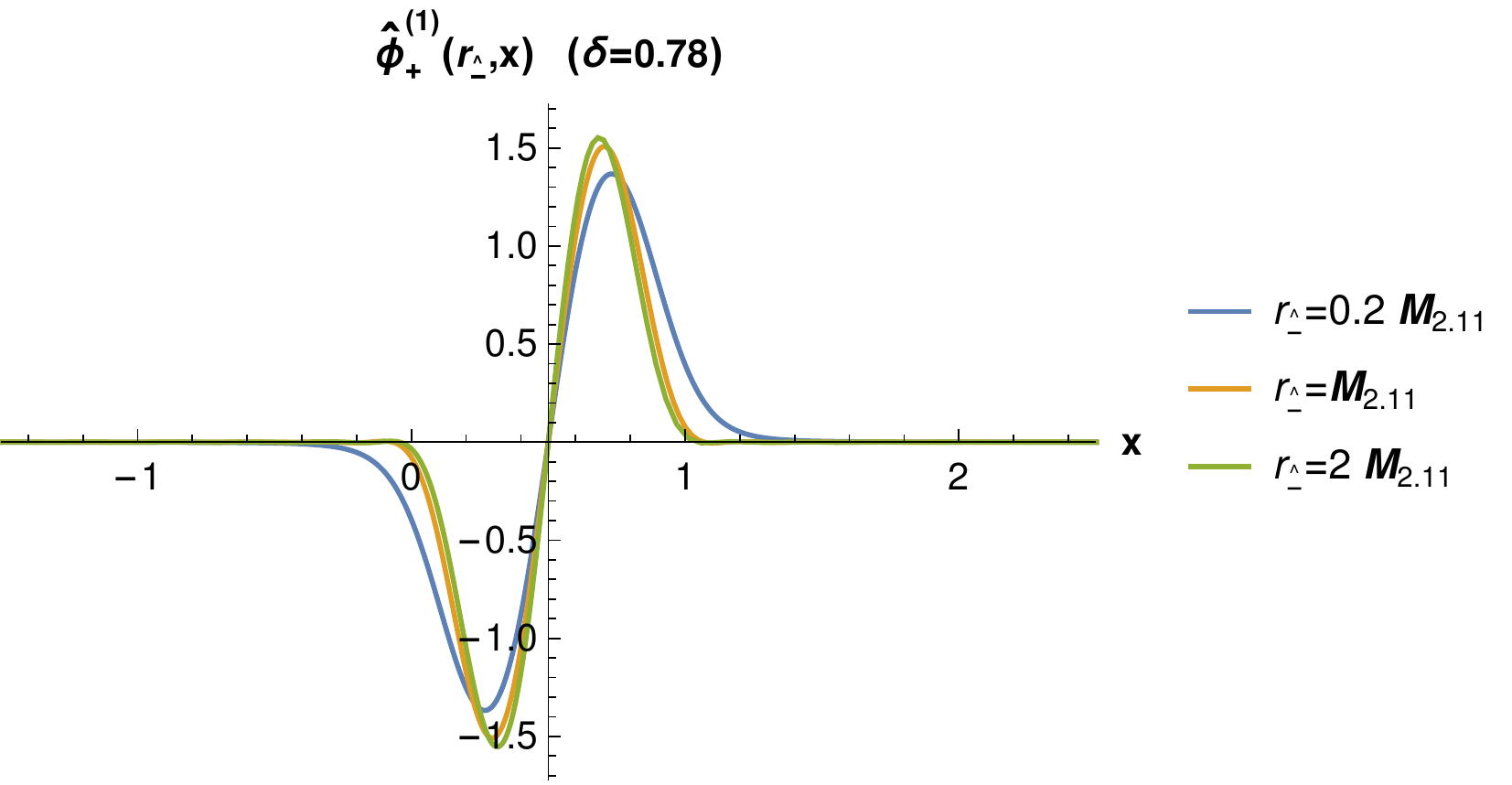}
		\label{fig:m211fezd078}
	}
	\centering
	\subfloat[]{
		\includegraphics[width=0.5\linewidth]{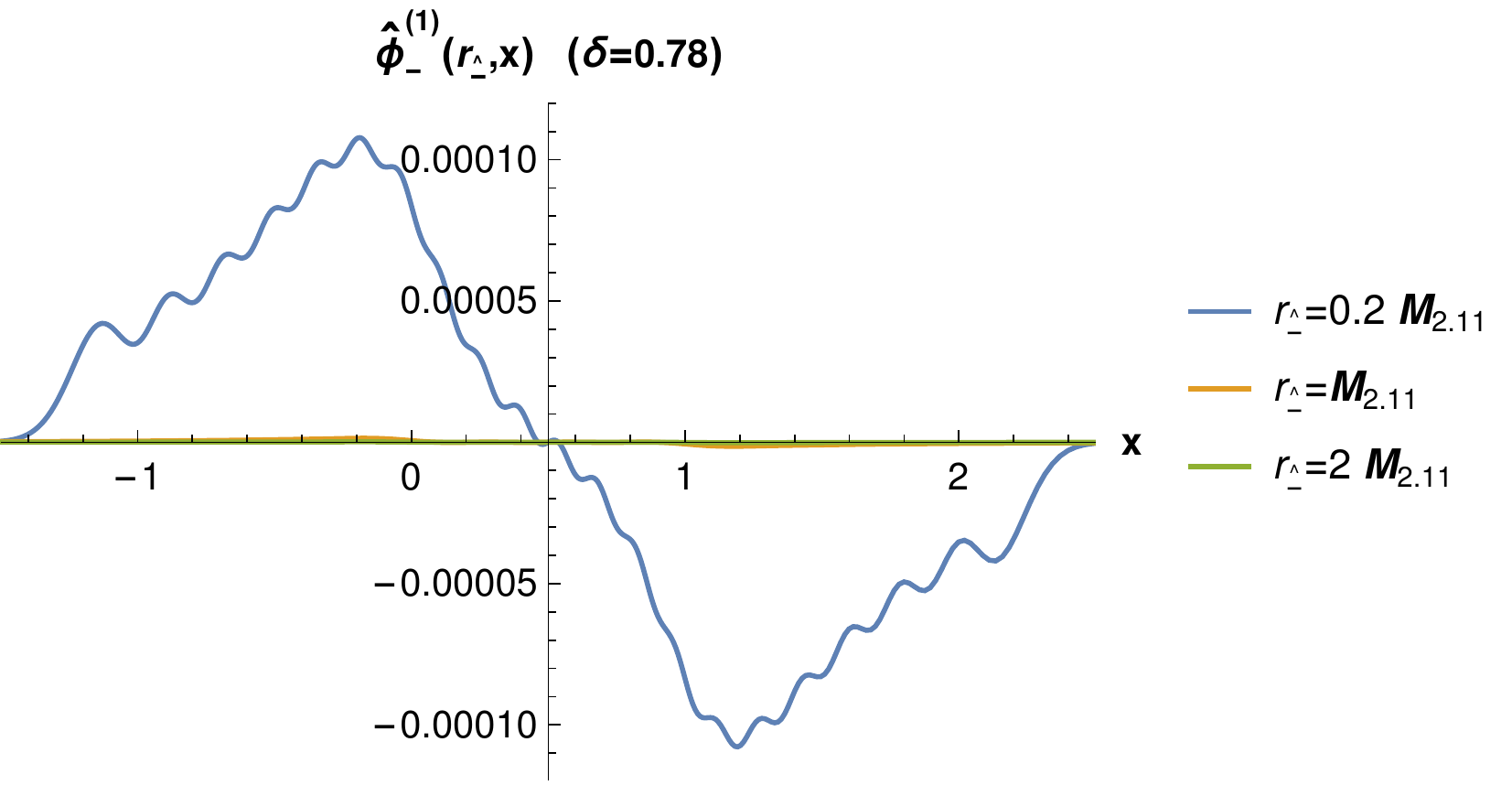}
		\label{fig:m211fefd078}
	}\\
	\caption{First excited state wave functions $\hat\phi_+^{(1)}(r_{\hat{-}},x)$ and $\hat\phi_-^{(1)}(r_{\hat{-}},x)$ for $ m=2.11 $. All quantities are in proper units of $ \sqrt{2\lambda} $.\label{fig:m211fez}}
\end{figure*}

In Sec.~\ref{sub:wavefunc}, we discussed the wavefunctions 
of the quark-antiquark bound-states for the cases of $m=0$ and $m=0.18$
in the unit of $\sqrt{2\lambda}$, i.e. $\bar{m} =$ 0 and 0.18. In this Appendix,  
we summarize the numerical results of $\hat\phi_{\pm}^{(0)}(r_{\hat{-}},x)$ and 
$\hat\phi_{\pm}^{(1)}(r_{\hat{-}},x)$ for a few other bare quark/antiquark mass cases; $m= 0.045, 1.0$ and 2.11 in the unit of $\sqrt{2\lambda}$, i.e. $\bar{m} =$ 0.045, 1.0 and 2.11. In particular, $\bar{m}=0.045$ corresponds to the physical pion mass $M_\pi = 0.41$ in the unit of $\sqrt{2\lambda}$ according to the reasoning~\cite{mov} mentioned in Sec.~\ref{sec:sol}. We also note that the free mass gap solution $\theta_f(\bar{p}'_{\hat{-}})$ given by Eq.~(\ref{eqn:freethetasolutionre}) for $\bar{m} = 1.0$ exhibits the straight line profile in the plot with respect to $\bar{\xi}' =  \tan^{-1} \bar{p}'_{\hat{-}}$ while the profile of the solution for $\bar{m} > 1.0$, e.g. $\bar{m} = 2.11$, gets bended toward the concave
shaped profile from the convex shaped profile for $\bar{m} < 1.0$ as one can see 
in Fig.~\ref{fig:solutionthetaintervsfreere}.   

For the case of $\bar{m}=0.045$, the numerical results of 
$\hat\phi_{\pm}^{(0)}(r_{\hat{-}},x)$ and 
$\hat\phi_{\pm}^{(1)}(r_{\hat{-}},x)$ are presented in Figs.~\ref{fig:m0045grz} and~\ref{fig:m0045fez}, respectively. In each figure, the results of 
$\delta = 0, 0.6$ and $0.78$ are shown in the 
top, middle and bottom panels, respectively. 
In each panel, the results of $r_{\hat{-}}=\mathbf{M}_{0.045}, 5 \mathbf{M}_{0.045}$ and $8 \mathbf{M}_{0.045}$, where $\mathbf{M}_{0.045} = 0.42$ (see Table~\ref{tab:mass}) is the ground state meson mass for the quark mass value $\bar{m}=0.045$, are depicted by the solid lines in blue, yellow, and green, respectively.
We note that the IFD ($\delta=0$) results shown in Figs.~\ref{fig:m0045grzd0},~\ref{fig:m0045grfd0},~\ref{fig:m0045fezd0} and~\ref{fig:m0045fefd0} coincide with the corresponding plots in Fig. 9 of Ref.~\cite{mov}. The results for the case of $\bar{m}=0.045$ exhibit the similar features that we discussed for the case of $\bar{m}=0$ in Sec.~\ref{sub:wavefunc}. 
Namely, the large-momentum IFD ($\delta=0$) numerical results approach to the LFD results quite slowly~\cite{pdf} as the momenta $r_{\hat{-}}$ get large (see Figs.~\ref{fig:m0045grzd0},~\ref{fig:m0045grfd0},~\ref{fig:m0045fezd0} and~\ref{fig:m0045fefd0}), while the results getting close to $\delta=\pi/4$ (e.g. $\delta=0.78$) yield very quickly the essential features of the LFD results fitting in the region $[0,1]$ regardless of the momenta $r_{\hat{-}}=\mathbf{M}_{0.045}$, $5\mathbf{M}_{0.045}$, or $8\mathbf{M}_{0.045}$ and the minus component disappears (see Figs.~\ref{fig:m0045grzd078},~\ref{fig:m0045grfd078},~\ref{fig:m0045fezd078} and~\ref{fig:m0045fefd078}) 
although the numerical sensitivity gets enhanced with some wiggles or bulges in $\hat\phi_{+}^{(0)}(r_{\hat{-}},x) (\delta=0.78)$ for $r_{\hat{-}} = 5 \mathbf{M}_{0.045}$ and $ 8 \mathbf{M}_{0.045}$ due to the enhanced demand of numerical accuracy as 
$\mathbb{C}$ gets close to zero and $r_{\hat{-}}$ gets large.
In Figs.~\ref{fig:m0045grz} and~\ref{fig:m0045fez}, the charge conjugation symmetry under the exchange of $x \leftrightarrow 1-x$ is manifest as we have discussed for the ground state and the first excited state previously in Sec.~\ref{sub:wavefunc}, i.e.
$\hat\phi_{+}^{(1)}(r_{\hat{-}},x)$ and $\hat\phi_{-}^{(1)}(r_{\hat{-}},x)$ reveal the antisymmetric profiles while $\hat\phi_{+}^{(0)}(r_{\hat{-}},x)$ and $\hat\phi_{-}^{(0)}(r_{\hat{-}},x)$ exhibit the symmetric profiles.

For the case of $\bar{m} = 1.0$ which we noted above its straight line profile for $\theta_f(\bar{p}'_{\hat{-}})$ in Fig.~\ref{fig:solutionthetaintervsfreere}, 
the numerical results of 
$\hat\phi_{\pm}^{(0)}(r_{\hat{-}},x)$ and $\hat\phi_{\pm}^{(1)}(r_{\hat{-}},x)$ 
are presented in Figs.~\ref{fig:m100grz} and~\ref{fig:m100fez}, respectively. The frames for $\bar{m}=1.0$ are chosen as $r_{\hat{-}}=0.2 \mathbf{M}_{1.0}, 2 \mathbf{M}_{1.0}$ and $5 \mathbf{M}_{1.0}$, where $\mathbf{M}_{1.0} = 2.70$ (see Table~\ref{tab:mass}) is the ground state meson mass for the quark mass value $\bar{m}=1.0$. The essential features that we discussed
for the lower masses ($\bar{m}=$ 0, 0.045, 0.18) in Sec.~\ref{sub:wavefunc} and above remain without much change. As mentioned earlier, we haven't increased the number of grid points beyond 600 while the numerical accuracy is much more demanded as $\mathbb{C}$ gets close to zero and $r_{\hat{-}}$ gets large.
However, we are not alarmed by the appearance of ``rabbit ear" for $\delta = 0.78$ and $r_{\hat{-}}=  5 \mathbf{M}_{1.0} = 13.5$ in Fig.~\ref{fig:m100grzd078} as 
we are convinced from our numerical analyses that such numerical noise would disappear as we keep pushing the number of grid points even higher.   

Finally, in Figs.~\ref{fig:m211grz} and~\ref{fig:m211fez}, we present our numerical results of $\hat\phi_{+}^{(0)}(r_{\hat{-}},x)$, $\hat\phi_{-}^{(0)}(r_{\hat{-}},x)$, $\hat\phi_{+}^{(1)}(r_{\hat{-}},x)$ and $\hat\phi_{-}^{(1)}(r_{\hat{-}},x)$ for the case of ${\bar m}=2.11$. In this case, the 
$\theta(\bar{p}'_{\hat{-}})$ solution gets close to the free mass gap solution $\theta_f(\bar{p}'_{\hat{-}})$ as shown in Fig.~\ref{fig:solutionthetaintervsfreere}, 
which may indicate that the binding effect gets lesser while the quark mass effect gets larger. In fact, the extreme heavy quark mass limit would yield the non-relativistic 
$\delta$-function type of ground-state meson wavefunction peaked highly at $x= 1/2$ to share the longitudinal momentum equally between the two equal mass quark and antiquark. In Fig.~\ref{fig:m211grz}, we see a kind of precursor for such tendency toward the heavy quark-antiquark bound-state system.  
In the case of $\bar{m}=2.11$, we take our frames as $r_{\hat{-}}=0.2 \mathbf{M}_{2.11}, \mathbf{M}_{2.11}$ and $2 \mathbf{M}_{2.11}$, where $\mathbf{M}_{2.11} = 4.91$ (see Table~\ref{tab:mass}). As $2 \mathbf{M}_{2.11}$ is already large enough
for our numerical computation, we do not go beyond $r_{\hat{-}}=2 \mathbf{M}_{2.11}$. Besides the tendency toward the heavy quark-antiquark bound-state system,
the essential features that we discussed previously including  
the charge conjugation symmetry under the exchange of $x \leftrightarrow 1-x$ for the ground state and the first excited state appear similar in  
Figs.~\ref{fig:m211grz} and~\ref{fig:m211fez}. We notice some wiggles in the $\hat\phi_{-}$ component of the wavefunction solution in e.g. Fig.~\ref{fig:m211grfd078}, but the overall magnitude of the $\hat\phi_{-}$ wavefunction is always negligible compared to $\hat\phi_{+}$ whenever this occurs, thus it does not cause concern to us. 

\section{``Quasi-PDFs" corresponding to mesonic wavefunctions for 
$m=$ 0.045, 1.0 and 2.11 in the unit of $\sqrt{2\lambda}$}
\label{quasi-PDF-other-masses}
Starting from the definition of the ``quasi-PDFs" interpolating between 
IFD and LFD given by Eq.~(\ref{intepolating-pdf-gauge-link}),
we obtained the ``quasi-PDFs" using the mesonic wavefunctions of the quark-antiquark bound-states in the interpolating axial gauge, $ A_{\hat{-}}^a=0$, as given by Eq.~(\ref{interpol-pdf}) and discussed the ``quasi-PDFs" for the cases of $m=0$ and $m=0.18$ in Sec.~\ref{sub:quasipdf}. As the mesonic wavefunctions for $m= 0.045, 1.0$ and 2.11 were presented in the previous appendix, Appendix~\ref{app:figures}, we now discuss the corresponding ``quasi-PDFs" in this Appendix~\ref{quasi-PDF-other-masses}. 

\begin{figure*}
	\centering
	\subfloat[]{
		\includegraphics[width=0.4\linewidth]{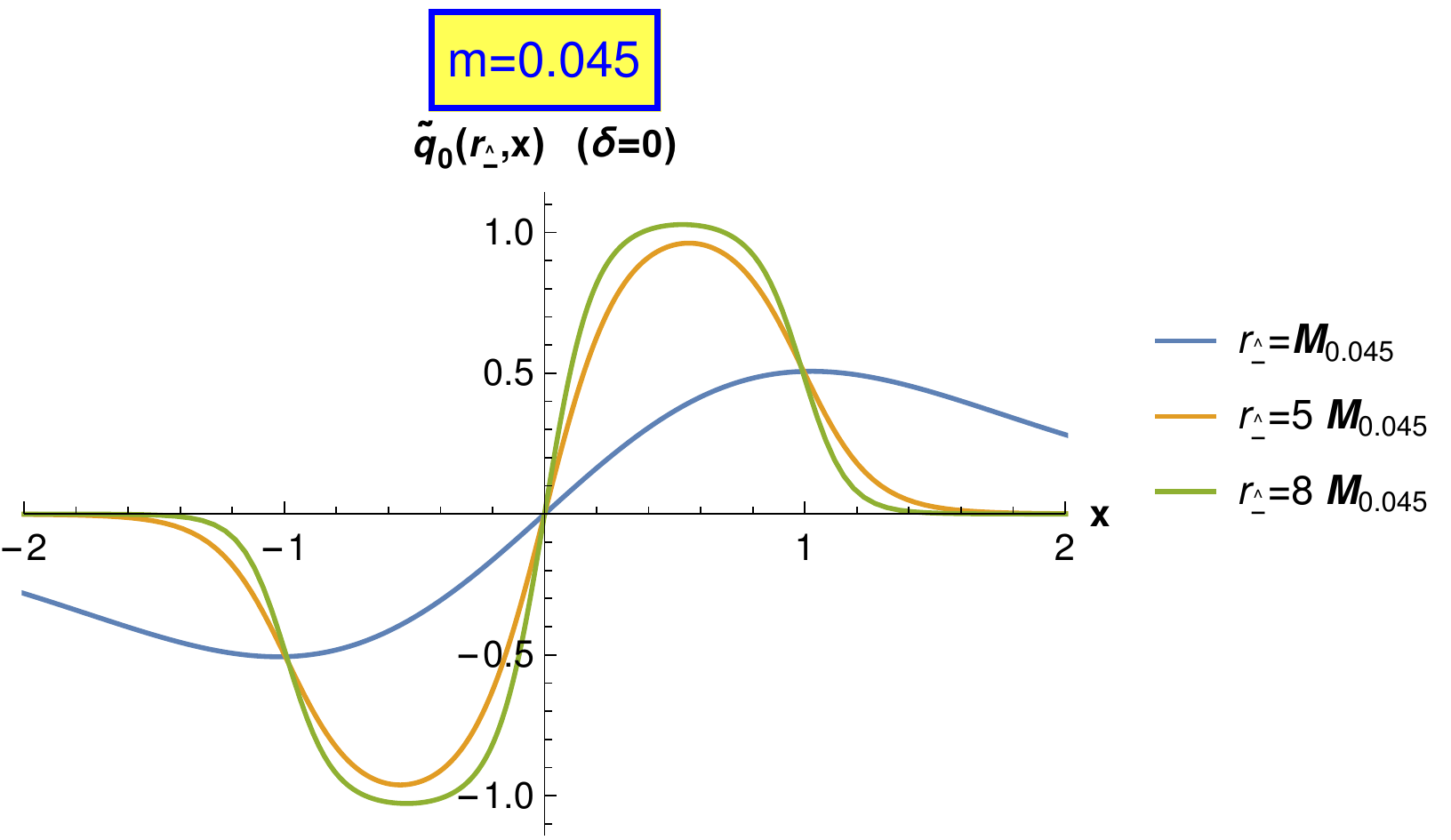}
		\label{fig:m0045_delta0_grquasipdf}
	}
	\centering
	\subfloat[]{
		\includegraphics[width=0.4\linewidth]{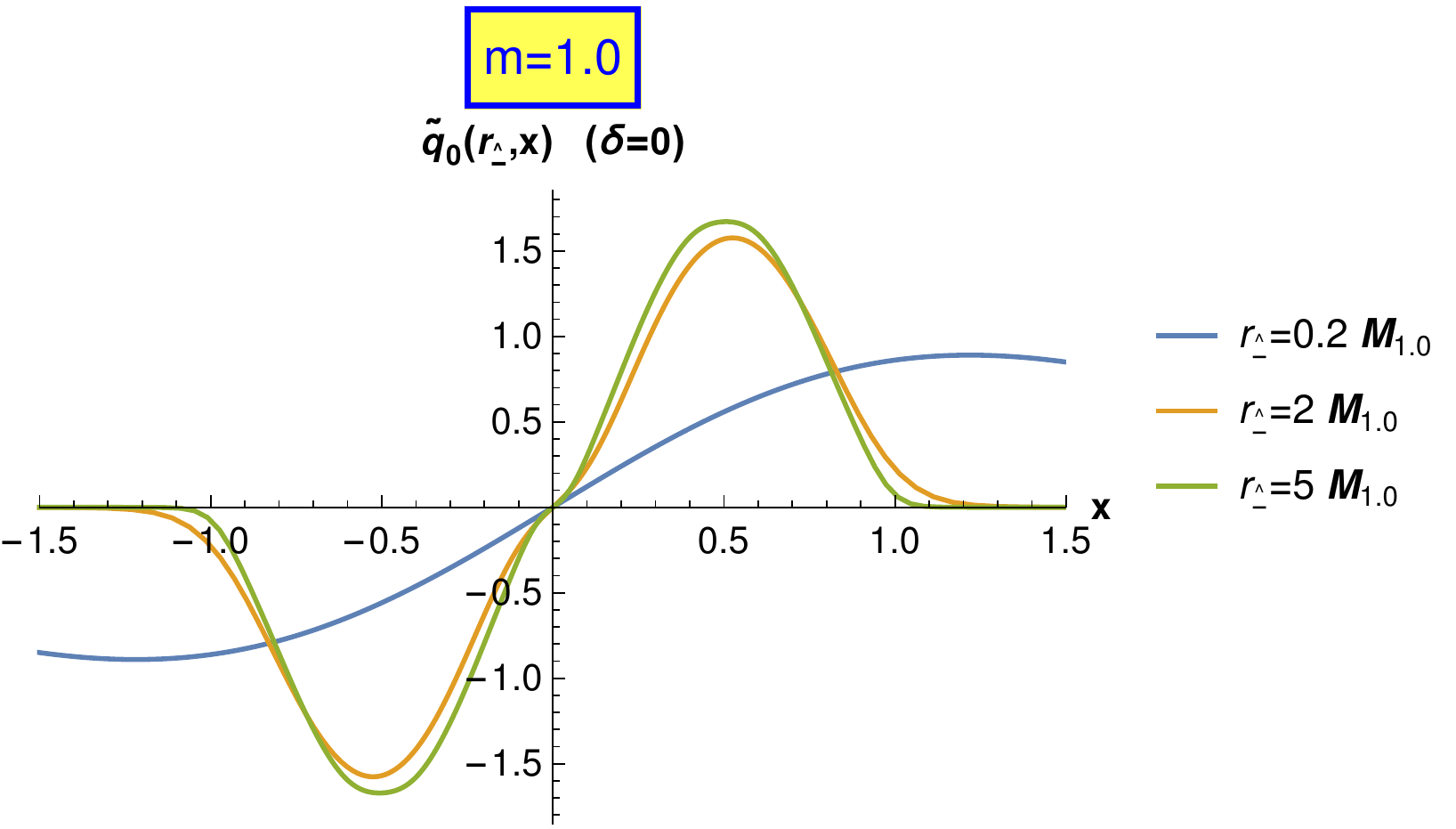}
		\label{fig:m100_delta0_grquasipdf}
	}\\
	\centering
	\subfloat[]{
		\includegraphics[width=0.4\linewidth]{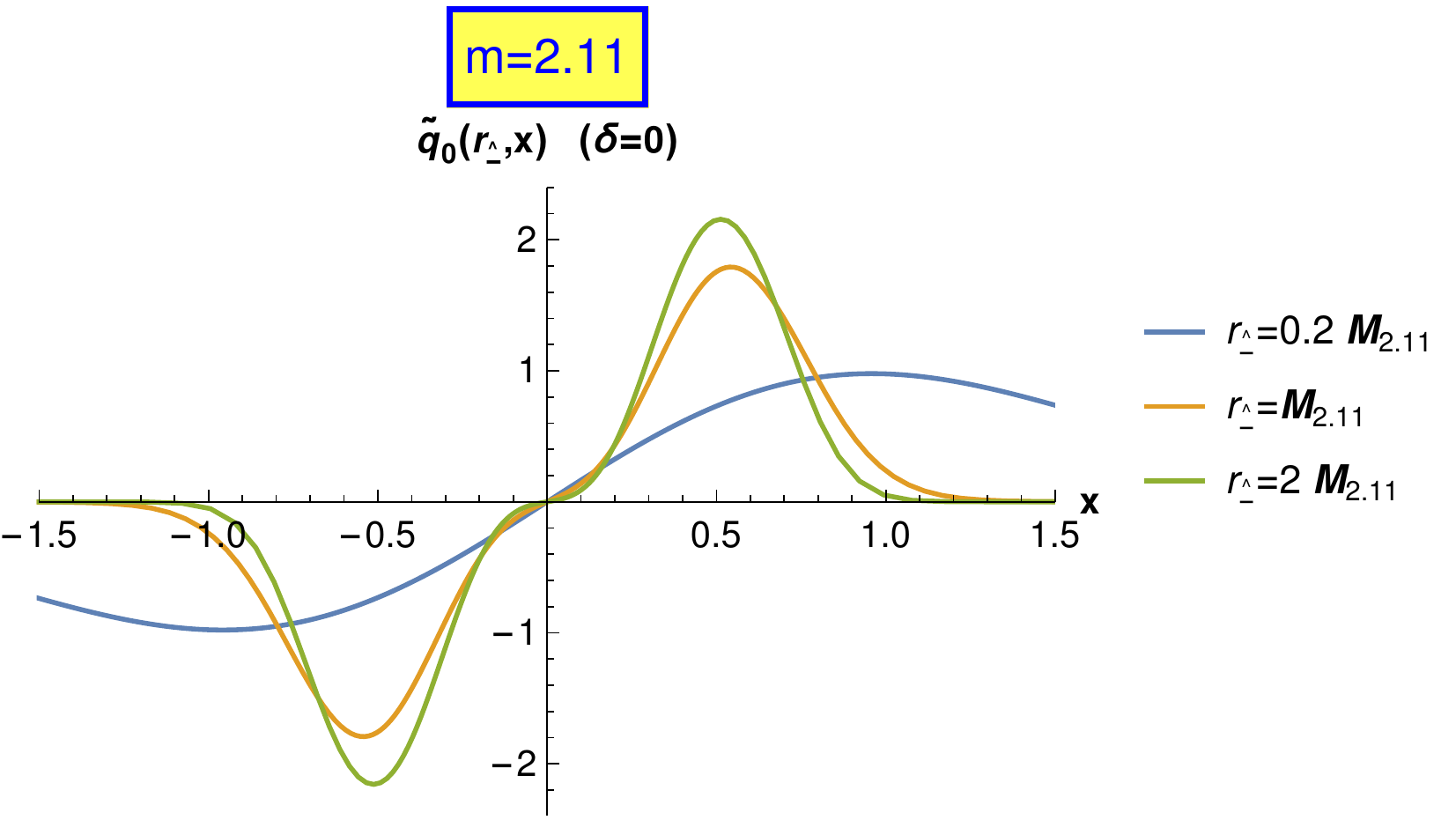}
		\label{fig:m211_delta0_grquasipdf}
	}
	\caption{Quasi-PDFs in IFD ($\delta =0$) for the ground state ($n=0$) wave functions of (a) $m=0.045$, (b) $m=1.00$, and (c) $m=2.11$  in the unit of $ \sqrt{2\lambda} $.
\label{figapp:delta0_grquasipdf}}
\end{figure*}

\begin{figure*}
	\centering
	\subfloat[]{
		\includegraphics[width=0.4\linewidth]{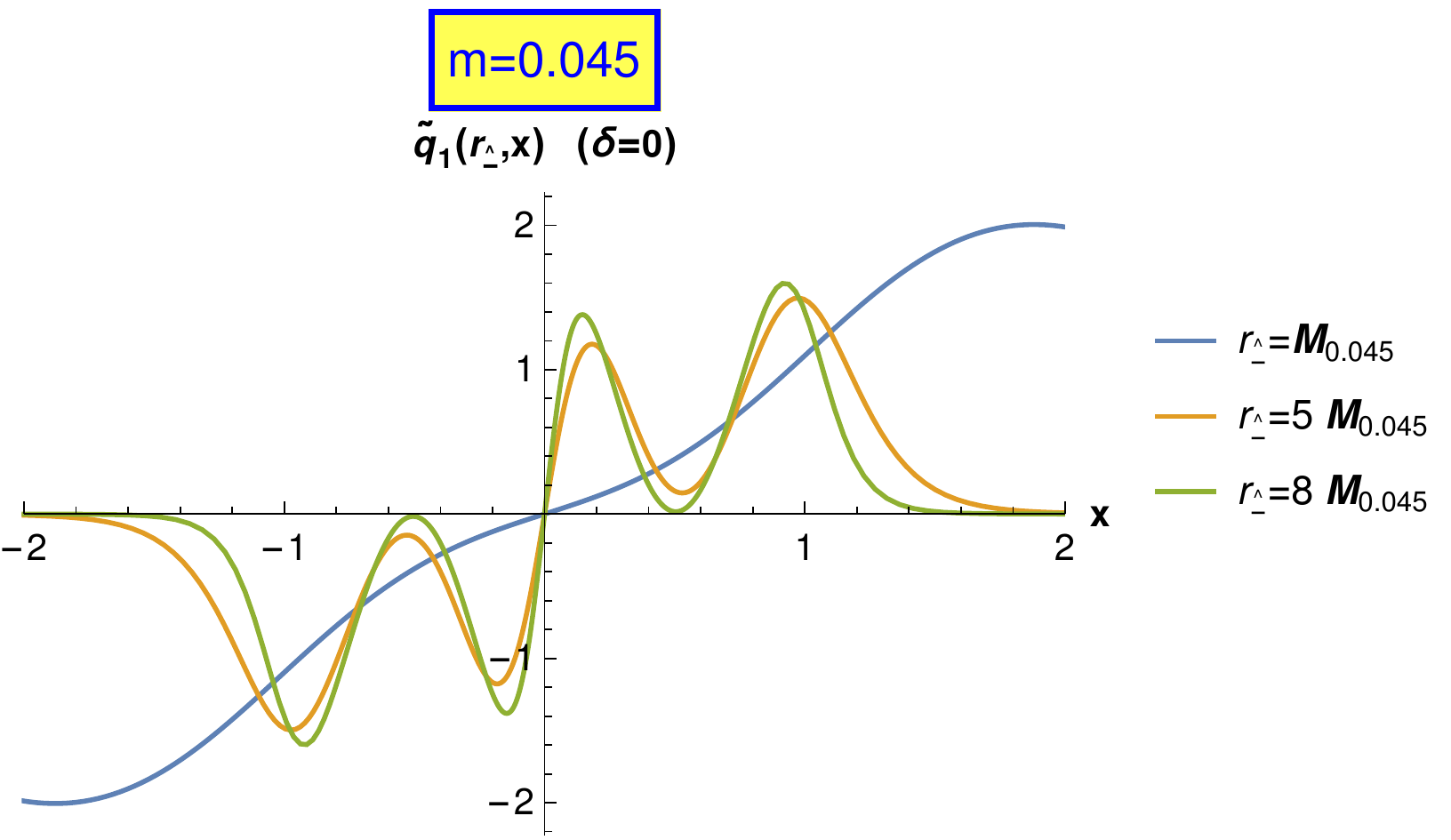}
		\label{fig:m0045_delta0_fequasipdf}
	}
	\centering
	\subfloat[]{
		\includegraphics[width=0.4\linewidth]{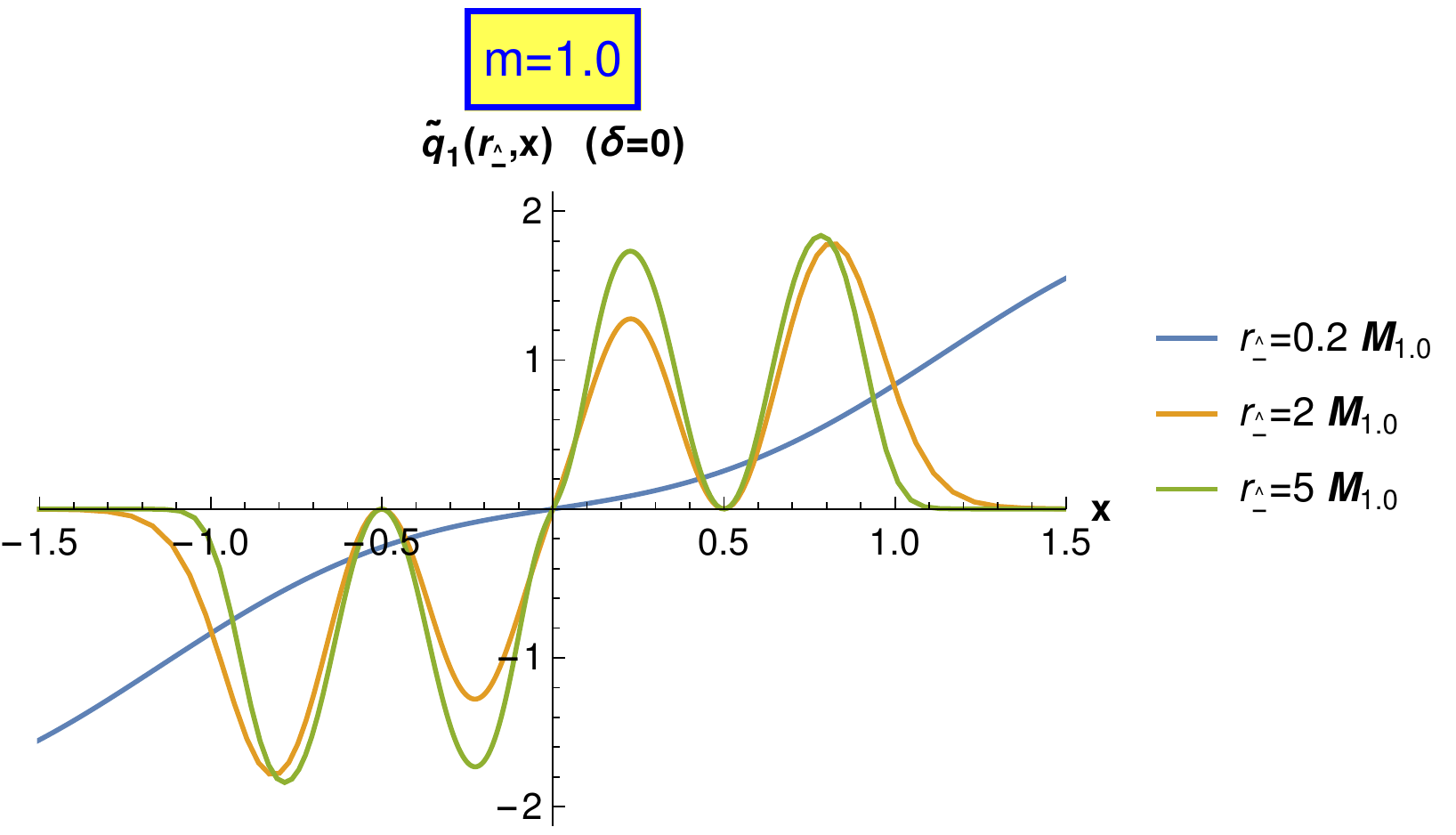}
		\label{fig:m100_delta0_fequasipdf}
	}\\
	\centering
	\subfloat[]{
		\includegraphics[width=0.4\linewidth]{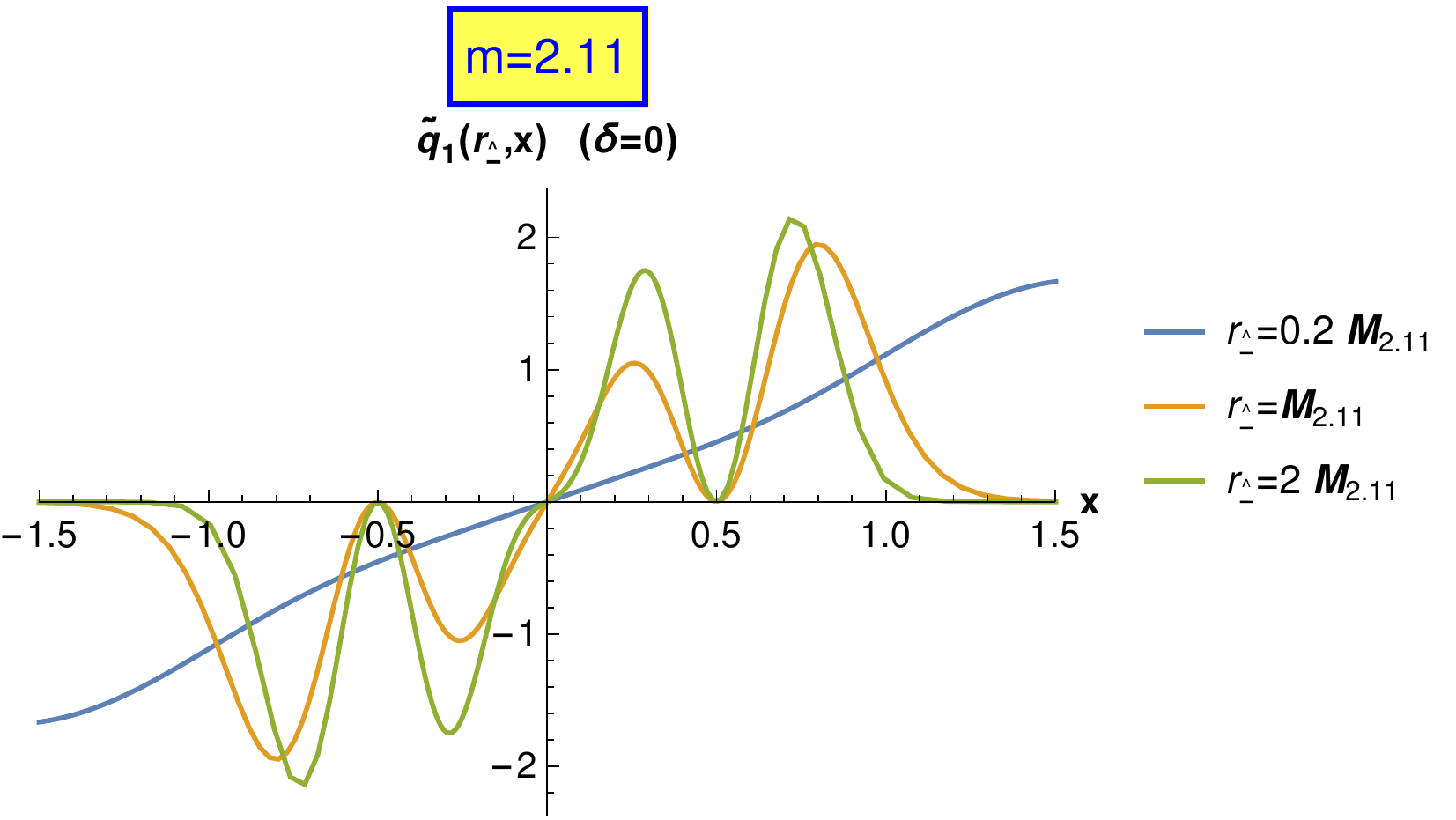}
		\label{fig:m211_delta0_fequasipdf}
	}
	\caption{Quasi-PDFs in IFD ($\delta=0$) for the first excited state ($n=1$) wave functions of (a) $m=0.045$, (b) $m=1.00$, and (c) $m=2.11$ in the unit of $ \sqrt{2\lambda} $.
\label{figapp:delta0_fequasipdf}}
\end{figure*}

First, the $\delta=0$ (IFD) results of the ground state and first excited state mesonic quasi-PDFs are shown in Figs.~\ref{figapp:delta0_grquasipdf} and \ref{figapp:delta0_fequasipdf}, respectively, for a few different quark mass values, not only $m=0.045$ as taken in Ref.~\cite{pdf} but also $m=1.0$ and $m=2.11$. The results of $m=0.045$ shown in Figs.~\ref{fig:m0045_delta0_grquasipdf} and \ref{fig:m0045_delta0_fequasipdf} agree very well with the top right panels of Figs. 2 and 3 of Ref.~\cite{pdf}. 
Due to the charge conjugation symmetry under the exchange of $x \leftrightarrow 1-x$
, $\hat\phi_{\pm}^{(0)}(r_{\hat{-}},x)= \hat\phi_{\pm}^{(0)}(r_{\hat{-}},1-x)$ and $\hat\phi_{\pm}^{(1)}(r_{\hat{-}},x)= - \hat\phi_{\pm}^{(1)}(r_{\hat{-}},1-x)$, we see the peak and valley at $x=1/2$ for ${\tilde q}_0(r_{\hat{-}},x)$ and 
${\tilde q}_1(r_{\hat{-}},x)$, respectively, as shown in Figs.~\ref{figapp:delta0_grquasipdf} and \ref{figapp:delta0_fequasipdf}. 
Although the peak and valley get a little more sharpened as $m$ gets larger, 
the essential features of the symmetry remain intact regardless of the $m$ values.
In each panel, the results of the moving frames with the longitudinal meson momentum
$r_{\hat{-}}=r^1=\mathbf{M}_{0.045}, 5 \mathbf{M}_{0.045}$ and $8 \mathbf{M}_{0.045}$, where $\mathbf{M}_{0.045} = 0.42$ (see Table~\ref{tab:mass}) is the ground state meson mass for the quark mass value $m=0.045$, are depicted by the solid lines in blue, yellow, and green, respectively.
As noted in Ref.~\cite{pdf}, the large-momentum IFD ($\delta=0$) numerical results approach to the LFD results quite slowly as the momentum $r_{\hat{-}}=r^1$ gets large.
The corresponding results for the larger $m$ values, $m=1.0$ and $m=2.11$, are shown in Figs.~\ref{fig:m100_delta0_grquasipdf} and \ref{fig:m100_delta0_fequasipdf} and 
Figs.~\ref{fig:m211_delta0_grquasipdf} and \ref{fig:m211_delta0_fequasipdf} for 
the ground state and the first excited state, respectively.

\begin{figure*}
	\centering
	\subfloat[]{
		\includegraphics[width=0.4\linewidth]{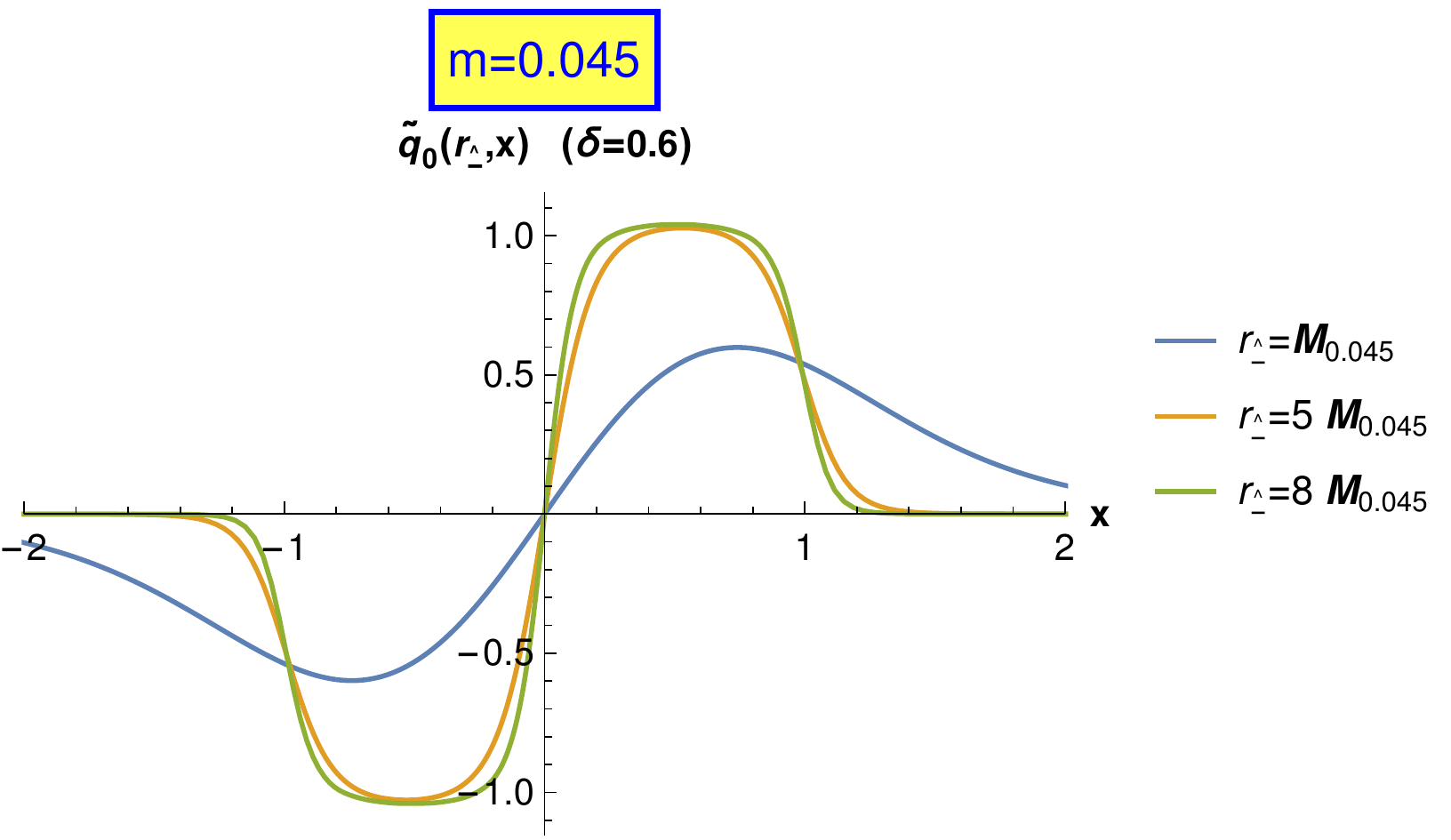}
		\label{fig:m0045_delta06_grquasipdf}
	}
	\centering
	\subfloat[]{
		\includegraphics[width=0.4\linewidth]{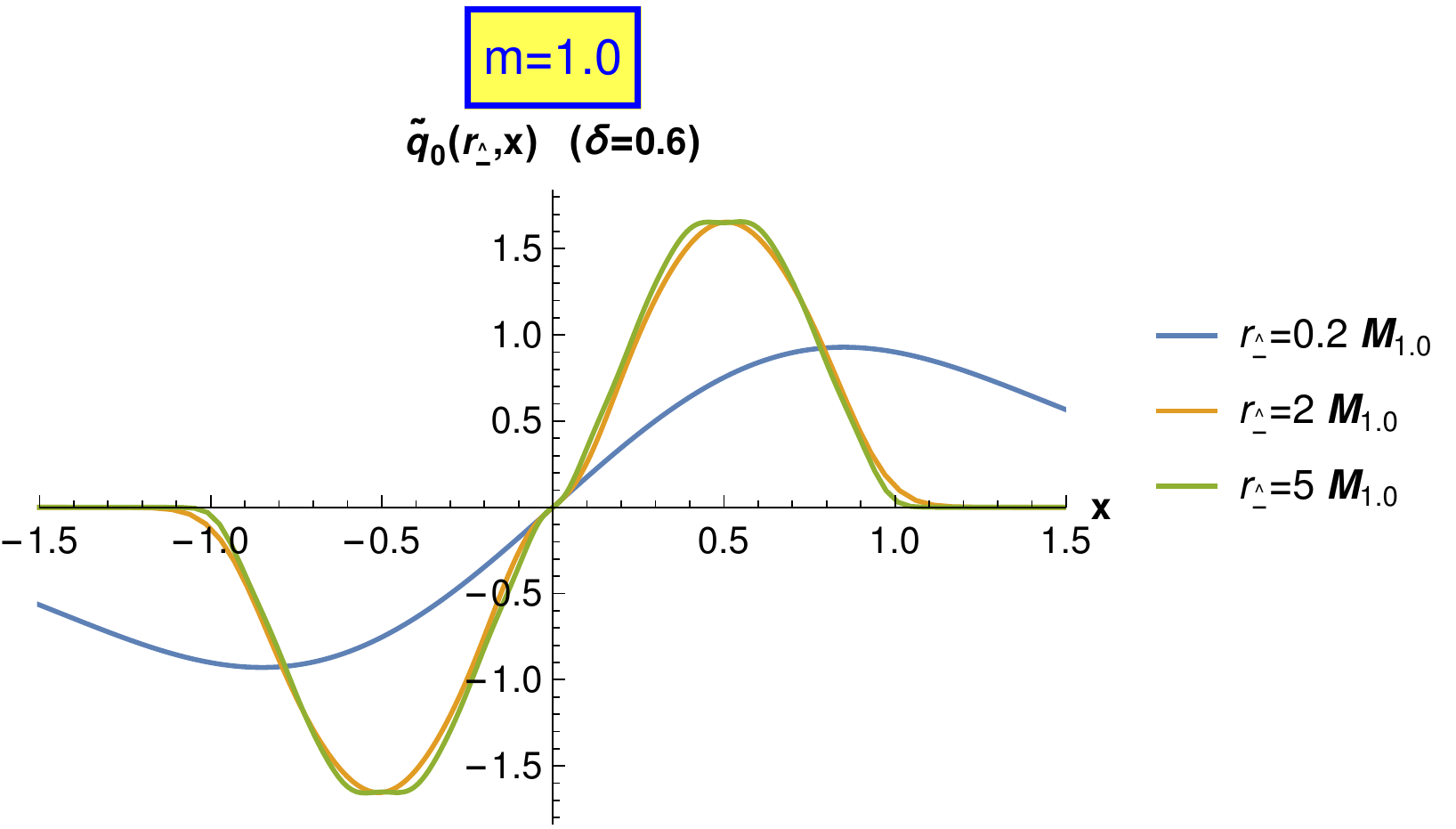}
		\label{fig:m100_delta06_grquasipdf}
	}\\
	\centering
	\subfloat[]{
		\includegraphics[width=0.4\linewidth]{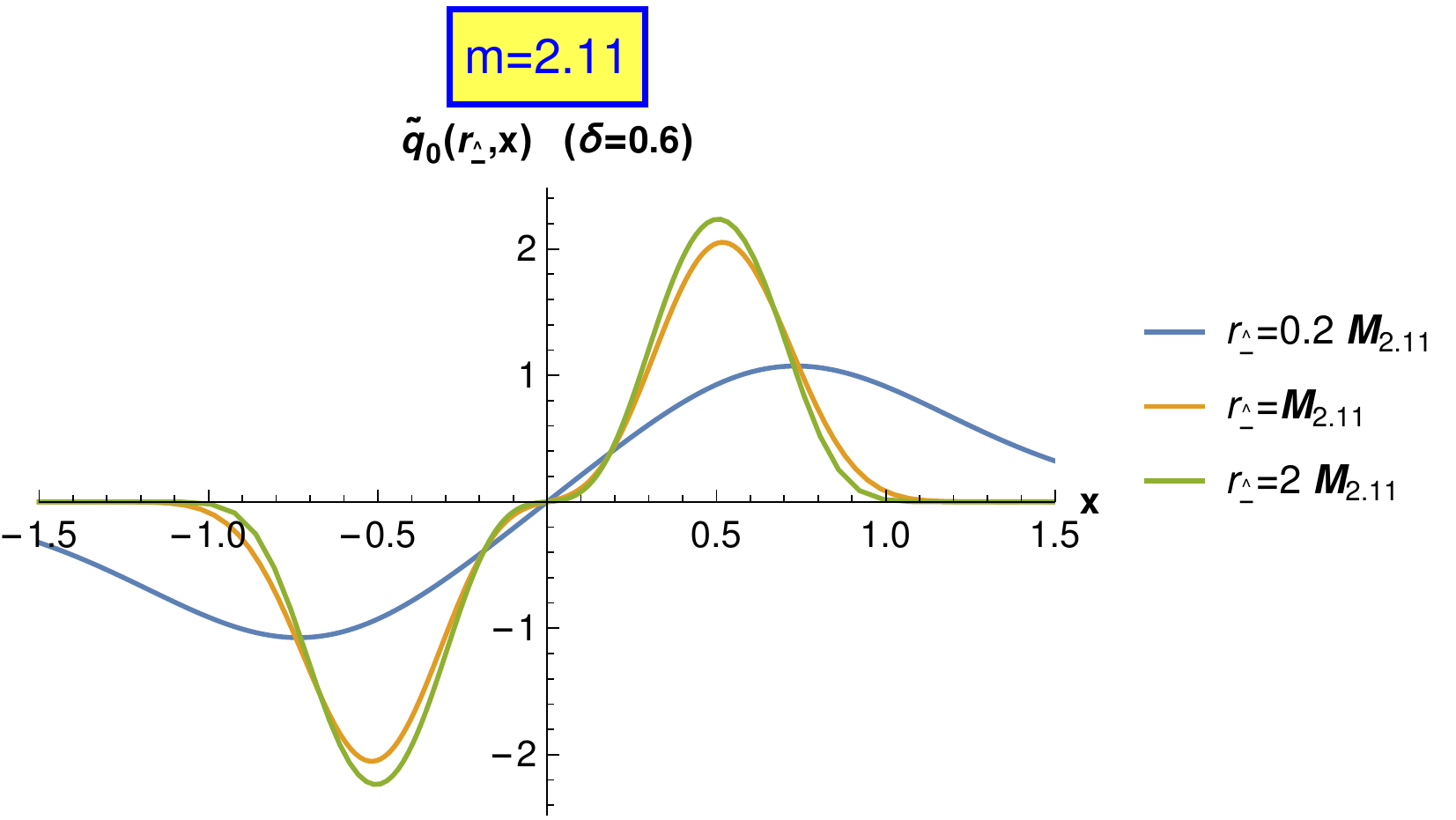}
		\label{fig:m211_delta06_grquasipdf}
	}
	\caption{$ \delta=0.6$ interpolating ``quasi-PDFs" for the ground state ($n=0$) wave functions of (a) $m=0.045$, (b) $m=1.00$, and (c) $m=2.11$  in the unit of $ \sqrt{2\lambda} $.\label{figapp:delta06_grquasipdf}}
\end{figure*}

\begin{figure*}
	\centering
	\subfloat[]{
		\includegraphics[width=0.4\linewidth]{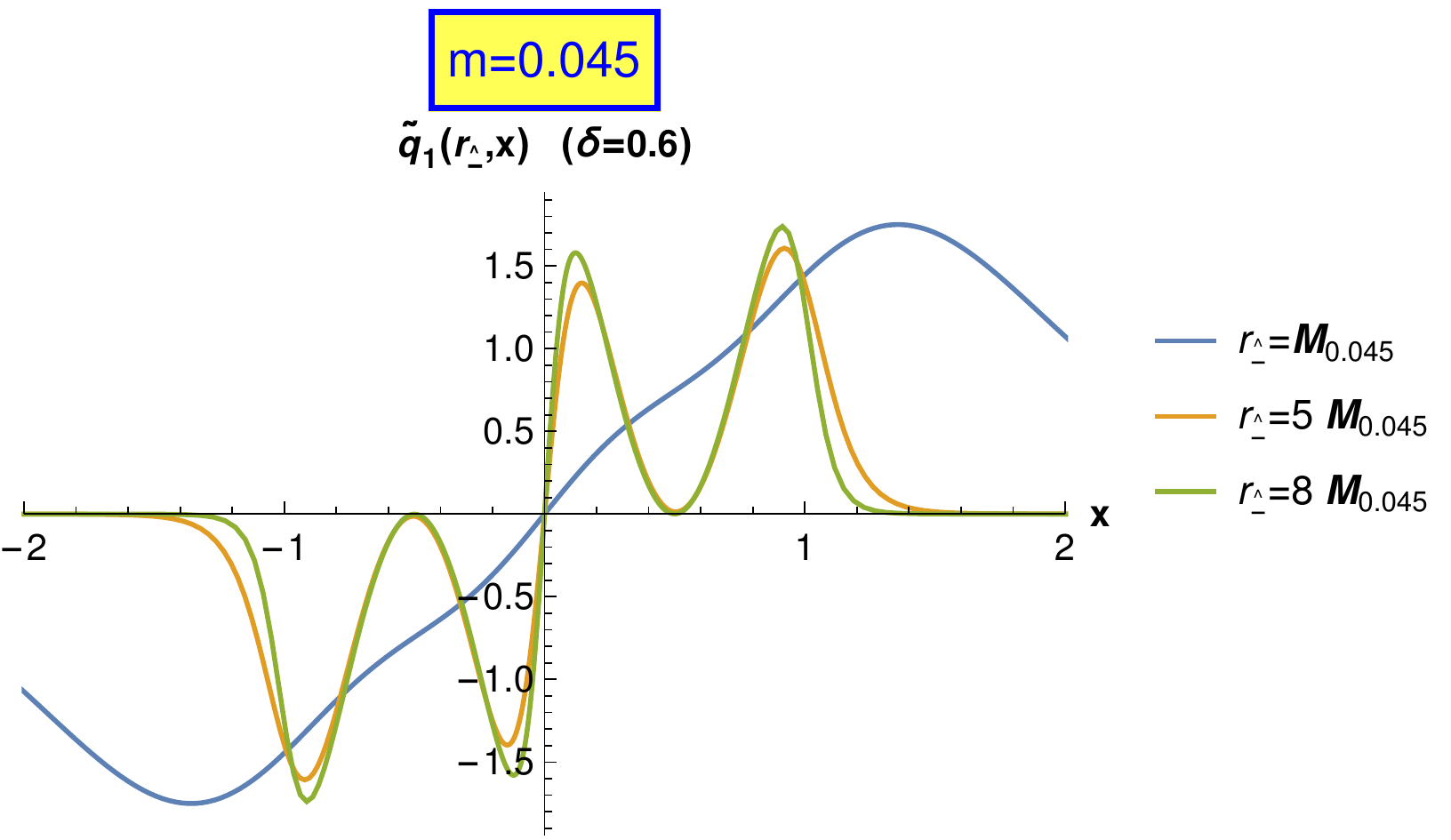}
		\label{fig:m0045_delta06_fequasipdf}
	}
	\centering
	\subfloat[]{
		\includegraphics[width=0.4\linewidth]{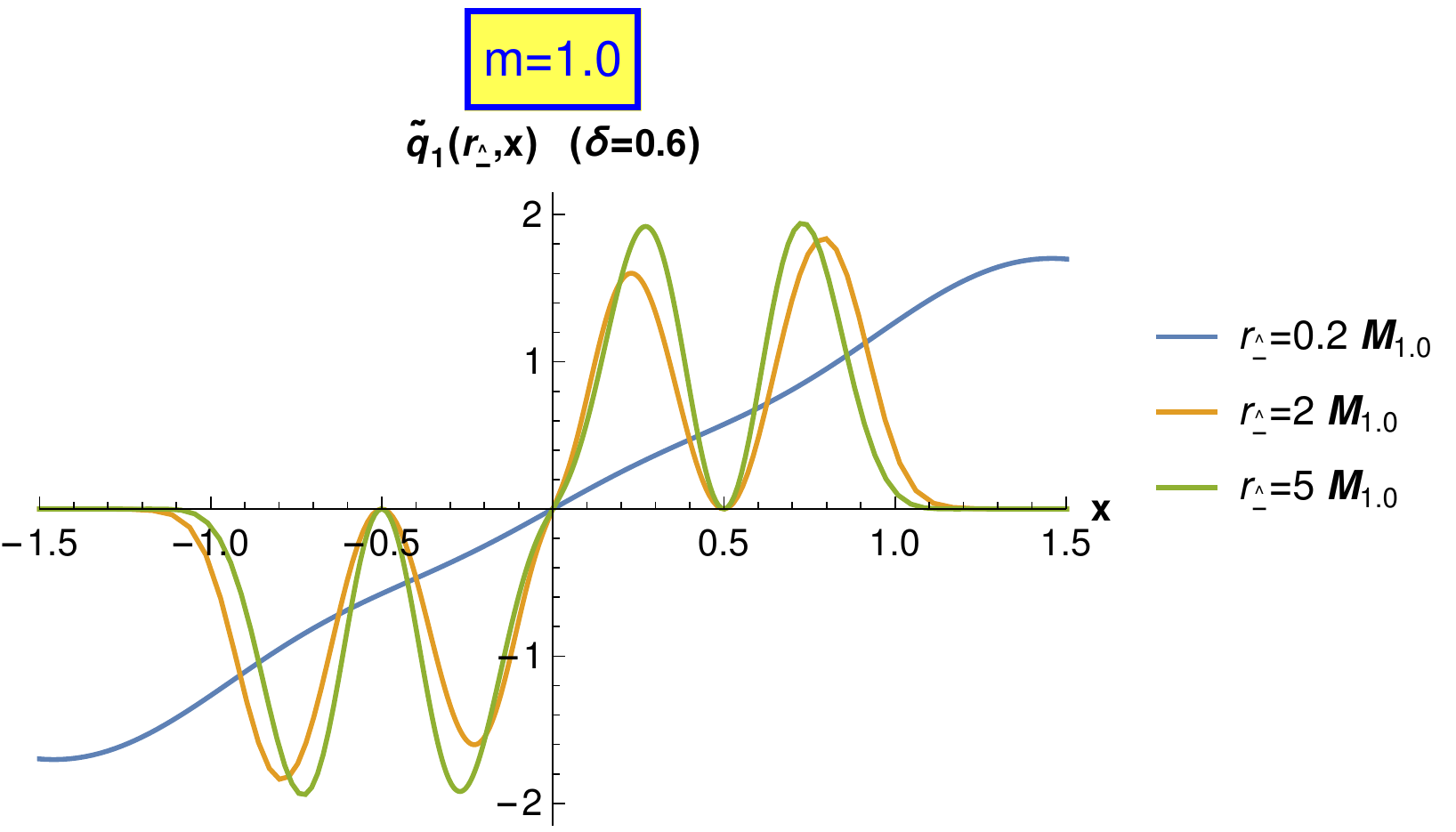}
		\label{fig:m100_delta06_fequasipdf}
	}\\
	\centering
	\subfloat[]{
		\includegraphics[width=0.4\linewidth]{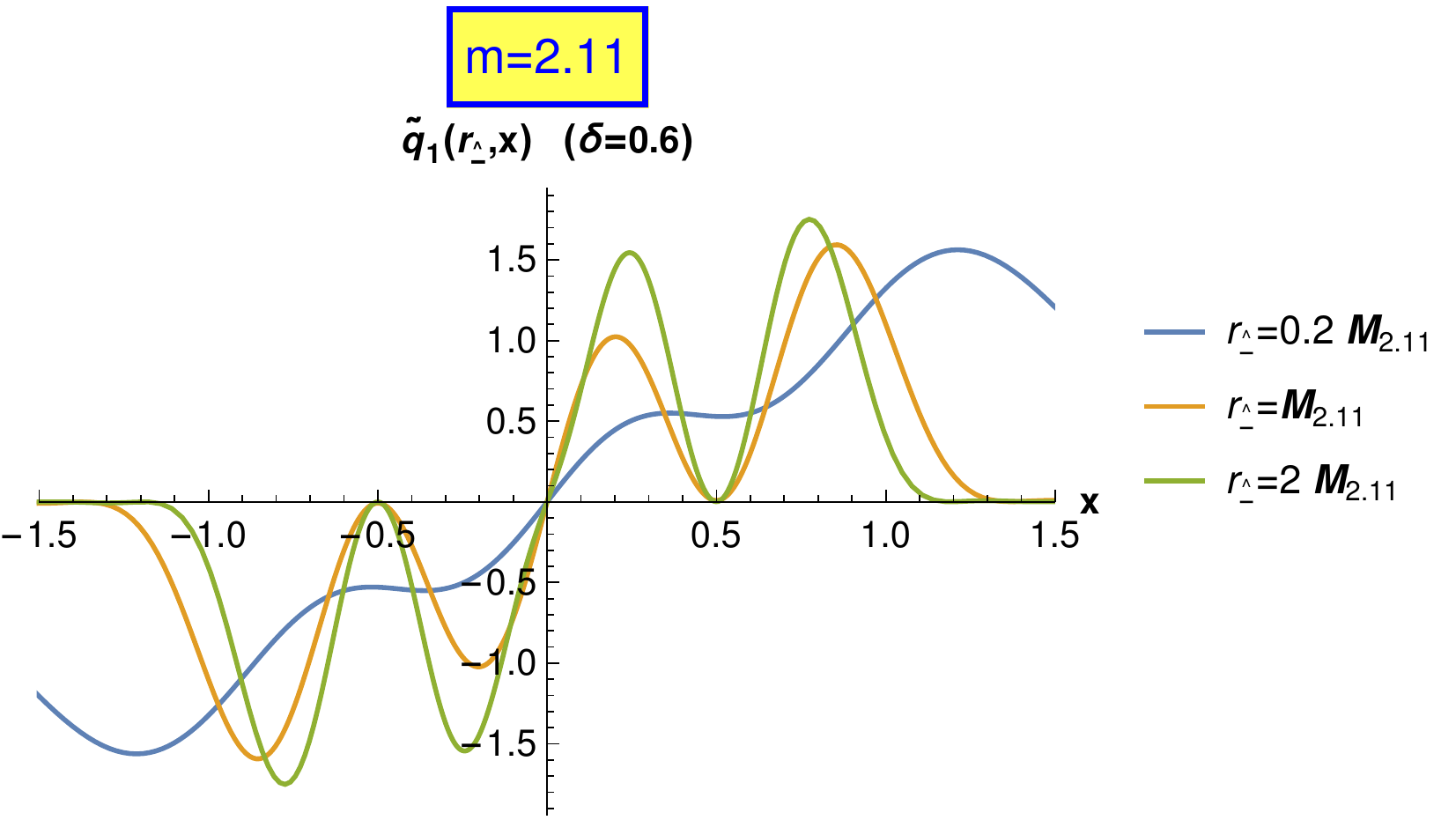}
		\label{fig:m211_delta06_fequasipdf}
	}
	\caption{$ \delta=0.6$ interpolating ``quasi-PDFs" for the first excited state ($n=1$) wave functions of (a) $m=0.045$, (b) $m=1.00$, and (c) $m=2.11$ in the unit of $ \sqrt{2\lambda} $.
\label{figapp:delta06_fequasipdf}}
\end{figure*}

\begin{figure*}
	\centering
	\subfloat[]{
		\includegraphics[width=0.4\linewidth]{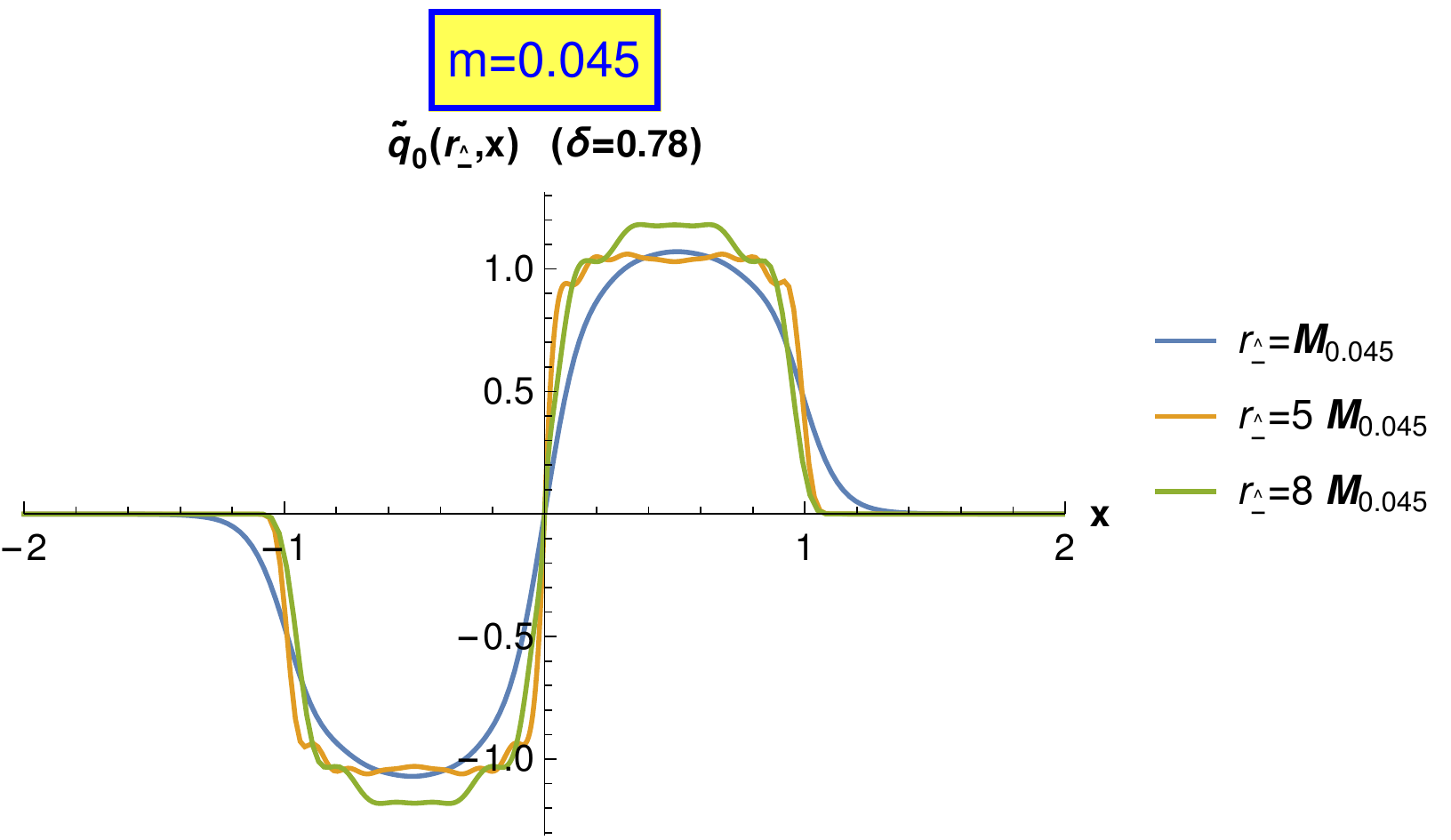}
		\label{fig:m0045_delta078_grquasipdf}
	}
	\centering
	\subfloat[]{
		\includegraphics[width=0.4\linewidth]{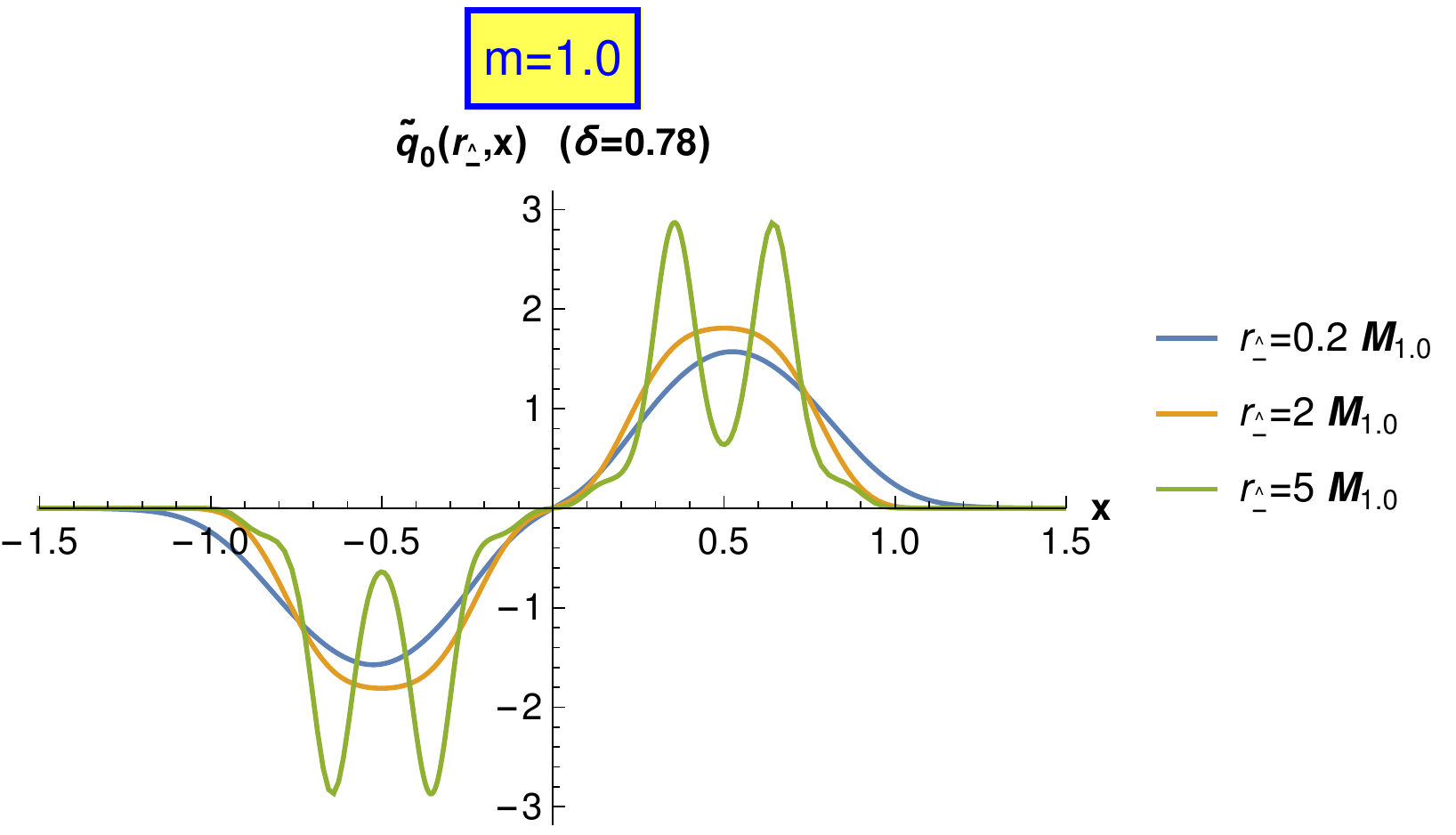}
		\label{fig:m100_delta078_grquasipdf}
	}\\
	\centering
	\subfloat[]{
		\includegraphics[width=0.4\linewidth]{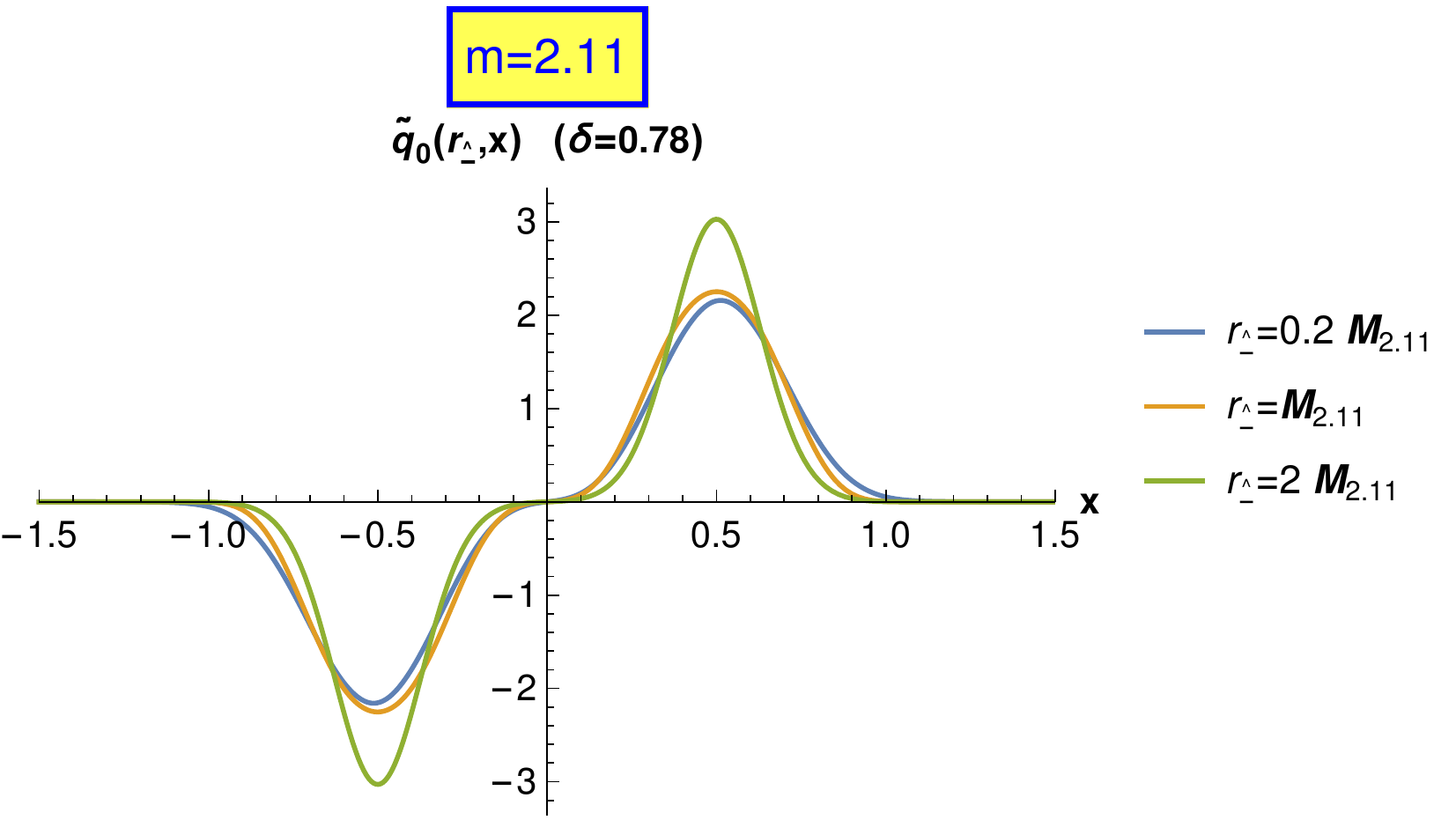}
		\label{fig:m211_delta078_grquasipdf}
	}
	\caption{$ \delta=0.78$ interpolating ``quasi-PDFs" for the ground state ($n=0$) wave functions of (a) $m=0.045$, (b) $m=1.00$, and (c) $m=2.11$ in the unit of $ \sqrt{2\lambda} $.\label{figapp:delta078_grquasipdf}}
\end{figure*}

\begin{figure*}
	\centering
	\subfloat[]{
		\includegraphics[width=0.4\linewidth]{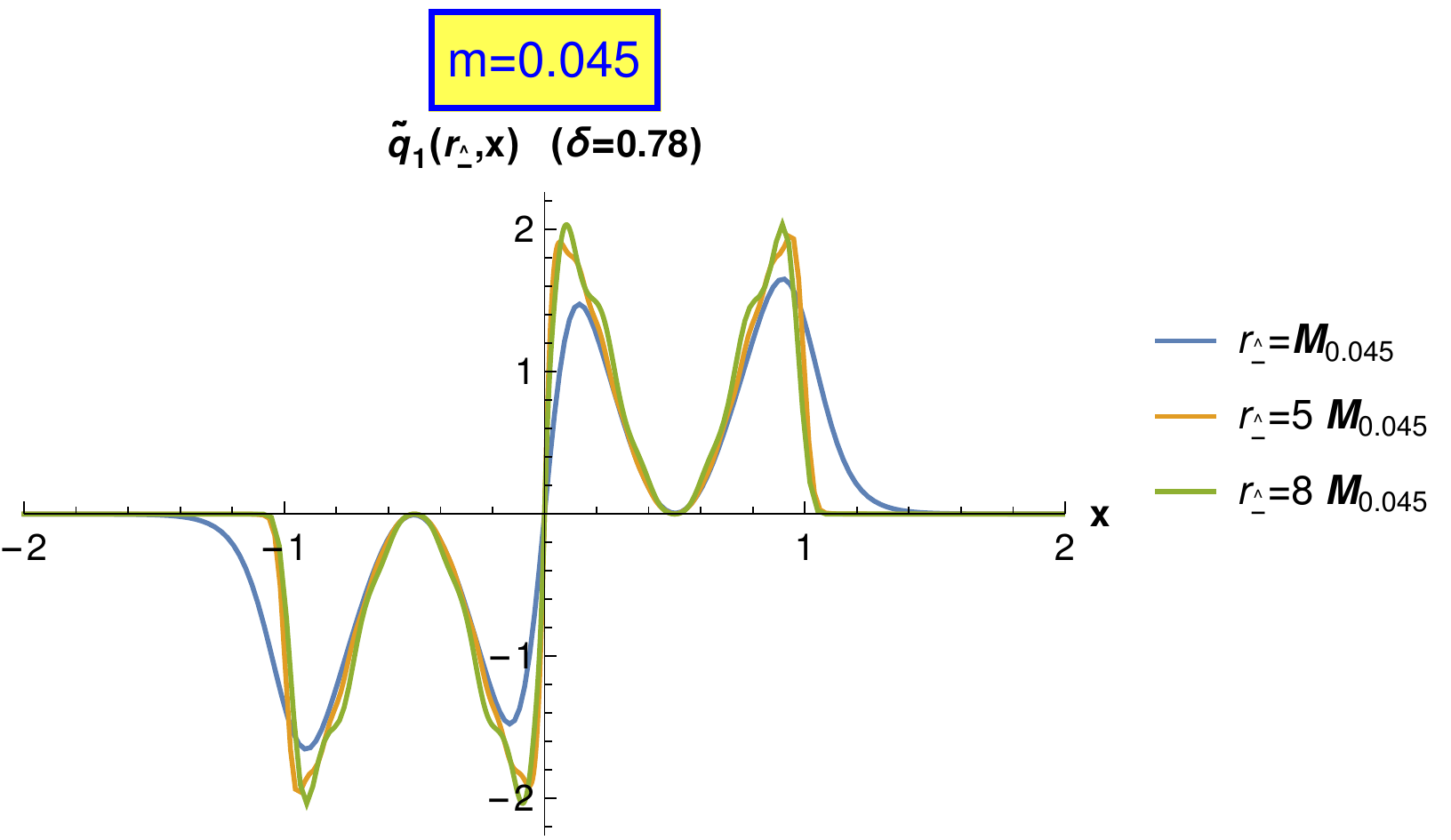}
		\label{fig:m0045_delta078_fequasipdf}
	}
	\centering
	\subfloat[]{
		\includegraphics[width=0.4\linewidth]{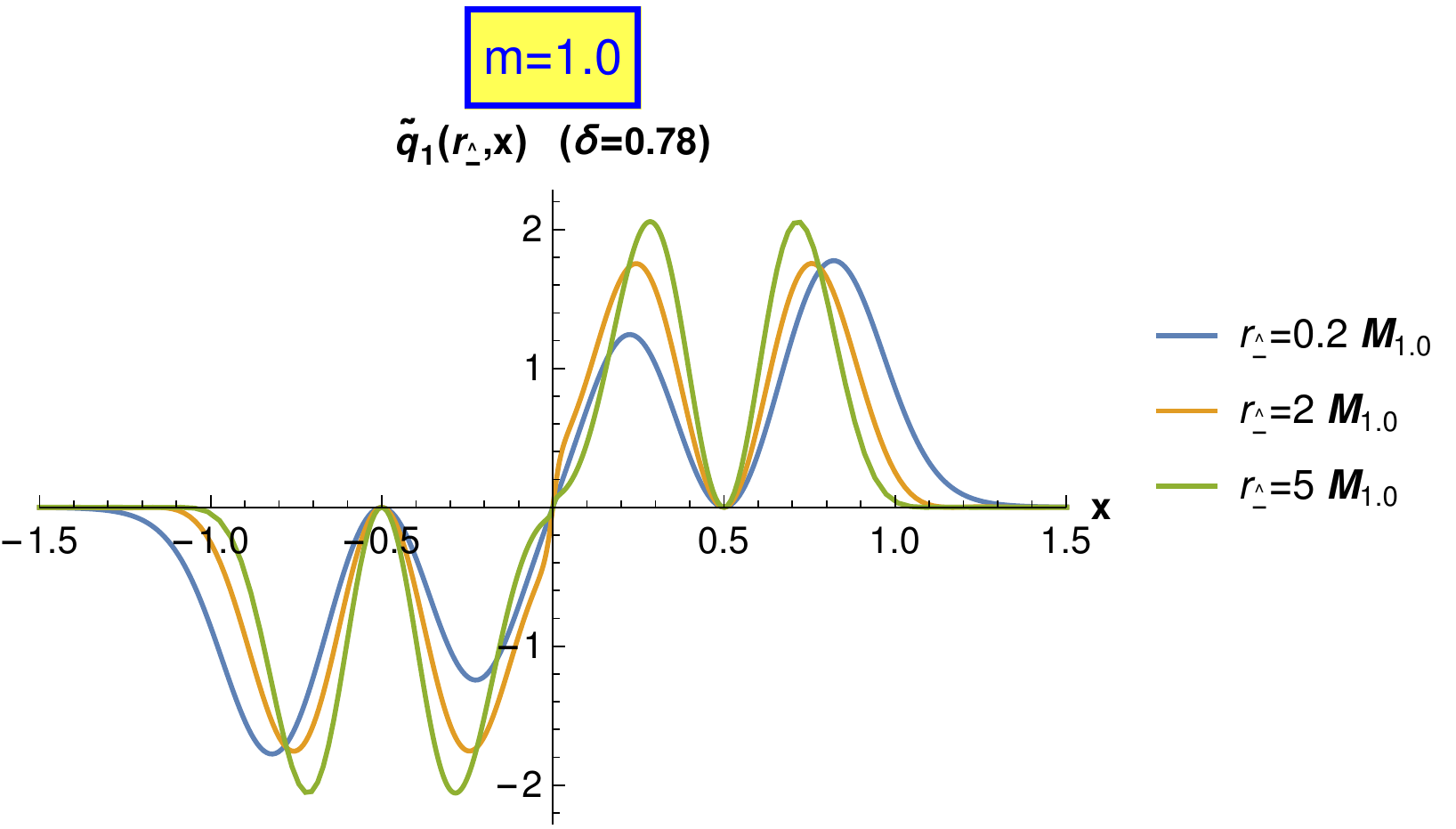}
		\label{fig:m100_delta078_fequasipdf}
	}\\
	\centering
	\subfloat[]{
		\includegraphics[width=0.4\linewidth]{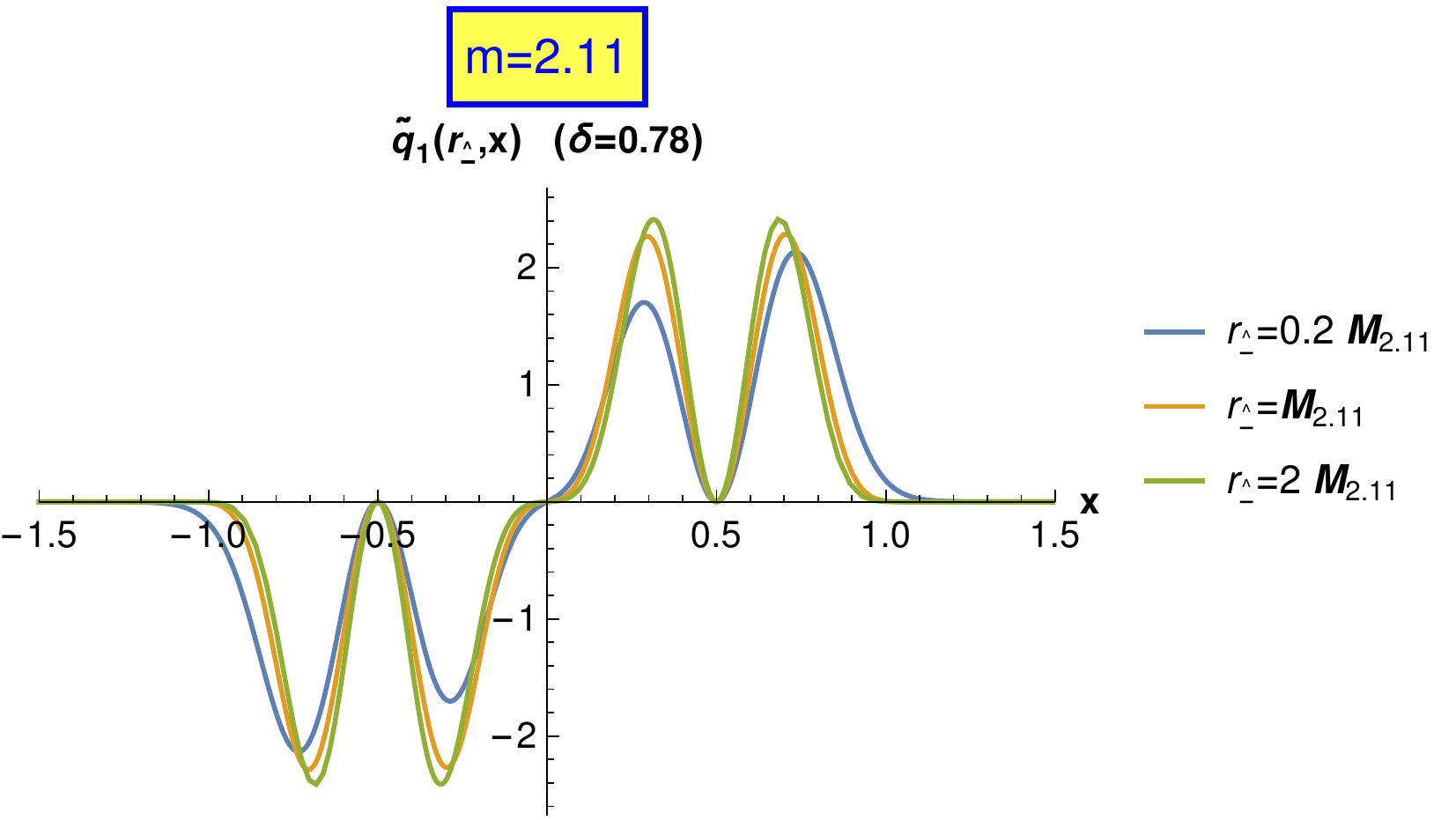}
		\label{fig:m211_delta078_fequasipdf}
	}
	\caption{$ \delta=0.78$ interpolating ``quasi-PDFs" for the first excited state ($n=1$) wave functions of (a) $m=0.045$, (b) $m=1.00$, and (c) $m=2.11$  in the unit of $ \sqrt{2\lambda} $.
\label{figapp:delta078_fequasipdf}}
\end{figure*}

As discussed in Sec.~\ref{sub:quasipdf}, the variation of the interpolating parameter $\delta$ may remedy the slow approach to the LFD results
in IFD ($\delta=0$). To exhibit this feature, we show 
the results of the ground state and first excited state mesonic quasi-PDFs for different $\delta$ values ($\delta =$ 0.6 and 0.78) in Figs.~\ref{figapp:delta06_grquasipdf} and \ref{figapp:delta06_fequasipdf} and Figs.~\ref{figapp:delta078_grquasipdf} and \ref{figapp:delta078_fequasipdf}, respectively, with the same arrangement of corresponding $m$ values ($m=$ 0.045, 1.0 and 2.11). For the $\delta=0.6$ case shown in Figs.~\ref{figapp:delta06_grquasipdf} and \ref{figapp:delta06_fequasipdf},
one may see already some improvement in the approach to the LFD results
by comparing the corresponding ${\tilde q}_0(r_{\hat{-}},x)$ and 
${\tilde q}_1(r_{\hat{-}},x)$ results in Figs.~\ref{figapp:delta0_grquasipdf} and \ref{figapp:delta0_fequasipdf} with the corresponding $r_{\hat{-}}$ values. 
For the $\delta=0.78$ case shown in Figs.~\ref{figapp:delta078_grquasipdf} and \ref{figapp:delta078_fequasipdf},
the results get improved much more dramatically yielding very quickly the essential features of the LFD results fitting in the region $x\in[0,1]$. 
The similar features have been noted earlier in Secs.~\ref{sub:wavefunc} and ~\ref{sub:quasipdf} as well as in Appendix~\ref{app:figures}, i.e. taking $\delta$ away from the IFD ($\delta=0$), the resemblance to the PDFs in LFD may appear more swiftly achieved with the boost of the meson longitudinal momentum $r_{\hat{-}}$ to the larger value.
Nevertheless, one should note here a numerical caveat demanding much higher numerical 
accuracy as the meson momentum $r_{\hat{-}}$ gets larger while the $\delta$ value gets close to $\pi/4$ (e.g. $\delta=0.78$). In such situation, the numerical sensitivity kicks in so strongly that the results cannot
be trusted unless they get tested for the improvement with much higher numerical accuracy. However, as discussed in Sec.~\ref{sub:quasipdf}, one doesn't 
need to boost the longitudinal momentum $r_{\hat{-}}$ too large if 
the $\delta$ value gets close to $\pi/4$. As the $\delta$ value gets close to $\pi/4$, relatively smaller $r_{\hat{-}}$ value can do the job. Thus, it would be  
worthwhile to search for a ``sweet spot" by varying both $\delta$ and $r_{\hat{-}}$ together to obtain the ``LFD-like" result. 

\section{\label{app:rest}Rest frame bound-state equation and its solution}
\begin{figure}
	\centering
	\subfloat[]{
		\includegraphics[width=1.0\columnwidth]{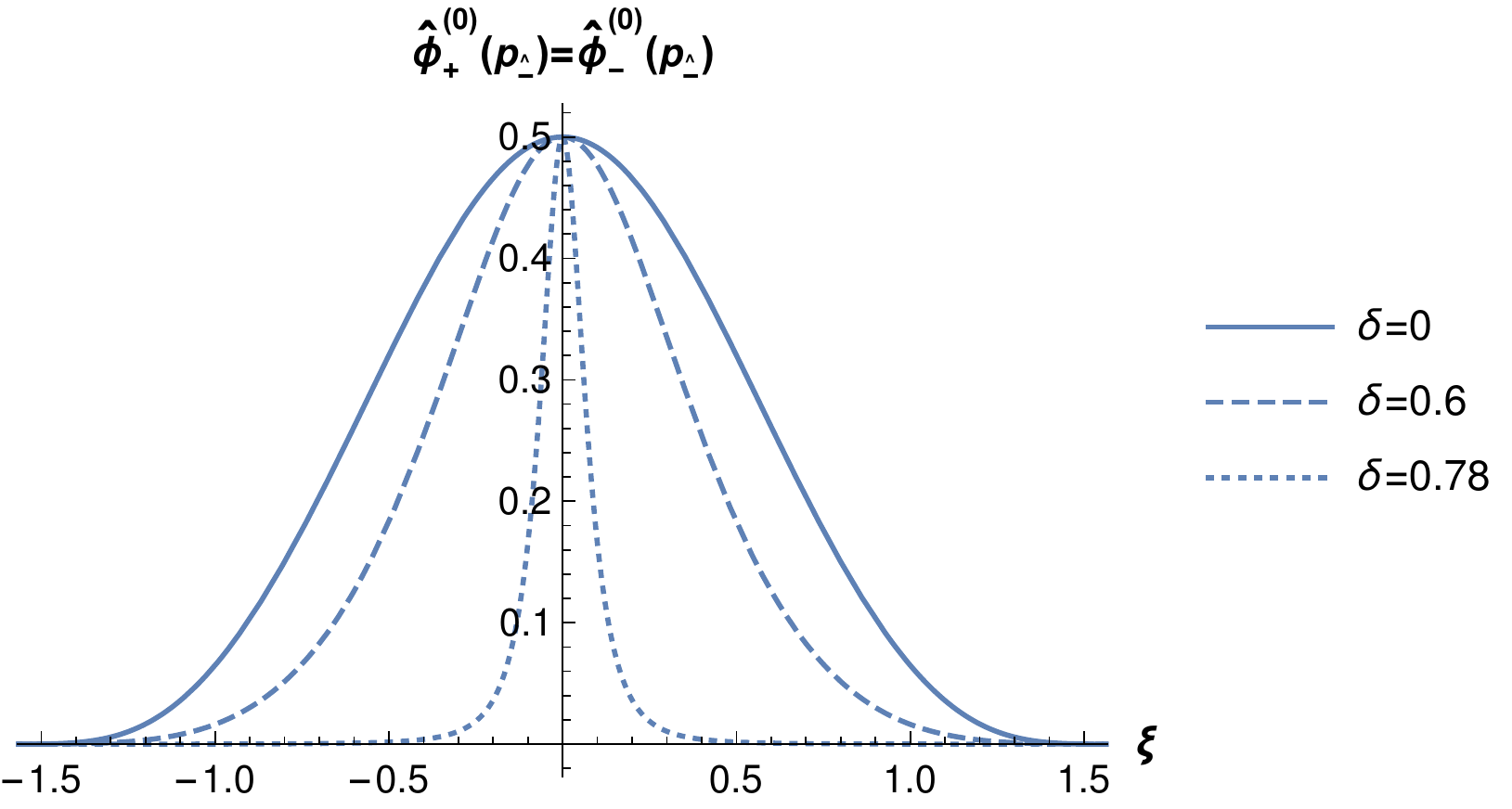}
	}\\
	\caption{Analytic solutions for the ground state wave functions in the rest frame for the bare quark mass $ m=0 $ for three different interpolation angles as functions of $\xi=\tan^{-1} p_{\hat{-}}$. All quantities are in proper units of $ \sqrt{2\lambda} $.\label{fig:m0rfn0ana}}
\end{figure}
\begin{figure*}
	\centering
	\subfloat[]{
		\includegraphics[width=0.5\linewidth]{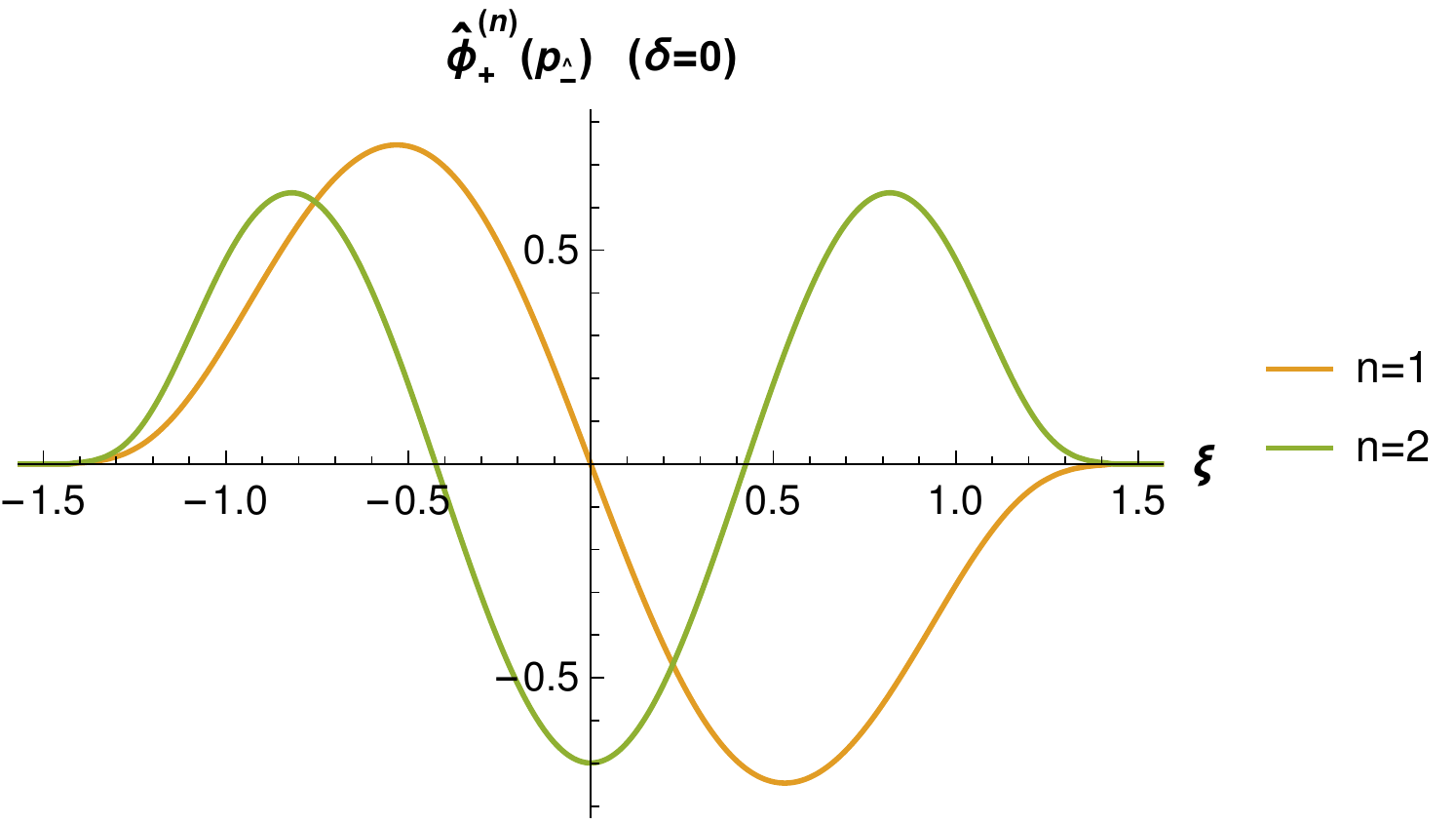}
		\label{fig:m0rfzd0}
	}
	\centering
	\subfloat[]{
		\includegraphics[width=0.5\linewidth]{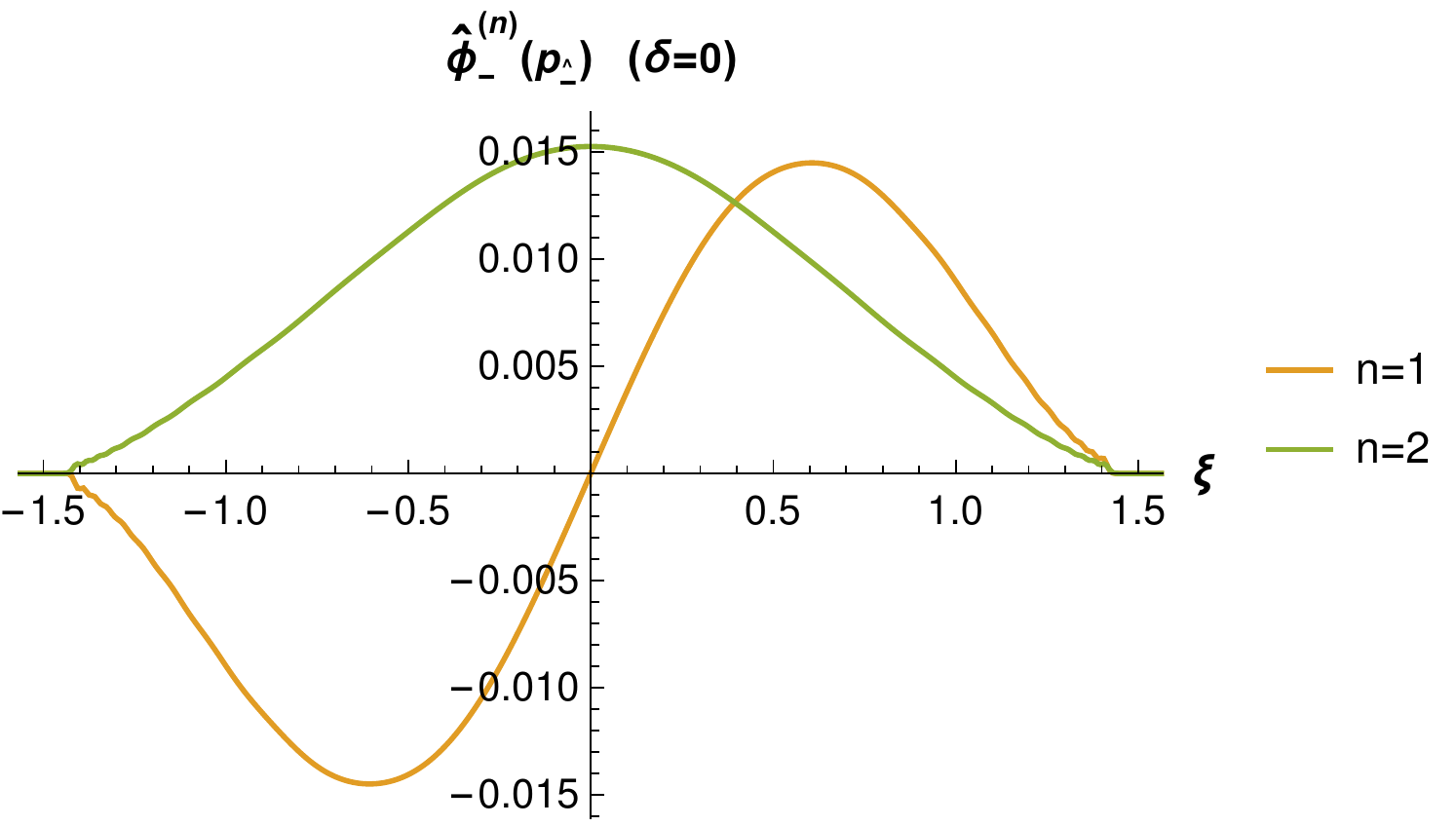}
		\label{fig:m0rffd0}
	}\\
	\centering
	\subfloat[]{		
		\includegraphics[width=0.5\linewidth]{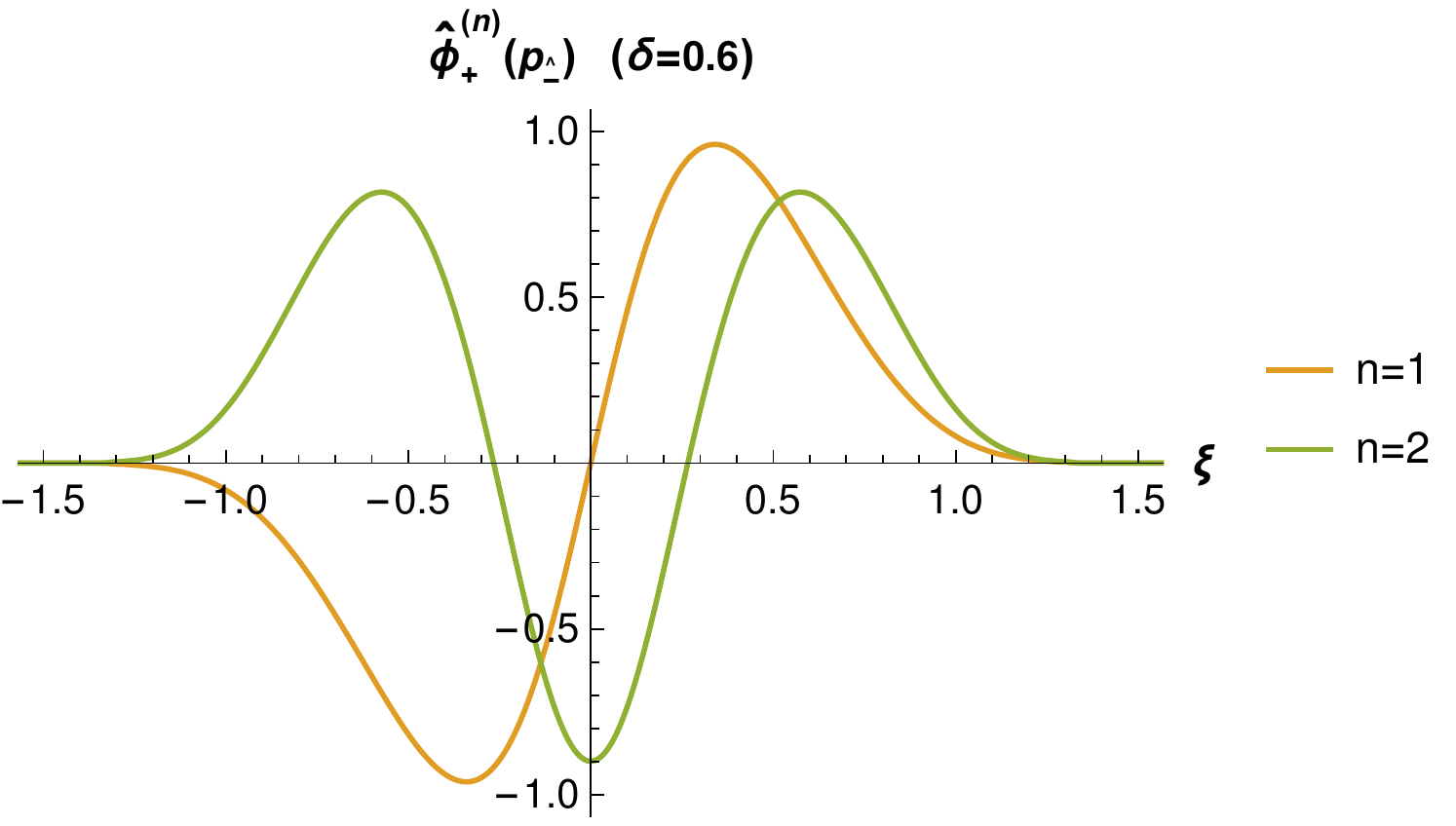}
		\label{fig:m0rfzd06}
	}
	\centering
	\subfloat[]{
		\includegraphics[width=0.5\linewidth]{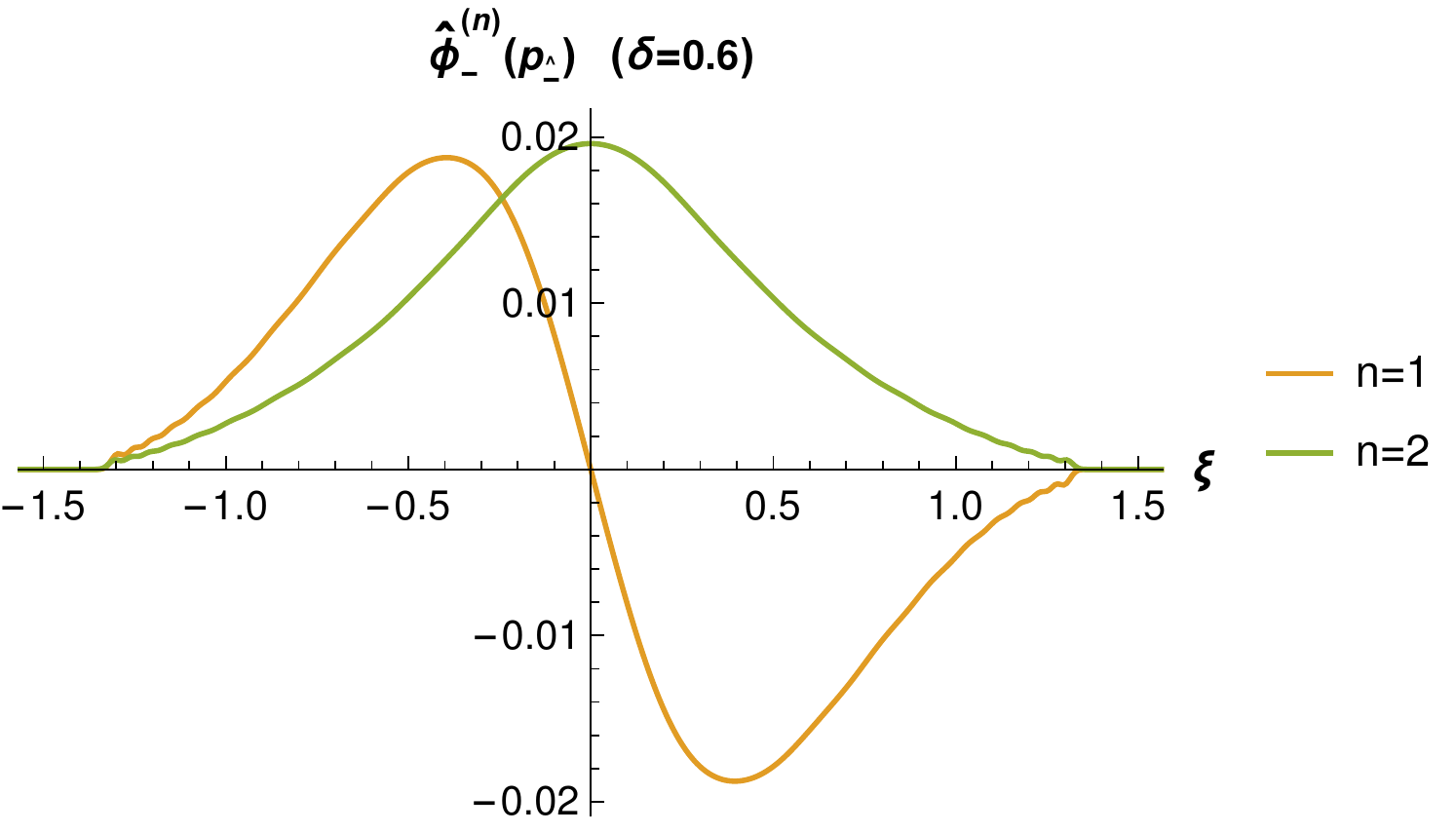}
		\label{fig:m0rffd06}
	}\\
	\centering
	\subfloat[]{
		\includegraphics[width=0.5\linewidth]{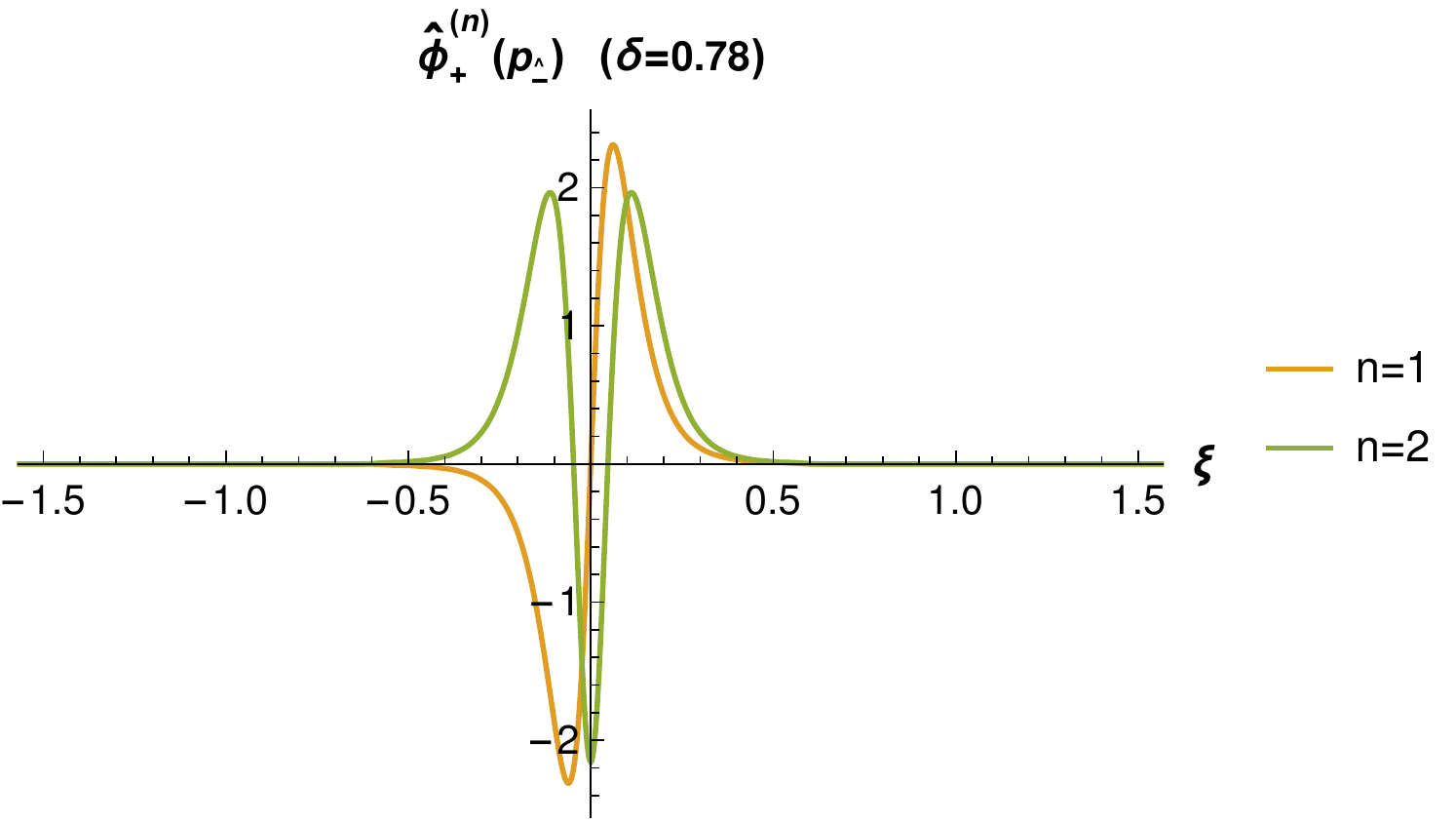}
		\label{fig:m0rfzd078}
	}
	\centering
	\subfloat[]{
		\includegraphics[width=0.5\linewidth]{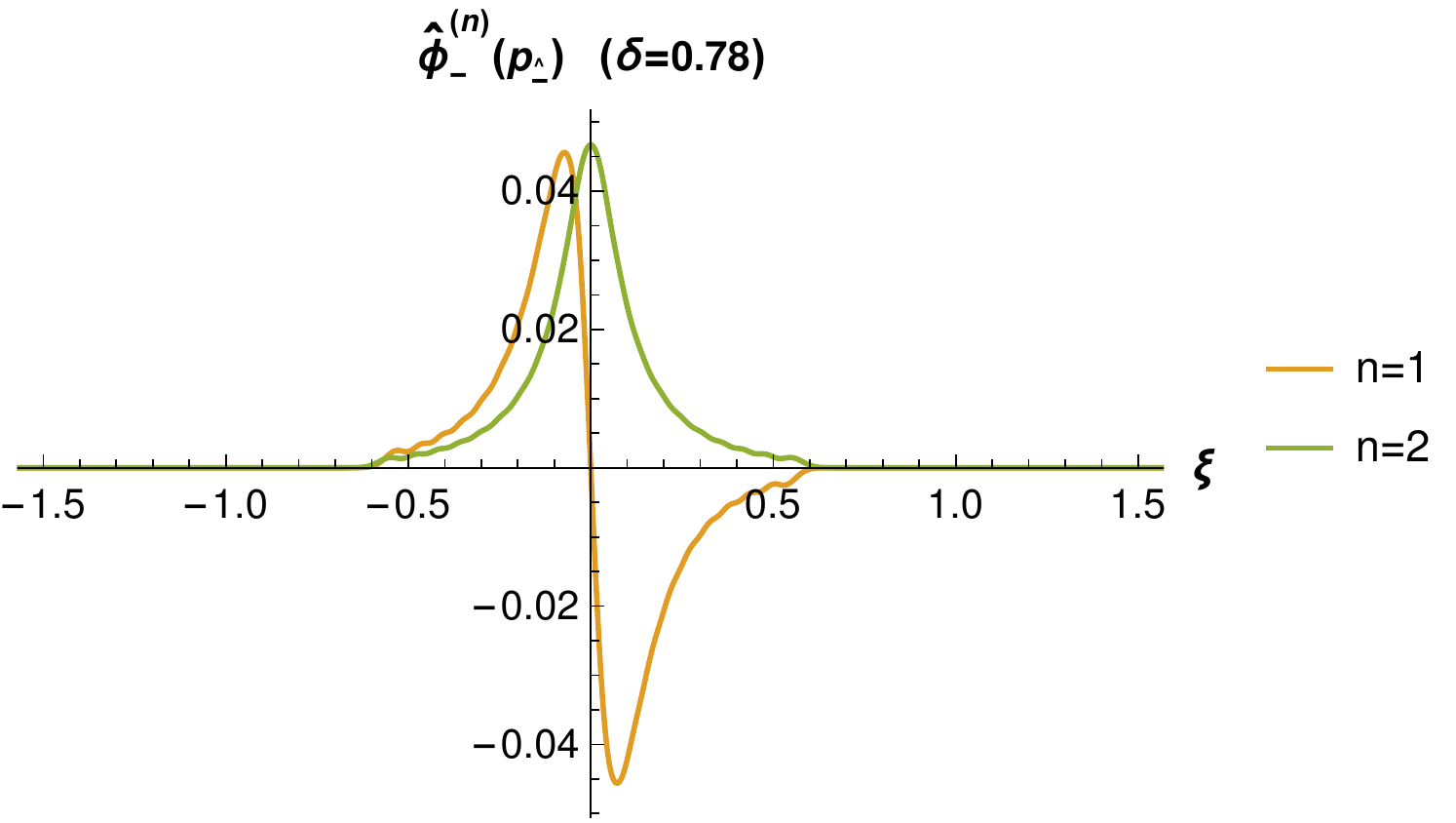}
		\label{fig:m0rffd078}
	}\\
	\caption{Rest frame wave functions $\hat\phi_+^{(n)}(p_{\hat{-}})$ and $\hat\phi_-^{(n)}(p_{\hat{-}})$ for $ m=0 $ ($n=1$ and $n=2$) as functions of $\xi=\tan^{-1} p_{\hat{-}}$. All quantities are in proper units of $ \sqrt{2\lambda} $.\label{fig:m0rfz}}
\end{figure*}

\begin{figure*}
	\centering
	\subfloat[]{
		\includegraphics[width=0.5\linewidth]{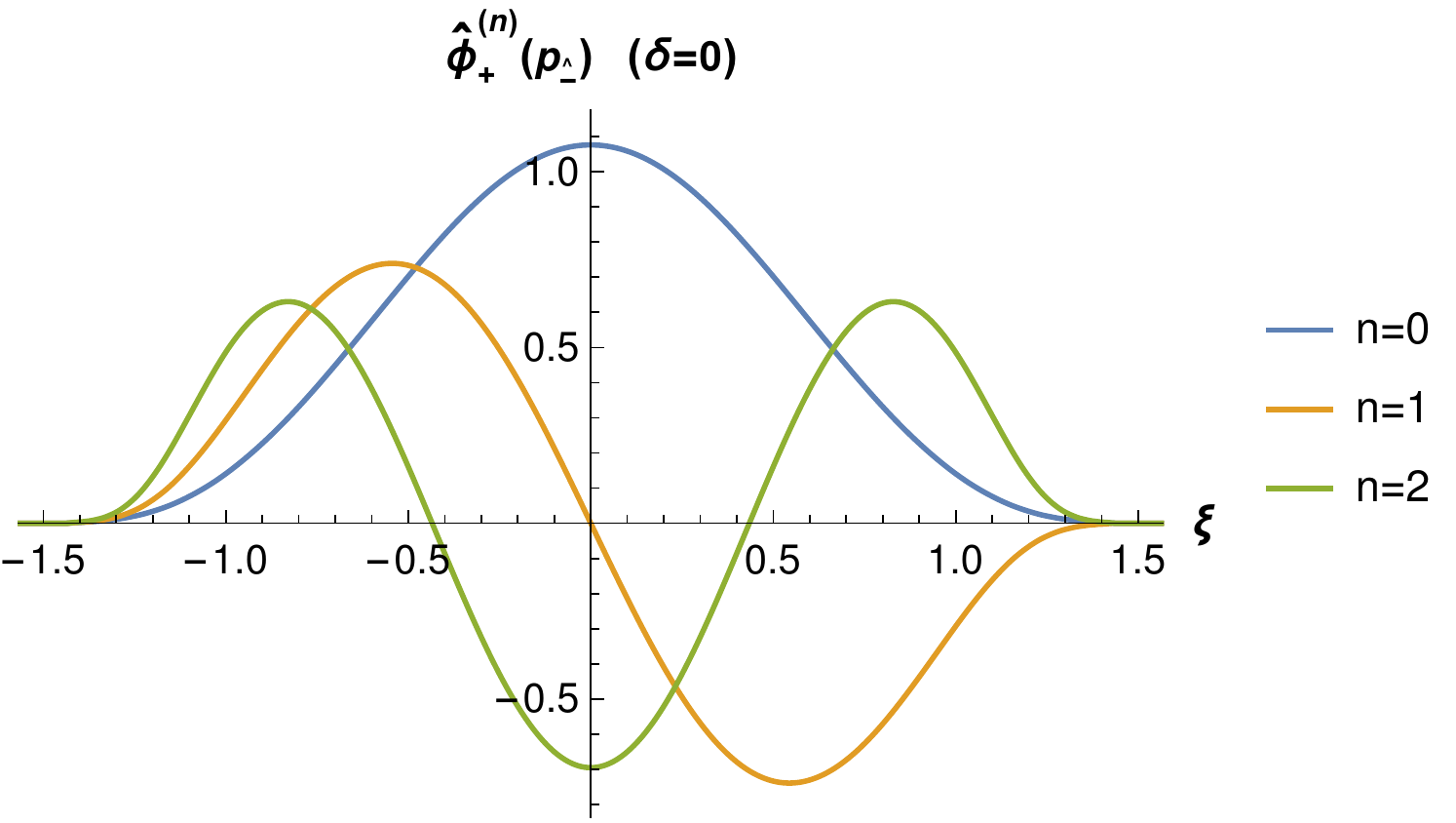}
		\label{fig:m0045rfzd0}
	}
	\centering
	\subfloat[]{
		\includegraphics[width=0.5\linewidth]{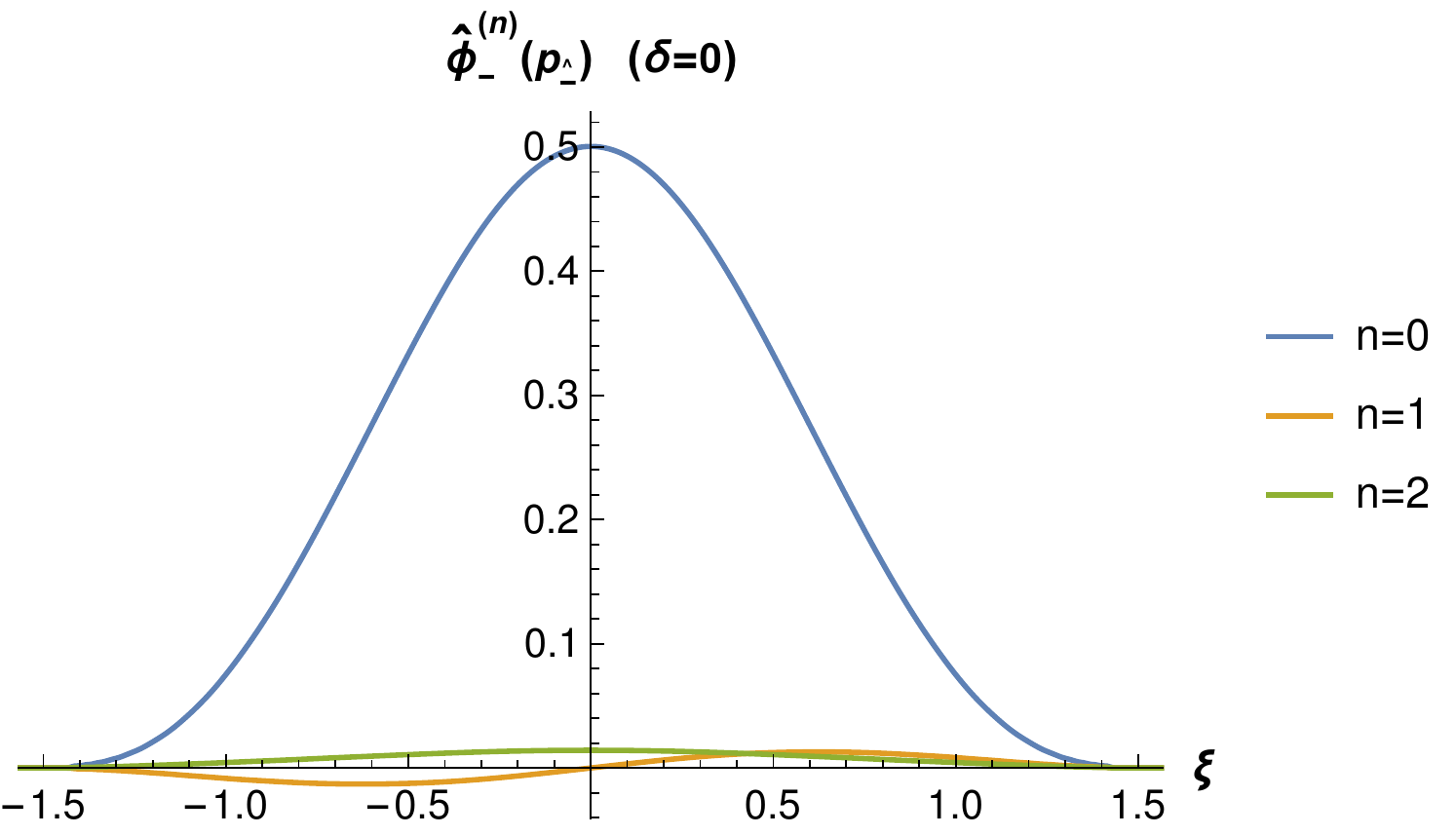}
		\label{fig:m0045rffd0}
	}\\
	\centering
	\subfloat[]{
		\includegraphics[width=0.5\linewidth]{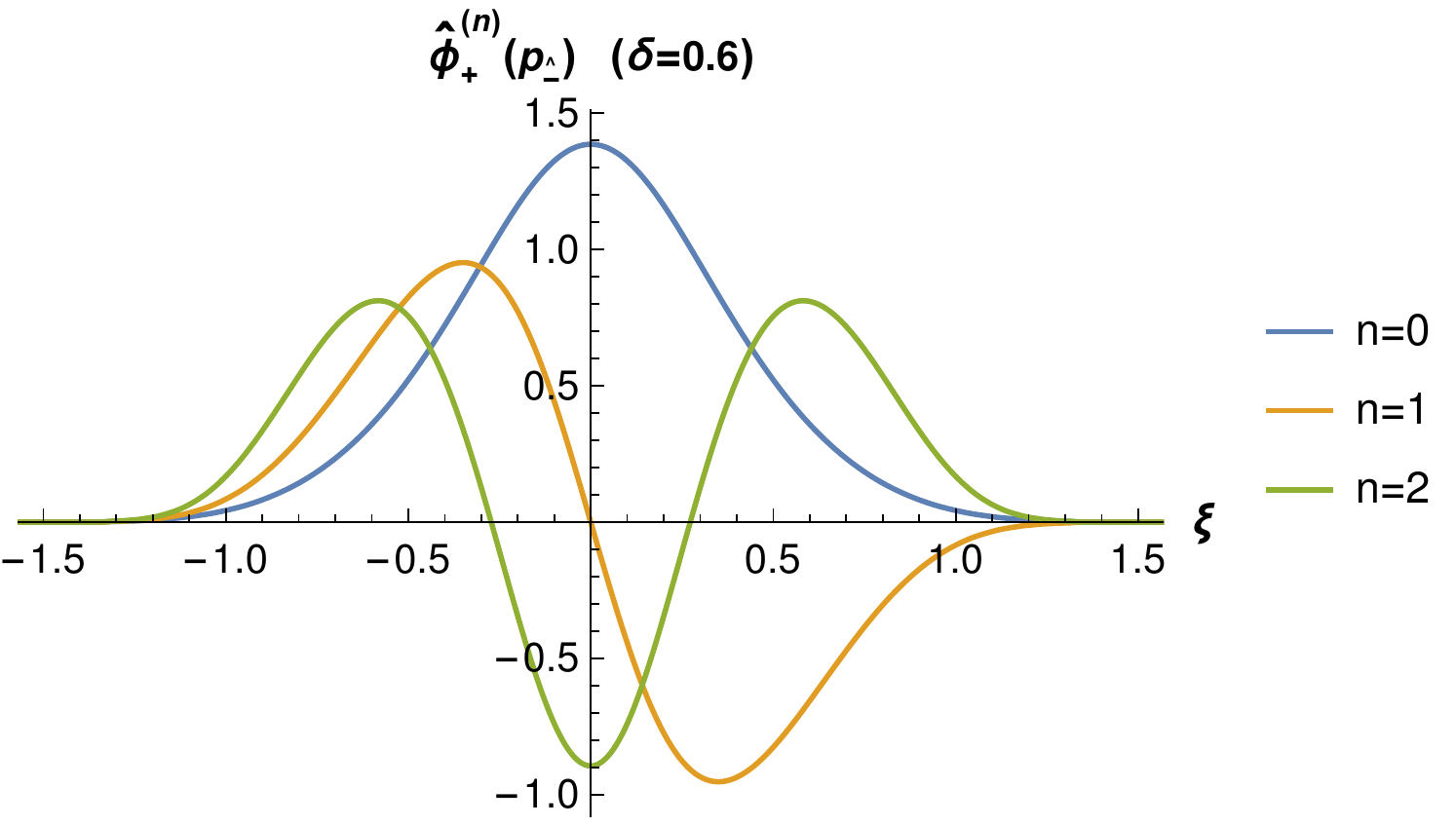}
		\label{fig:m0045rfzd06}
	}
	\centering
	\subfloat[]{
		\includegraphics[width=0.5\linewidth]{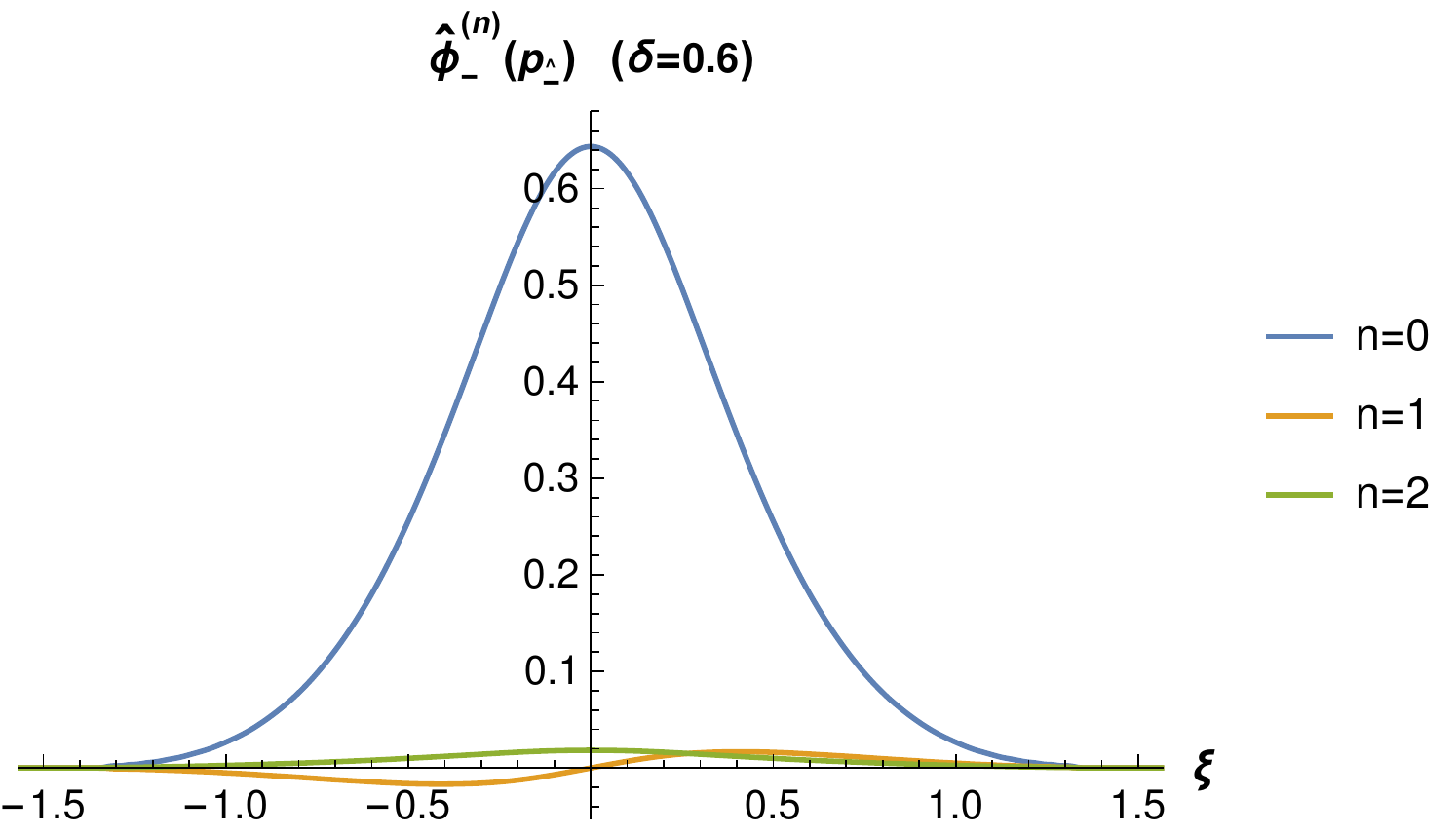}
		\label{fig:m0045rffd06}
	}\\
	\centering
	\subfloat[]{
		\includegraphics[width=0.5\linewidth]{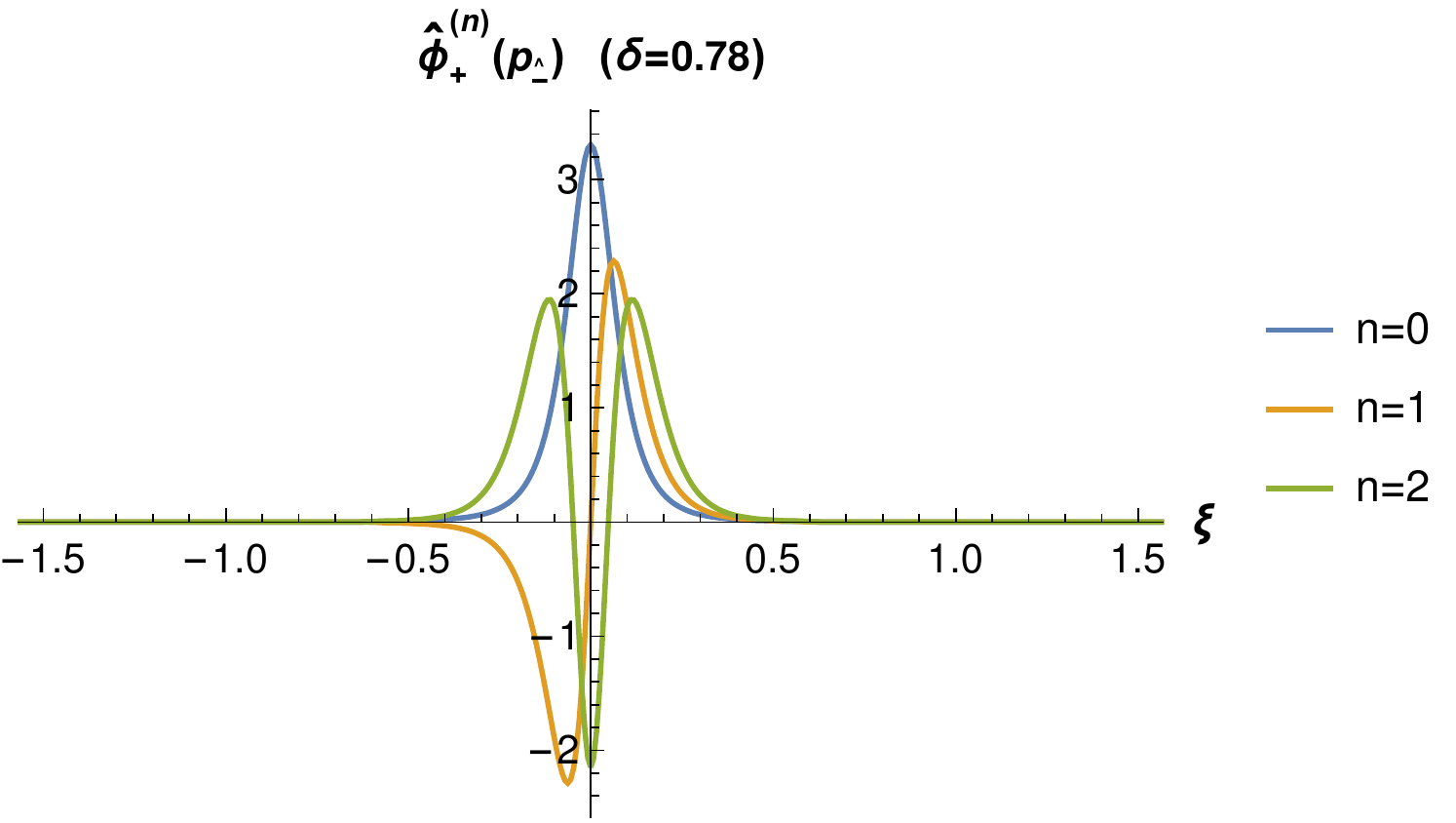}
		\label{fig:m0045rfzd078}
	}
	\centering
	\subfloat[]{
		\includegraphics[width=0.5\linewidth]{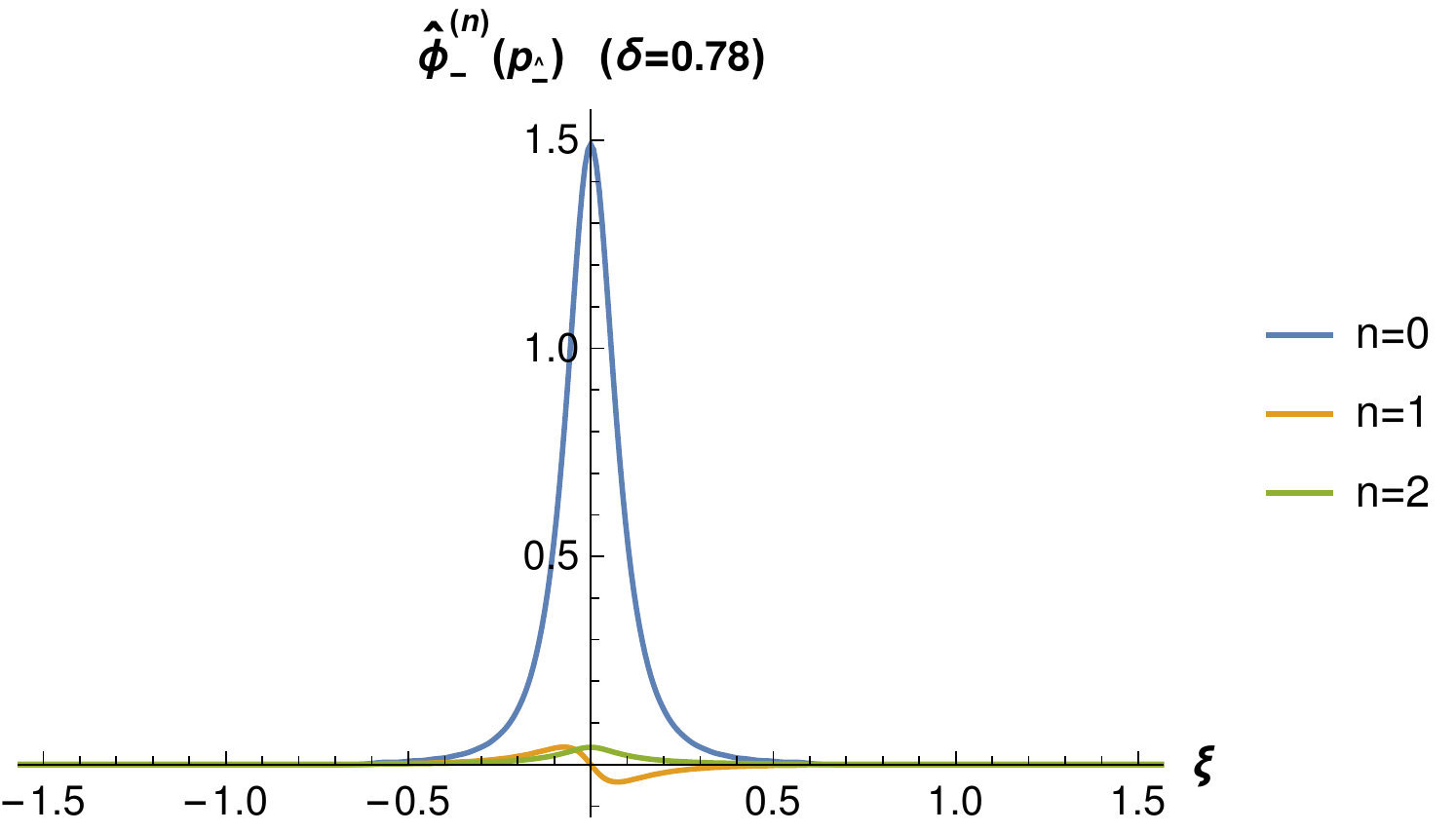}
		\label{fig:m0045rffd078}
	}\\
	\caption{Rest frame wave functions $\hat\phi_+^{(n)}(p_{\hat{-}})$ and $\hat\phi_-^{(n)}(p_{\hat{-}})$ for $ m=0.045 $
as functions of $\xi=\tan^{-1} p_{\hat{-}}$. All quantities are in proper units of $ \sqrt{2\lambda} $.\label{fig:m0045rfz}}
\end{figure*}

\begin{figure*}
	\centering
	\subfloat[]{
		\includegraphics[width=0.5\linewidth]{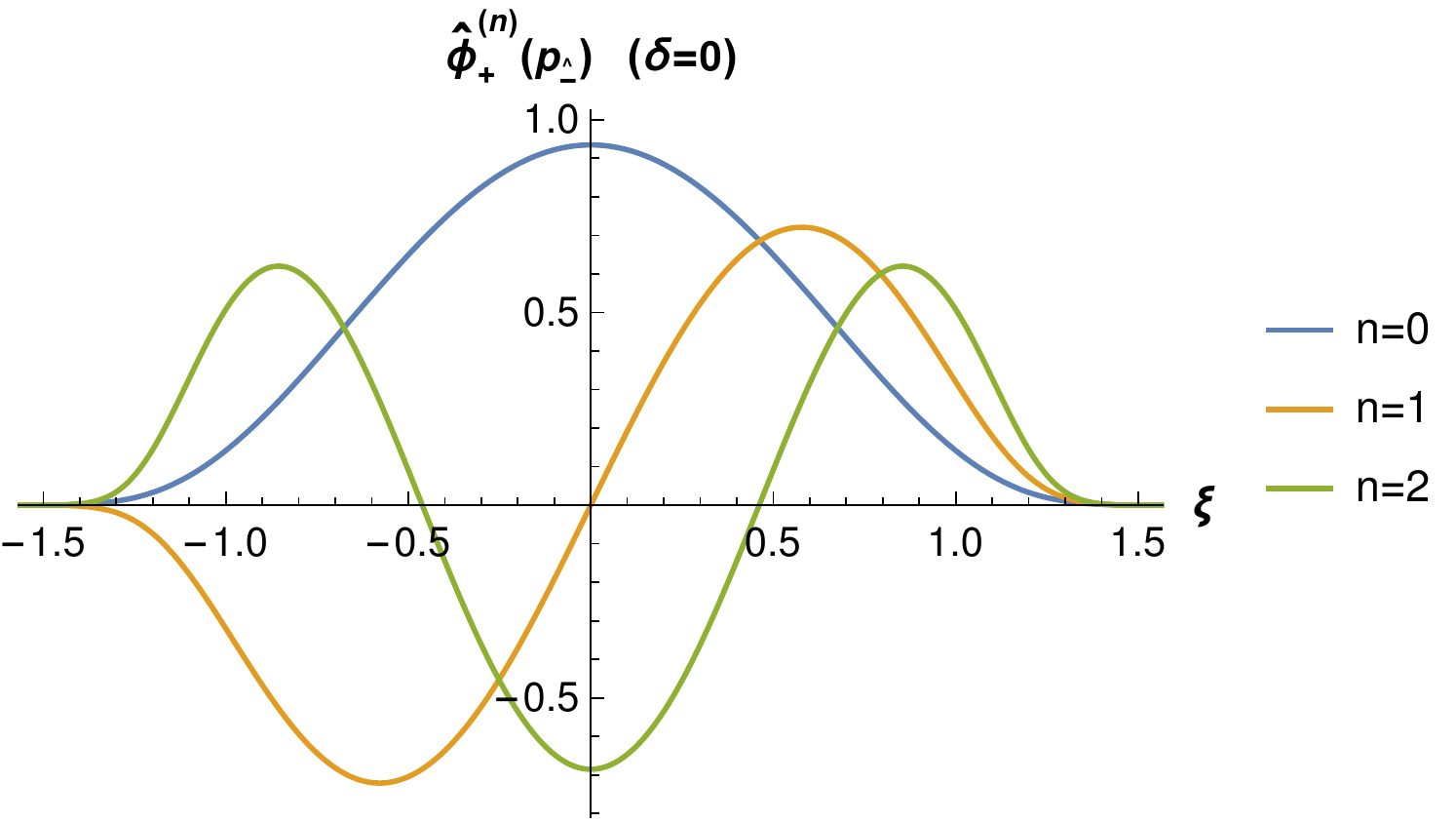}
		\label{fig:m018rfzd0}
	}
	\centering
	\subfloat[]{
		\includegraphics[width=0.5\linewidth]{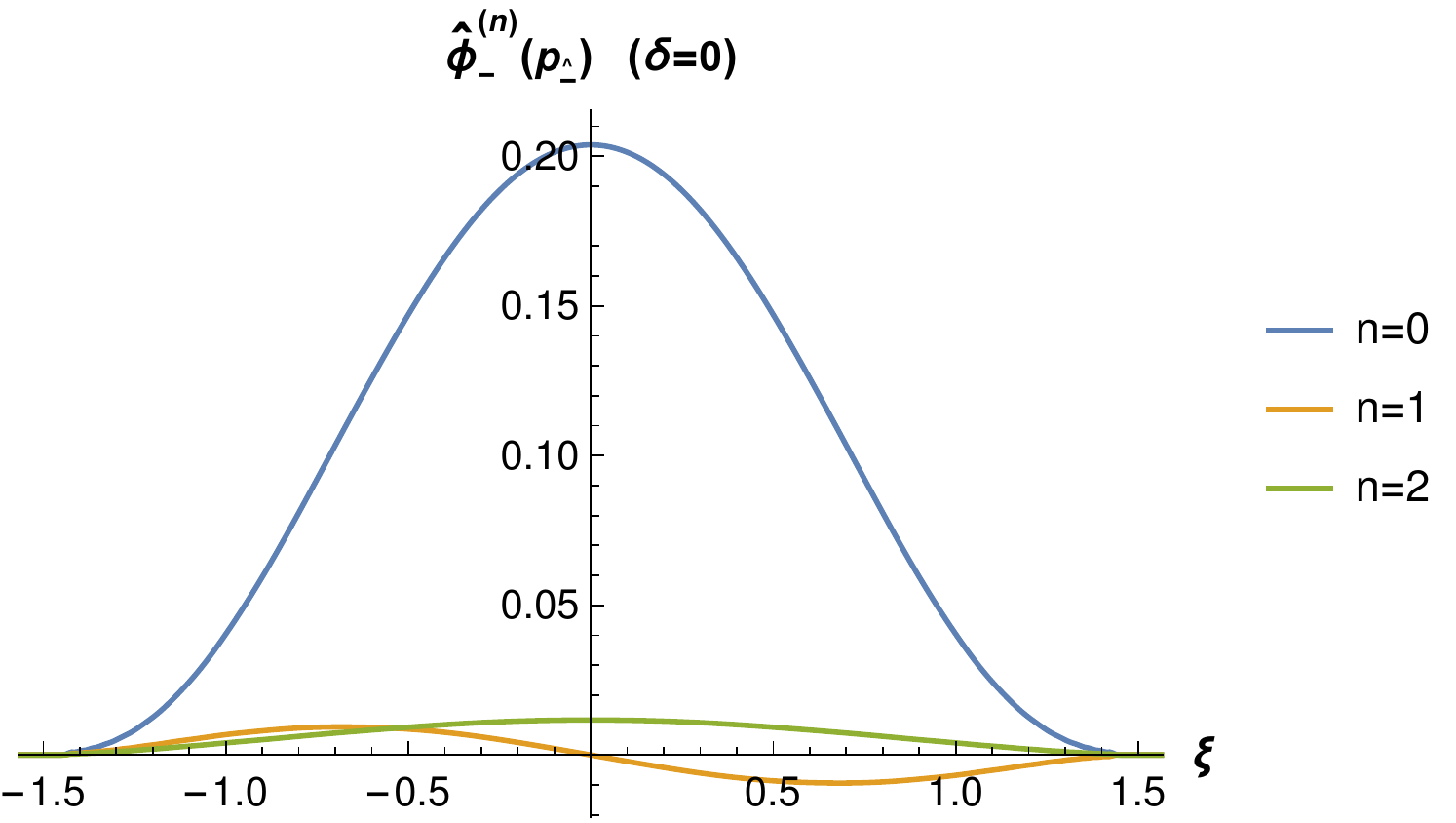}
		\label{fig:m018rffd0}
	}\\
	\centering
	\subfloat[]{
		\includegraphics[width=0.5\linewidth]{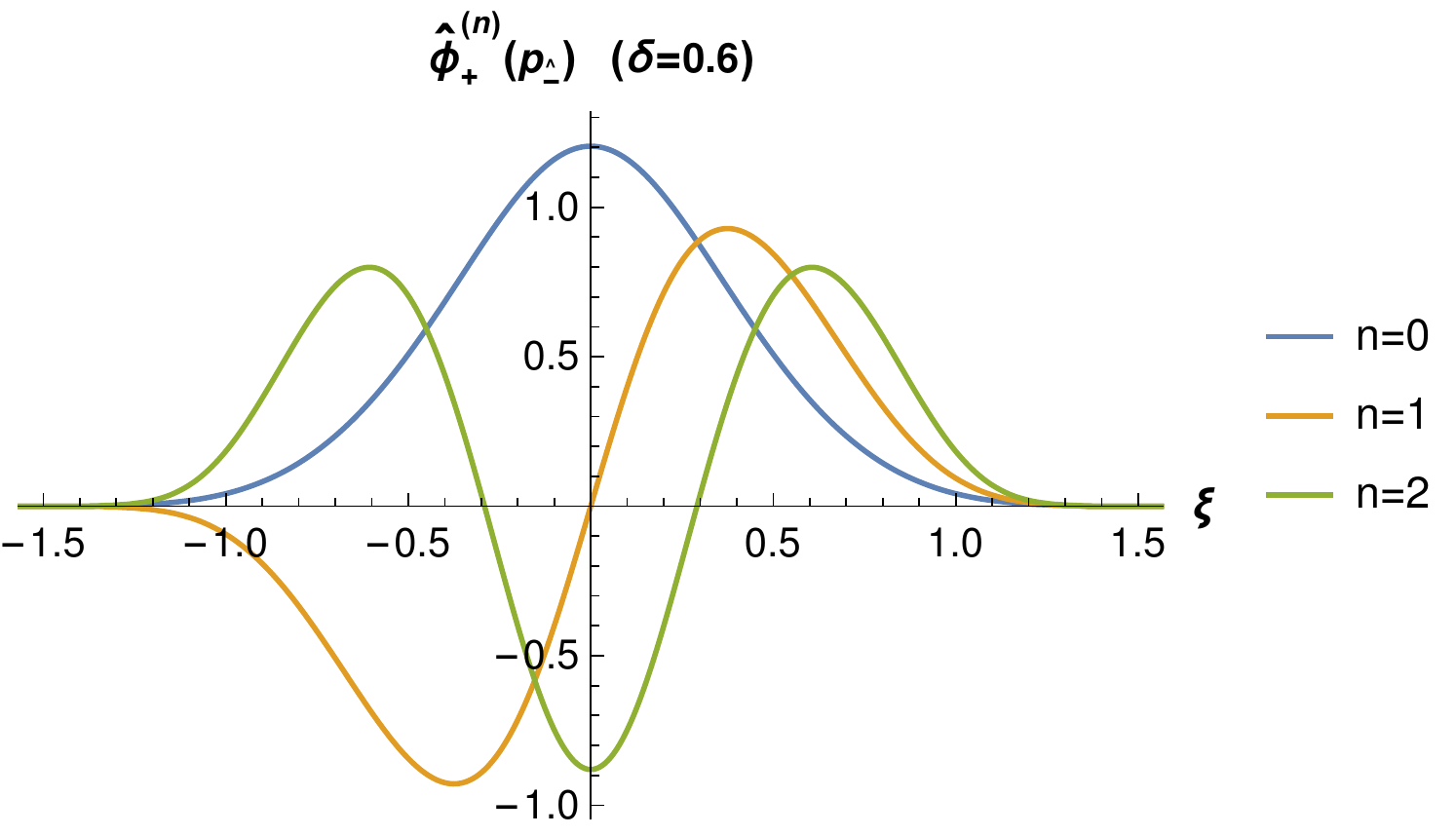}
		\label{fig:m018rfzd06}
	}
	\centering
	\subfloat[]{
		\includegraphics[width=0.5\linewidth]{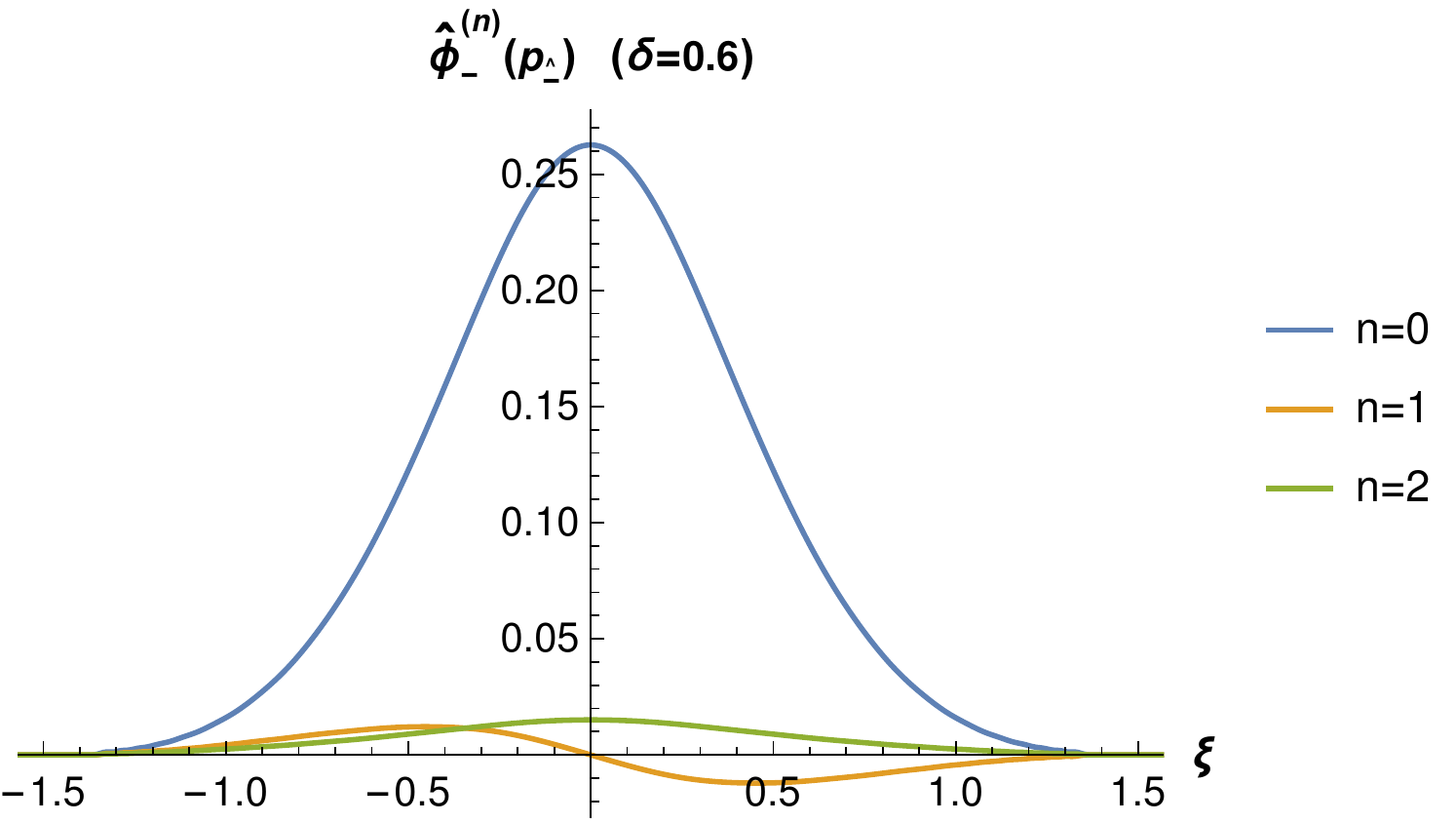}
		\label{fig:m018rffd06}
	}\\
	\centering
	\subfloat[]{
		\includegraphics[width=0.5\linewidth]{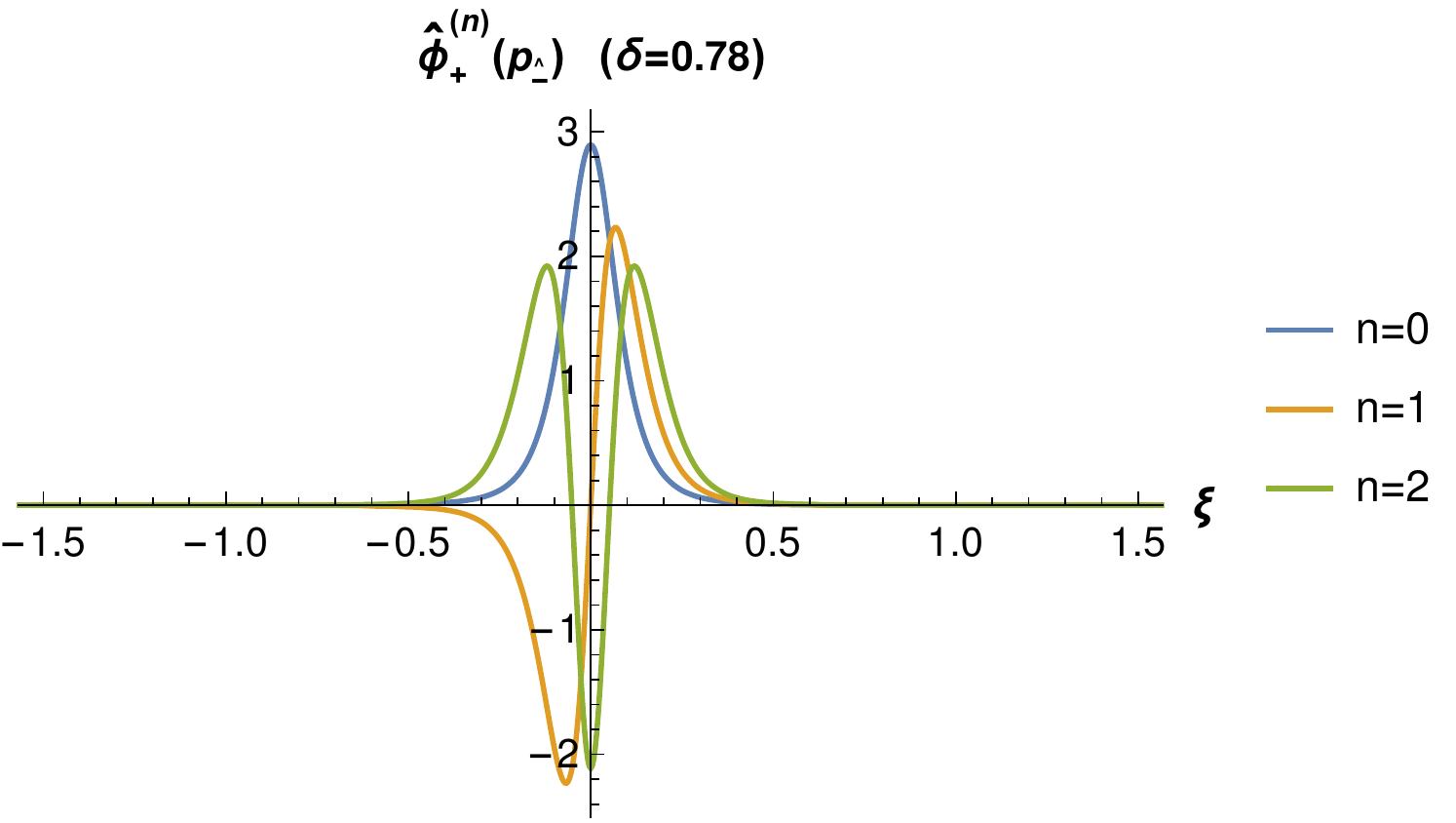}
		\label{fig:m018rfzd078}
	}
	\centering
	\subfloat[]{
		\includegraphics[width=0.5\linewidth]{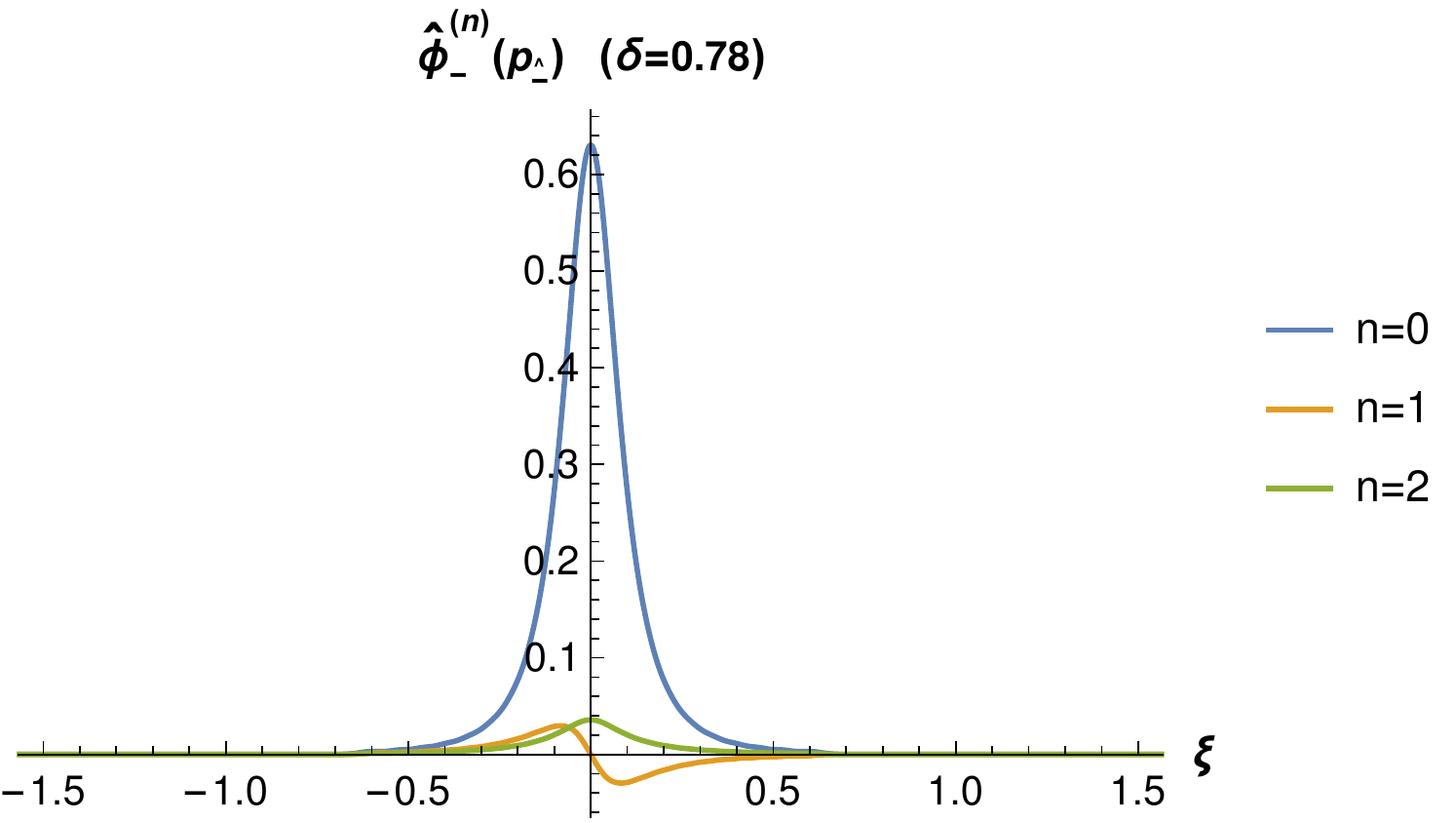}
		\label{fig:m018rffd078}
	}\\
	\caption{Rest frame wave functions $\hat\phi_+^{(n)}(p_{\hat{-}})$ and $\hat\phi_-^{(n)}(p_{\hat{-}})$ for $ m=0.18 $
as functions of $\xi=\tan^{-1} p_{\hat{-}}$. All quantities are in proper units of $ \sqrt{2\lambda} $.\label{fig:m018rfz}}
\end{figure*}

\begin{figure*}
	\centering
	\subfloat[]{
		\includegraphics[width=0.5\linewidth]{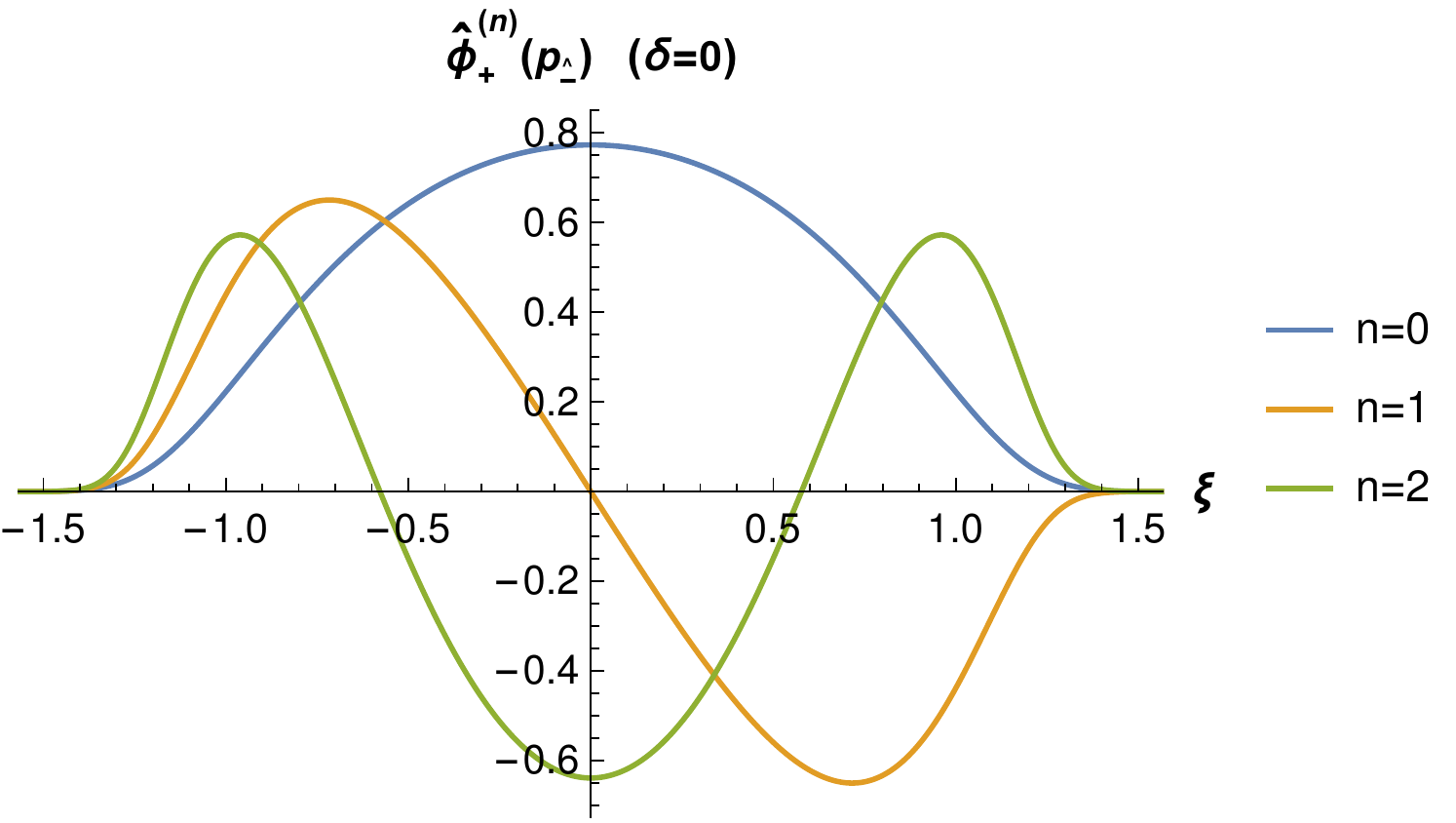}
		\label{fig:m100rfzd0}
	}
	\centering
	\subfloat[]{
		\includegraphics[width=0.5\linewidth]{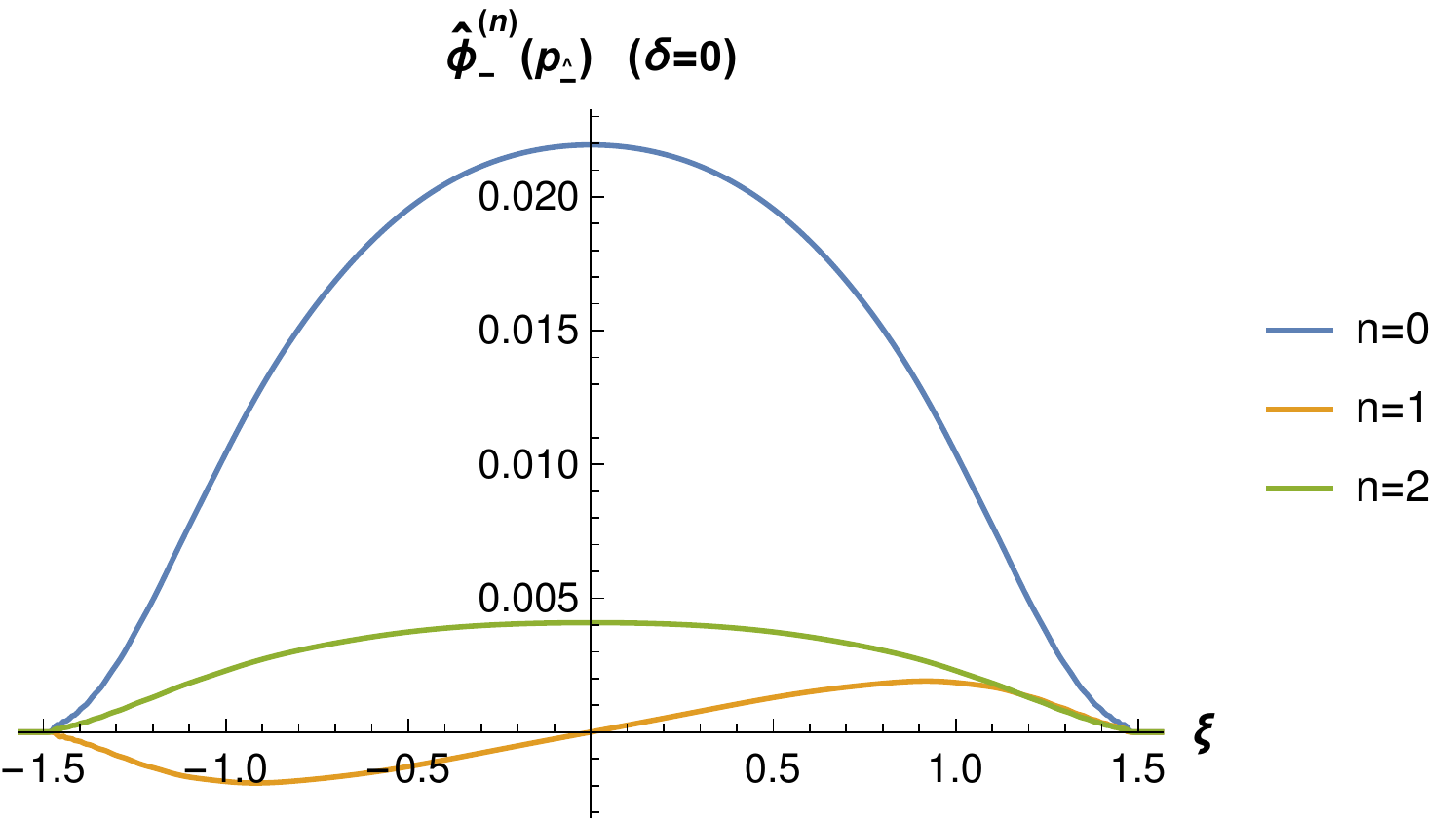}
		\label{fig:m100rffd0}
	}\\
	\centering
	\subfloat[]{		
		\includegraphics[width=0.5\linewidth]{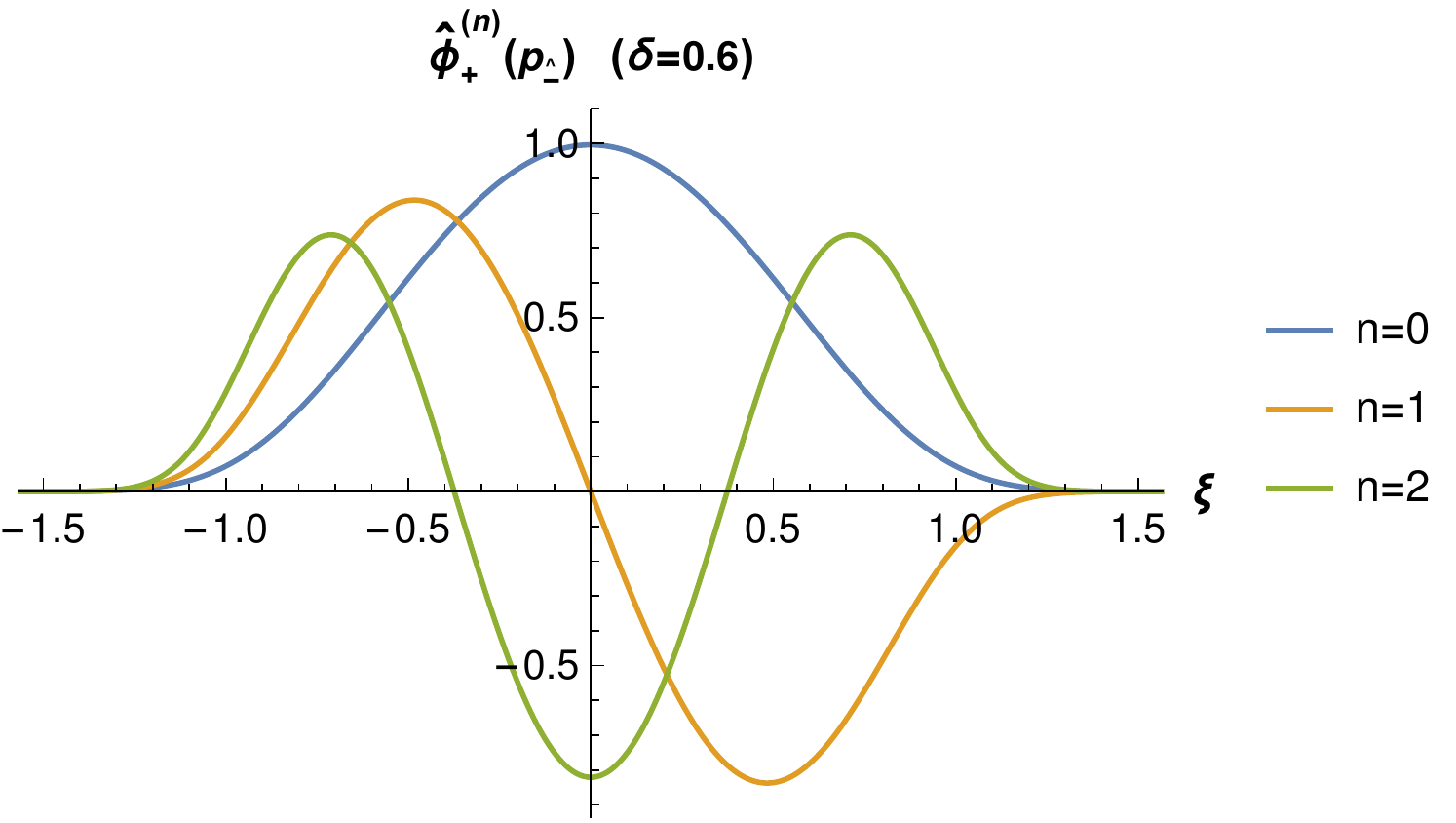}
		\label{fig:m100rfzd06}
	}
	\centering
	\subfloat[]{
		\includegraphics[width=0.5\linewidth]{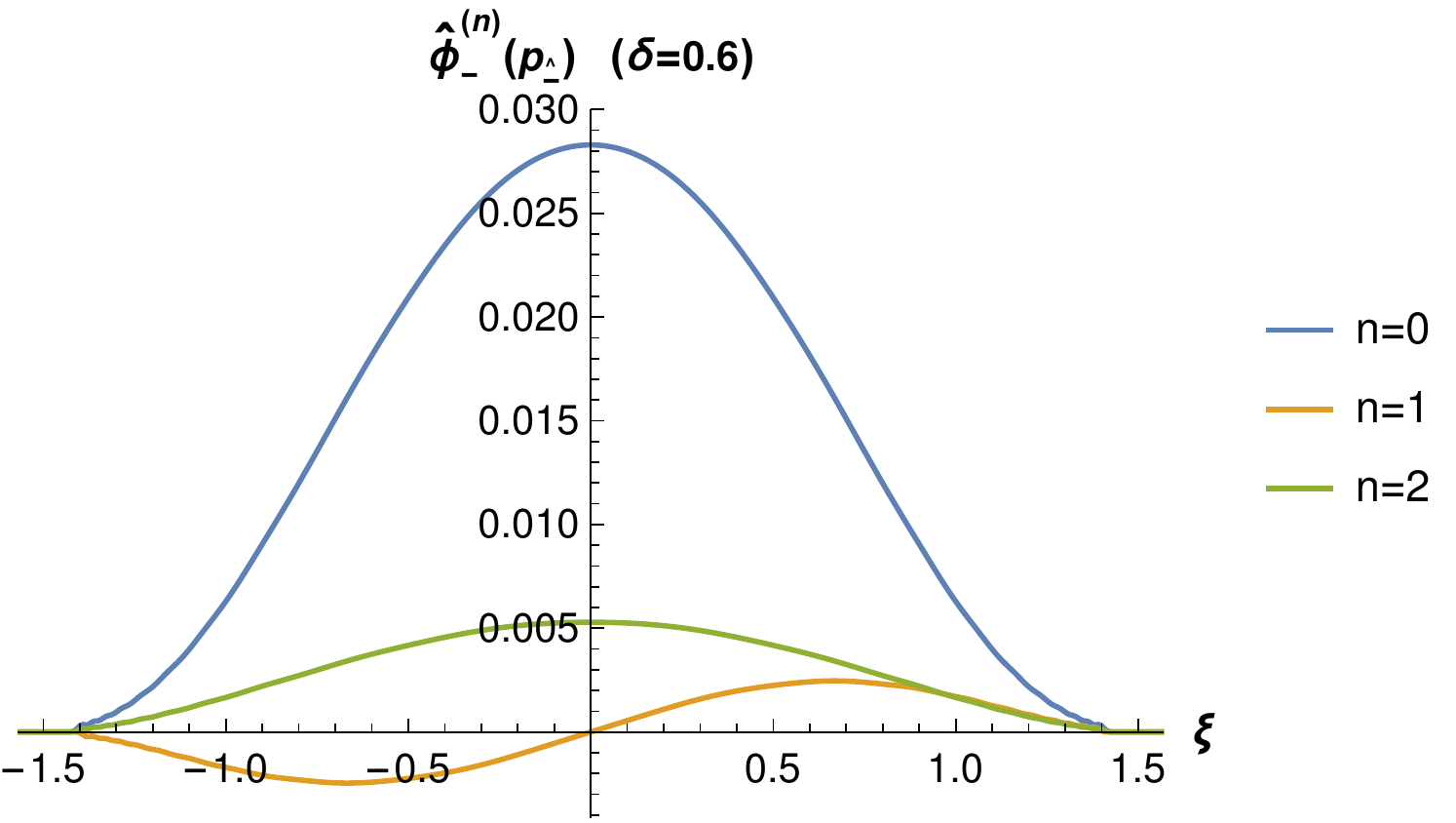}
		\label{fig:m100rffd06}
	}\\
	\centering
	\subfloat[]{
		\includegraphics[width=0.5\linewidth]{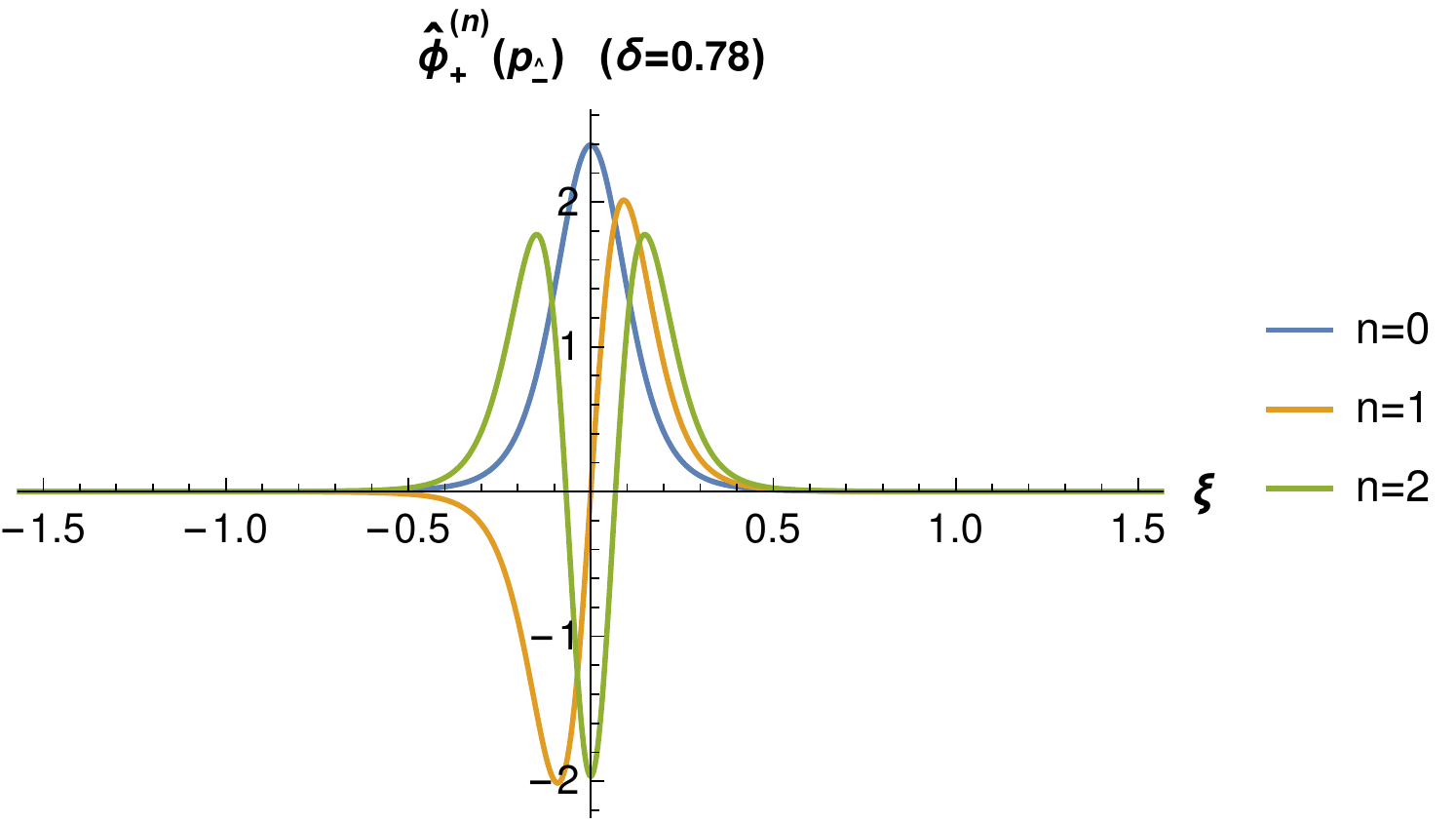}
		\label{fig:m100rfzd078}
	}
	\centering
	\subfloat[]{
		\includegraphics[width=0.5\linewidth]{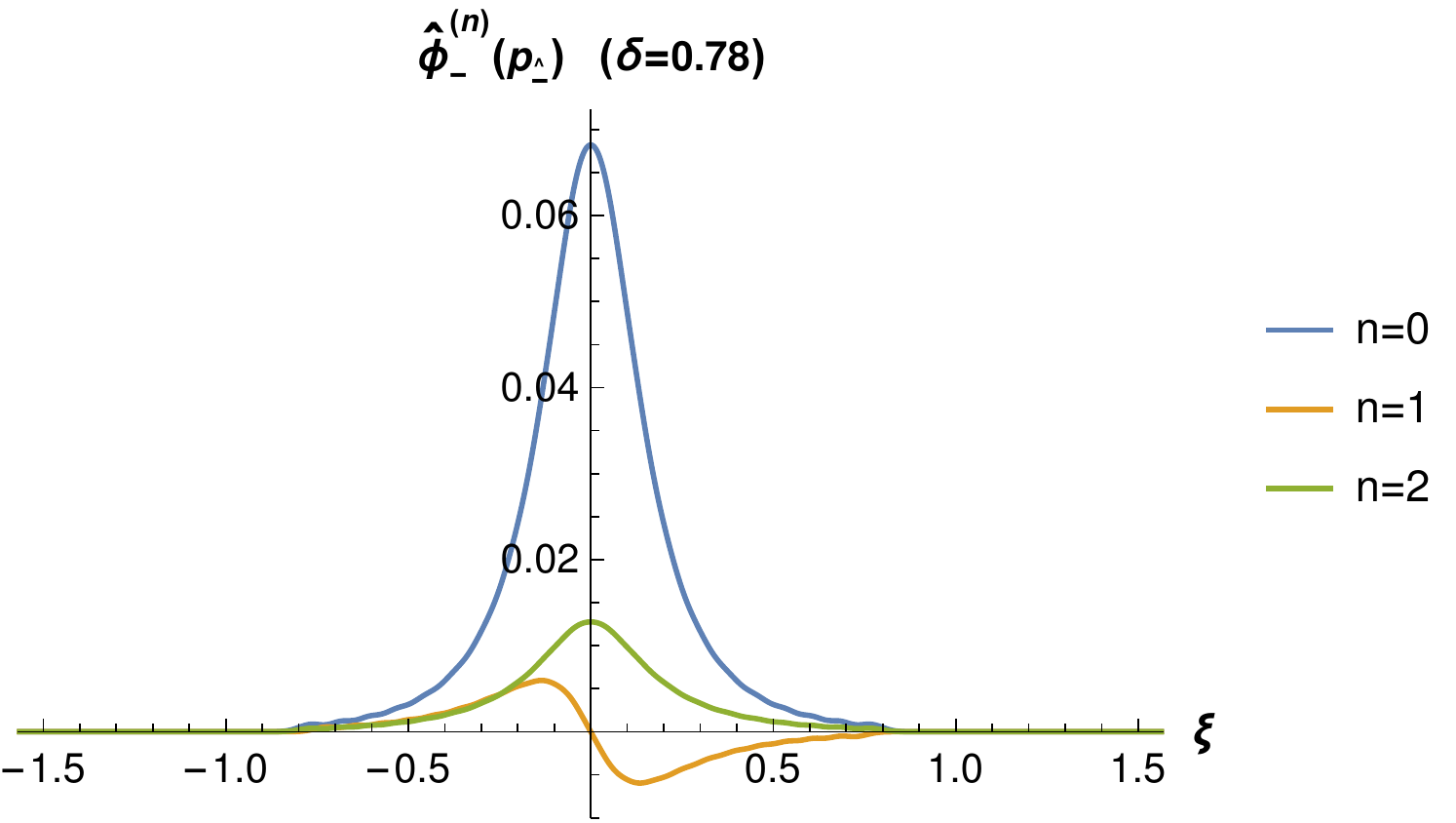}
		\label{fig:m100rffd078}
	}\\
	\caption{Rest frame wave functions $\hat\phi_+^{(n)}(p_{\hat{-}})$ and $\hat\phi_-^{(n)}(p_{\hat{-}})$ for $ m=1.0 $
as functions of $\xi=\tan^{-1} p_{\hat{-}}$. All quantities are in proper units of $ \sqrt{2\lambda} $.\label{fig:m100rfz}}
\end{figure*}

\begin{figure*}
	\centering
	\subfloat[]{
		\includegraphics[width=0.5\linewidth]{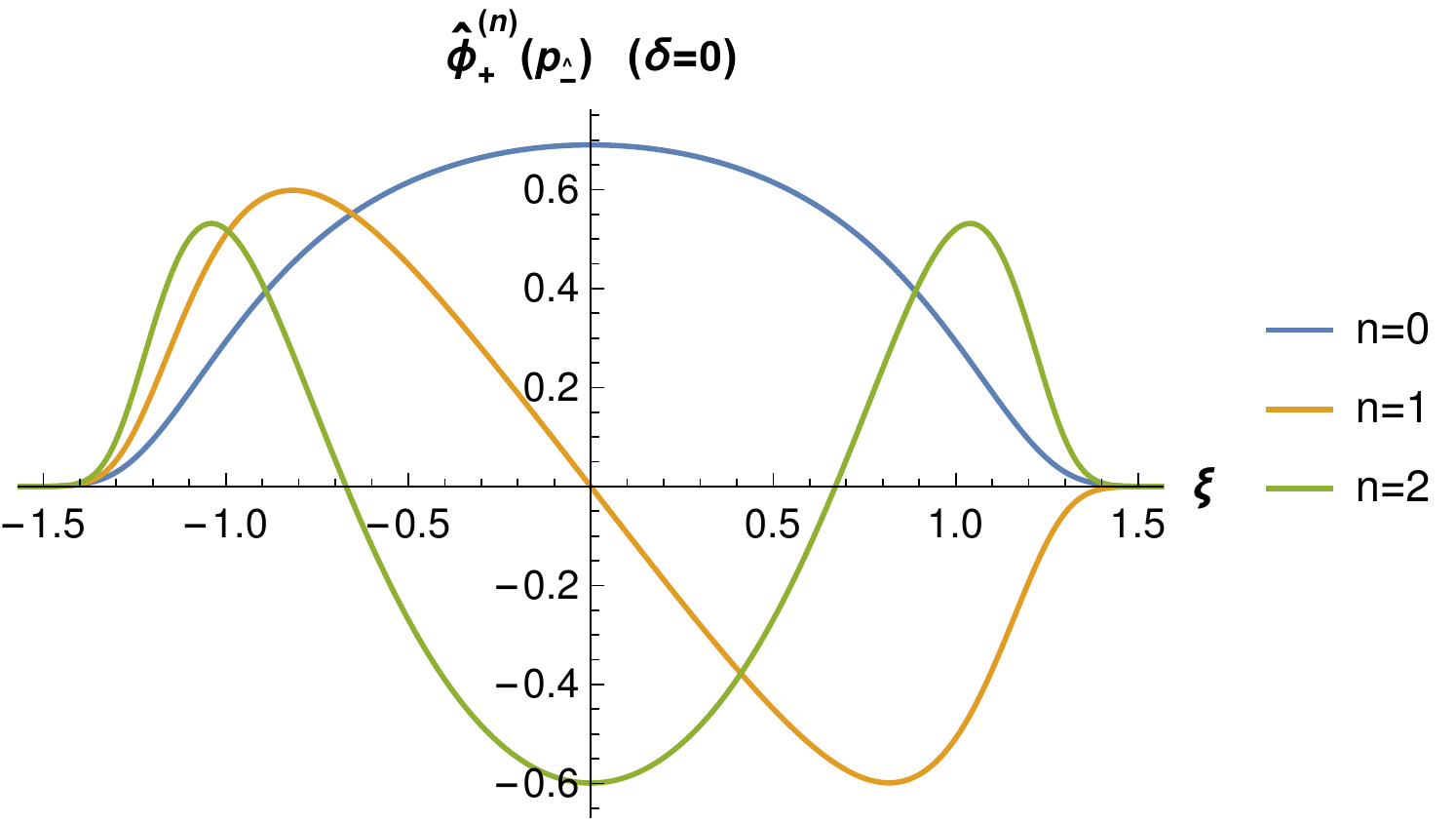}
		\label{fig:m211rfzd0}
	}
	\centering
	\subfloat[]{
		\includegraphics[width=0.5\linewidth]{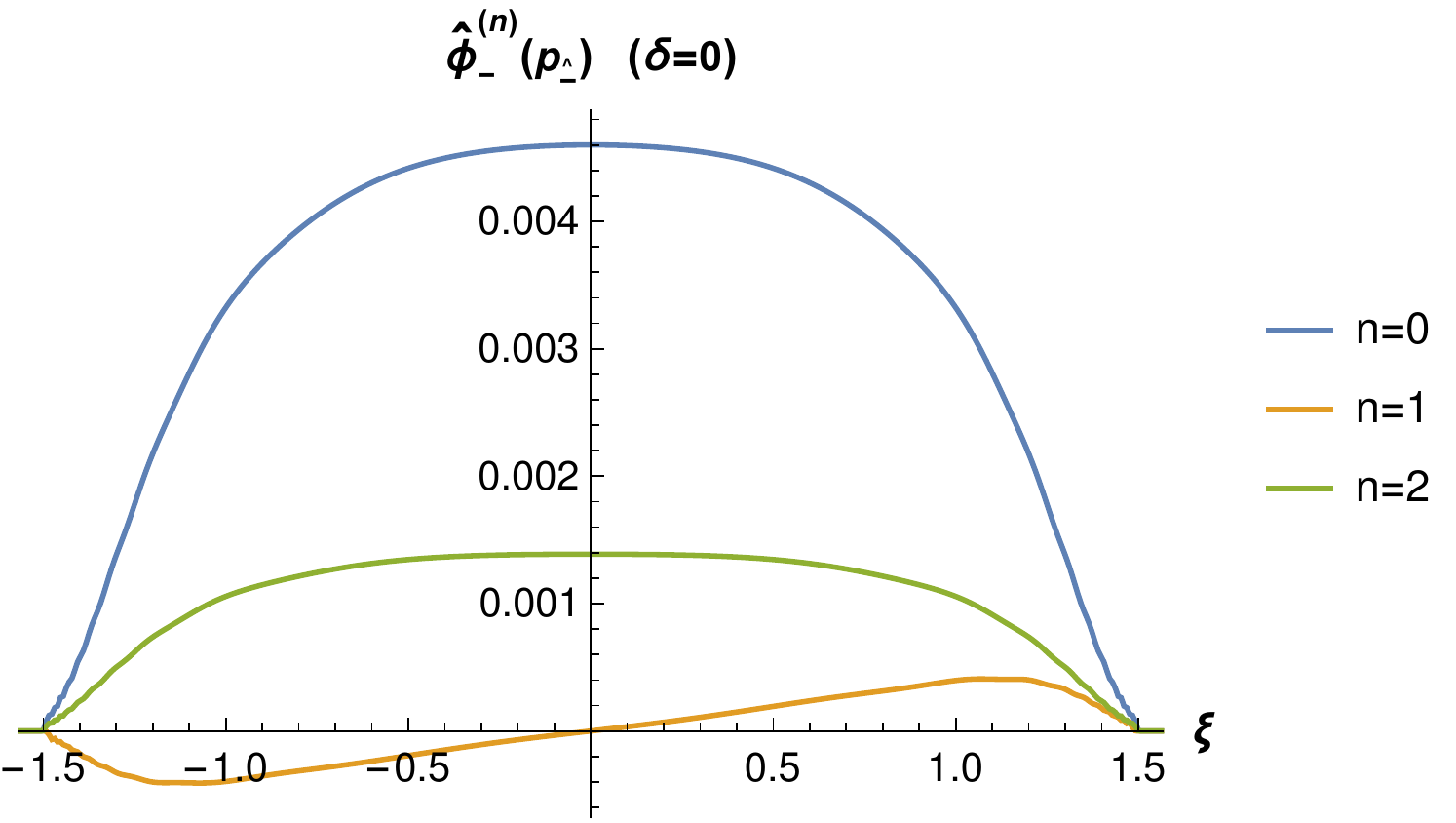}
		\label{fig:m211rffd0}
	}\\
	\centering
	\subfloat[]{	
		\includegraphics[width=0.5\linewidth]{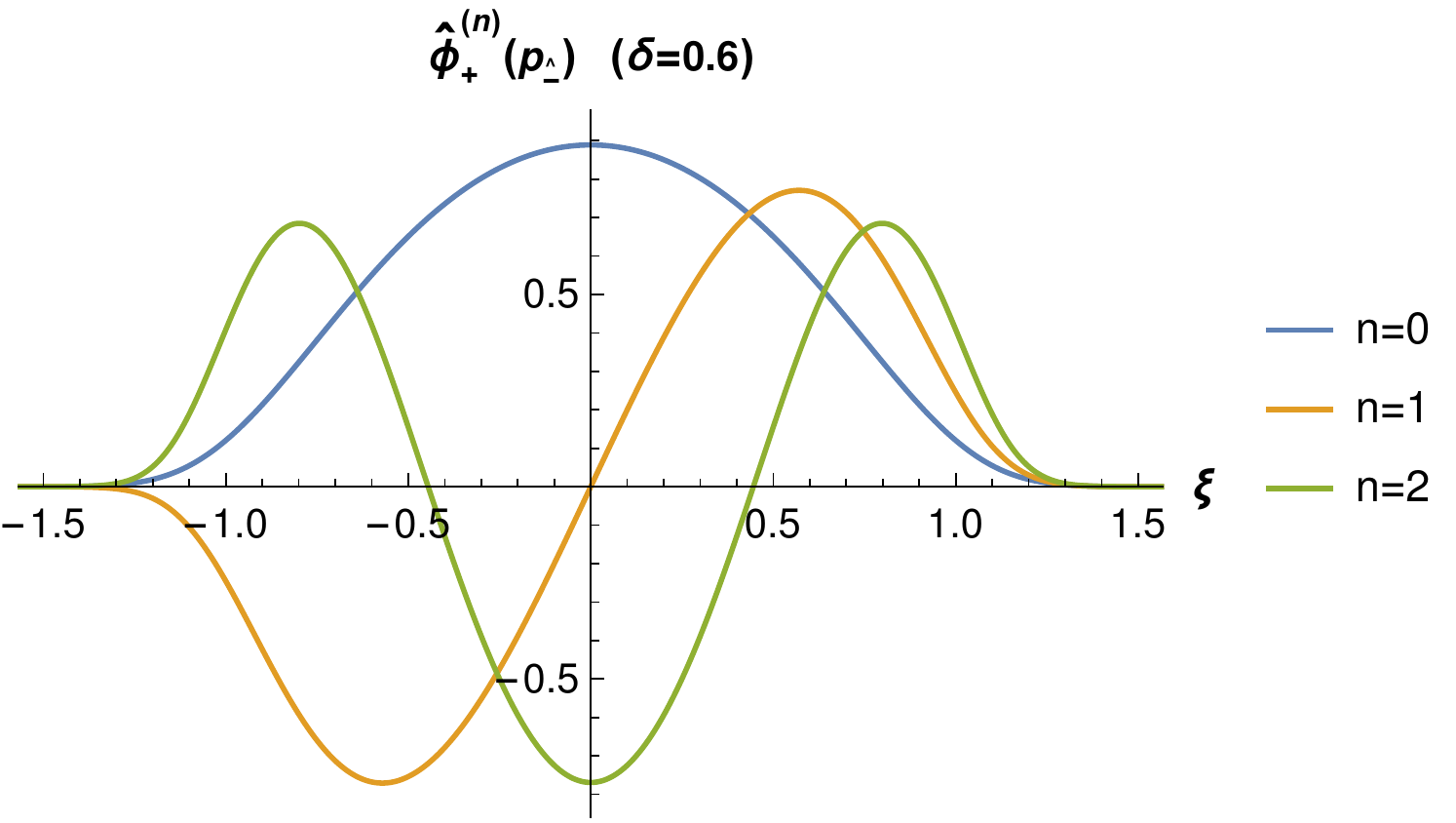}
		\label{fig:m211rfzd06}
	}
	\centering
	\subfloat[]{		
		\includegraphics[width=0.5\linewidth]{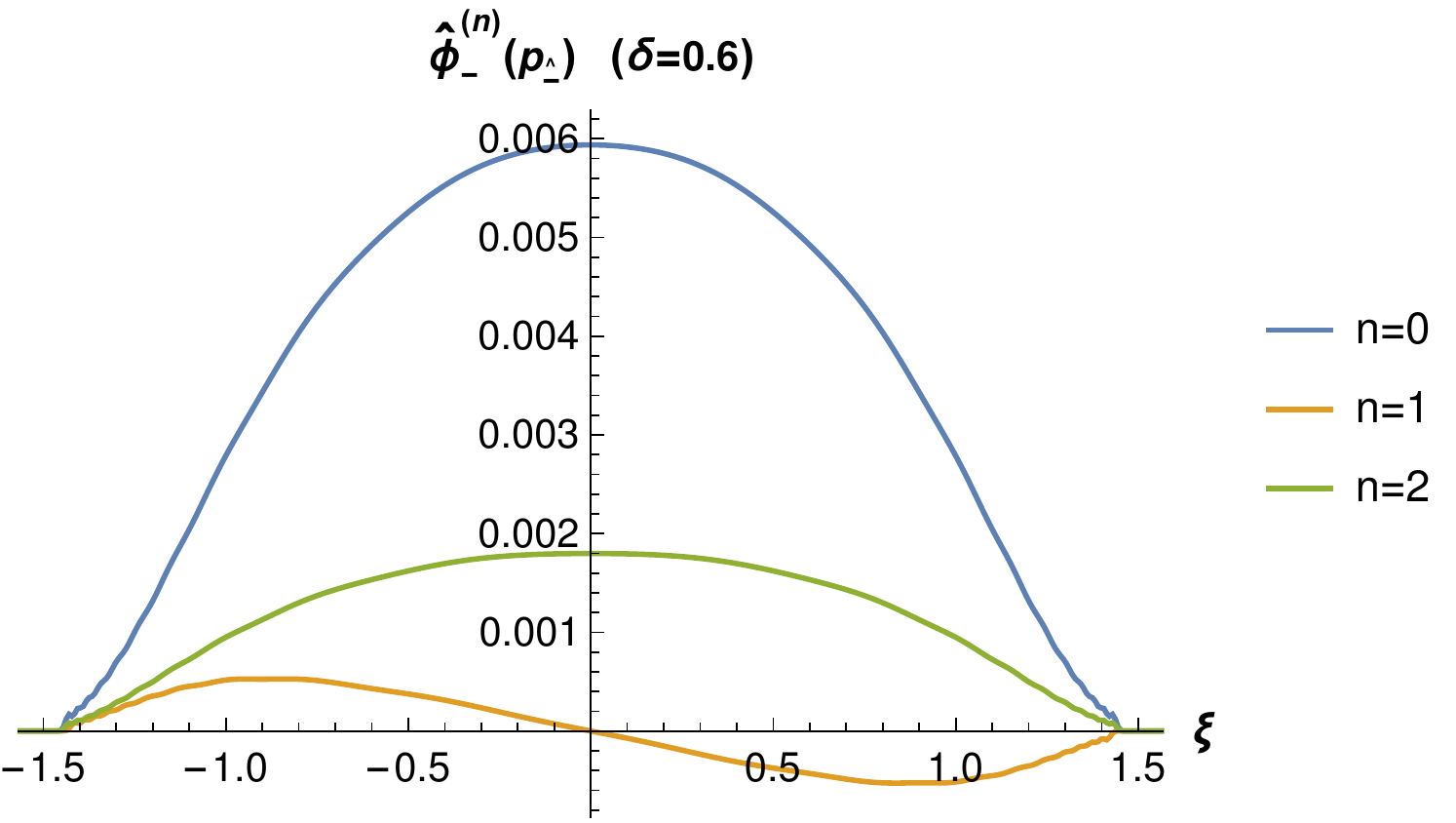}
		\label{fig:m211rffd06}
	}\\
	\centering
	\subfloat[]{
		\includegraphics[width=0.5\linewidth]{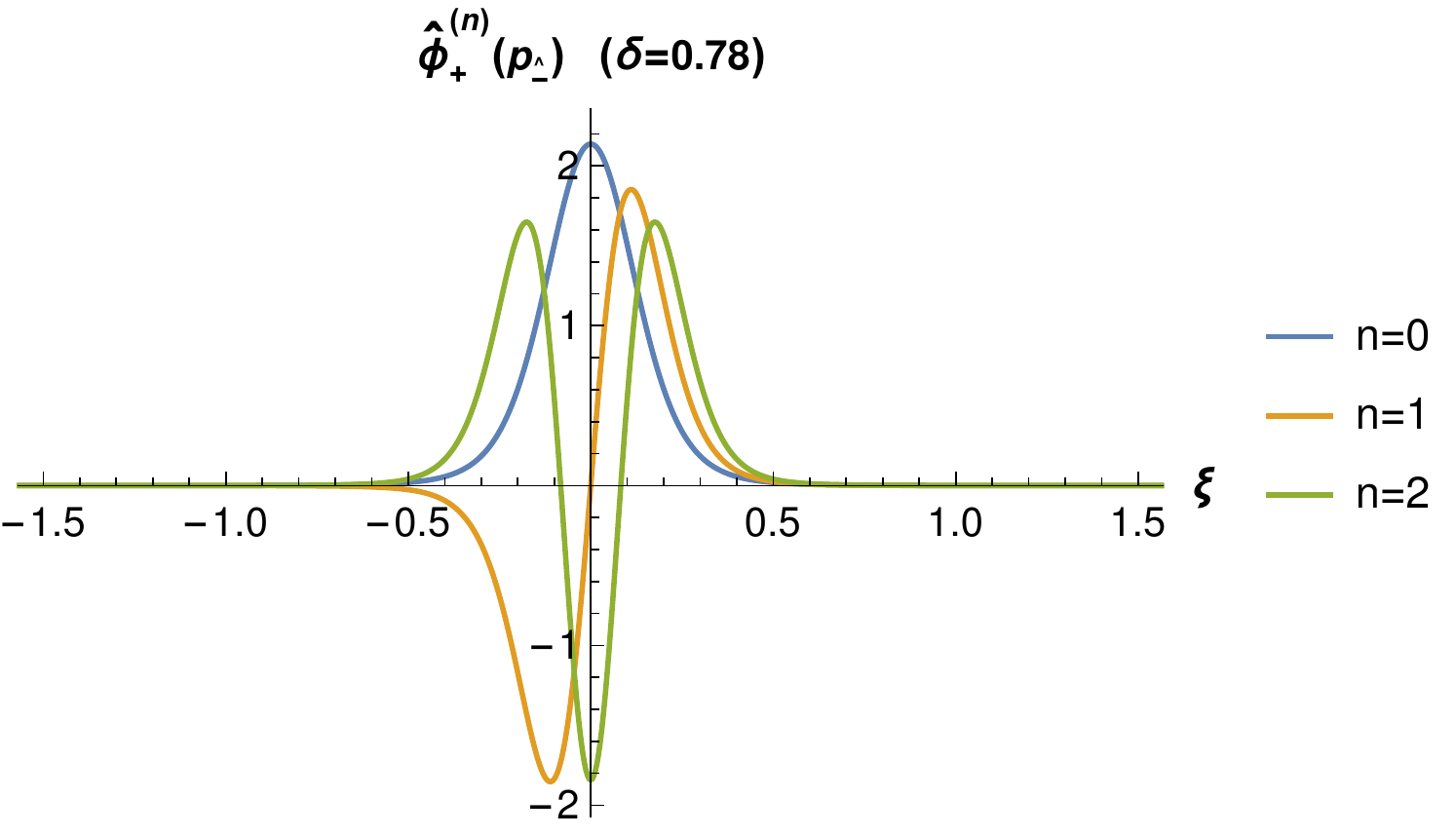}
		\label{fig:m211rfzd078}
	}
	\centering
	\subfloat[]{	
		\includegraphics[width=0.5\linewidth]{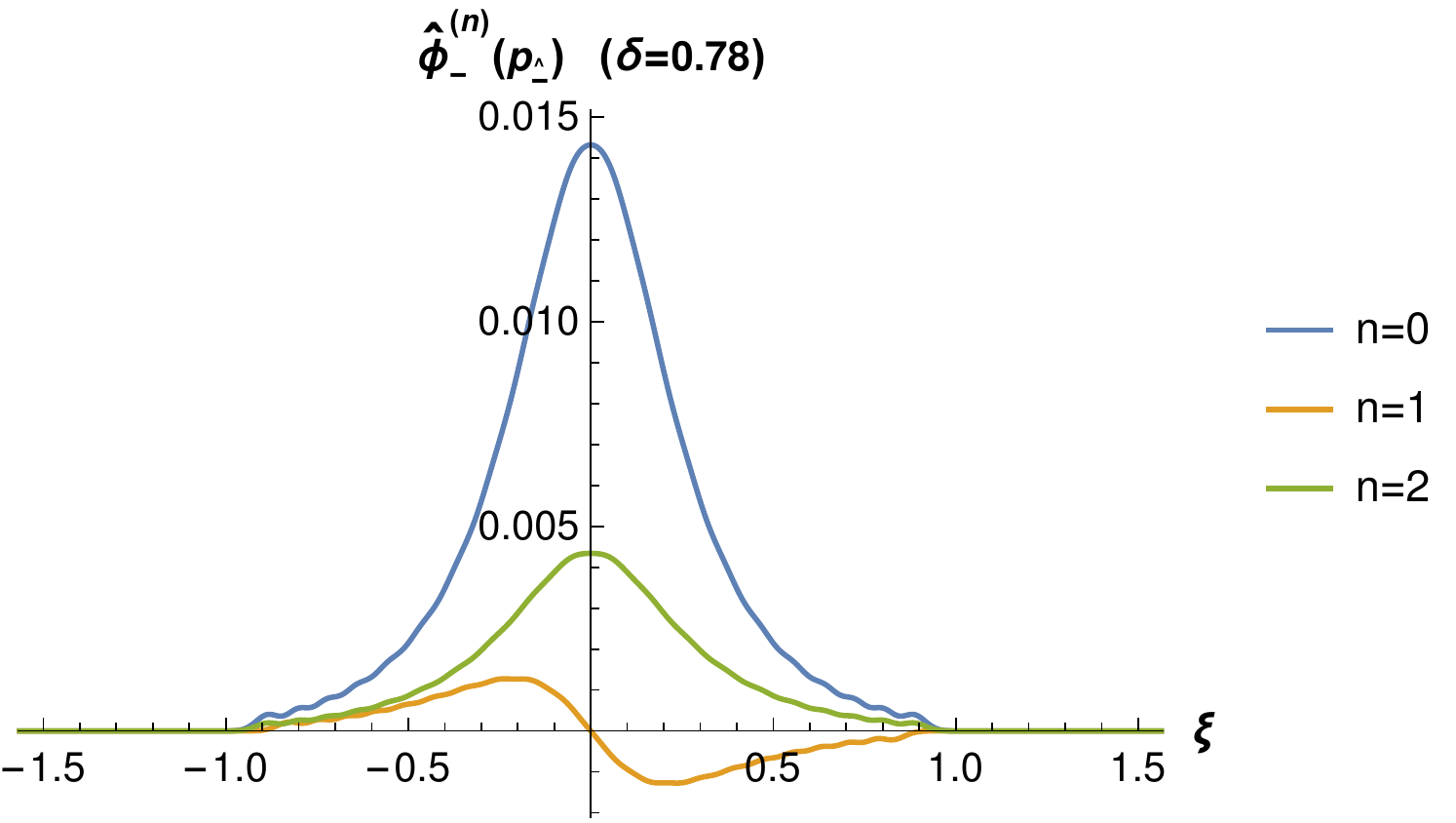}
		\label{fig:m211rffd078}
	}\\
	\caption{Rest frame wave functions $\hat\phi_+^{(n)}(p_{\hat{-}})$ and $\hat\phi_-^{(n)}(p_{\hat{-}})$ for $ m=2.11 $
as functions of $\xi=\tan^{-1} p_{\hat{-}}$. All quantities are in proper units of $ \sqrt{2\lambda} $.\label{fig:m211rfz}}
\end{figure*}

While we presented our numerical solutions of the bound-state wavefunctions 
$\hat\phi_{\pm}^{(n)}(r_{\hat{-}},x)$ in terms of the interpolating longitudinal momentum fraction variable $x=p_{\hat{-}}/r_{\hat{-}}$ in Sec.~\ref{sub:wavefunc} and 
Appendix~\ref{app:figures} to discuss the moving frame dependence of the interpolating wavefunctions between IFD and LFD, 
the rest frame is special and deserves separate description/discussion. In particular, the massless particles can't exist in the rest frame according to the relativity although 
the GOR relation ${\cal M}_{(0)}^2 \sim m\sqrt{\lambda} \to 0$
indicates that the meson mass ${\cal M}_{(0)} \to 0$ as $m \to 0$ in the chiral limit. As the massless Goldstone boson moves
with the speed of light, it can't exist in the rest frame.
We thus devote this final Appendix for the discussion of
the rest frame bound-state equation and 
its solution. 

Taking $r_{\hat{-}}=0$ in Eq.~(\ref{boundeq}), we get
	\begin{subequations}\label{boundeqrf}
		\begin{align}\label{boundeqrf1}
		&\left[ -r^{\hat{+}}+2E(p_{\hat{-}})\right] \hat\phi_+(p_{\hat{-}})=\lambda\mathbb{C}\dashint\frac{dk_{\hat{-}}}{(p_{\hat{-}}-k_{\hat{-}})^2}\notag\\
		&\ \ \times\left[C(p_{\hat{-}},k_{\hat{-}}) \hat\phi_+(k_{\hat{-}})-S(p_{\hat{-}},k_{\hat{-}}) \hat\phi_-(k_{\hat{-}})\right], \\
		&\left[ r^{\hat{+}}+2E(p_{\hat{-}})\right] \hat\phi_-(p_{\hat{-}})=\lambda\mathbb{C}\dashint\frac{dk_{\hat{-}}}{(p_{\hat{-}}-k_{\hat{-}})^2}\notag\\
		&\ \ \times\left[C(p_{\hat{-}},k_{\hat{-}}) \hat\phi_-(k_{\hat{-}})-S(p_{\hat{-}},k_{\hat{-}}) \hat\phi_+(k_{\hat{-}})\right],\label{boundeqrf2}
		\end{align}
	\end{subequations}
	where 
	\begin{equation}\label{Cpk}
	C(p_{\hat{-}},k_{\hat{-}})=C(p_{\hat{-}},k_{\hat{-}},r_{\hat{-}}=0)=\cos^2 \left(\frac{\theta(p_{\hat{-}})-\theta(k_{\hat{-}})}{2}\right),
	\end{equation}
and
	\begin{equation}\label{Spk}
	S(p_{\hat{-}},k_{\hat{-}})=S(p_{\hat{-}},k_{\hat{-}},r_{\hat{-}}=0)=-\sin^2 \left(\frac{\theta(p_{\hat{-}})-\theta(k_{\hat{-}})}{2}\right).
	\end{equation}
The basis wavefunction is also provided without scaling the interpolating momentum variable $p_{\hat{-}}$ with respect to $r_{\hat{-}}$ (in contrast to Eq.~(\ref{eqn:wfbasis}))
\begin{equation}\label{wfbasisrf}
\Psi_m(\alpha,p_{\hat{-}})=\sqrt{\frac{\alpha}{2^m m! \sqrt{\pi}}}{\rm e}^{-\frac{1}{2}\alpha^2p_{\hat{-}}^2}H_m\left( \alpha p_{\hat{-}}\right),
\end{equation}
and the $n$-th bound-state wavefunctions ${\hat\phi}_{\pm}^{(n)}(p_{\hat{-}})$ are normalized as
\begin{equation}\label{normrf}
\int dp_{\hat{-}}\left\lbrace  |\hat\phi_{+}^{(n)}(p_{\hat{-}})|^2-|\hat\phi_{-}^{(n)}(p_{\hat{-}})|^2\right\rbrace =1,
\end{equation}
where one should note the caveat of $n=0$ solution for $m=0$ that becomes null
in the rest frame due to the relativity.

In the frame $r_{\hat{-}} = 0$, the meson mass square $ {\cal M}_{(n)}^2=\frac{(r^{\hat{+}}_{(n)})^2}{\mathbb{C}} $ and the corresponding wavefunctions $ \hat\phi_{\pm}^{(n)}(p_{\hat{-}}) $ are obtained by solving the coupled bound-state equations, Eqs.(\ref{boundeqrf1}) and (\ref{boundeqrf2}), using essentially the same technique that we described in Sec.\ref{sub:spec}. 

Due to the GOR relation ${\cal M}_{(0)}^2 \sim m\sqrt{\lambda} \to 0$, 
the ground-state meson mass ${\cal M}_{(0)} \to 0$ as $m \to 0$. 
As mentioned earlier, the massless Goldstone boson then moves with the speed of light and can't exist in the rest frame
according to the relativity. How can one understand this 
distinction of the massless Goldstone boson in the rest frame?
To realize it, one may take a look more closely 
the analytic solution of the ground-state ($n=0$) wavefunction 
($\hat\phi_\pm^{(0)}(p_{\hat{-}})$) given by Eq.(\ref{pionic-wavefunction}) and find that the rest frame $r_{\hat{-}} = 0$ yields $\hat\phi_\pm^{(0)}(p_{\hat{-}})$ as
\begin{equation}
\label{GoldstoneRest}
\hat\phi_{+}^{(0)}(p_{\hat{-}}) = \hat\phi_{-}^{(0)}(p_{\hat{-}}) =  
\frac{1}{2}\cos\theta(p_{\hat{-}}).
\end{equation}
Then, indeed, the normalization condition given by Eq.(\ref{norm}) becomes null
indicating the absence of the massless Goldstone boson in the rest frame. However, the individual interpolating wavefunctions 
$\hat\phi_{+}^{(0)}(p_{\hat{-}})$ and 
$\hat\phi_{-}^{(0)}(p_{\hat{-}})$ do not vanish as plotted in Fig.~\ref{fig:m0rfn0ana}. Here, we use the variable of the horizontal axis $\xi=\tan^{-1}(p_{\hat{-}})$. In Fig.~\ref{fig:m0rfn0ana}, the interpolation angle $\delta=0$, $0.6$, and $0.78$ results are depicted in solid, dashed, and dotted lines, respectively. 

In contrast to the ground state, the excited states ($n \neq 0$) for $m=0$ acquire the non-zero bound-state mass as shown in Table~\ref{tab:mass}, e.g. ${\cal M}_{(1)}$ = 2.43, ${\cal M}_{(2)}$ = 3.76, etc. in the unit of $\sqrt{2\lambda}$, and they can take the rest frame $r_{\hat{-}} = 0$. In Fig.~\ref{fig:m0rfz}, the bound-state wavefunctions $\hat\phi_+^{(n)}(p_{\hat{-}})$ and $\hat\phi_-^{(n)}(p_{\hat{-}})$ 
for the first ($n=1$) and second ($n=2$) excited states obtained by numerically solving Eq.~(\ref{boundeqrf}) are plotted with yellow and green lines, respectively. 
From Fig.~\ref{fig:m0rfz}, we can see the 
odd and even parities respectively for $n=1$ and $n=2$ state wavefunctions 
$\hat\phi_{\pm}^{(n)}(p_{\hat{-}})$ under the exchange of $p_{\hat{-}}\leftrightarrow-p_{\hat{-}}$. For the $\delta$ value close to $\pi/4$, the wavefunctions are very sharply peaked and constrained in a relatively small $|p_{\hat{-}}|$ region, indicating 
that ultimately only the zero-mode $p^+ = 0$ survives for $r^+$ = 0 frame in LFD. 
Not only does the supporting momentum region $\xi=\tan^{-1}(p_{\hat{-}})$
for $\hat\phi_{\pm}^{(n)}(p_{\hat{-}})$ get shrunken to the zero-mode
$\xi=\tan^{-1}(p^+) = 0$ but also the magnitude of $\hat\phi_{-}^{(n)}(p_{\hat{-}})$ gets much 
more suppressed compare to $\hat\phi_{+}^{(n)}(p_{\hat{-}})$ as 
$\delta$ value close to $\pi/4$, indicating the absence of 
$\hat\phi_{-}^{(n)}(p^+)$ solutions in LFD. 

For the non-zero mass cases, we plot $n=$ 0, 1, and 2 together for 
the rest frame wavefunctions $\hat\phi_+^{(n)}(p_{\hat{-}})$ and $\hat\phi_-^{(n)}(p_{\hat{-}})$ as shown in Fig.~\ref{fig:m0045rfz} for 
$m=0.045$, Fig.~\ref{fig:m018rfz} for $m=0.18$, 
Fig.~\ref{fig:m100rfz} for $m=1.0$, and 
Fig.~\ref{fig:m211rfz} for $m=2.11$, respectively. 

For $m=0.045$, although we have compared our results for the moving frames
shown in Figs.~\ref{fig:m0045grz} and~\ref{fig:m0045fez} with the corresponding results in Ref.~\cite{mov} as discussed
in Appendix~\ref{app:figures}, we couldn't compare our results shown in Fig.~\ref{fig:m0045rfz} with Ref.~\cite{mov} as the rest frame was not considered in Ref.~\cite{mov}. However, the results shown in 
Fig.~\ref{fig:m0045rfz} are not much different from 
the corresponding results for $m=0$ discussed above as the difference in the 
mass $m$ is rather marginal. 

For $m=$ 0.18, 1.0 and 2.11, all of our $\delta=0$ results shown in Figs.~\ref{fig:m018rfz}, \ref{fig:m100rfz}, and \ref{fig:m211rfz} can be respectively compared with the 
corresponding rest frame results shown in Figs. 7, 8, 9 and 10 of Ref.~\cite{Li}.
Although the normalization of the bound-state wavefunction was mistaken 
in Ref.~\cite{Li} by taking the + sign between $|\hat\phi_{+}^{(n)}(p_{\hat{-}})|^2$
$|\hat\phi_{-}^{(n)}(p_{\hat{-}})|^2$ in Eq.~(\ref{normrf}), our results look quite consistent with theirs as the magnitude of $\hat\phi_{+}^{(n)}(p_{\hat{-}})$ is much larger
than the magnitude of $\hat\phi_{-}^{(n)}(p_{\hat{-}})$ to reveal any sizable difference
for the comparison. All of our results look consistent with the characteristics of 
 $\hat\phi_{+}^{(n)}(p_{\hat{-}})$ and $\hat\phi_{-}^{(n)}(p_{\hat{-}})$ discussed 
 previously.

\end{document}